\documentclass[fleqn,usenatbib]{mnras}
\usepackage{newtxtext,newtxmath}
\usepackage[T1]{fontenc}
\DeclareRobustCommand{\VAN}[3]{#2}
\let\VANthebibliography\thebibliography
\def\thebibliography{\DeclareRobustCommand{\VAN}[3]{##3}\VANthebibliography}

\usepackage{graphicx}	
\usepackage{amsmath}	
\usepackage[most]{tcolorbox}
\usepackage{caption}
\usepackage{subcaption}
\usepackage{tikz}
\usepackage{bm}
\usepackage{overpic}
\usepackage{siunitx}
\usetikzlibrary{arrows.meta,positioning,calc,fit,shapes.misc}

\newcommand{\atonhe}{\textsc{aton-he}}
\newcommand{\pgt}{\textsc{p-gadget3}}

\title[Tomographic reconstruction of Reionization]{TORRCH: Tomographic reconstruction of the reionization of cosmic hydrogen with Ly$\bm{\alpha}$ emitters and non-Ly$\bm{\alpha}$-selected galaxies}

\author[Maitra et al.]{
Soumak Maitra$^{1}$\thanks{E-mail: soumak.maitra@theory.tifr.res.in},
Girish Kulkarni$^{1}$,
Vipul Arora$^{2,3}$,
Matteo Viel$^{4,5,6,7,8}$,
Shikhar Asthana$^{9}$,
\newauthor\mbox{} James S. Bolton$^{10}$, Martin G. Haehnelt$^{9}$ and
Laura Keating$^{11}$\\
$^{1}$Tata Institute of Fundamental Research, Homi Bhabha Road, Mumbai 400005, India\\
$^{2}$Department of Electrical Engineering, IIT~Kanpur, Kanpur 208016, India\\
$^{3}$Department of Electrical Engineering (ESAT), KU Leuven, 3001 Leuven, Belgium\\
$^{4}$Istituto Nazionale di Astrofisica -Osservatorio Astronomico di Trieste, Via Tiepolo 11, Trieste, Italy\\
$^{5}$SISSA International School for Advanced Studies, Via Bonomea 265, 34136, Trieste, Italy\\
$^{6}$INFN - Sezione di Trieste, via Valerio 2, 34127, Trieste, Italy\\
$^{7}$IFPU, Institute for Fundamental Physics of the Universe, Via Beirut 2, 34014 Trieste, Italy\\
$^{8}${ICSC - Centro Nazionale di Ricerca in High Performance Computing, Big Data e Quantum Computing, Via Magnanelli 2, Bologna, Italy}\\
$^{9}$Kavli Institute for Cosmology and Institute of Astronomy, Madingley Road, Cambridge, CB3 0HA, UK\\
$^{10}$School of Physics and Astronomy, University of Nottingham, University Park, Nottingham, NG7 2RD, UK\\
$^{11}$Institute for Astronomy, University of Edinburgh, Blackford Hill, Edinburgh, EH9 3HJ, UK
}

\date{Accepted ---. Received ---; in original form ---}

\pubyear{\the\year{}}

\begin{document}
\label{firstpage}
\pagerange{\pageref{firstpage}--\pageref{lastpage}}
\maketitle

\begin{abstract}
Tomographic reconstruction of reionization is a long-sought goal. It would move the field beyond global summary statistics, such as the volume-averaged ionised fraction, to direct, field-level constraints on the ionization topology. With this in mind, we present \textsc{TORRCH} (TOmographic Reconstruction of the Reionization of Cosmic Hydrogen), a deep-learning framework that reconstructs the neutral-hydrogen fraction field during the epoch of reionization from the spatial distributions of Ly$\alpha$ emitters (LAEs) and non-Ly$\alpha$-selected galaxies (NLSGs) at luminosity limits comparable to current surveys. Using hydrodynamical simulations post-processed with radiative transfer, we train a deterministic 3D U-Net on mock surveys spanning diverse reionization scenarios and predict the neutral-fraction field. We find that \textsc{TORRCH} recovers the large-scale ionization morphology from synthetic data comparable to current surveys with high fidelity, and reproduces both the one-point distribution and the 2D power spectrum of projected neutral fractions. The predicted galaxy-IGM cross-correlation is also captured well, including the expected small-scale anti-correlation and its decline towards zero at large separations. Reconstruction quality depends on tracer completeness, with deep joint LAE+NLSG samples yielding the most accurate morphology, while LAE-only selections retain bubble-scale topology but with reduced fidelity. Robustness tests show that the method is stable to variations in ionization conditions between training and test data, and to realistic redshift uncertainties. Our results suggest that galaxy-based tomography can potentially deliver reliable reionization maps across realistic survey redshift windows.
\end{abstract}


\section{Introduction}

The epoch of reionization (EoR) marks a pivotal phase in the thermal and ionization history of the Universe, during which ultraviolet (UV) photons from the first generation of galaxies and black holes ionized the diffuse neutral hydrogen in the intergalactic medium (IGM). The \emph{global} timing of this transition is increasingly well constrained by several probes: the Thomson optical depth from cosmic microwave background (CMB) polarization \citep{planck2018cp}, damping-wing absorption in high-redshift quasars \citep{Mortlock2011,Banados2018,davies2018,becker2024,Spina2024,Zhu2024,Kist2025}, Ly$\alpha$ forest opacity fluctuations at $z\sim 5$--6 \citep{Daloisio2015,chardin2017,kulkarni2019,Bosman2018,Bosman2022}, and galaxy-based inferences of the volume-averaged neutral fraction $\langle x_{\mathrm{HI}}\rangle(z)$ from Ly$\alpha$ emission statistics \citep{Ouchi2010,konno2014,2015MNRAS.446..566M,Ouchi2018,Konno2018,2018PASJ...70...55I, hoag2019, mason2019,morales2021, 2021ApJ...923..229G, 2023ApJ...949L..40B, 2023ApJ...947L..24M,     2024ApJ...971..124U}. However, these measurements are largely insensitive to the detailed spatial arrangement of ionized and neutral regions. As a result, different ionization morphologies and source populations can yield similar values of the mean neutral hydrogen fraction, $\langle x_{\mathrm{HI}}\rangle(z)$, leaving the morphology and topology of reionization only weakly constrained. To break these degeneracies, it is necessary to extract higher-dimensional information that directly traces the spatial structure of the ionized and neutral IGM \citep[e.g.][]{Gnedin2022}.

The ionization field is shaped by the clustering of ionizing sources and the coupling between radiation and large-scale structure, giving rise to a network of ionized bubbles and neutral islands whose sizes, shapes, and connectivity encode information about reionization physics \citep{furlanetto2004, Mesinger2016, naidu2020}. Excursion-set and semi-analytic schemes such as \textsc{21cmfast} efficiently generate large ionization maps and bubble statistics \citep{mesinger2007, Mesinger2011}. GPU-accelerated radiative-transfer frameworks such as \textsc{aton} and \textsc{aton-he} \citep{aubert2008, aubert2010, Asthana2024} provide detailed, large-volume ionization fields that track recombination-regulated ionization fronts and IGM thermal evolution, while radiation-hydrodynamics simulations such as \textsc{thesan} self-consistently couple galaxy formation, radiative transfer, and IGM thermochemistry \citep{Kannan2022, Yeh2023, Neyer2024}. These simulations have been used to quantify bubble-size distributions, percolation behaviour, and the morphology of ionized structures \citep[e.g.][]{Lin2016, Giri2018bsd, Doussot2022, Shimabukuro2022}, as well as the population of neutral ``islands’’ that characterise the late stages of reionization \citep{Giri2019islands, Giri2025islands}. Morphological diagnostics such as the Minkowski functionals, Betti numbers \citep{Giri2021betti}, and higher-order correlation functions \citep{majumdar2018,Gorce2019,Jennings2020,Raste2023} show that the ionization field contains a rich set of observables beyond two-point clustering. Together, these theoretical efforts demonstrate how much information on reionization is encoded at the \emph{field level}, which current global constraints largely compress away.

High-redshift Ly$\alpha$ emitters (LAEs) occupy a unique position among different probes of reionization. Whereas the CMB Thomson optical depth constrains only the redshift-integrated ionization history, quasar damping-wing absorption and Ly$\alpha$ forest opacity fluctuations sample individual sightlines, and forthcoming 21-cm measurements are challenged by Galactic and extragalactic foreground contamination, LAEs uniquely provide simultaneous access to both global and spatial information within a single, foreground-clean galaxy sample. Because Ly$\alpha$ photons get resonantly attenuated by the presence of neutral hydrogen, the visibility of LAEs is strongly modulated by the surrounding ionization state: galaxies in large ionized regions are preferentially detected, while those embedded in neutral zones suffer substantial damping-wing absorption \citep{Dijkstra2014, Sadoun2017}. As a result, the same LAE samples that constrain $\langle x_{\mathrm{HI}}\rangle$ through number counts and luminosity functions \citep{Ouchi2018, Harish2022} also carry information about bubble sizes and topology through their spatial configuration \citep{McQuinn2007, Mesinger2011, Hutter2015, Weinberger2018,mason2018,Konno2018,hu2019,2024ApJ...971..124U,lu2024,Chen2025,Lu2025}.  Simulation-based studies explicitly link LAEs to ionized bubbles, finding that LAEs preferentially occupy overdense, highly ionized regions \citep{Mesinger2008lae,Sobacchi2015} and that LAE associations or overdensities can mark extended ionized patches embedded in a more neutral IGM \citep{Trapp2023}. The LAEs also act as effective signposts of bubble sizes, with their three-dimensional spatial distribution encoding information about the local ionization topology \citep{Neyer2024, NeyerLAE2025, Maitra2025}. JWST observations already hint at such connections between LAE overdensities, protocluster-scale structures, and extended ionized regions \citep[e.g.][]{Witstok2024env, Saxena2024}. This dual role of LAEs as probes of both $\langle x_{\mathrm{HI}}\rangle$ and ionization morphology naturally motivates moving from scalar to field-level analyses.

A first step in this direction is to exploit higher-order and explicitly morphological statistics of the LAE field. The LAE bispectrum has been proposed and applied as a sensitive probe of the non-Gaussian, bubble-dominated topology of reionization that is invisible to two-point statistics \citep{Greig2013, Maitra2025}. Complementary measures, including bubble-size distributions, neutral-island statistics, and topological descriptors, have been explored primarily using simulated 21-cm and ionization maps \citep[e.g.][]{Lin2016, Giri2018bsd, Giri2019islands, Doussot2022}, and can in principle be applied to reconstructed fields traced by galaxies. Collectively, these works show that LAEs and related galaxy populations are not only sensitive to the global ionization fraction $\langle x_{\mathrm{HI}}\rangle$ but also encode detailed information about the underlying \emph{spatial} neutral-hydrogen field, $x_{\mathrm{HI}}(\mathbf{r})$, that defines the geometry of reionization.

Field-level reconstruction of $x_{\mathrm{HI}}(\mathbf{r})$ has so far been developed most extensively in the context of 21-cm cosmology. Forecasts for direct imaging and mock observational pipelines propagate instrumental and foreground effects to assess the recoverability of bubble morphology from interferometric data \citep{liu2016, Beardsley2016, Ghara2020}. Early modelling work explored spatial correlations between 21-cm maps and LAE distributions, demonstrating their sensitivity to ionized structure and reionization parameters \citep[e.g.][]{Wiersma2013, Sobacchi2016, Hutter2017, Heneka2017, Kubota2018, Vrbanec2020}. More recently, machine-learning approaches have been proposed to infer astrophysical and cosmological parameters directly from mock 21-cm observables \citep{Gillet2019, LaPlante2019, Hassan2020}, and to identify ionized regions or neutral islands in noisy 21-cm images \citep{Bianco2021}. These data-driven methods demonstrate that complex, non-linear mappings between observable fields and the underlying ionization structure can be learned from simulations, but their practical deployment remains tightly coupled to overcoming low-frequency foregrounds and calibration systematics in real 21-cm data, particularly the chromatic, direction-dependent instrumental response and ionosphere-induced errors that leak foregrounds into otherwise clean Fourier modes \citep{koopmans2015,LiuShaw2020,Barry2016,EwallWice2017,Patil2016,Mertens2020,Gan2023,Brackenhoff2024}. 

By contrast, a galaxy-based tomographic framework that combines LAEs with a broader high-redshift galaxy population is immediately applicable to existing and near-future spectroscopic samples. Such galaxy-based reconstructions provide a foreground-free route to field-level EoR information that is naturally complementary to 21-cm imaging and cross-correlation studies \citep[e.g.][]{koopmans2015,robertson2022}. A key ingredient in this framework is the combination of multiple galaxy populations. LAEs respond sensitively to the local ionization state, whereas galaxies selected via their rest-UV continuum but lacking detectable Ly$\alpha$ emission primarily trace the underlying stellar- and halo-mass distributions.  This multi-tracer logic is already embedded in existing inference approaches. In the Bayesian framework of \citet{mason2018} and its extensions \citep[e.g.][]{Bolan2022}, Ly$\alpha$ equivalent width measurements and upper limits for UV-continuum-selected galaxies are combined with reionization simulations to infer $\langle x_{\mathrm{HI}}\rangle$ while marginalising over ISM radiative transfer and intrinsic line-profile variations. In this approach, galaxies with detected Ly$\alpha$ emission and those without detectable Ly$\alpha$ provide complementary information on the IGM neutral fraction: the former are directly modulated by IGM transmission, while the UV-continuum selection is comparatively insensitive to IGM attenuation. In what follows, we refer to galaxies selected independently of Ly$\alpha$ (e.g., via their rest-UV continuum or other tracers) generically as non-Ly$\alpha$-emission-selected galaxies, or in short, ``NLSGs''. These galaxies may or may not have detectable Ly$\alpha$ emission. Inference frameworks that jointly model Ly$\alpha$ emission statistics, continuum-selected NLSGs, and reionization simulations have therefore begun to exploit such multi-tracer datasets to tighten constraints on $\langle x_{\mathrm{HI}}\rangle$ \citep{mason2018,Bolan2022} and to probe reionization patchiness via local bubble requirements around individual sources (e.g. a local ionized bubble of characteristic radius $R_{\rm ion}\approx 0.2\,\mathrm{pMpc}$ required by the analysis of \citet{Witstok2025} to enable Ly$\alpha$ transmission at $z\approx 13$). Extending these analyses from global or one-dimensional summaries to genuinely three-dimensional, multi-tracer reconstructions of the ionization field is the next logical step.

This step is strongly motivated by the rapid growth of high-redshift galaxy datasets, which now probe well inside the reionization era. Ground-based narrowband programmes have already assembled wide-field samples of Ly$\alpha$ emitters. The SILVERRUSH survey identifies $\sim 2230$ LAEs at $z\simeq 5.7$--6.6 over $13.8$--$21.2~\mathrm{deg}^2$ \citep{Ouchi2018,Shibuya2018}, while the LAGER programme targets higher redshifts, reporting $79$ LAEs at $z\simeq 6.9$ over a combined effective area of $\sim 4.5~\mathrm{deg}^2$ and aiming to expand to $\sim 24~\mathrm{deg}^2$ upon completion \citep{hu2019,Khostovan2020}. In the same wide-field narrowband ecosystem, the CHORUS survey extends Subaru/HSC-based LAE mapping over $\lesssim 24~\mathrm{deg}^2$ across multiple redshift slices, delivering large, homogeneous LAE samples and clustering measurements that directly connect post-reionization evolution to the heart of the EoR \citep{Inoue2020}. Integral-field and multi-object spectrographs provide the three-dimensional positional information and Ly$\alpha$ spectroscopy required for tomographic analyses. Optical IFU surveys such as MUSE enable dense Ly$\alpha$ mapping up to $z\lesssim 6.7$ \citep{Bacon2017,Bacon2023}, while near-infrared facilities extend spectroscopic confirmation deeper into the reionization era. In particular, JWST/NIRSpec and NIRISS have already delivered confirmed LAEs at $z>7$, including four LAEs at $7.5\le z\le 9.5$ from the PASSAGE survey and a Ly$\alpha$ detection at $z\simeq 13$ \citep[e.g.][]{Runnholm2025,Witstok2025}. Upcoming wide-field near-IR spectrographs such as MOONS will further expand the accessible survey volumes and multiplexing capabilities for Ly$\alpha$ and rest-frame UV spectroscopy at these redshifts \citep{Cirasuolo2014,Cirasuolo2020}.
In parallel, JWST imaging and spectroscopy from campaigns such as JADES, CEERS, COSMOS-Web, PASSAGE and related programmes reveal large samples of LAEs and continuum-selected NLSGs extending to $z\gtrsim 8$--10, including $717$ $z>8$ candidates over $125~\mathrm{arcmin}^2$ (JADES) and $88$ candidates at $z\sim 8.5$--14.5 over $\sim 90~\mathrm{arcmin}^2$ (CEERS), while COSMOS-Web maps a contiguous $\sim 0.54~\mathrm{deg}^2$ \citep[e.g.][]{Casey2023,Hainline2024,Finkelstein2024}. Building directly on this COSMOS-wide legacy, the COSMOS-3D JWST programme adds wide-area NIRCam/WFSS slitless spectroscopy (with parallel MIRI imaging) to secure redshifts for large samples into the EoR, enabling panoramic three-dimensional large-scale structure and IGM-galaxy connection studies in the same field \citep{Kakiichi2024}. Future wide-field facilities such as the Roman Space Telescope and Subaru Prime Focus Spectrograph (PFS; $2394$ fibres over a $1.3$ deg diameter field) will further increase the sample sizes and survey volumes of both populations, enabling statistically robust clustering and morphology studies across the heart of the EoR \citep{Wang2022HLSS,Sugai2015}. Cross-correlation studies with background quasars in projects such as ASPIRE (25 quasar fields) and EIGER (six quasar fields) provide complementary constraints on small- and large-scale IGM structure, including neutral islands and transmission corridors \citep{Wang2023ASPIRE,Matthee2023EIGER}. A broad compilation of current and upcoming efforts toward three-dimensional LAE mapping is summarised in Table~1 of \citet{Maitra2025}. Together, these heterogeneous datasets demand reconstruction methods that can leverage multiple tracer populations to map the three-dimensional ionization field.

In this work, we present \textsc{TORRCH} (TOmographic Reconstruction of the Reionization of Cosmic Hydrogen), a tomographic deep-learning framework that reconstructs the hydrogen ionization field from the three-dimensional distribution of galaxies. Building upon the \textsc{DeepCHART} architecture introduced in \citet{Maitra2025b}, we do this via a modified 3D U-Net model that learns non-linear mappings from the observed galaxy field, represented by the positions and Ly$\alpha$ luminosities of LAEs together with NLSGs, to the underlying ionization structure. Our training data are drawn from the \textsc{Sherwood-Relics} cosmological hydrodynamical simulations \citep{puchwein2023}, post-processed with the GPU-accelerated radiative-transfer code \textsc{aton-he} \citep{Asthana2024}, and span a variety of plausible reionization scenarios that differ in source populations and timing. In this paper, we focus primarily on tomographic reconstruction at $z=7.14$ and further validate the robustness of our approach by applying models trained at $z=7.14$ to an independent slice at $z=6.6$. Our goal is to move beyond traditional scalar probes by performing field-level inference of ionization morphology incorporating both LAEs and NLSGs as complementary tracers. In doing so, \textsc{TORRCH} aims to elevate galaxy clustering from a purely statistical probe to a full-fledged tomographic tracer of reionization, aligned with the capabilities of ongoing and upcoming surveys in the JWST and SKA eras. More broadly, such galaxy-informed reconstructions naturally provide informative priors on the ionization field that can be incorporated into joint analyses of 21cm intensity mapping and CMB-based reionization observables, enabling a coherent, multi-probe view of the epoch of reionization.

This paper is organised as follows. In Section~\ref{Sec:Methodology} we describe the simulation suite and radiative-transfer post-processing, and outline the forward modelling used to generate the mock LAE and NLSG tracer catalogues. In Section~\ref{sec:vae3dunet_reg} we present our tomographic reconstruction framework, including the 3D U-Net architecture, loss function, and training strategy. In Section~\ref{sec:results} we quantify reconstruction performance using field-level comparisons and one- and two-point statistical diagnostics of the reconstructed $x_{\rm HI}$ field. In Section~\ref{sec:robustness} we perform robustness tests, including generalisation across ionization conditions and the impact of line-of-sight redshift uncertainties. We summarise our main findings and discuss implications in Section~\ref{sec:summary}. Throughout this work we adopt the $\Lambda$CDM cosmology of the Sherwood-Relics simulations, with cosmological parameters
$(\Omega_{\rm m},\Omega_{\Lambda},\Omega_{\rm b},h,\sigma_8,n_{\rm s})
= (0.308,\,0.692,\,0.0482,\,0.678,\,0.829,\,0.961)$ from \cite{planck2014}.

\section{Models for Reionization and LAE\lowercase{s}}
\label{Sec:Methodology}

To model the spatial distribution of LAEs during the epoch of reionization, we combine large-scale cosmological simulations with a semi-empirical prescription for assigning LAEs to dark matter haloes. Our simulations are drawn from the Sherwood-Relics suite \citep{bolton17,puchwein2023}, which captures the thermal and dynamical evolution of the high-redshift intergalactic medium (IGM). Radiative transfer of ionizing photons is computed in post-processing using the \atonhe\ code. Our analysis, as mentioned before, focuses on snapshots at $z=7.14$, with additional validation performed at $z=6.6$. LAEs are populated using a forward-modeling prescription based on \citet{weinberger2019} and further developed in \citet{Maitra2025}. We describe each component in detail below.

\subsection{Hydrodynamical Simulation of the IGM}

We begin with one of the cosmological hydrodynamical runs from the Sherwood-Relics suite, evolved using \pgt\ code (a modified version of {\sc Gadget-2} code described by \citealt{springel2005}). The simulation tracks structure formation within a periodic box of comoving volume $(160\,h^{-1}\,{\rm cMpc})^3$, using $2048^3$ gas and $2048^3$ dark matter particles, and evolves from redshift $z=99$ to $z=4$. Simulation snapshots are stored at regular intervals of 40 Myr, enabling fine temporal resolution of the IGM’s evolution during reionization.
To reduce computational cost while preserving the large-scale gas dynamics, the run employs the \texttt{QUICK\_LYALPHA} approximation \citep{viel2004a}. This scheme converts dense, cool gas (with $T < 10^5$~K and overdensities $\delta > 1000$) into collisionless star particles, effectively bypassing the computationally expensive formation of dense galactic gas clumps.

A spatially uniform ultraviolet background (UVB) from \citet{puchwein2019} is included during the hydrodynamical evolution to mimic the cumulative effects of photoionization heating and pressure smoothing prior to full reionization. Although the Sherwood-Relics suite includes simulations with coupled radiative transfer, our analysis uses the hydro-only run, allowing us to explore a range of ionization histories via post-processing with \atonhe.

For radiative transfer calculations, we interpolate the SPH fields: gas density, temperature, and velocity, onto a uniform Cartesian grid of $2048^3$ cells. This gridded representation of the IGM provides the physical input for solving the time-dependent radiative transfer and thermochemistry in subsequent stages.

\subsection{Radiative Transfer with \atonhe}

To model the inhomogeneous propagation of ionizing radiation during the epoch of reionization, we post-process the hydrodynamical simulation outputs using the \atonhe\ code \citep{Asthana2024}, an enhanced version of the hydrogen-only radiative transfer code \textsc{aton} \citep{aubert2008}. \atonhe\ extends the original code by including helium physics and solving the radiative transfer equation using a multi-frequency, moment-based scheme with M1 closure \citep{levermore1984}, allowing for a self-consistent evolution of gas temperature and ionization state.

Ionizing sources are associated with dark matter halos above a threshold mass of $10^9\,\mathrm{M}_\odot/h$, with emissivities assumed to scale linearly with halo mass \citep{kulkarni2019, keating2020}, motivated by the empirical correlation between halo mass and star formation rate at high redshift. Each source is assigned a blackbody spectral energy distribution with an effective temperature of $T = 4 \times 10^4$~K. All simulations are performed in multi-frequency mode, except for the {Extremely Early} model, where a single-frequency, hydrogen-only approximation is adopted to reduce computational cost.

Each reionization model is calibrated against observational constraints by adjusting the emissivities of ionizing sources to reproduce the observed mean Ly$\alpha$ forest transmission at $z<6$, ensuring that all models remain consistent with the established thermal and ionization state of the post-reionization IGM. We explore four distinct reionization scenarios, each producing a different topology of ionized regions:

\begin{itemize}
    \item \textbf{Fiducial (Late) Reionization Model:} In this baseline scenario, hydrogen reionization progresses gradually, with a midpoint at $z_{\mathrm{mid}} \sim 6.5$ and completion by $z \sim 5.5$. Ionizing sources are massive stars in galaxies, with emissivities  proportional to their halo masses above the $10^9\, M_{\odot}h^{-1}$ threshold. This model is tuned to match both the Ly$\alpha$ forest mean transmission and temperature measurements and serves as our default.

    \item \textbf{Early Reionization Model:} This model adopts the same source prescription as the Fiducial model but assumes higher ionizing emissivities at earlier times, shifting the midpoint of reionization to $z_{\mathrm{mid}} \sim 7.5$. It reflects scenarios where star formation is more efficient or begins earlier, and remains consistent with Ly$\alpha$ forest data.

    \item \textbf{Extremely Early Reionization Model:} Reionization in this case proceeds significantly earlier, reaching its midpoint at $z_{\mathrm{mid}} \sim 9.5$. This model is run in hydrogen-only, single-frequency mode and features a sharply peaked ionizing emissivity at high redshift that declines steeply with time. While it is in tension with CMB measurements of the optical depth $\tau$, it serves as a useful upper bound for assessing the impact of early ionization on high-redshift galaxy observables. 

    \item \textbf{Oligarchic source model:} In this scenario, the ionizing photon budget is dominated by a relatively small population of massive galaxies. Only halos with masses above $8.5 \times 10^9\,M_\odot/h$ are included, thereby privileging more massive halos than in the fiducial source model and producing a more patchy reionization morphology. The ionizing emission is assumed to arise from stellar sources with a stellar spectrum, and the model is calibrated against Ly$\alpha$\ forest constraints at $z<6$.

\end{itemize}

Together, these three scenarios bracket a broad range of reionization timelines and morphologies, allowing us to evaluate the sensitivity of LAE observables to the timing and patchiness of reionization.

\subsection{LAE Modeling}

To connect the reionization-era ionization field with observable galaxy populations, we construct mock LAE catalogs using a forward-modeling framework that combines empirical galaxy properties with radiative transfer-informed attenuation. This approach is semi-empirical in nature and designed to reflect both the physics of galaxy formation and the complexities of Ly$\alpha$ radiative transfer in a patchy, evolving IGM. Our methodology builds on the framework of \citet{weinberger2019} and \citet{Maitra2025} to account for reionization models, UV luminosity function data, and realistic CGM/IGM transmission calculations. The implementation of this methodology is provided through \textsc{SiMPLE-Gen}\footnote{\textsc{SiMPLE-Gen}: Simulated Mock Population of Lyman-Alpha Emitters Generator. Code available at \url{https://github.com/soumak-maitra/SiMPLE-Gen}.}\citep{Maitra2025}. Below, we describe the three core steps of our LAE modeling pipeline.

\subsubsection{UV Galaxy Assignment via Abundance Matching}

We begin by associating UV-bright galaxies with the dark matter halos (collapsed structures that serve as sites of early galaxy formation) resolved in our simulation. This is achieved through an abundance matching technique that ensures the cumulative number density of galaxies above a given UV luminosity matches the cumulative number density of halos above a corresponding mass threshold.
Specifically, we use the redshift-dependent UV luminosity functions compiled by \citet{Bouwens2015}, parameterized as Schechter functions across $z \sim 4$-8, and map them onto the halo mass function derived from our simulation. To capture the stochastic nature of high-redshift star formation, we include a mass- and redshift-dependent duty cycle, which quantifies the probability that a halo of a given mass is actively hosting a star-forming galaxy at the time of observation. This duty cycle is computed based on recent halo mass growth over a 50~Myr time interval, following the formalism of \citet{trenti2010}. The resulting galaxy population thus reflects the bursty and intermittent star formation activity that is expected to characterize early galaxies.

\subsubsection{Intrinsic Ly$\alpha$ Emission: Assigning Equivalent Widths and Luminosities}

Once galaxies are assigned to halos, we estimate their intrinsic Ly$\alpha$ output based on their UV properties. Each galaxy is assigned a rest-frame Ly$\alpha$ equivalent width (REW), drawn from a conditional probability distribution that depends on its UV magnitude. We adopt the REW distribution proposed by \citet{dijkstra2012}, which models the observed trend that fainter galaxies tend to have higher Ly$\alpha$ equivalent widths.
Combining the drawn REW with the UV luminosity of each galaxy yields an estimate of its intrinsic Ly$\alpha$ luminosity. This approach incorporates the observed diversity of Ly$\alpha$ emission strengths at fixed UV magnitude and redshift. \citet{dijkstra2012} demonstrated that this prescription successfully reproduces the observed rest-frame equivalent-width distributions at $z \simeq 3.1$, 3.7, and 5.7, as measured by \citet{ouchi2008}. Although this model may overpredict strong REW values at higher redshift, we find that CGM and IGM absorption largely mitigates this overprediction, in agreement with earlier findings by \citet{weinberger2019}.

\subsubsection{Transmission through the CGM and IGM}

The intrinsic Ly$\alpha$ emission from galaxies must propagate through both the circumgalactic medium (CGM) and the intergalactic medium (IGM), where it is subject to resonant scattering by neutral hydrogen. The degree of attenuation depends sensitively on the local ionization environment and the velocity structure of the gas along the line of sight.
To capture this, we extract one-dimensional skewers through the simulated ionization and density fields centered on each galaxy. Along each skewer, we compute the Ly$\alpha$ optical depth as a function of velocity offset $v$ from the line center, integrating the transmission over a Gaussian line profile. The line profile is assumed to be redshifted by a velocity offset $\Delta v = 1.5\,v_{\rm circ}$, where $v_{\rm circ}$ is the halo's circular velocity, and has a fixed velocity dispersion of $\sigma_v = 88\,\mathrm{km\,s^{-1}}$. These values are chosen to broadly match observations of Ly$\alpha$ line shapes in high-redshift galaxies.

This simplified emission model captures key features of ISM radiative transfer, such as bulk redshifting and line broadening, while the CGM and IGM transmission is computed fully self-consistently from the simulated ionization fields. This step is critical as the patchiness of the IGM and the local velocity gradients can strongly modulate LAE visibility, even for intrinsically bright galaxies.

\subsubsection{Selection of Observable LAEs}

We define galaxies as observable LAEs if they meet a minimum Ly$\alpha$ luminosity threshold. These selection criteria are designed to mimic those used in actual narrow-band surveys, particularly those conducted with the Hyper Suprime-Cam (HSC) on Subaru. For each redshift, we apply survey-specific REW and luminosity cuts. When constructing mock tomographic surveys in this work, we consider three representative configurations drawn from these generic selection cuts:
(i) a deep LAE+NLSG sample with $\log (L_{\rm Ly\alpha}/\mathrm{erg\,s^{-1}}) > 41$ and $M_{\rm UV} < -18$,
(ii) a shallow LAE+NLSG sample with $\log (L_{\rm Ly\alpha}/\mathrm{erg\,s^{-1}}) > 42$ and $M_{\rm UV} < -19$, and
(iii) an LAE-only sample with $\log (L_{\rm Ly\alpha}/\mathrm{erg\,s^{-1}}) > 42$. This design allows for a direct and survey-specific comparison between our mock catalogs and observed LAE luminosity functions. For comparisons to JWST-selected samples, it may also be useful to account for potential aperture/slit losses, given that Ly$\alpha$ emission can be spatially extended.

This LAE modeling framework provides a physically grounded yet flexible approach to linking theoretical reionization models with observable galaxy populations. It incorporates both empirical inputs and simulation-derived quantities, enabling a robust exploration of how different ionization histories and topologies affect LAE observability and clustering. For full algorithmic details, performance tests, and parameter choices, we refer the reader to \citet{Maitra2025}.

\begin{figure*}
    \centering
    \includegraphics[width=17cm, trim={0cm 1.5cm 0cm 1.5cm}, clip]{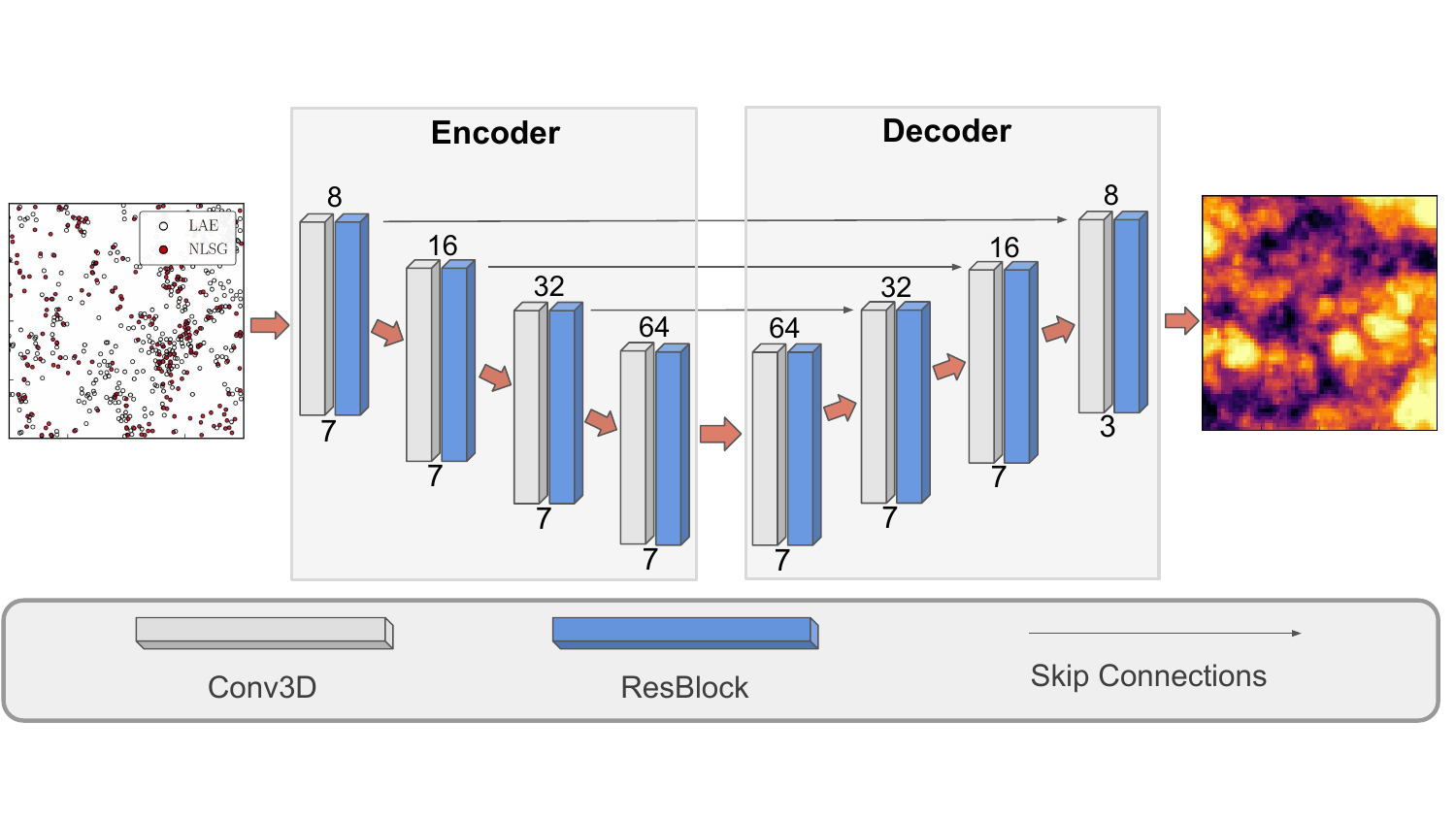}
    \caption{
Schematic architecture of the 3D U-Net used for tomographic reconstruction in this work. 
The network takes as input the three-dimensional distribution of Ly$\alpha$ emitters (LAEs) 
and non-Ly$\alpha$ selected galaxies (NLSGs) in a $80 \,h^{-1}\mathrm{cMpc} \times 80 \,h^{-1}\mathrm{cMpc} \times 30 \,h^{-1}\mathrm{cMpc}$ subvolume, which is processed through a hierarchy of 3D convolutional 
and residual blocks in the encoder (left). Each stage applies convolutional filters with 
kernel size $7\times 7\times 7$ and increases the feature dimensionality, beginning from 8 channels in 
the first layer to 64 in the final layer. Skip connections link encoder outputs to their corresponding decoder 
layers (right), enabling the recovery of fine-scale spatial structure otherwise lost during 
deep feature extraction. The decoder mirrors the encoder with a sequence of channel-reducing 
3D convolutions and residual blocks, reconstructing the three-dimensional ionization 
field in a $80 \,h^{-1}\mathrm{cMpc} \times 80 \,h^{-1}\mathrm{cMpc} \times 30 \,h^{-1}\mathrm{cMpc}$ subvolume from the learned hierarchical representation. Kernel sizes ($7\times 7\times 7$ for the first 3 layers and $3\times 3\times 3$ for the final layer) and feature 
channel counts (shown above each block) indicate the effective receptive field and 
representation capacity at every stage. This U-Net formulation, without a latent 
compression step, focuses purely on deterministic reconstruction and preserves maximal 
spatial information throughout the network.
}

    \label{fig:unet}
\end{figure*}

\section{Neural Network Architecture and Training Strategy}\label{sec:vae3dunet_reg}

Reconstructing spatially distributed physical fields from sparse and noisy data constitutes a classical inverse problem encountered across a wide range of scientific disciplines, from medical imaging \citep{McCann2017ReviewInverseProblems} and geophysics \citep{Tarantola2005InverseProblemTheory} to astrophysical large-scale structure inference \citep{Jasche2013BayesianLSS}. Traditional solutions typically rely on forward-modelling within a Bayesian framework, which provides physically interpretable results and principled uncertainty quantification \citep{Stuart2010Inverse, Jasche2013BayesianLSS}. However, these methods require high-dimensional sampling through techniques that become computationally expensive for large or high-resolution datasets \citep[see][for example]{BuiThanh2013Scale}.

The advent of deep learning has transformed this landscape by providing data-driven alternatives to classical inverse solvers. Convolutional neural networks (CNNs), in particular, exploit spatial locality and hierarchical feature learning, making them highly effective for recovering spatially correlated quantities from incomplete or noisy observations. They have been applied across a wide range of domains, including computed tomography and MRI reconstruction \citep{Jin2017DeepCT, Wang2016DeepMRI}, seismic inversion \citep{li2023seismicinversion}, and cosmological parameter estimation from large-scale structure maps \citep{ribli2019, ribli2019b, Krachmalnicoff2019, Lu2023}. CNNs learn statistical correlations directly from data, bypassing the need for handcrafted priors or explicit forward models, and offer fast, differentiable, end-to-end inference once trained. Nonetheless, standard CNN architectures are typically optimized for dense and isotropic inputs, which limits their effectiveness in regimes such as astrophysical tomography where sampling is sparse, anisotropic, or highly irregular.

These challenges motivate architectures that can simultaneously aggregate information across scales while preserving fine-grained spatial structure and robustness to missing data. Among modern architectures, the U-Net \citep{Ronneberger2015} has emerged as a particularly powerful framework for inverse problems. Its symmetric encoder-decoder design allows the encoder to compress information hierarchically into multi-scale feature maps, while skip connections transmit fine spatial detail directly to the decoder, ensuring that small-scale structure is not lost during compression. Three-dimensional extensions \citep{Cicek2016} have successfully generalized this approach to volumetric data, enabling applications ranging from medical 3D segmentation \citep{Milletari2016VNet, Isensee2021nnUNet} to cosmological inference of large-scale structure, including density-field reconstruction and structural classification\citep{Aragon-Calvo2019, Garcia2023, Du2025}. The combination of local receptive fields and cross-scale feature fusion makes the U-Net particularly well suited for tomographic inference problems where the goal is to recover continuous spatial fields from incomplete observations.

Our work is inspired by the {\sc DeepCHART} framework \citep{Maitra2025b}, which introduced a modified 3D U-Net with an additional variational latent space, representing both local and global statistical structure. Although a variational latent bottleneck can capture non-local correlations across the reconstructed volume, we find that in our setting where the training set is limited and each subvolume spans a relatively small spatial extent such global modelling provides little practical benefit. In this work, we therefore adopt a fully deterministic U-Net architecture, removing the stochastic latent space and focusing on a direct convolutional mapping between inputs and outputs. The model relies on the multi-scale receptive fields of the encoder-decoder structure to capture spatial correlations locally, prioritizing reconstruction accuracy and data efficiency in the small sample regime considered here.

In addition, our implementation introduces several architectural refinements tailored to stable training under sparse and noisy conditions:
(i) \emph{Group normalization} \citep{Wu2018GroupNorm} replaces batch normalization to ensure consistent feature scaling even for small mini-batches;
(ii) \emph{3D dropout} \citep{Srivastava2014Dropout, Tompson2015SpatialDropout} encourages spatial regularization by randomly masking entire feature maps during training; and
(iii) a \emph{Heteroscedastic output head} predicts both the voxel-wise mean $\hat{\mu}(\mathbf{r})$ and variance $\hat{\sigma}^2(\mathbf{r})$ of the reconstructed field at spatial location $\mathbf{r}$ in real space.
In this work, our primary objective is the accurate recovery of the mean ionization field $\hat{\mu}(\mathbf{r})$. The predicted variance $\hat{\sigma}^2(\mathbf{r})$ is therefore employed mainly as a regularization mechanism that stabilizes training and improves the fidelity of the mean reconstruction, rather than as a physically calibrated uncertainty estimate. While the variance does not yet provide a fully reliable measure of predictive uncertainty in our current setup, understanding its quantitative interpretation and calibration is left for future investigation. The complete architecture is shown in Figure~\ref{fig:unet}.

The following subsections describe the Dataset preparation, network architecture, loss formulation, and training strategy in detail.

\subsection{Training and Validation Dataset Preparation}

The neural network was trained on simulated tomographic data derived from four radiative-transfer reionization models: { Fiducial}, { Extremely Early}, { Early}, and { Oligarchic}. Each model provides a full three-dimensional distribution of the neutral hydrogen fraction, $x_{\mathrm{HI}}(\mathbf{r})$, and the associated galaxy and LAE populations. From these large-scale cosmological boxes, we constructed compact training and validation samples designed to capture both the local morphology of ionized regions and the statistical diversity across models.

We divided the full simulation cube ($160\,h^{-1}\,\mathrm{cMpc}$ on a side) into eight equal octants, each treated as an independent realization of the ionization field. Within each octant, we extracted subvolumes of size $64\times64\times24$ voxels ($80 \times 80 \times 30\,h^{-1}\,\mathrm{cMpc}$) using a sliding window with a stride of two voxels, yielding 21 subvolumes per octant. The physical size of each subvolume is well matched to the effective volumes probed by current and near-term high-redshift Lyman-$\alpha$ emitter surveys. In particular, the transverse extent is comparable to the comoving scales covered by deep JWST spectroscopic surveys such as COSMOS-3D \citep{Kakiichi2024}, as well as by wide-field narrow-band and multi-object spectroscopic LAE surveys including SILVERRUSH \citep{Ouchi2018}, the Subaru Prime Focus Spectrograph survey (PFS-SSP; \citealt{Takada2014, Tamura2016}), and COSMOS-Web \citep{Casey2023}. Meanwhile, the line-of-sight depth closely matches the $\sim 20$--$40\,h^{-1}\,\mathrm{cMpc}$ redshift slices probed by typical narrow-band and spectroscopic LAE selections.
We augmented each subvolume by reflections along the transverse directions (no reflection, $x$, $y$, and $x+y$), resulting in 84 subvolumes per octant. One octant was reserved for validation, yielding 84 validation samples, while the remaining seven octants were used for training, producing 588 training samples per reionization history. With four reionization models, the final training set contains 2352 subvolumes. This octant-based split ensures spatial independence between training and validation sets while sampling statistically similar environments.

Each subvolume carries, in addition to the $x_{\mathrm{HI}}$ field, the corresponding distribution of simulated LAEs and NLSGs. The LAE catalogues were constructed by assigning intrinsic Ly$\alpha$ luminosities to dark-matter haloes and modulating them by the local ionization state and the IGM transmission fraction. Similarly, NLSG galaxies were included by associating $M_{\mathrm{UV}}$ values to the same halo population. For each subvolume, the relative spatial positions and luminosities of these sources were stored alongside the three-dimensional ionization field, providing the inputs and targets for the tomographic reconstruction task.

The training and validation datasets were generated for three representative survey configurations designed to emulate realistic observational depths. For clarity, we classify these configurations into three categories {Deep}, {Shallow}, and {LAE-only}, summarised in Table~\ref{tab:survey_configs}. In the remainder of this paper, we adopt this terminology when referring to the different survey selections.

\begin{table}
    \centering
    \caption{Configurations of the three galaxy surveys considered in this work.}
    \label{tab:survey_configs}
    \begin{tabular}{lccc}
        \hline
        Survey type & $\log (L_{\rm Ly\alpha}/\mathrm{erg\,s^{-1}})$ cut & $M_{\mathrm{UV}}$ cut & Tracers included \\
        \hline
        Deep        & $>41$ & $<-18$ & LAEs + NLSGs \\
        Shallow     & $>42$ & $<-19$ & LAEs + NLSGs \\
        LAE-only    & $>42$ & --      & LAEs only   \\
        \hline
    \end{tabular}
\end{table}

The Deep configuration is representative of JWST spectroscopic programmes, such as COSMOS-3D \citep{Kakiichi2024}, which combine deep NIRCam imaging with NIRSpec spectroscopy over transverse scales of $\sim100\,h^{-1}\,\mathrm{cMpc}$. The Shallow configuration corresponds to wide-field narrow-band and multi-object spectroscopic LAE surveys such as SILVERRUSH, supplemented by UV-continuum-selected galaxy samples from GOLDRUSH \citep{Ono2018} and forthcoming spectroscopic programmes including the PFS-SSP. The LAE-only configuration represents pure line-selected samples drawn from SILVERRUSH and PFS-like spectroscopic catalogues.

In summary, we construct survey-matched, spatially independent training and validation datasets from multiple radiative-transfer reionization models by extracting and augmenting subvolumes whose physical scales and source selections are directly comparable to current and forthcoming LAE surveys. It is to be noted that MUSE- and JWST-selected samples are subject to different selection functions than Subaru/HSC narrow-band surveys, and would therefore require dedicated forward-modelling for application to real data. Each survey configuration is trained independently, yielding a physically grounded training sample that captures a wide range of reionization morphologies while maintaining statistical independence between the training and validation subsets.

\subsection{Architecture Description}

The network is implemented in \textsc{PyTorch} \citep{Paszke2019} and processes volumetric inputs of shape $64\times 64\times 24$, as mentioned above, corresponding to three-dimensional subvolumes extracted from the simulation boxes. It adopts a symmetric encoder-decoder architecture, wherein the input is first processed through an encoder path that extracts increasingly abstract feature representations, followed by a decoder path that incrementally refines and restores spatial detail to reconstruct the output field. Skip connections directly link corresponding encoder and decoder layers, preserving fine-grained information lost during deep feature transformations. Each block also includes residual connections \citep{he2016}, enabling the network to learn identity mappings and improving gradient flow during training.

\subsubsection{Encoder: hierarchical feature extraction.}
The encoder consists of four convolutional stages that progressively increase the feature depth while maintaining spatial resolution. Each stage performs:
\begin{enumerate}
    \item A 3D convolution with kernel size $7\times7\times7$, aggregating spatial information across neighbouring voxels.
    \item Group Normalization \citep{Wu2018}, which normalizes channel statistics within small groups of size 8, stabilizing training for small mini-batches.
    \item A ReLU activation \citep{Nair2010} that introduces non-linearity.
    \item A residual block \citep{he2016} containing two convolutional layers, each followed by Group Normalization and 3D Dropout ($p=0.1$). The residual addition allows the block to learn refinements on top of the input representation.
\end{enumerate}
At each successive layer, the number of channels doubles: $(8, 16, 32, 64)$. Unlike classical downsampling U-Nets, no strided convolution or pooling is applied. The model maintains the original spatial resolution of $64\times64\times24$, relying instead on deeper receptive fields and residual refinement to capture multi-scale spatial structure.

\subsubsection{Decoder: volumetric reconstruction.}
The decoder mirrors the encoder but inverts the feature hierarchy. It sequentially reduces channel depth while refining spatial features through convolutional and residual operations. Each decoder block performs:
\begin{enumerate}
    \item A 3D convolution followed by Group Normalization and ReLU activation to refine spatial details.
    \item A residual block with Group Normalization and Dropout to promote regularization.
    \item An additive skip connection that merges features from the corresponding encoder stage, ensuring both global and local information are preserved.
\end{enumerate}
The final layer is a $3\times3\times3$ convolution that outputs two channels per voxel: the predicted mean $\hat{\mu}(\mathbf{r})$ and the log-variance $\log\hat{\sigma}^2(\mathbf{r})$. Together they define a spatially resolved Gaussian likelihood,
\[
p(y|\mathbf{r}) = \mathcal{N}\!\bigl(y\,;\,\hat{\mu}(\mathbf{r}),\,\hat{\sigma}^2(\mathbf{r})\bigr).
\]
Our primary objective in this work is the accurate recovery of the mean ionization field $\hat{\mu}$. The network outputs both the voxel-wise mean $\hat{\mu}$ and the log-variance $\log \hat{\sigma}^2$ during inference. The variance models the aleatoric uncertainty arising from intrinsic ambiguity and noise in the tracer-field mapping. Here, $\hat{\sigma}^2$ acts as a stabilising regularization term during training, encouraging smoother and more stable reconstructions. Although our analysis focuses on the mean field, treating the variance output as an auxiliary regularizer, it will be interesting to see it as a physically interpretable uncertainty measure for our tomographic estimates; we leave this for future investigation.

\subsubsection{Loss formulation, optimization, and inference.}
Training employs a heteroscedastic negative log-likelihood (NLL) loss \citep[][]{Seitzer2022}, which assumes that the neutral hydrogen fraction at each voxel, $x_{\mathrm{HI}}(\mathbf r)$, is drawn from the modeled Gaussian distribution $p(x_{\mathrm{HI}}|\mathbf r)$:
\begin{equation}
    \mathcal{L}_{\beta\text{-NLL}} =
    \lfloor\sigma^{2\beta}\rfloor \!\left[\tfrac{1}{2}\log\sigma^2 + 
    \tfrac{1}{2}\frac{\bigl(x_{\mathrm{HI}}-\mu\bigr)^2}{\sigma^2}\right],
\end{equation}
where $\lfloor\cdot\rfloor$ denotes the stop-gradient operator and $\beta \in (0,1)$ modulates the relative weighting of uncertain regions. In this work, we adopt $\beta=0.5$.
This formulation enables the network to accommodate spatially varying effective noise levels by adaptively downweighting uncertain voxels during optimization, thereby improving convergence stability and generalization under sparse observational constraints. In contrast to earlier deterministic formulations such as \textsc{DeepCHART} \citep{Maitra2025b}, which optimize a point estimate of the ionization field through a deterministic loss, the present framework incorporates an explicit variance term within the loss function.

Model parameters are optimized using the Adam algorithm \citep{KingmaWelling2013} with a learning rate of $10^{-4}$, weight decay $10^{-4}$, and gradient clipping with a maximum norm of 1. Training is performed in mixed precision on a single NVIDIA V100 GPU with a batch size of 16.
Although the network was trained for a total of 200 epochs, the final model was selected at epoch~100 based on multiple diagnostic criteria. The $\beta$-NLL loss on the validation set reached its minimum near epoch$\sim$20 and began to increase thereafter, indicating mild overfitting in the uncertainty component $\sigma$. This behaviour is expected in heteroscedastic formulations, where the variance term can adaptively inflate to account for residual noise in the data. A rising $\beta$-NLL therefore does not necessarily signify deterioration in the predictive mean, but rather reflects the model’s growing tendency to assign higher uncertainty to regions it finds difficult to reconstruct. 

Since our primary objective in this work is accurate recovery of the mean ionization field $\hat{\mu}$, we additionally monitored the root-mean-square error (RMSE) and the Pearson correlation coefficient between the predicted and true $x_{\mathrm{HI}}$ fields, alongside $\mathcal{L}_{\beta\text{-NLL}}$. The RMSE reached a stable minimum around epoch$\sim$50, implying that the mean-field predictions had already converged, even though the $\beta$-NLL continued to fluctuate due to the adaptive variance term. The Pearson correlation coefficient, which quantifies large-scale structural agreement, improved gradually until epoch~100 and then plateaued, with no further gain in either RMSE or correlation. We therefore adopted the epoch$\sim$100 checkpoint as the optimal trade-off between reconstruction accuracy, smoothness regularization, and stability.

Each training run required approximately 7--8\,GPU-hours in total, with each epoch taking roughly 2--2.5\,minutes. Model checkpoints and detailed logs were recorded at every validation interval to ensure reproducibility and to facilitate quantitative comparison across survey configurations.
A single forward pass on a $64\times64\times24$ subvolume requires less than $\SI{0.5}{\second}$, enabling efficient large-volume tomographic reconstruction using a sliding-window or patch-based inference strategy.

In summary, the neural network in this work integrates hierarchical convolutional encoding, residual refinement, and voxel-level heteroscedastic uncertainty estimation into a unified 3D reconstruction framework. Its design yields stable and accurate reconstructions across varying tracer selections and noise levels, making it particularly suitable for tomographic studies of reionization and large-scale cosmic structure.

\tcbset{
    colback=white,
    colframe=black!40,
    fonttitle=\bfseries,
    coltitle=black,
    boxsep=3pt,
    arc=2pt,
    outer arc=2pt,
    left=4pt,
    right=4pt,
    top=2pt,
    bottom=2pt,
    width=\textwidth,
    enhanced,
    before skip=1pt,   % reduced top spacing
    after skip=1pt     % reduced bottom spacing
}

\begin{figure*}
\centering

% ==================== EARLY MODEL (NO SMOOTHING) ====================
\begin{tcolorbox}[title={\textbf{Early Model:} Deep Survey}]
\includegraphics[width=12.5cm, trim={3.5cm 0.5cm 3cm 1cm}, clip]{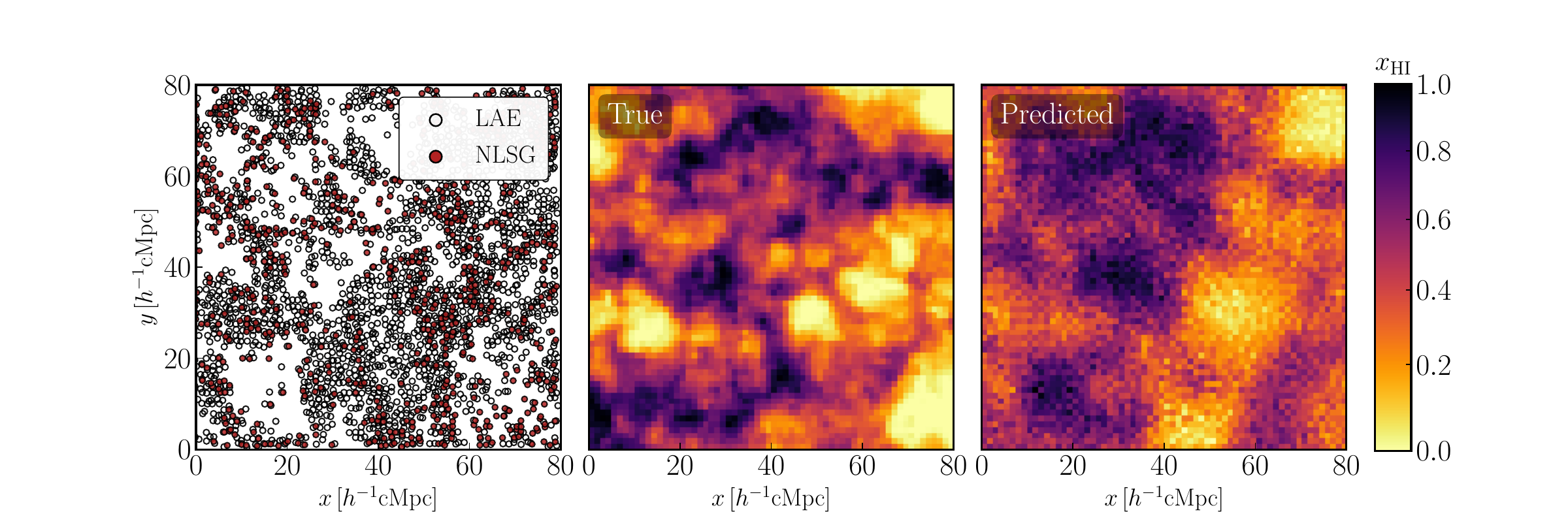}%
\hspace{0.15cm}%
\includegraphics[width=4.4cm, trim={0cm 0.7cm 0.67cm 0.2cm}, clip]{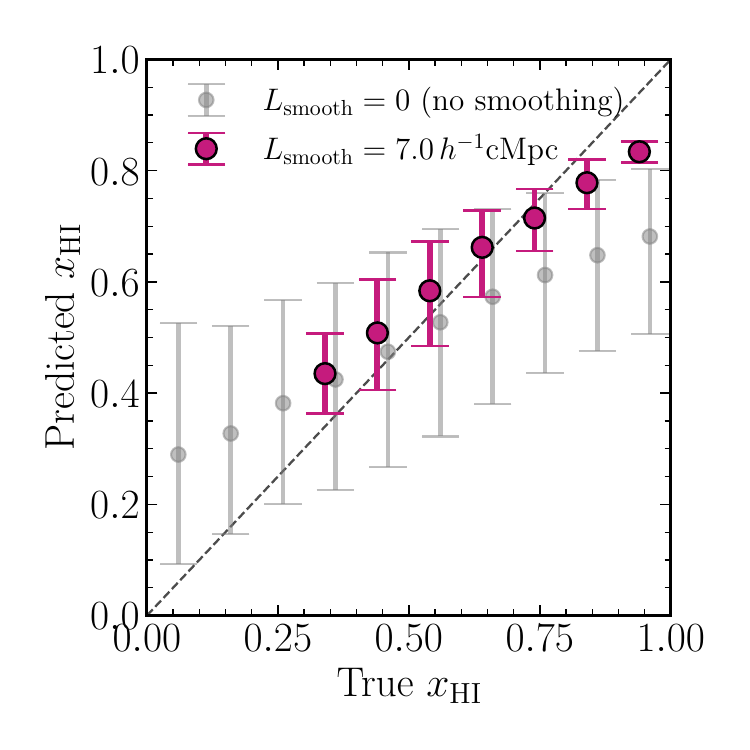}
\end{tcolorbox}

\begin{tcolorbox}[title={\textbf{Early Model:} Shallow Survey}]
\includegraphics[width=12.5cm, trim={3.5cm 0.5cm 3cm 1cm}, clip]{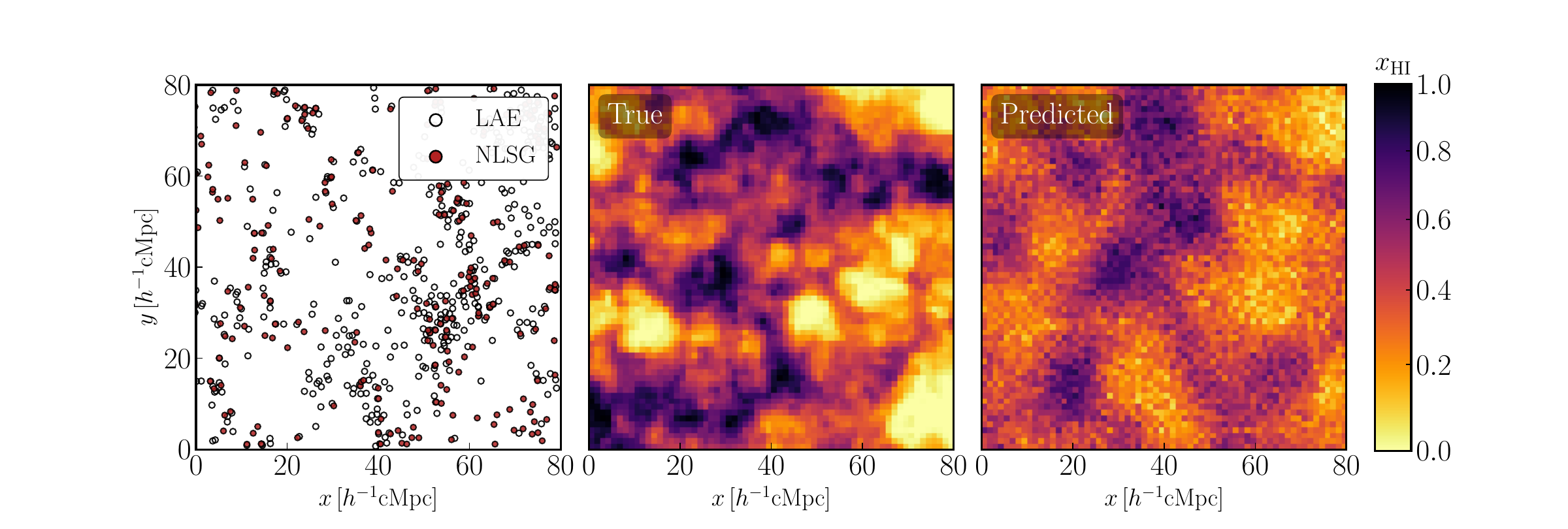}%
\hspace{0.15cm}%
\includegraphics[width=4.4cm, trim={0cm 0.7cm 0.67cm 0.2cm}, clip]{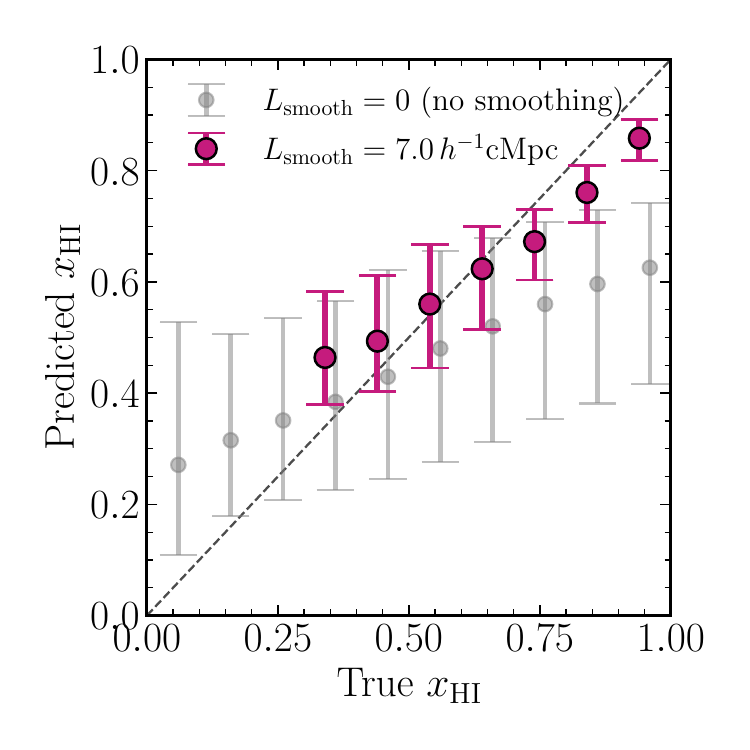}
\end{tcolorbox}

\begin{tcolorbox}[title={\textbf{Early Model:} LAE-only Survey}]
\includegraphics[width=12.5cm, trim={3.5cm 0.5cm 3cm 1cm}, clip]{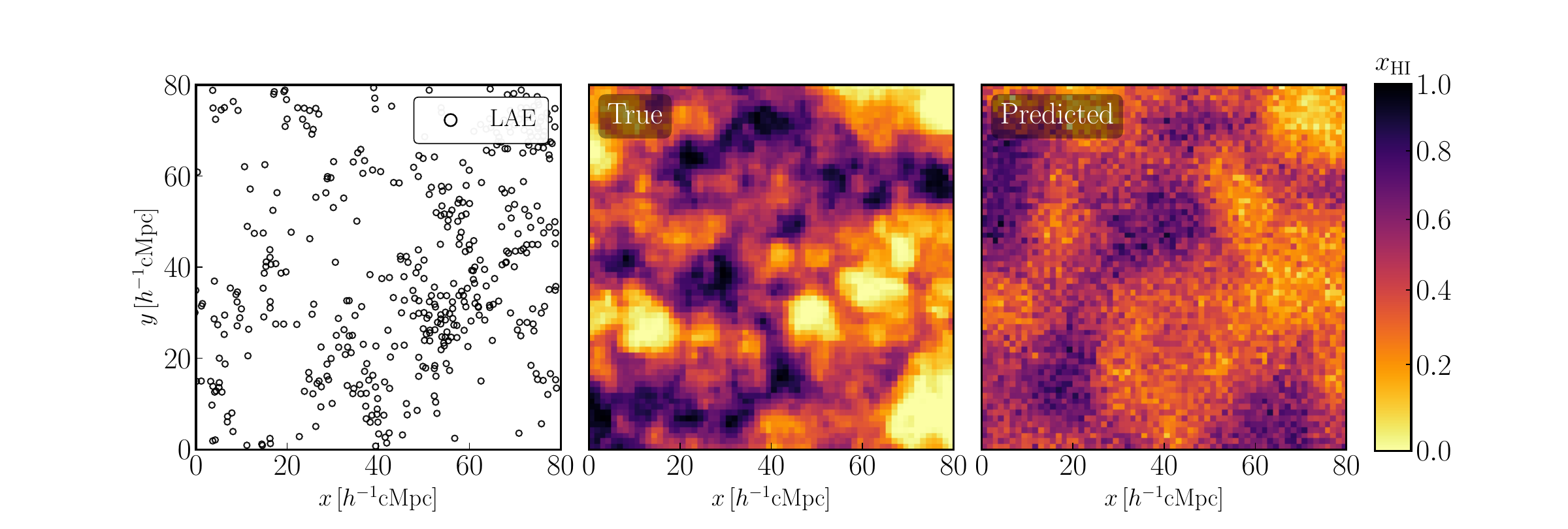}%
\hspace{0.15cm}%
\includegraphics[width=4.4cm, trim={0cm 0.7cm 0.67cm 0.2cm}, clip]{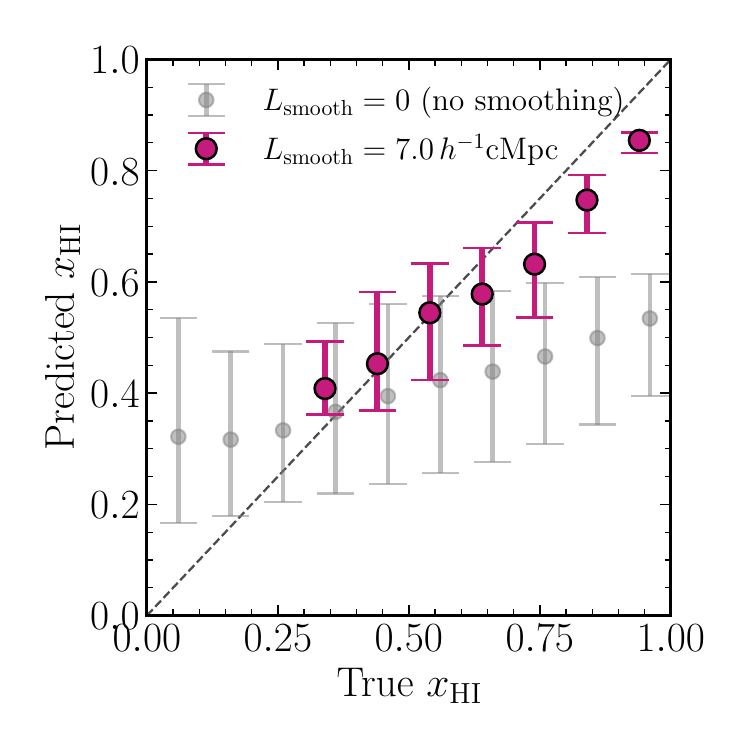}
\end{tcolorbox}

\caption{
Comparison of reconstructed neutral hydrogen fraction fields ($x_{\mathrm{HI}}$) projected along the redshift axis over $30~h^{-1}\mathrm{cMpc}$ and true-predicted relations for the {Early reionization model} at $z=7.14$. 
Each row corresponds to a different survey configuration: {Deep}, {Shallow}, and {LAE-only} (see Table~\ref{tab:survey_configs}). 
For each case, the left three panels show the spatial distribution of LAEs and NLSGs, 
the true ionization map, and the reconstructed map. 
The rightmost panels display the voxel-wise comparison between true and reconstructed 
$x_{\mathrm{HI}}$, showing the median relation and 68\% confidence intervals for both the 
unsmoothed ($L_{\mathrm{smooth}}=0$; grey) and smoothed 
($L_{\mathrm{smooth}}=7~h^{-1}\,\mathrm{cMpc}$; magenta) fields. 
Smoothing reduces small-scale fluctuations and improves the correlation with the true field, 
while the overall trends remain consistent across survey configurations.
}
\label{fig:Early_NoSmooth}
\end{figure*}

\begin{figure*}
\centering

% ==================== OLIGARCHIC MODEL (NO SMOOTHING) ====================
\begin{tcolorbox}[title={\textbf{Oligarchic Model:} Deep Survey}]
\includegraphics[width=12.5cm, trim={3.5cm 0.5cm 3cm 1cm}, clip]{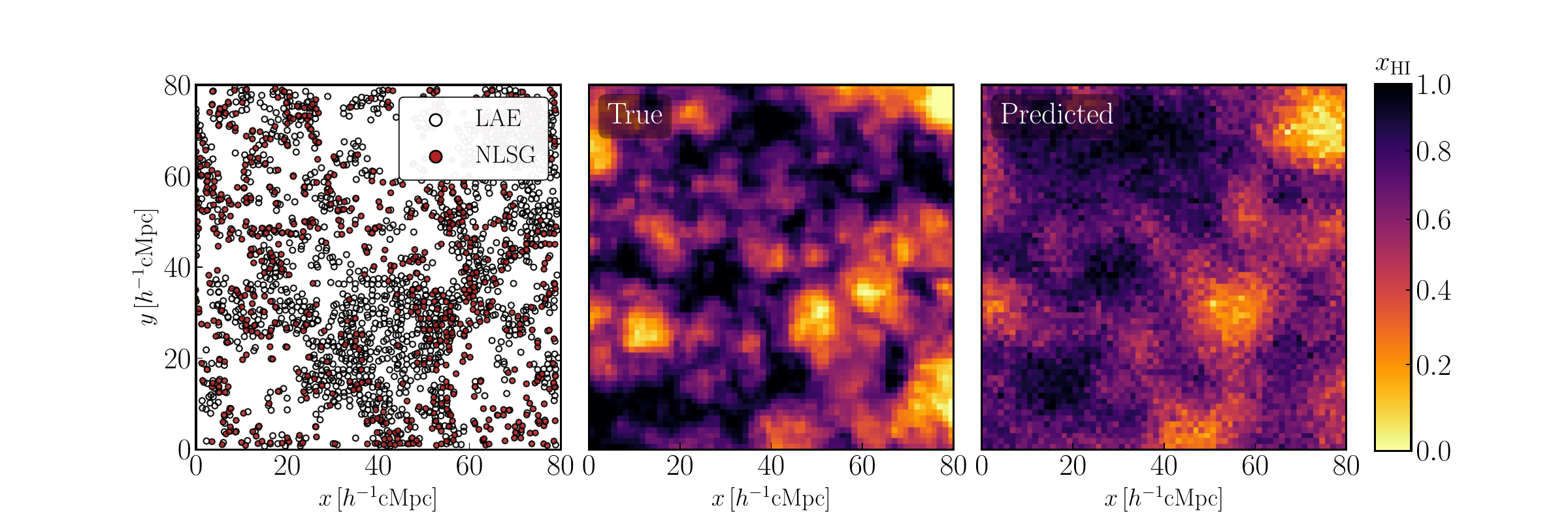}%
\hspace{0.15cm}%
\includegraphics[width=4.4cm, trim={0cm 0.7cm 0.67cm 0.2cm}, clip]{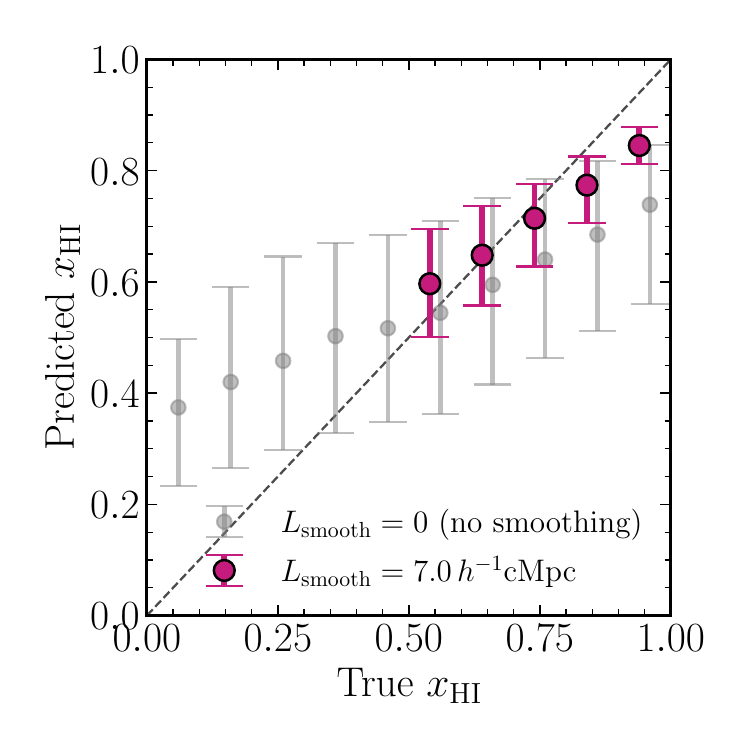}
\end{tcolorbox}

\begin{tcolorbox}[title={\textbf{Oligarchic Model:} Shallow Survey}]
\includegraphics[width=12.5cm, trim={3.5cm 0.5cm 3cm 1cm}, clip]{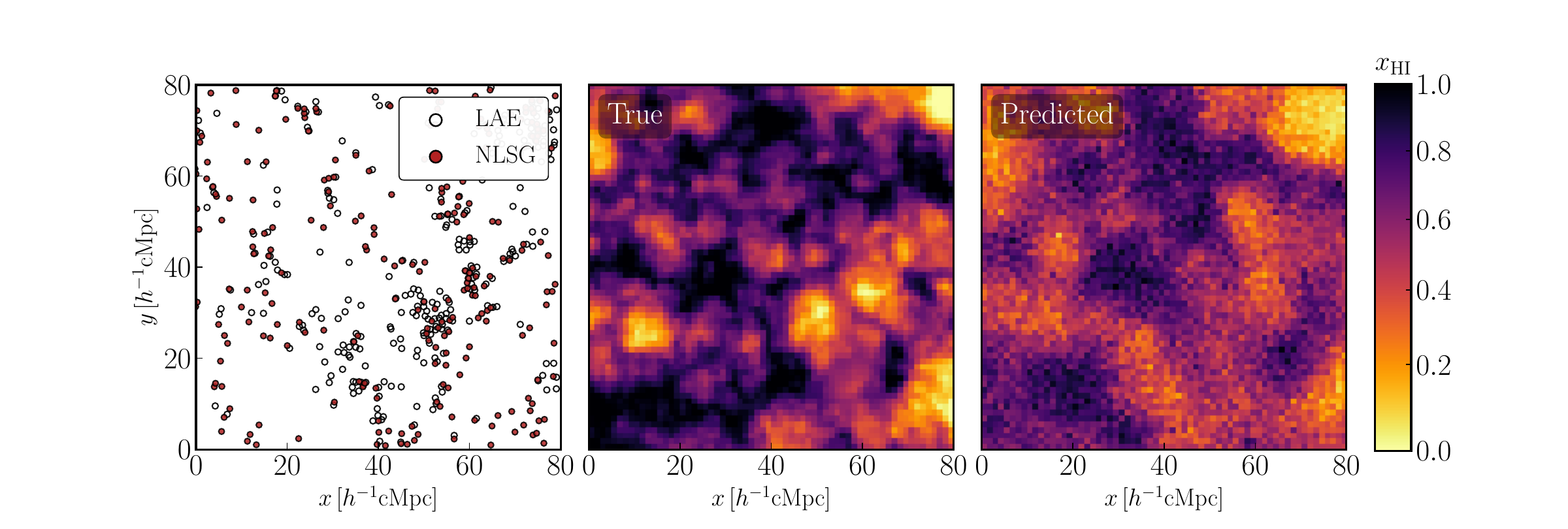}%
\hspace{0.15cm}%
\includegraphics[width=4.4cm, trim={0cm 0.7cm 0.67cm 0.2cm}, clip]{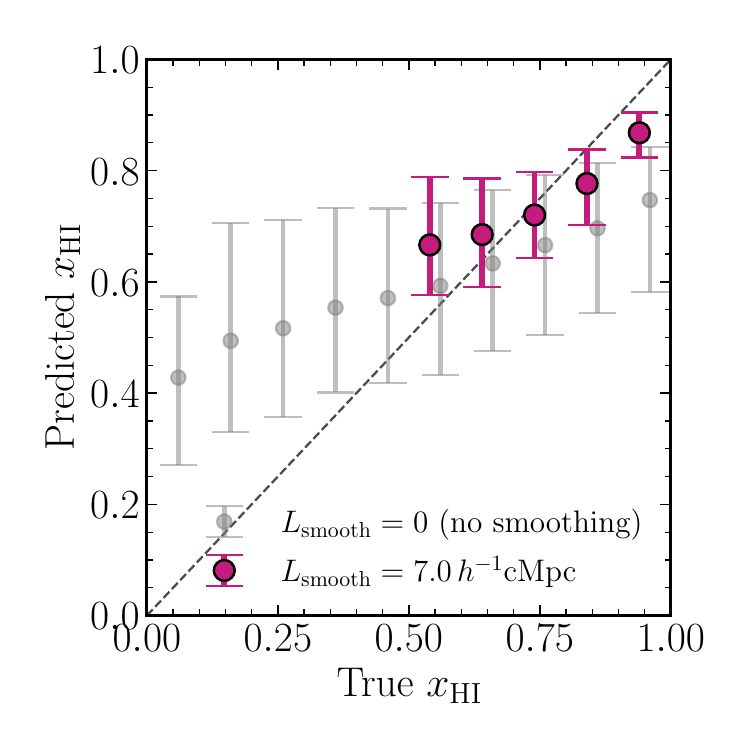}
\end{tcolorbox}

\begin{tcolorbox}[title={\textbf{Oligarchic Model:} LAE-only Survey}]
\includegraphics[width=12.5cm, trim={3.5cm 0.5cm 3cm 1cm}, clip]{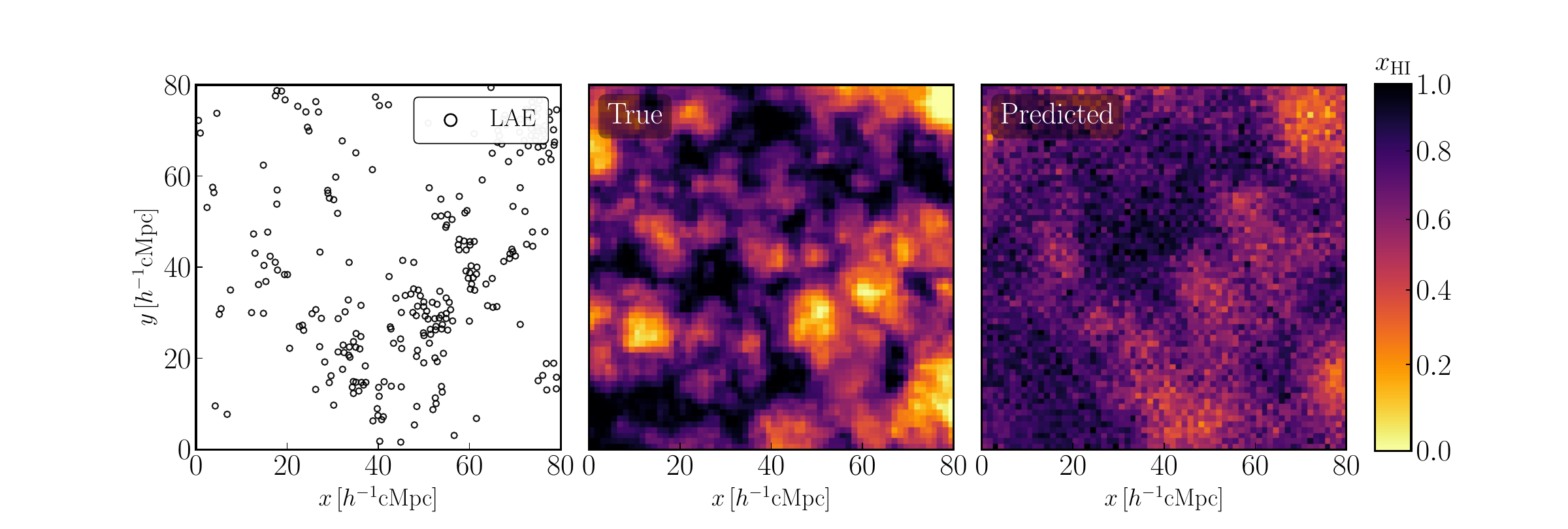}%
\hspace{0.15cm}%
\includegraphics[width=4.4cm, trim={0cm 0.7cm 0.67cm 0.2cm}, clip]{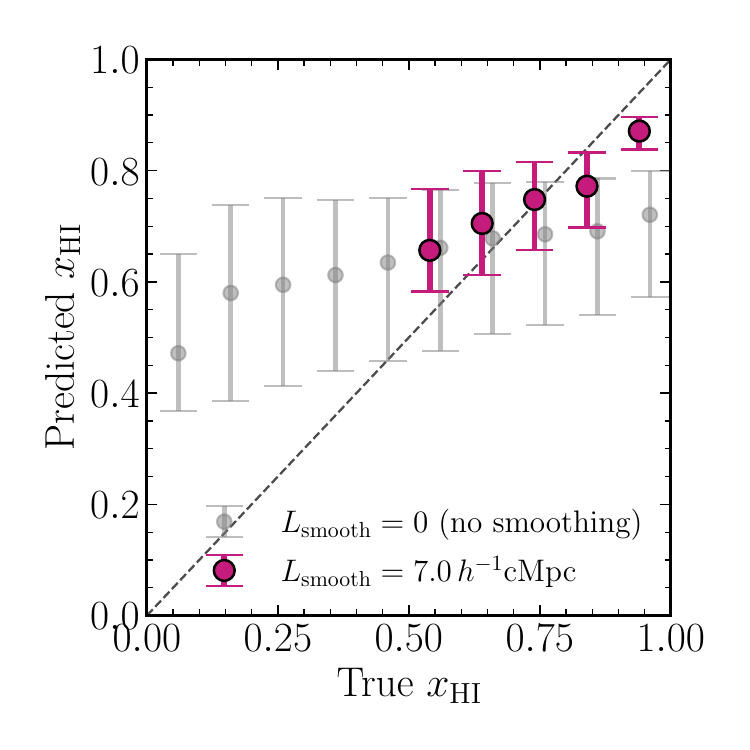}
\end{tcolorbox}

\caption{
Comparison of reconstructed neutral hydrogen fraction fields ($x_{\mathrm{HI}}$) projected along the redshift axis over $30~h^{-1}\mathrm{cMpc}$ and true–predicted relations for the {Oligarchic reionization model} at $z=7.14$. 
Each row corresponds to a different survey configuration: {Deep}, {Shallow}, and {LAE-only} (see Table~\ref{tab:survey_configs}).
For each case, the left three panels show the spatial distribution of LAEs and NLSGs, 
the true ionization map, and the reconstructed map. 
The rightmost panels display the voxel-wise comparison between true and reconstructed 
$x_{\mathrm{HI}}$, showing the median relation and $68\%$ confidence intervals for both the 
unsmoothed ($L_{\mathrm{smooth}}=0$; grey) and smoothed 
($L_{\mathrm{smooth}}=7~h^{-1}\,\mathrm{cMpc}$; magenta) fields. 
}
\label{fig:Oligarchic_NoSmooth}
\end{figure*}

\section{Results}
\label{sec:results}

We now examine the performance of our deep-learning framework in reconstructing the neutral-hydrogen fraction field during the EoR from mock observations of LAEs and NLSGs. 
We examine the model’s behaviour across multiple reionization scenarios that differ in their source prescriptions and ionization morphologies, and under varying survey selection thresholds that control the tracer population. 
Our analysis proceeds from direct, field-level comparisons to statistical diagnostics (including one-point and two-point statistics of the reconstructed and true ionization fields), in order to assess the physical scales over which the tomographic inference remains predictive. All statistical diagnostics, unless otherwise mentioned, are evaluated on $x_{\mathrm{HI}}$ projected along the redshift axis over slabs of thickness $30~h^{-1}\mathrm{cMpc}$. This choice is motivated by the typical comoving depth probed by narrowband Lyman-$\alpha$ surveys and the associated redshift-space uncertainties, making such a projection a natural and observationally relevant representation of the reconstructed ionization field.

\subsection{Reconstruction Fidelity}
\label{subsec:pearson_and_maps}

We first examine the fidelity with which the network recovers the spatial morphology of the neutral-hydrogen fraction field, $x_{\mathrm{HI}}$, from the observed distributions of Lyman-$\alpha$ emitters (LAEs) and non-Lyman-$\alpha$-selected galaxies (NLSGs). 
Throughout this subsection, the term ``Predicted'' denotes the network-inferred field, and ``True'' refers to the corresponding radiative-transfer simulation used as ground truth. 
All analyses are performed on statistically independent test subvolumes that were not used during training. 
Our goal here is to quantify, both visually and statistically, how accurately the model reproduces the ionization morphology on different physical scales.

\subsubsection{Visual comparison of reconstructed and true neutral-hydrogen maps}
\label{subsubsec:visual_maps}

Figures~\ref{fig:Early_NoSmooth} and \ref{fig:Oligarchic_NoSmooth} (\ref{fig:Crazy_NoSmooth} and \ref{fig:Fiducial_NoSmooth}, in the appendix) present representative 2D projections of the reconstructed and true neutral-hydrogen fraction fields, $x_{\mathrm{HI}}$, for the {Early} and {Oligarchic} ({Extremely Early} and {Fiducial}) reionization scenarios. Each row corresponds to a distinct tracer-selection threshold: 
(i)~$\log L_{\mathrm{Ly}\alpha}>41$, $M_{\mathrm{UV}}<-18$; 
(ii)~$\log L_{\mathrm{Ly}\alpha}>42$, $M_{\mathrm{UV}}<-19$; and 
(iii)~$\log L_{\mathrm{Ly}\alpha}>42$, reflecting progressively sparser galaxy samples. 
In all panels, the spatial distribution of LAEs (open black circles) and NLSGs (filled red circles) is shown alongside the true and reconstructed $x_{\mathrm{HI}}$ maps, projected over a $\Delta z = 30~h^{-1}\mathrm{cMpc}$ slab.

Figure~\ref{fig:Early_NoSmooth} illustrates the reconstructions for the {Early} model. The true ionization field exhibits large-scale ionized regions surrounding dense galaxy concentrations, with interspersed neutral islands. We see that the reconstructions recover these features with good fidelity across all tracer thresholds, correctly identifying the dominant ionized regions and their approximate extent. We also see that the reconstruction is much better in capturing the large-scale features; it fails to capture the small-scale features to the same level of fidelity. We also see that the reconstruction quality varies systematically with tracer selection. The deepest survey ($\log L_{\mathrm{Ly}\alpha}>41$, $M_{\mathrm{UV}}<-18$) yields the most detailed and accurate reconstructions, successfully capturing bubble shapes and identifying even some small-scale neutral features. As the survey becomes shallower ($\log L_{\mathrm{Ly}\alpha}>42$, $M_{\mathrm{UV}}<-19$), the morphology becomes more blurred, with some bubble boundaries misaligned or merged. In the sparsest LAE-only case ($\log L_{\mathrm{Ly}\alpha}>42$), reconstructions retain only large-scale topological features, and the boundaries of ionized regions become significantly distorted due to the lack of faint LAEs and supporting NLSGs. 

%Voxel-wise comparisons between the true and predicted projected $x_{\mathrm{HI}}$ (right panels) quantify this agreement and simultaneously illustrate the impact of smoothing. For each bin of $\Delta x_{\mathrm{HI}}=0.1$, we plot the median reconstructed value of $x_{\mathrm{HI}}$ and the $68\%$ confidence interval across the test subvolumes, for both the unsmoothed fields ($L_{\mathrm{smooth}}=0$) and versions smoothed with a Gaussian kernel of $L_{\mathrm{smooth}} = 7~h^{-1}\mathrm{cMpc}$. This smoothing scale is chosen to lie above the threshold at which the Pearson correlation coefficient exceeds $r=0.5$ for all reionization scenarios for the deepest survey configuration (see Sec.~\ref{subsubsec:pearson} for details). Across the dynamic range, the smoothed reconstructions show reduced scatter and improved alignment with the one-to-one relation in comparison to the unsmoothed case, particularly at the extremes of $x_{\mathrm{HI}}$. This is consistent with the expectation that averaging over bubble-scale regions suppresses stochastic voxel-level fluctuations and yields a more faithful recovery of the contrast between ionized and neutral regions. The agreement improves marginally with survey depth, with the deepest sample showing the best recovery of extreme $x_{\mathrm{HI}}$ values.

Voxel-wise comparisons between the true and predicted projected $x_{\mathrm{HI}}$ (right panels in Figure~\ref{fig:Early_NoSmooth}) quantify this agreement. We do these comparisons in two ways. First, we compare the two maps at the simulation resolution. Second, we smooth the maps with a Gaussian kernel at $L_{\mathrm{smooth}} = 7~h^{-1}\mathrm{cMpc}$ before comparing. (This smoothing scale is chosen to lie above the threshold at which the Pearson correlation coefficient exceeds $r=0.5$ for all reionization scenarios for the deepest survey configuration. See Sec.~\ref{subsubsec:pearson} for details.) For each bin of $\Delta x_{\mathrm{HI}}=0.1$, we plot the median reconstructed value of $x_{\mathrm{HI}}$ and the $68\%$ confidence interval across the test subvolumes, for both the unsmoothed and the smoothed fields. Across the dynamic range, the smoothed reconstructions show reduced scatter and improved alignment with the one-to-one relation in comparison to the unsmoothed case, particularly at the extremes of $x_{\mathrm{HI}}$. This is consistent with the expectation that averaging over bubble-scale regions suppresses stochastic voxel-level fluctuations and yields a more faithful recovery of the contrast between ionized and neutral regions. The agreement improves marginally with survey depth, with the deepest sample showing the best recovery of extreme $x_{\mathrm{HI}}$ values.

Figure~\ref{fig:Oligarchic_NoSmooth} presents the corresponding reconstructions for the {Oligarchic} model, where ionized regions form small, disconnected bubbles embedded in a largely neutral background. The fragmented morphology is more challenging to reconstruct, particularly under sparse sampling. While all three survey thresholds allow the network to recover the overall segregation between ionized and neutral environments, bubble boundaries appear smeared, and small-scale neutral features are often missed. As in the {Early} case, denser sampling yields improved visual fidelity, while the large-scale correspondence between galaxy overdensities and ionized regions remains broadly preserved across the different survey configurations. The voxel-wise statistics for the {Oligarchic} model likewise show overall consistency with the identity relation, albeit with reduced accuracy in the unsmoothed fields compared to the {Early} case. Applying Gaussian smoothing suppresses scatter and brings the reconstructed field into closer agreement with the one-to-one relation. However, small residual deviations persist at the extremes of $x_{\mathrm{HI}}$, particularly for the sparsest surveys, similar to the behaviour seen in the {Early} model.

Taken together, these visual comparisons demonstrate that reconstruction fidelity depends sensitively on both tracer sampling density and the intrinsic complexity of the reionization morphology. Models characterised by extended, coherent ionized regions are more readily recovered than those dominated by fragmented bubbles, and deep, multi-tracer surveys substantially enhance the quality of tomographic inference.

\begin{figure*}
    \centering
    \includegraphics[width=0.5\linewidth]{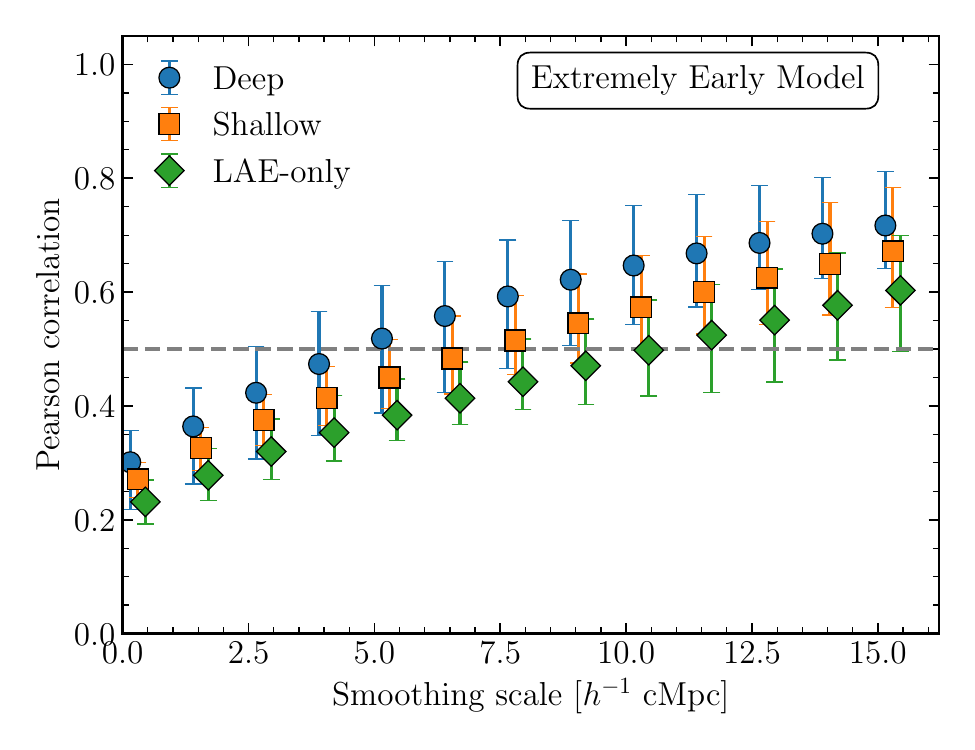}%
    \includegraphics[width=0.5\linewidth]{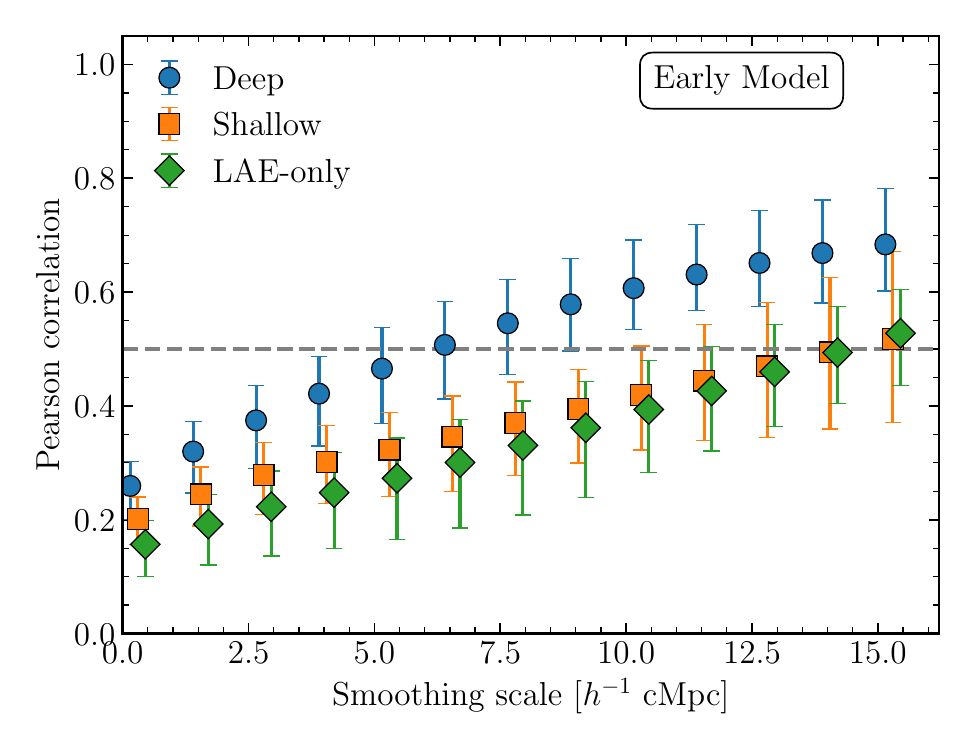}%
    
    \includegraphics[width=0.5\linewidth]{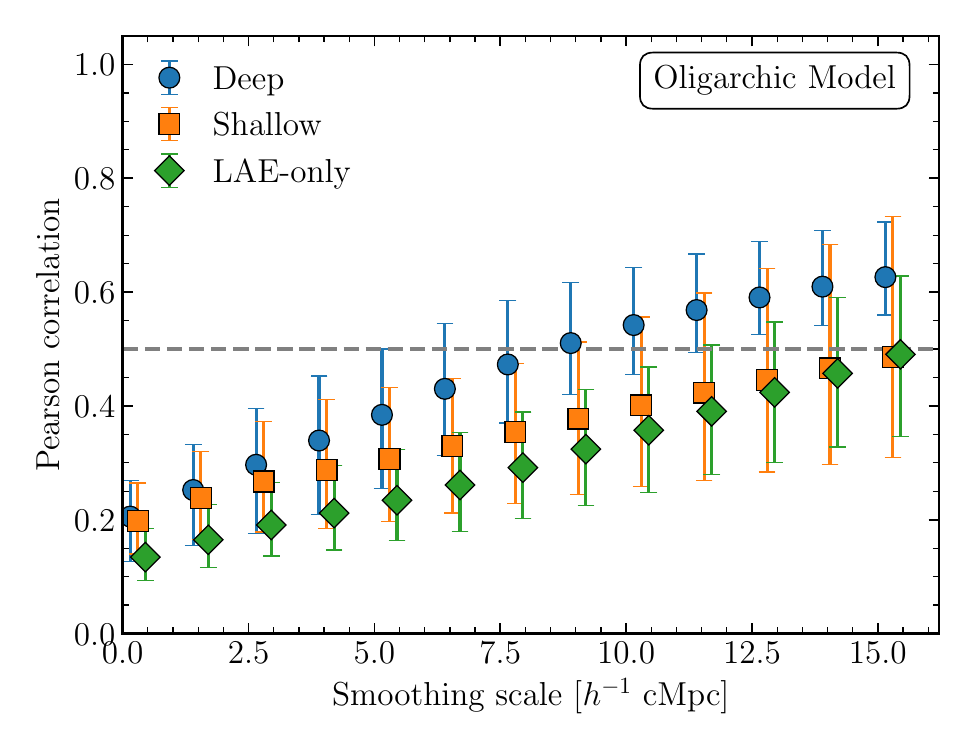}%
    \includegraphics[width=0.5\linewidth]{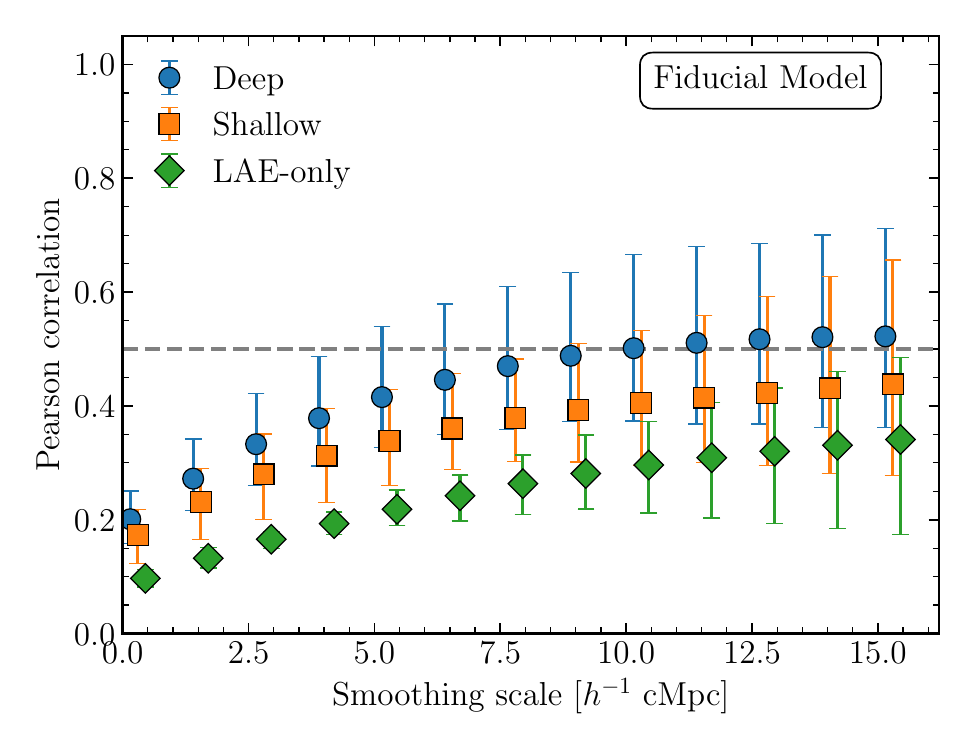}%
    \caption{
Pearson correlation coefficient between the true and reconstructed three-dimensional $x_{\mathrm{HI}}$ field as a function of Gaussian smoothing scale for different reionization models at $z=7.14$. 
Each panel corresponds to one reionization scenario ({Extremely Early}, {Early}, {Oligarchic}, and {Fiducial}). 
Markers with $1\sigma$ error bars indicate the inter-slice variation in correlation for the three survey configurations: Deep, Shallow, and LAE-only (see Table~\ref{tab:survey_configs}). 
A horizontal dashed line at $r=0.5$ marks the approximate threshold above which the reconstruction begins to recover meaningful large-scale structure. 
Overall, the correlation strength increases with smoothing scale, with the {Early} and {Extremely Early} models exhibiting systematically higher fidelity than the {Oligarchic} and {Fiducial} models.
}    
\label{fig:Pearson}
\end{figure*}

\subsubsection{Voxel-wise correlation across smoothing scales}
\label{subsubsec:pearson}

To quantify the qualitative trends seen in the maps in a statistically robust manner, we next compute the voxel-wise Pearson correlation coefficient between the reconstructed and true neutral-hydrogen fraction fields as a function of spatial smoothing scale. Unlike the projected-field statistics discussed above, this diagnostic is evaluated directly on the full three-dimensional $x_{\mathrm{HI}}$ fields, providing a scale-dependent measure of how faithfully the network recovers the ionization morphology.

The Pearson correlation coefficient is defined as
\begin{equation}
r(L_{\mathrm{smooth}}) \;=\;
\frac{\left\langle \big(x_{\mathrm{HI}} - \bar{x}_{\mathrm{HI}}\big)\,\big(\hat{x}_{\mathrm{HI}} - \overline{\hat{x}_{\mathrm{HI}}}\big) \right\rangle}
{\sigma_{x_{\mathrm{HI}}}\,\sigma_{\hat{x}_{\mathrm{HI}}}}\,,
\end{equation}
where $x_{\mathrm{HI}}(\mathbf{r})$ and $\hat{x}_{\mathrm{HI}}(\mathbf{r})$ denote the true and reconstructed fields, respectively; $\bar{x}_{\mathrm{HI}}$ and $\overline{\hat{x}_{\mathrm{HI}}}$ are their volume-weighted means; and $\sigma_{x_{\mathrm{HI}}}$ and $\sigma_{\hat{x}_{\mathrm{HI}}}$ are the corresponding standard deviations. The angular brackets denote an average over all voxels within a test subvolume after applying an isotropic Gaussian filter of comoving width $L_{\mathrm{smooth}}$.

We vary the smoothing length from $L_{\mathrm{smooth}}=0$ to $15~h^{-1}\mathrm{cMpc}$ to probe reconstruction fidelity from voxel scales to the characteristic sizes of ionized bubbles. Because $r$ is insensitive to global amplitude offsets, it serves as a robust diagnostic of morphological agreement, probing whether the network correctly recovers the spatial phases of ionized and neutral regions irrespective of mean-field calibration. For each reionization scenario and survey configuration, we compute $r(L_{\mathrm{smooth}})$ across all test subvolumes and report the ensemble mean, with uncertainties estimated from the inter-slice variance and quoted as $68\%$ confidence intervals.

Figure~\ref{fig:Pearson} presents the resulting scale-dependent Pearson correlation coefficients for the four reionization scenarios ({Extremely Early}, {Early}, {Oligarchic}, and {Fiducial}) and for three tracer selections of increasing sparsity. Across all models, the correlation increases monotonically with smoothing scale, consistent with the visual impression that large-scale ionization morphology is recovered more robustly than small-scale structure. At small scales ($L_{\mathrm{smooth}}\lesssim3~h^{-1}\mathrm{cMpc}$), correlations remain weak ($r\simeq0.2$–$0.4$), reflecting the combined effects of voxel-level noise, sparse sampling, and stochastic tracer bias. As the smoothing scale increases, these small-scale discrepancies are averaged out, leading to steadily improving alignment between reconstructed and true fields.

For $L_{\mathrm{smooth}}\gtrsim10~h^{-1}\mathrm{cMpc}$, the correlation approaches an asymptotic plateau that depends on the underlying reionization morphology. The {Extremely Early} and {Early} models, characterised by large, coherent ionized regions, reach $r\simeq0.7$–$0.8$, whereas the more fragmented {Fiducial} and {Oligarchic} models saturate at $r\simeq0.5$–$0.6$. This systematic offset reflects the greater difficulty of reconstructing complex, small-scale ionization topologies.

A horizontal dashed line at $r=0.5$ marks the empirical threshold adopted to indicate statistically meaningful morphological agreement. This level is typically reached at $L_{\mathrm{smooth}}\gtrsim7~h^{-1}\mathrm{cMpc}$, identifying the physical scale above which tomographic reconstructions reliably recover the ionization topology. The transition around $L_{\mathrm{smooth}}\approx8$-$10~h^{-1}\mathrm{cMpc}$ resembles the characteristic bubble sizes during the mid-stages of reionization, reinforcing the conclusion that LAE/NLSG-based inference is primarily sensitive to bubble-scale structure, where spatial coherence dominates over stochastic fluctuations.

\begin{figure*}
    \centering
    \includegraphics[width=0.4\linewidth, trim=15 50 10 15, clip]{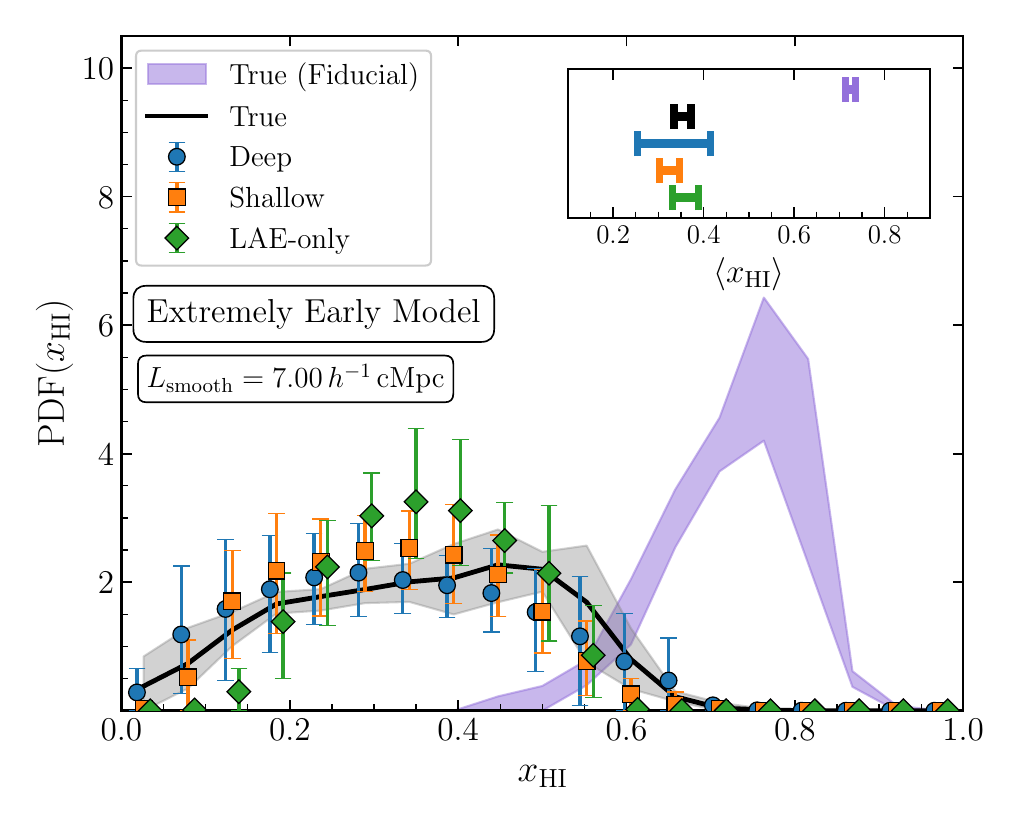}%
    \includegraphics[width=0.365\linewidth, trim=55 50 10 15, clip]{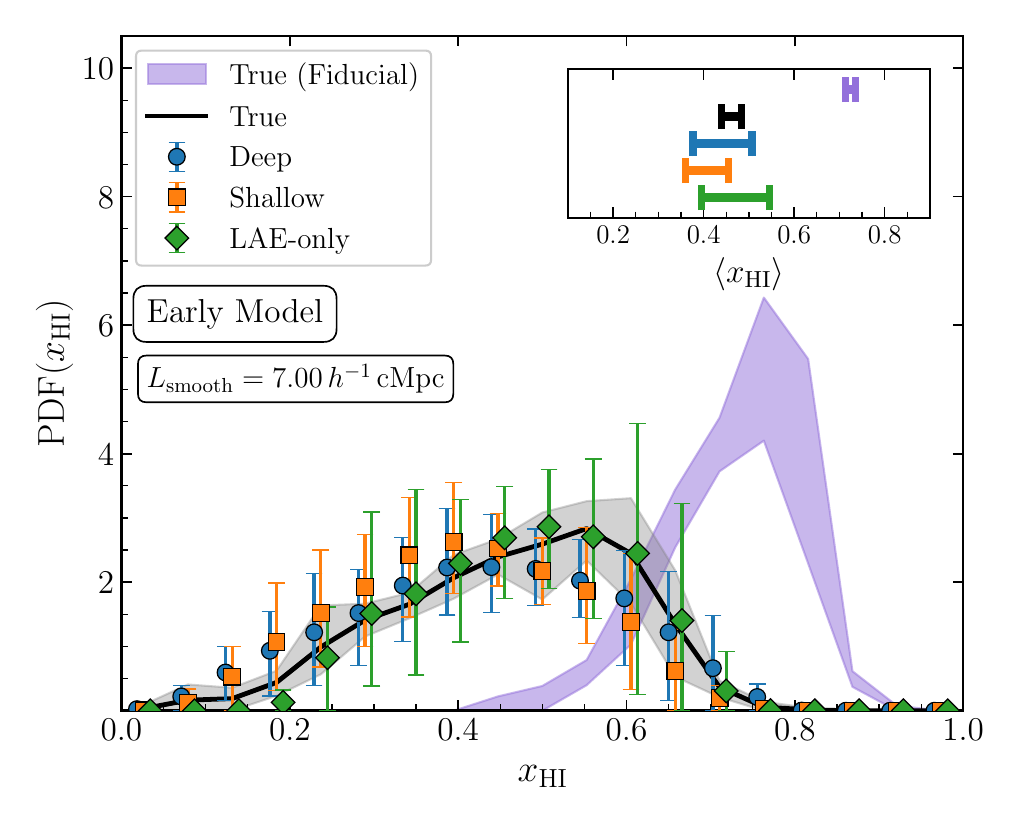}
    
    \includegraphics[width=0.4\linewidth, trim=15 10 10 10, clip]{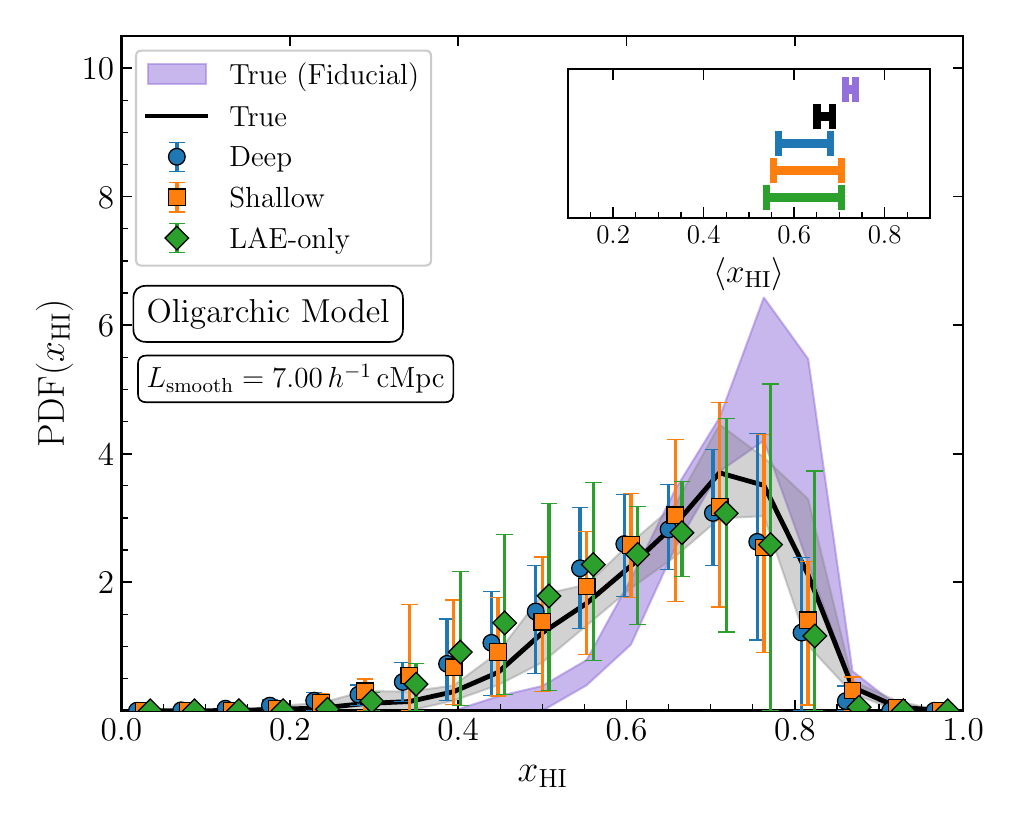}%
    \includegraphics[width=0.365\linewidth, trim=55 10 10 10, clip]{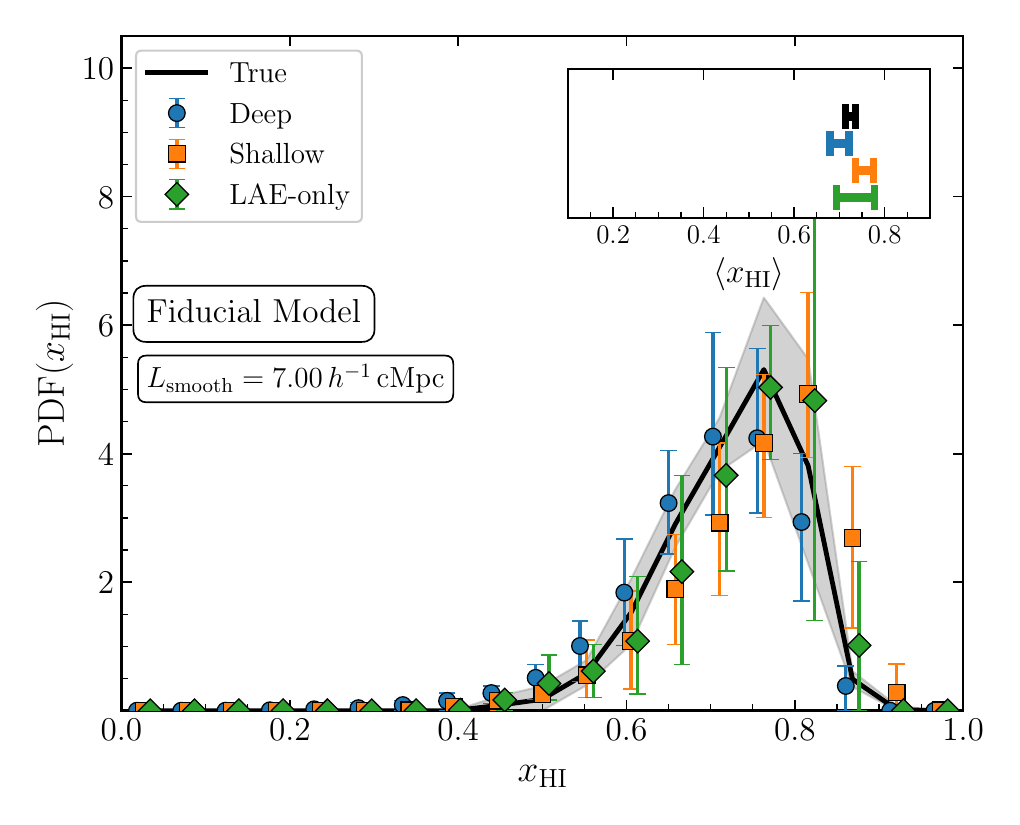}
    \caption{ Probability distribution functions (PDFs) of the projected (over $30~h^{-1}\mathrm{cMpc}$ in the redshift direction) neutral hydrogen fraction, $x_{\mathrm{HI}}$, reconstructed under four reionization scenarios at $z=7.14$: {Extremely Early}, {Early}, {Oligarchic}, and {Fiducial}, after Gaussian smoothing with $L_{\mathrm{smooth}}=7~h^{-1}\,\mathrm{cMpc}$. In each panel, the {Fiducial} model is shown as a common reference via the purple shaded band, enabling a direct comparison between alternative reionization histories and the same baseline. The solid black curve and grey band indicate the mean and 68\% confidence interval of the true $x_{\mathrm{HI}}$ distribution for the model shown in each panel. Colored symbols with error bars show reconstructed PDFs for the three survey configurations: Deep (blue), Shallow (orange) and LAE-only (green) (see Table~\ref{tab:survey_configs}). The inset panel summarises the mean neutral fraction $\langle x_{\mathrm{HI}}\rangle$ for each case, shown as thick horizontal error bars representing the 68\% confidence intervals.
}

    \label{fig:PDF}
\end{figure*}

\subsection{Probability Distribution Function of Reconstructed Fields}
\label{subsec:pdf_analysis}

Having examined voxel-wise correlations and visual morphology, we now turn to the one-point probability distribution function (PDF) of the projected $x_{\mathrm{HI}}$, which provides a complementary global diagnostic of the reconstructed fields. The PDF encodes the relative volume fractions of ionized and neutral regions and is sensitive not only to the global ionization level but also to the topology of the ionization field. Because observational samples suffer from redshift uncertainties which removes fine line-of-sight structures, projected statistics offer a natural comparison space. Accurate recovery of this distribution therefore constitutes a stringent test of whether the reconstruction preserves the correct mixture of ionization states across scales.

Figure~\ref{fig:PDF} shows the PDFs of the Gaussian-smoothed ($L_{\mathrm{smooth}}=7h^{-1}\mathrm{cMpc}$) $x_{\mathrm{HI}}$ fields for the four reionization scenarios: {Extremely Early}, {Early}, {Oligarchic}, and {Fiducial}. In each panel, the solid black curve and grey shaded band indicate the mean and 68\% inter-volume scatter of the true PDF for the model shown, computed over testing subvolumes. The {Fiducial} model is additionally overplotted in all panels as a common reference (purple shaded band), enabling direct comparison between alternative reionization histories. Colored symbols with error bars denote the reconstructed PDFs for three survey selections, with the error bars representing the 68\% inter-volume scatter across reconstructed realizations: (i)~log$L_{\mathrm{Ly}\alpha}>41$, $M_{\mathrm{UV}}<-18$ (blue); (ii)~log$L_{\mathrm{Ly}\alpha}>42$, $M_{\mathrm{UV}}<-19$ (orange); and (iii)~log$L_{\mathrm{Ly}\alpha}>42$ (green; LAE-only). The corresponding mean neutral fractions, $\langle x_{\mathrm{HI}}\rangle$, are summarised in the inset panel as horizontal coloured error bars, whose extents represent the 68\% inter-volume uncertainty on the mean.

The true PDFs display strong model dependence driven by differences in ionization morphology. The {Extremely Early} and {Early} models exhibit broad distributions peaking at $x_{\mathrm{HI}}\sim0.4$, consistent with extended reionization and early bubble overlap. In contrast, the {Oligarchic} and {Fiducial} models show narrower, more skewed PDFs with enhanced low-$x_{\mathrm{HI}}$ tails, characteristic of later-stage, patchier ionization dominated by compact bubbles. Notably, while the mean neutral fractions of the {Fiducial} and {Oligarchic} models partially overlap, their PDFs differ systematically in shape: the relative weight of the low-$x_{\mathrm{HI}}$ tail provides additional discriminatory power that is not captured by $\langle x_{\mathrm{HI}}\rangle$ alone. The explicit comparison to the fiducial reference PDF (purple band) highlights these shape differences across the full distribution.

For completeness, Fig.~\ref{fig:PDF_nosmooth} shows the PDFs of the unsmoothed fields ($L_{\mathrm{smooth}}=0$). Although the reconstructed PDFs recover the qualitative trends of the true distributions, they exhibit a somewhat compressed dynamic range, with both low- and high-$x_{\mathrm{HI}}$ tails suppressed. This behaviour is primarily driven by incomplete sampling of the IGM by discrete galaxy tracers for deeper survey selections with higher tracer densities. Residual compression nonetheless persists for the {Oligarchic} and {Fiducial} models and is further amplified in the LAE-only case.

Applying Gaussian smoothing at $L_{\mathrm{smooth}}=7\,h^{-1}\mathrm{cMpc}$ mitigates these limitations in a physically interpretable manner. Smoothing suppresses voxel-scale fluctuations while preserving large-scale ionization topology, reducing inter-volume scatter and significantly improving agreement between reconstructed and true PDFs. For the two denser tracer selections (blue and orange), both the peak position and width of the PDFs are recovered within the 68\% confidence interval, with mean neutral fractions converging to within $\Delta x_{\mathrm{HI}}\lesssim0.03$ for the {Extremely Early} and {Early} models and $\lesssim0.05$ for the more complex {Oligarchic} and {Fiducial} cases. The LAE-only reconstructions remain systematically narrower and biased toward higher $\langle x_{\mathrm{HI}}\rangle$, underscoring that PDF fidelity is ultimately limited by tracer completeness rather than smoothing alone.

Overall, the comparison between Figs.~\ref{fig:PDF} and \ref{fig:PDF_nosmooth} demonstrates that while unsmoothed reconstructions underestimate the intrinsic variance of the ionization field, smoothing on a characteristic scale substantially restores the correct global statistics. Crucially, the full PDF, particularly the low-$x_{\mathrm{HI}}$ tail, retains discriminatory power between reionization histories that are otherwise difficult to distinguish using the mean neutral fraction alone, highlighting the PDF as a valuable diagnostic for differentiating ionization morphologies during the epoch of reionization.

\begin{figure*}
    \centering
    \includegraphics[width=0.4\linewidth, trim=15 55 10 15, clip]{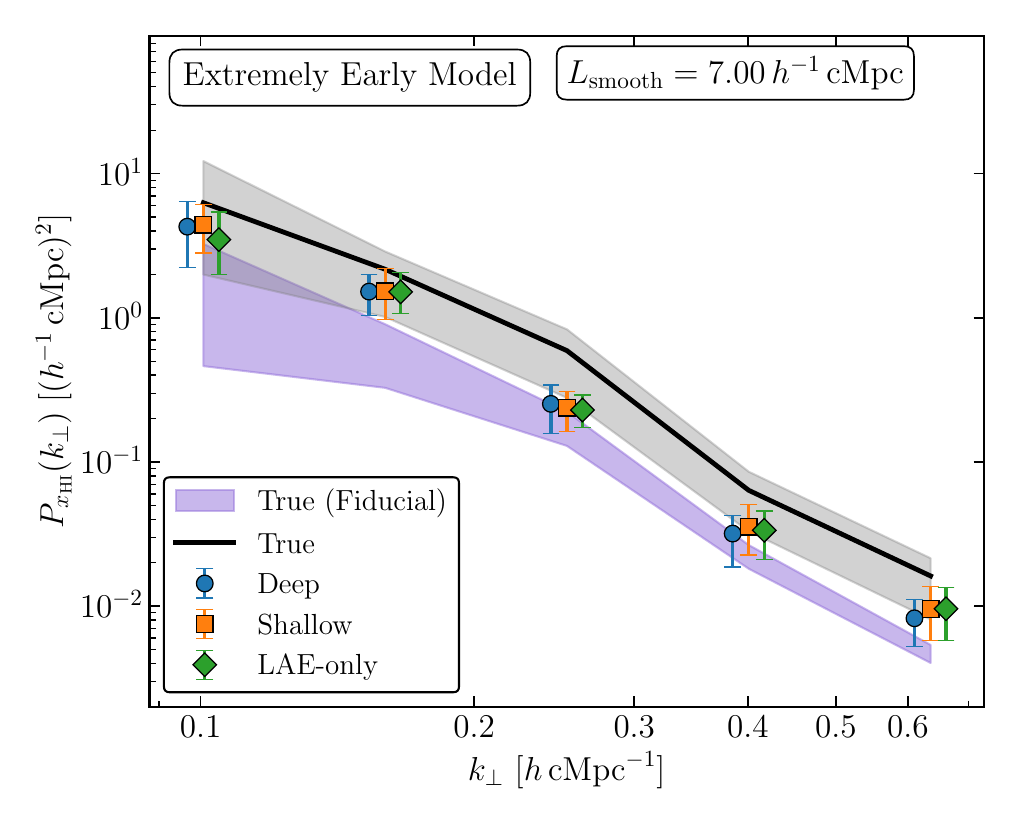}%
    \includegraphics[width=0.353\linewidth, trim=70 55 10 15, clip]{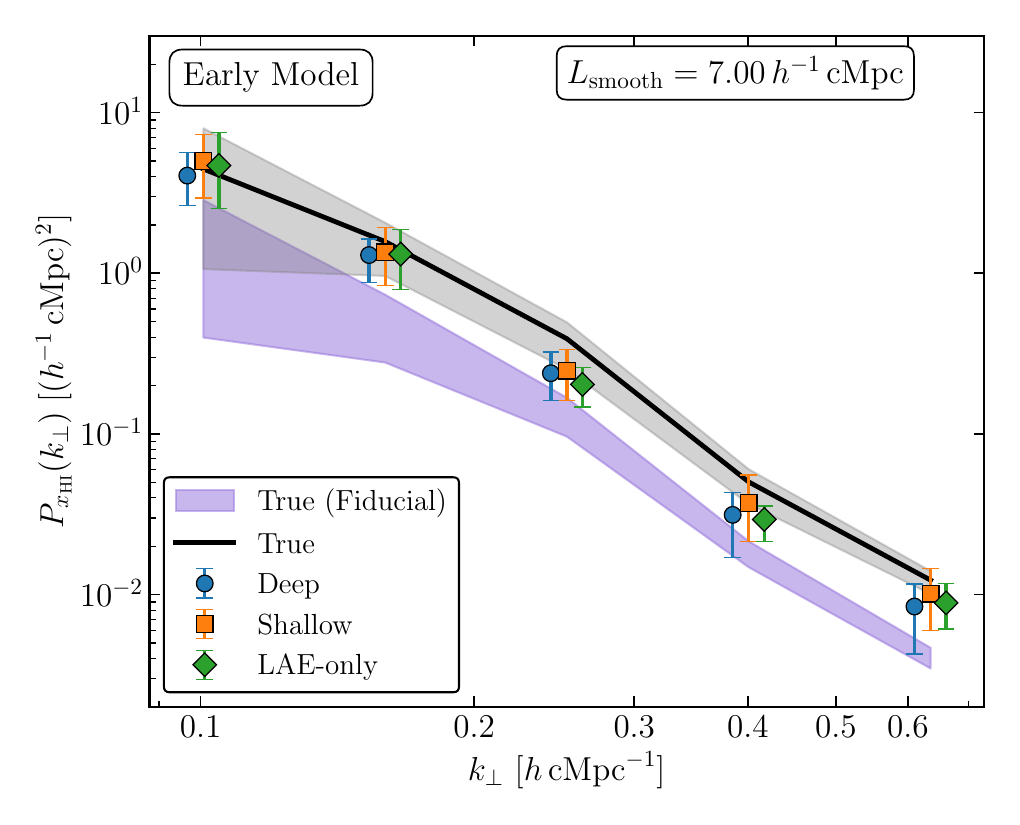}
    
    \includegraphics[width=0.4\linewidth, trim=15 10 10 10, clip]{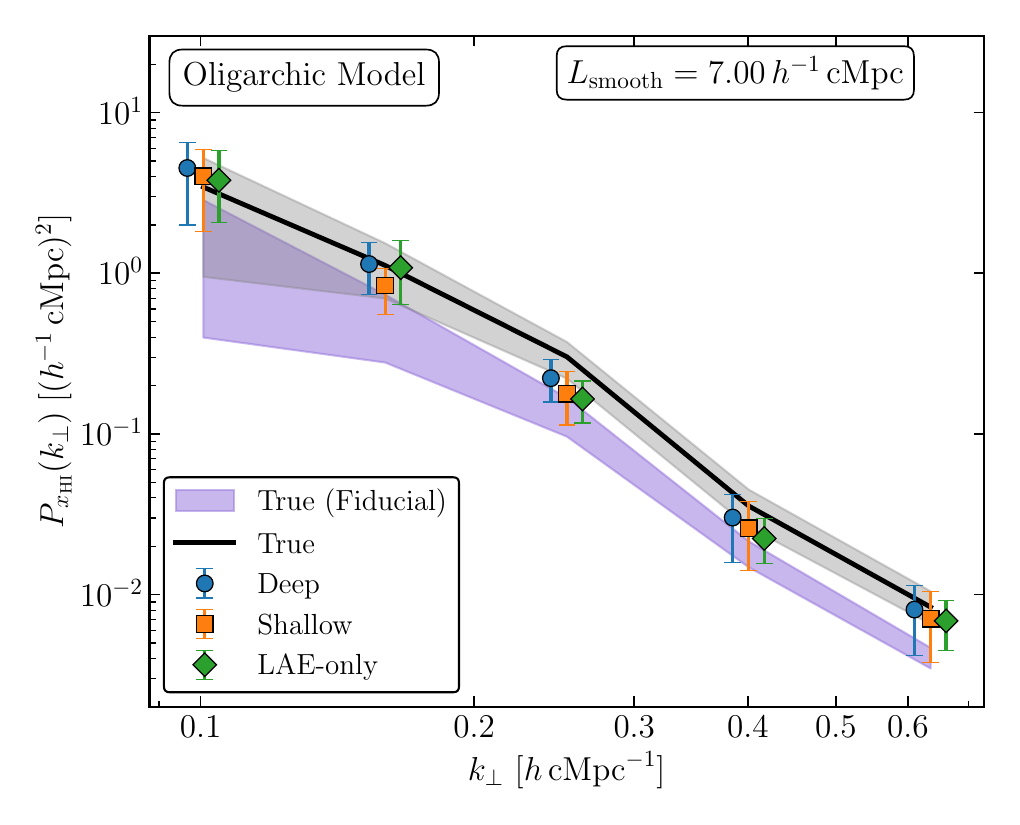}%
    \includegraphics[width=0.353\linewidth, trim=70 10 10 10, clip]{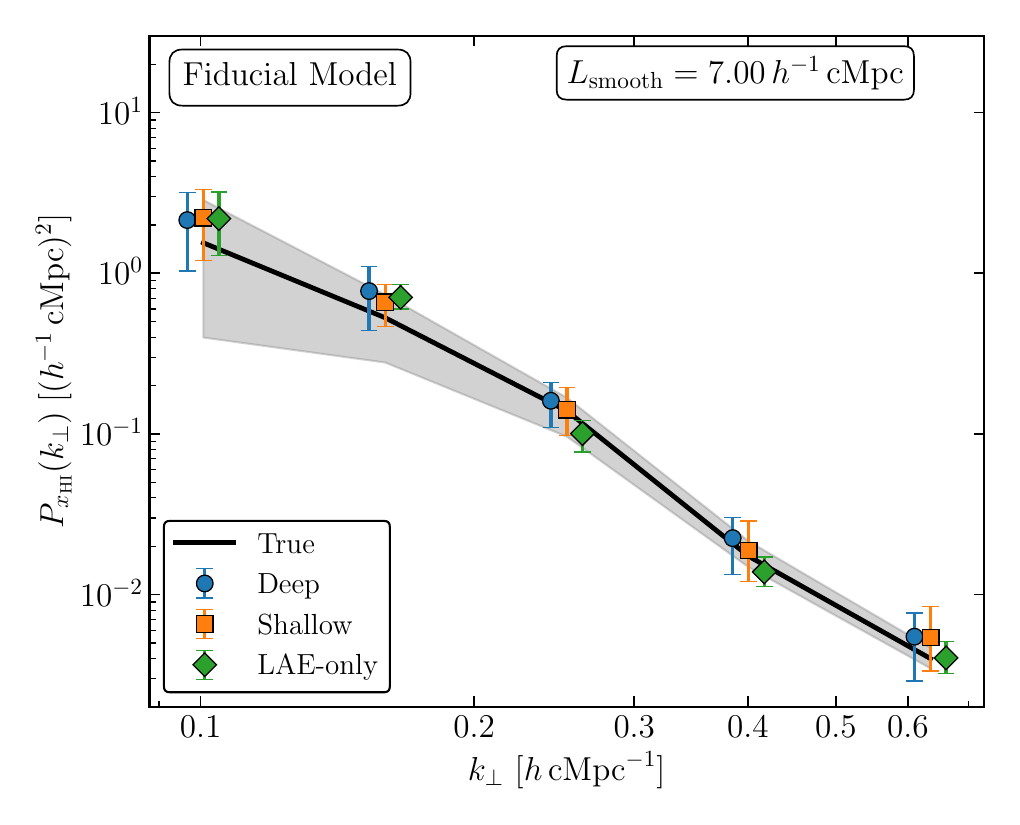}
    \caption{
Projected (over $30~h^{-1}\mathrm{cMpc}$ in the redshift direction) power spectra of the neutral-hydrogen fraction, 
$P_{x_{\mathrm{HI}}}(k_\perp)$, for true (black curves) and reconstructed 
fields across four reionization scenarios at $z=7.14$: 
{Extremely Early}, {Early}, {Oligarchic}, and {Fiducial} for gaussian-smoothed 
($L_{\mathrm{smooth}}=7\,h^{-1}\mathrm{cMpc}$) maps. 
Grey bands denote the $68\%$ scatter of the true spectra across 
the test subvolumes, while colored symbols with error bars show reconstructed PDFs for the three survey configurations: Deep (blue), Shallow (orange) and LAE-only (green) (see Table~\ref{tab:survey_configs}). In each panel, the {Fiducial} model is shown as a common reference via the purple shaded band.
}

    \label{fig:PS}
\end{figure*}

\subsection{Projected Power Spectrum of Reconstructed Fields}
\label{subsec:pk_analysis}

While the one-point PDF characterizes the distribution of ionization states, the projected power spectrum probes the scale-dependent clustering of neutral hydrogen. We compute the two-dimensional projected power spectrum $P_{x_{\mathrm{HI}}}(k_\perp)$ by projecting the three-dimensional neutral-fraction field over $30\,h^{-1}\mathrm{cMpc}$ along the line of sight and azimuthally averaging the squared Fourier amplitudes of the resulting transverse field. The spectrum is evaluated over the range $k_\perp = 0.08\text{--}0.8\,h\,\mathrm{cMpc}^{-1}$ using five logarithmically spaced bins and is analysed on logarithmic axes to capture its dynamic range. The mean and central 68\% confidence interval are estimated from multiple independent subvolumes. Agreement between the true and reconstructed $P_{x_{\mathrm{HI}}}(k_\perp)$ therefore provides a stringent test of whether the tomographic model preserves the correct scale-dependent morphology of the reionization topology.

Figure~\ref{fig:PS} compares the projected power spectra of the true and reconstructed $x_{\mathrm{HI}}$ fields for the four reionization scenarios after Gaussian smoothing with $L_{\mathrm{smooth}}=7\,h^{-1}\mathrm{cMpc}$. Each panel corresponds to a distinct model. The solid black curves denote the true power spectra averaged over the test subvolumes, with the shaded grey bands indicating the corresponding 68\% inter-volume scatter. The {Fiducial} model is additionally shown in all panels as a common reference via the purple shaded band, enabling a direct comparison between alternative reionization histories. Coloured markers represent the reconstructed spectra for three survey configurations. The blue symbols correspond to the Deep survey configuration, the orange symbols to the Shallow configuration, and the green symbols to the LAE-only configuration (see Table~\ref{tab:survey_configs}).
The behaviour of the corresponding unsmoothed power spectra ($L_{\mathrm{smooth}}=0$) is shown separately in Fig.~\ref{fig:PS_nosmooth}. Across all models, these reconstructions reproduce the large-scale ($k_\perp \lesssim 0.3\,h\,\mathrm{cMpc}^{-1}$) clustering of the true field with reasonable fidelity, while deviations become more pronounced at smaller scales ($k_\perp \gtrsim 0.3\,h\,\mathrm{cMpc}^{-1}$). The effect is strongest for the LAE-only selection, where sparse tracer coverage limits sensitivity to small-scale structure.

Smoothing both the reconstructed and true fields with $L_{\mathrm{smooth}}=7\,h^{-1}\mathrm{cMpc}$, as shown in Fig.~\ref{fig:PS}, systematically reduces inter-volume variance and improves agreement across all models. The smoothed reconstructions recover both the normalization and scale dependence of the true spectra to within the statistical uncertainty. This behaviour indicates that the network robustly captures the large-scale clustering of neutral hydrogen once sub-bubble fluctuations are averaged out. 

The relative differences among reionization models are also preserved. The {Extremely Early} and {Early} scenarios exhibit the highest large-scale power and the smoothest spectral slopes, consistent with extended reionization driven by numerous overlapping ionized regions. These models also display a larger inter-volume scatter in the power spectrum, reflecting enhanced sensitivity to cosmic variance associated with the stochastic overlap and percolation of large ionized bubbles. The {Oligarchic} model also shows higher power than the {Fiducial} case but reduced scatter compared to the earlier reionization histories, consistent with ionization driven by a smaller number of dominant sources. The {Fiducial} model exhibits the lowest overall amplitude and the smallest scatter, indicative of a more uniform and regulated ionization topology. The reconstructed spectra successfully reproduce these relative trends in both amplitude and slope, despite modest suppression at the highest $k_\perp$ due to implicit spatial regularization.
Dependence on tracer density is comparatively weak in the projected power spectrum. Reconstructions from all survey selections reproduce the true spectra within the inter-volume scatter across most scales. This behaviour contrasts with the stronger tracer-density dependence seen in the PDF analysis and indicates that the projected power spectrum is relatively robust to sampling sparsity, with large-scale clustering well captured even by the shallowest surveys.

In summary, the projected power-spectrum analysis demonstrates that the network reliably recovers the large-scale clustering and model-dependent scale dependence of the neutral hydrogen field. Gaussian smoothing on bubble scales enhances agreement by suppressing residual noise, while tracer density governs the accuracy of small-scale power recovery. Together with the PDF results, this confirms that the reconstruction preserves both global and scale-dependent ionization statistics, with remaining deviations driven primarily by observational sparsity.

\begin{figure*}
    \centering
    \includegraphics[width=0.5\linewidth]{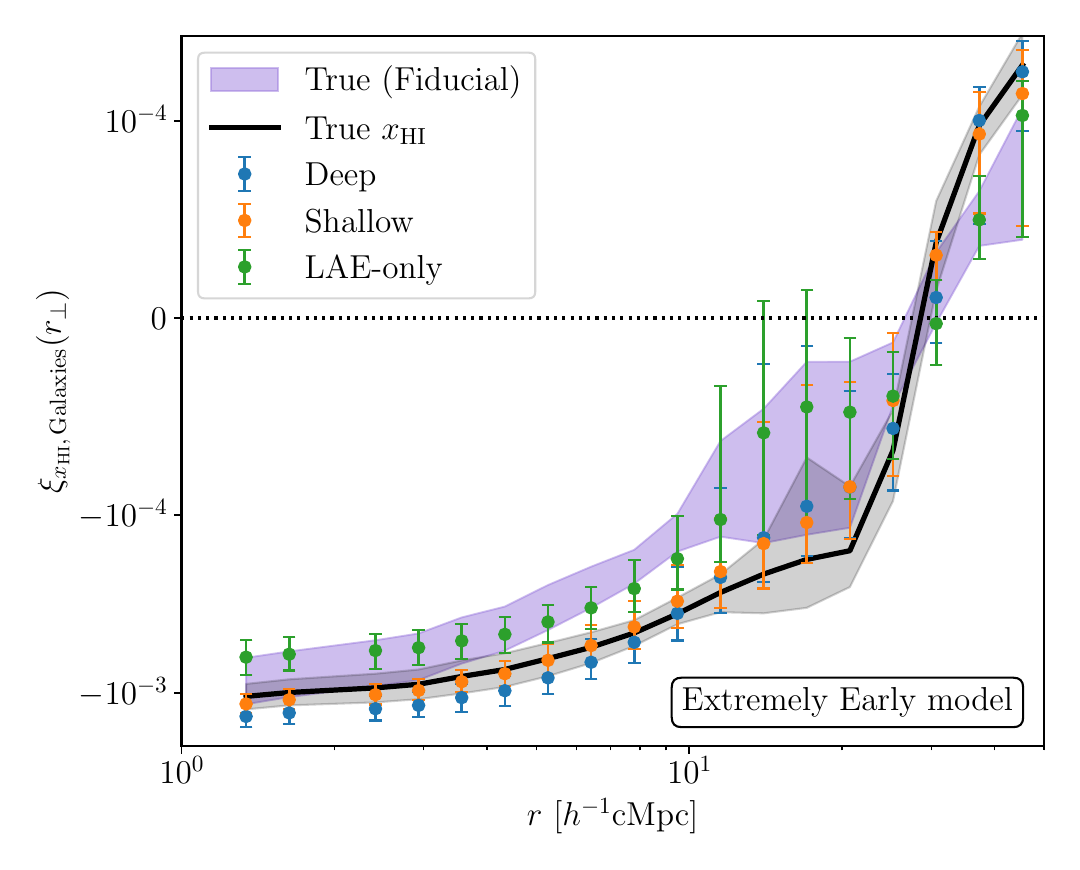}%    
    \includegraphics[width=0.5\linewidth]{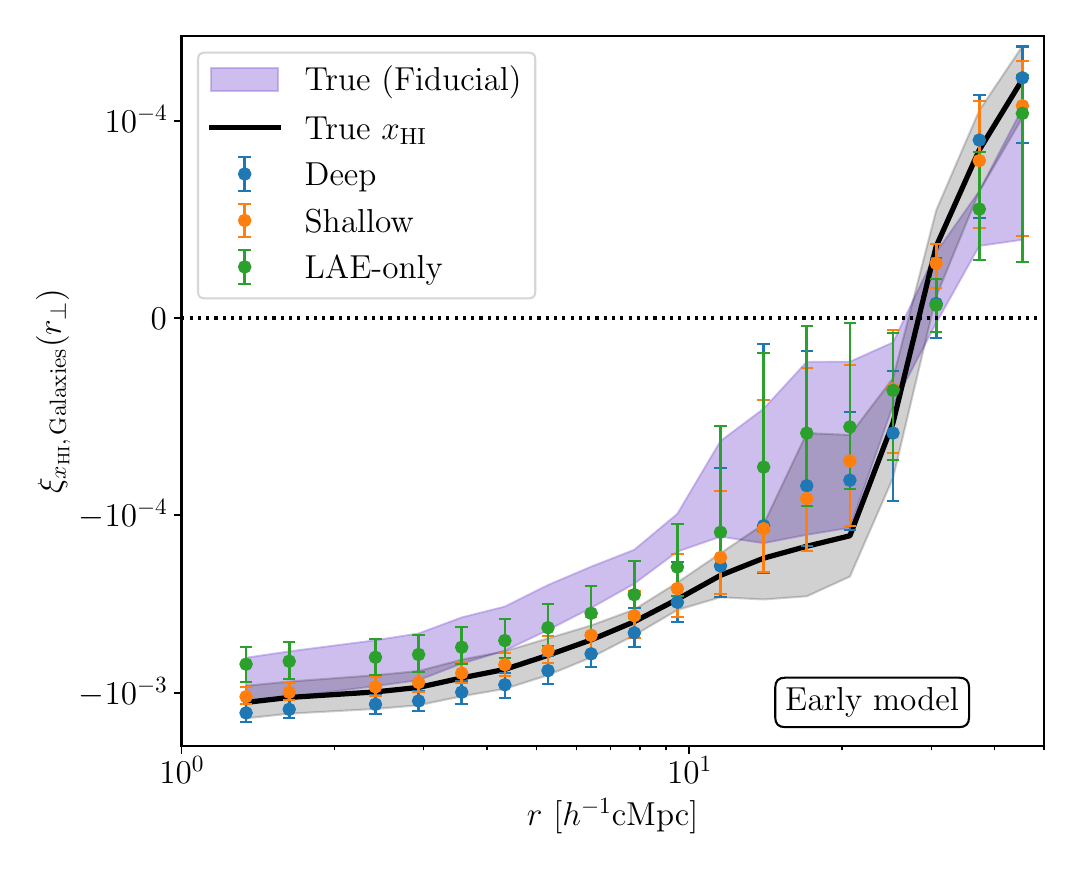}
    
    \includegraphics[width=0.5\linewidth]{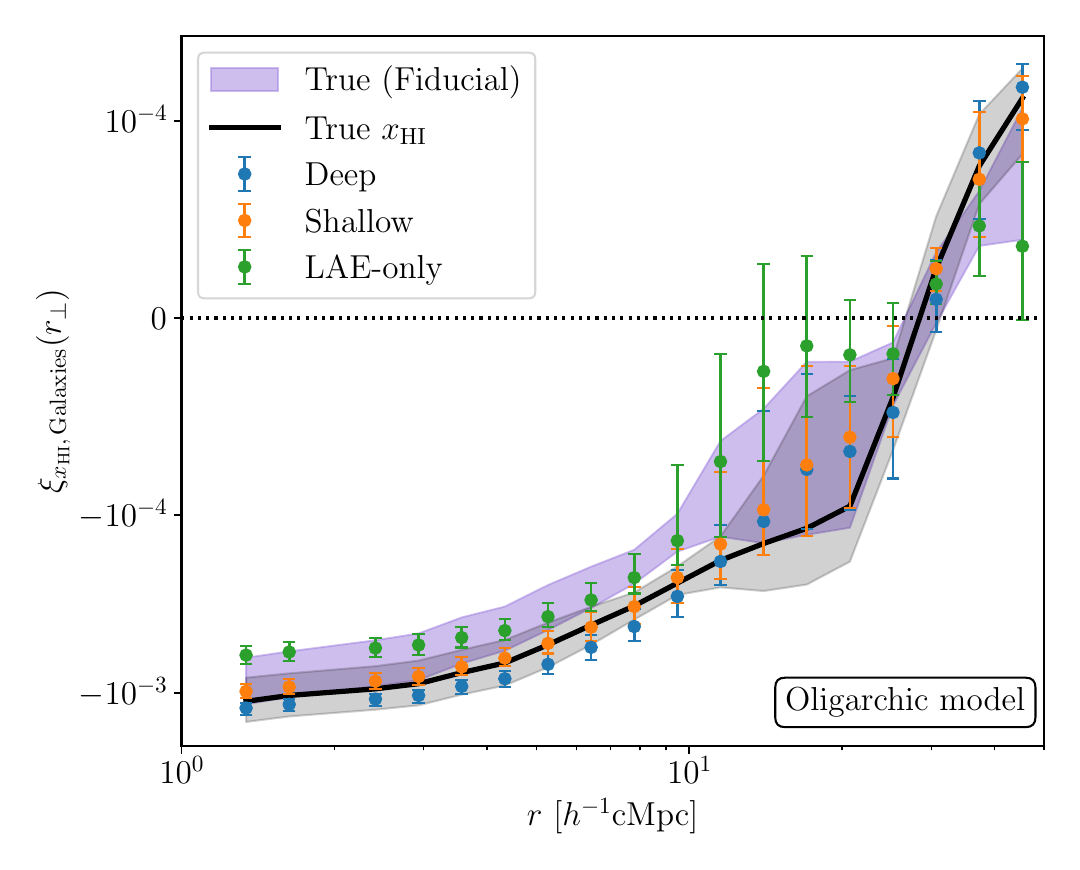}%
    \includegraphics[width=0.5\linewidth]{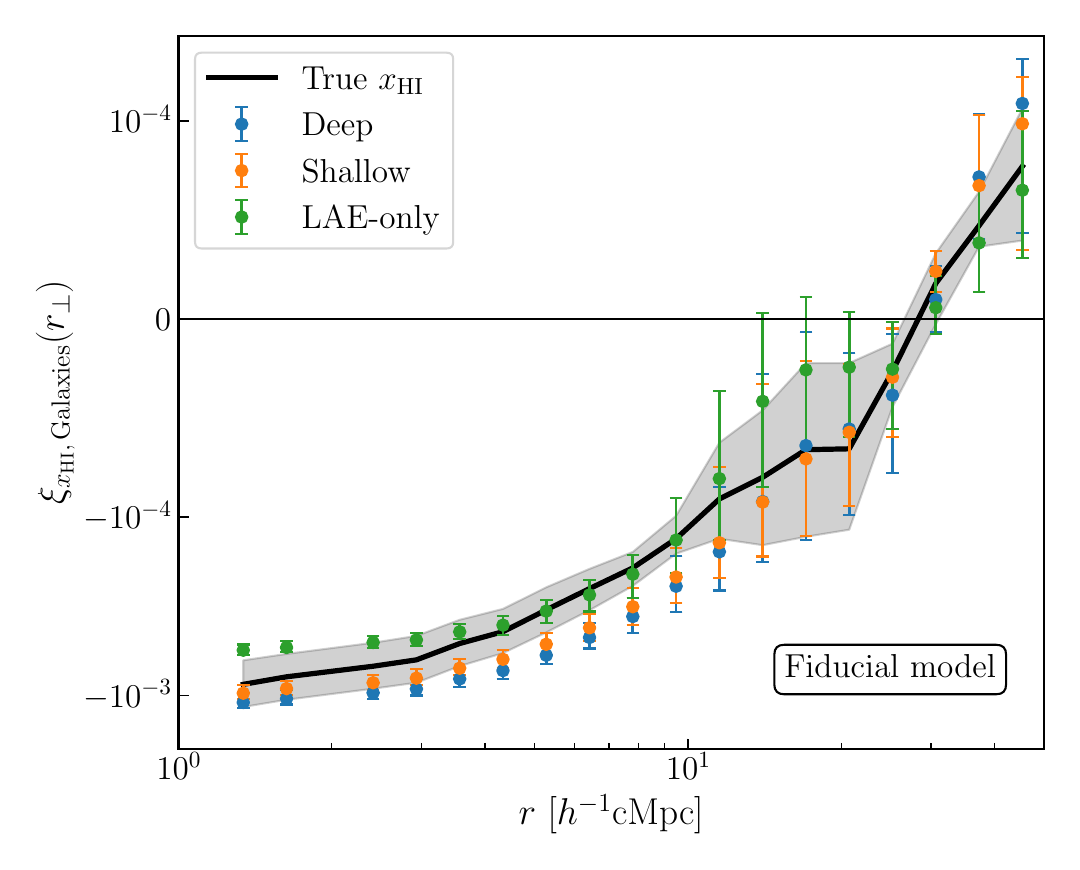}
    \caption{
Cross-correlation between the neutral hydrogen fraction and galaxy tracers,
$\xi_{x_{\mathrm{HI}},\,\mathrm{Galaxies}}(r_\perp)$, measured from the 
projected $2$D fields (over $30~h^{-1}\mathrm{cMpc}$ in the redshift direction) for four reionization scenarios at $z=7.14$:
{Extremely Early}, {Early}, {Oligarchic}, and {Fiducial}.
In each panel, the solid black curve and grey shaded band show the mean and
68\% confidence interval of the true $x_{\mathrm{HI}}$-galaxy
correlation measured across all test subvolumes, where the galaxy field is defined by a fixed population of NLSG galaxies selected by $M_{\mathrm{UV}}<-18$.  The {Fiducial} model is shown as a common reference via the purple shaded band.
Coloured markers denote the reconstructed correlations obtained from three
survey configurations: Deep (blue), Shallow (orange) and LAE-only (green) (see Table~\ref{tab:survey_configs}).  
Error bars indicate the 68\% inter-volume scatter for each survey case.
Across all models, the reconstructions reproduce the scale-dependent rise in
$\xi_{x_{\mathrm{HI}},\,\mathrm{Galaxies}}$ on large transverse separations
($r_\perp \gtrsim 5\,h^{-1}\mathrm{cMpc}$), reflecting the underlying topology
of ionized regions.
The apparent sharp transition from negative to positive values near 
$\xi_{x_{\mathrm{HI}},\,\mathrm{Galaxies}}=0$ is a visual artifact of the symmetric logarithmic scaling and does not correspond to a physical discontinuity in the correlation signal.
Differences between tracer selections highlight the sensitivity of the
cross-correlation to galaxy sampling density and Ly$\alpha$ visibility thresholds.
}

\label{fig:Cross_corr}
\end{figure*}

\subsection{Galaxy-IGM cross-correlations and their dependence on tracer completeness}
\label{subsec:crosscorr_analysis}

In galaxy-dominated reionization models, ionized regions preferentially form around galaxy overdensities. The ionizing sources carve \ion{H}{ii} regions and suppress $x_{\mathrm{HI}}$ in their immediate environments, leading to a scale-dependent anti-correlation between the galaxy field and the neutral hydrogen fraction. We quantify this using the projected cross-correlation function $\xi_{x_{\mathrm{HI}},\,\mathrm{Galaxies}}(r_\perp)$. For each subvolume, we construct a galaxy tracer field on the same grid as the IGM, defined as a binary mask of NLSG galaxies selected by a fixed UV magnitude threshold ($M_{\rm UV}<-18$), and correlate it with  $x_{\mathrm{HI}}(\textbf{r})$ after subtracting the respective means. The cross-correlation is estimated in Fourier space as the inverse transform of the cross-power, and is then projected by averaging along the line-of-sight direction to obtain $\xi(r_\perp)$. Finally, we azimuthally average the projected correlation in 20 logarithmically spaced transverse separation bins over $r_\perp=1$--$50\,h^{-1}\,\mathrm{cMpc}$, and we present the result on a log $r_\perp$ axis with a symmetric-logarithmic ordinate to retain both the anti-correlated small-scale signal and the large-scale approach to zero. The mean and central 68\% confidence intervals are estimated across the test subvolumes, enabling a direct comparison between the true and reconstructed $\xi_{x_{\mathrm{HI}},\,\mathrm{Galaxies}}(r_\perp)$ as a scale-resolved test of whether the reconstruction preserves the causal galaxy-IGM coupling.

Figure~\ref{fig:Cross_corr} presents $\xi_{x_{\mathrm{HI}},,\mathrm{Galaxies}}(r_\perp)$ for the four reionization scenarios, comparing the true cross-correlation (black curve with 68\% inter-volume scatter shown in grey) to the reconstructed correlations (coloured markers with 68\% inter-volume scatter). In all panels, the statistic is evaluated against the same underlying galaxy tracer field; differences between the coloured symbols therefore reflect only the survey configuration used to generate the reconstructed ionization maps. The blue symbols correspond to the Deep survey configuration, the orange symbols to the Shallow configuration, and the green symbols to the LAE-only configuration (see Table~\ref{tab:survey_configs}). For reference, the purple shaded band shows the true Fiducial correlation in each panel. The model-dependent offset of the black curve from this band (most clearly at intermediate separations $r_\perp \sim 10~h^{-1} \mathrm{cMpc}$ , with the precise range varying by model), highlights the scale on which different reionization topologies are distinguishable. The Deep and Shallow LAE+NLSG reconstructions closely track these departures and recover the correct sign and radial dependence across all models, remaining consistent with the true signal within the 68\% confidence interval over most of the measured range.

The dependence on tracer completeness follows a clear and physically intuitive hierarchy. Reconstructions built from the densest joint LAE+NLSG sample (blue) reproduce the amplitude and slope of the small-scale anti-correlation down to the smallest separations with minimal scatter, indicating that the network captures the strong galaxy–ionization coupling. At intermediate depth (orange), the reconstructed signal shows a mild suppression. The LAE-only case (green) exhibits the weakest and noisiest anti-correlation, reflecting the loss of information when NLSGs and sub-threshold LAEs are excluded.

Model-to-model variations are retained as well. The {Extremely Early} and {Early} scenarios, characterised by extended and spatially coherent ionized structures, yield broader anti-correlations, whereas the {Oligarchic} and {Fiducial} models show steeper small-scale trends. For the LAE+NLSG surveys, the reconstructed correlations reproduce the model-dependent separations from the Fiducial reference around $r_\perp\sim 10~h^{-1}\mathrm{cMpc}$, demonstrating discriminatory power on these intermediate scales. 
At large scales ($r_\perp\gtrsim 20~h^{-1},\mathrm{cMpc}$), all reconstructions converge toward the true mean, confirming that the global ionization balance and decorrelation scale between galaxies and the neutral fraction are accurately recovered. 

Overall, the cross-correlation analysis demonstrates that the reconstructions preserve the essential physical coupling between galaxies and the ionized IGM. The recovered $\xi_{x_{\mathrm{HI}},,\mathrm{Galaxies}}(r_\perp)$ reproduces the sign, amplitude, scale dependence, and model-specific variations of the true signal, with the degree of fidelity directly correlated with tracer completeness. The systematic weakening of the small-scale anti-correlation with decreasing survey depth quantitatively encapsulates the impact of observational limitations on recovering ionized-bubble contrast, while the agreement of Deep/Shallow reconstructions with the model-dependent offsets relative to the Fiducial reference indicates that joint LAE+NLSG surveys can distinguish reionization scenarios on $\sim 10 h^{-1}\mathrm{cMpc}$ scales.

This behaviour therefore appears to reflect the information content of the tracers rather than a fundamental limitation of the reconstruction method. If future surveys were to access a broader or deeper set of tracers, they might encode additional aspects of the small-scale ionization structure, potentially allowing the reconstructed fields to recover finer details of the neutral IGM than is currently possible.

\begin{figure*}

\centering
\includegraphics[width=0.55cm, trim={3.5cm 0.5cm 35.71cm 1.2cm}, clip]{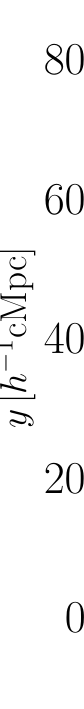}%
\begin{overpic}[width=8.15cm, trim={15cm 0.5cm 5.2cm 1.2cm}, clip]{Plots/Fiducial_L41_MUV_18_grid_019.pdf}
  \put(74,56.3){\makebox(0,0)[c]{\large\bfseries Deep Survey}}
\end{overpic}%
\begin{overpic}[width=4.05cm, trim={25.25cm 0.5cm 5.2cm 1.2cm}, clip]{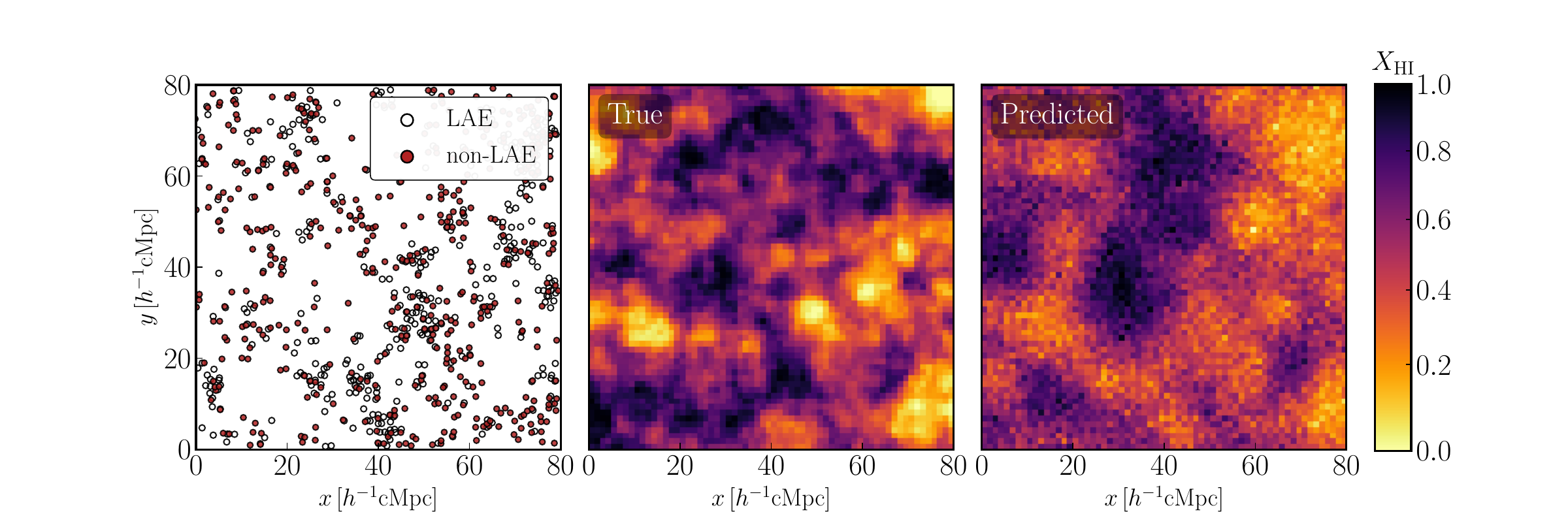}
  \put(41,96){\makebox(0,0)[c]{\large\bfseries Shallow Survey}}
\end{overpic}%
\begin{overpic}[width=4.95cm, trim={25.25cm 0.5cm 3cm 1.2cm}, clip]{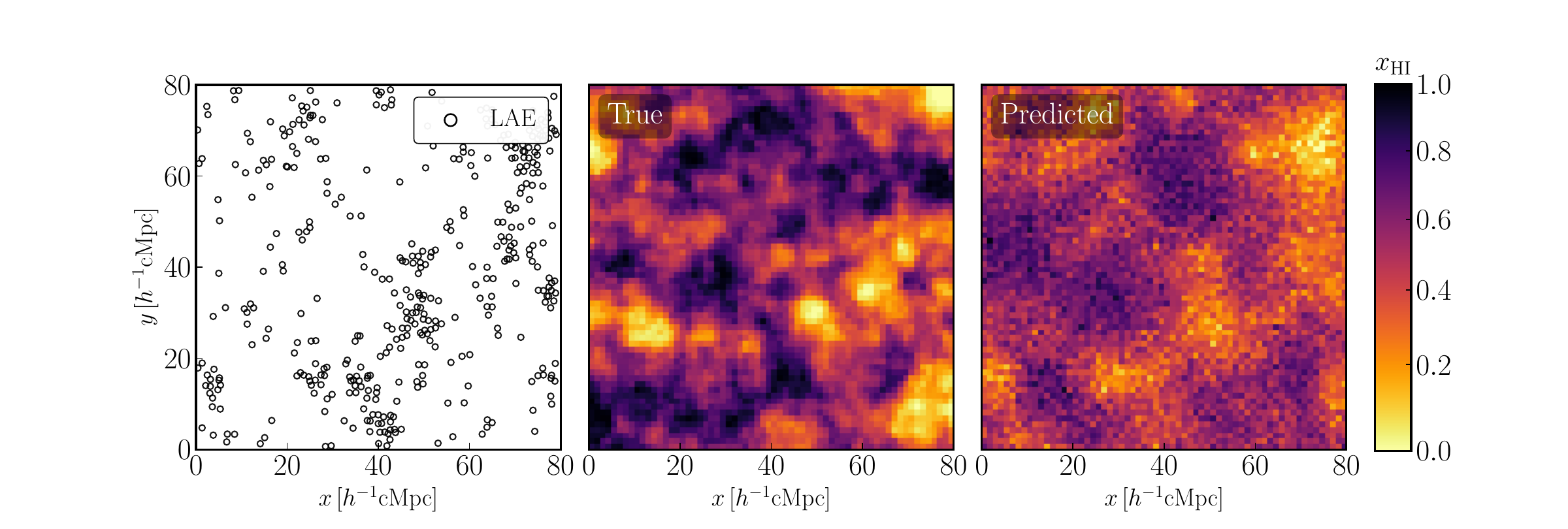}
  \put(40,93){\makebox(0,0)[c]{\large\bfseries LAE-only Survey}}
\end{overpic}%

    \includegraphics[width=0.45\linewidth, trim=15 10 10 0, clip]{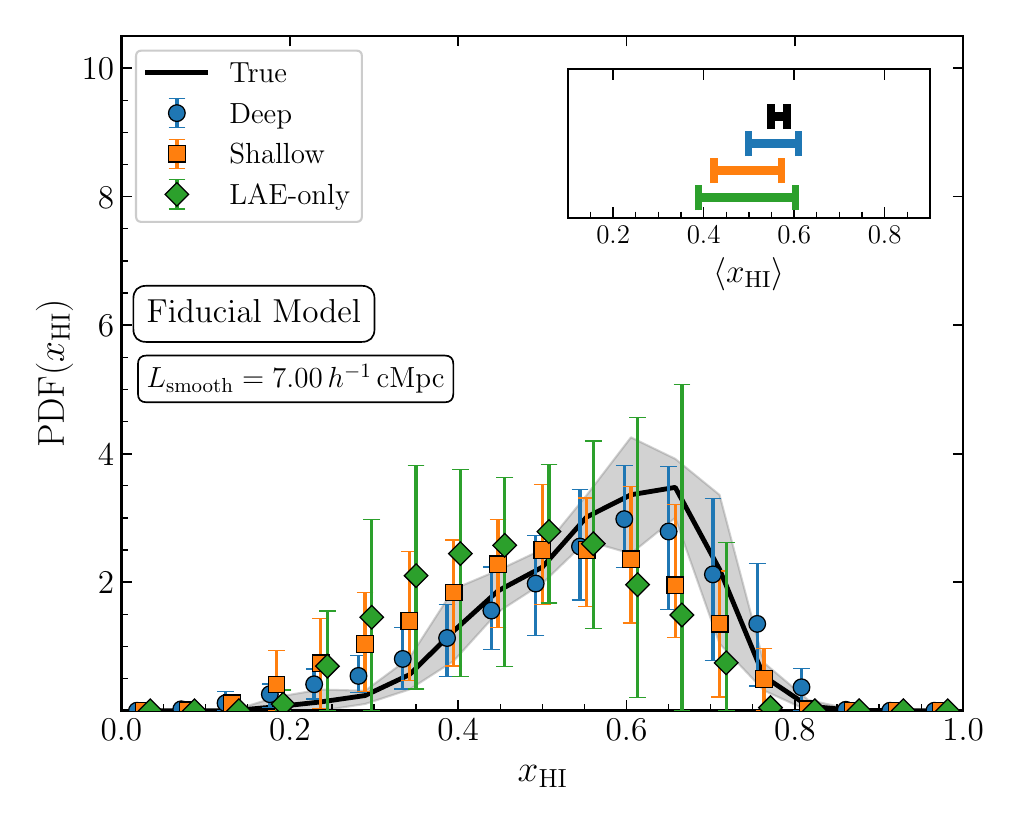}%
\includegraphics[width=0.46\linewidth, trim=15 18 10 0, clip]{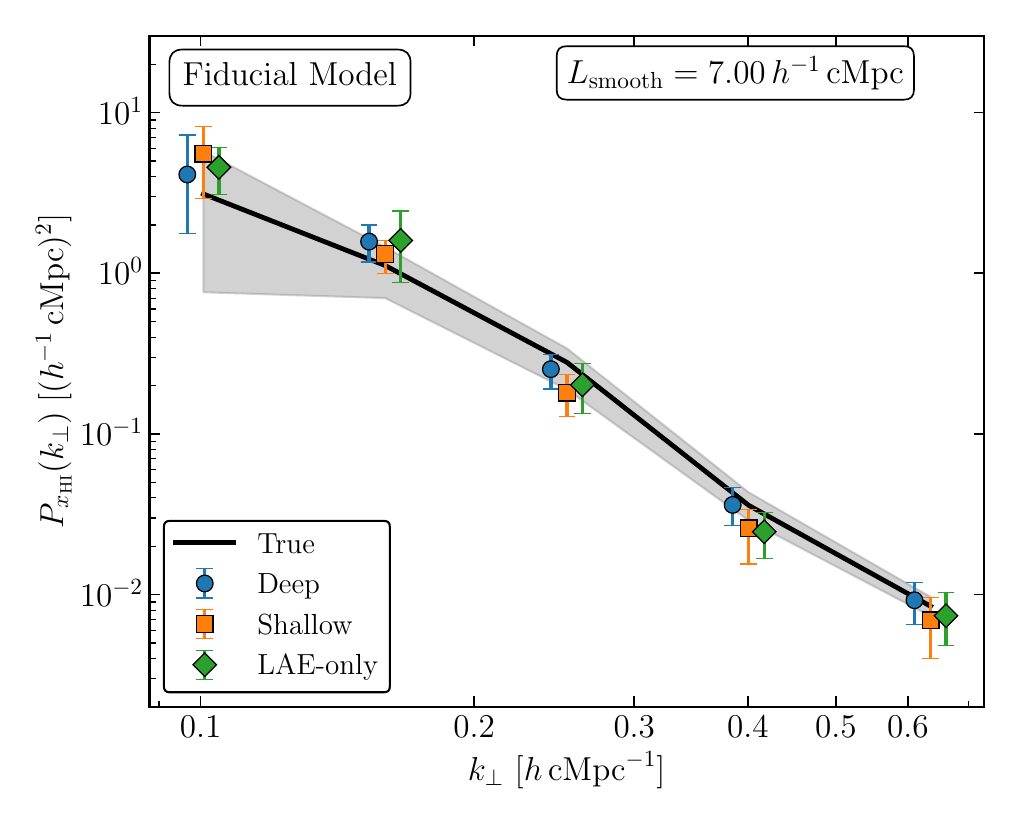}%
\caption{
Robustness of the tomographic reconstruction when applying the neural network trained at $z=7.14$ to mock observations of the {Fiducial Reionization model} at $z=6.6$. 
\textit{Top panels}: True and reconstructed neutral-hydrogen fraction fields, $x_{\mathrm{HI}}$, projected over $30~h^{-1}\mathrm{cMpc}$ in the redshift direction) for three representative survey selections: Deep (blue), Shallow (orange) and LAE-only (green) (see Table~\ref{tab:survey_configs}).  
\textit{Bottom-left}: Probability distribution function (PDF) of the projected $x_{\mathrm{HI}}$ fields at $z=6.6$, comparing the true distribution (black curve; shaded region shows inter-volume scatter) with reconstructions for the three survey selections (colored markers; 68\% uncertainties). The inset panel summarises the mean neutral fraction $\langle x_{\mathrm{HI}}\rangle$ for each case, shown as horizontal error bars representing the 68\% confidence intervals. \textit{Bottom-right}: Dimensionless 2D power spectrum, $P_{2\mathrm{D}}(k_\perp)$, of the true and reconstructed fields. Shaded bands show the field-to-field variance of the true statistics at $z=6.6$. }
\label{fig:tomo_z=6.6}
\end{figure*}

\begin{figure*}
\centering

% ==================== EARLY MODEL (NO SMOOTHING) ====================
\begin{tcolorbox}[title={\textbf{Early Model: Deep Survey}}]
\includegraphics[width=12.5cm, trim={3.5cm 0.5cm 3cm 1cm}, clip]{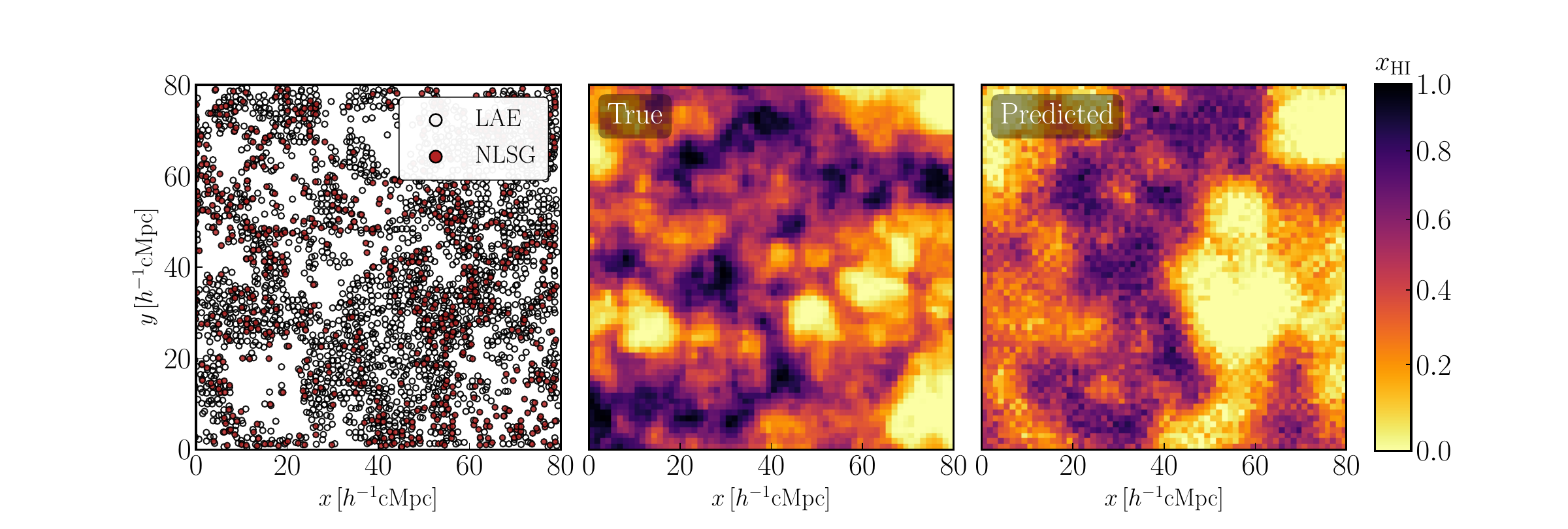}%
\hspace{0.15cm}%
\includegraphics[width=4.4cm, trim={0cm 0.7cm 0.67cm 0.2cm}, clip]{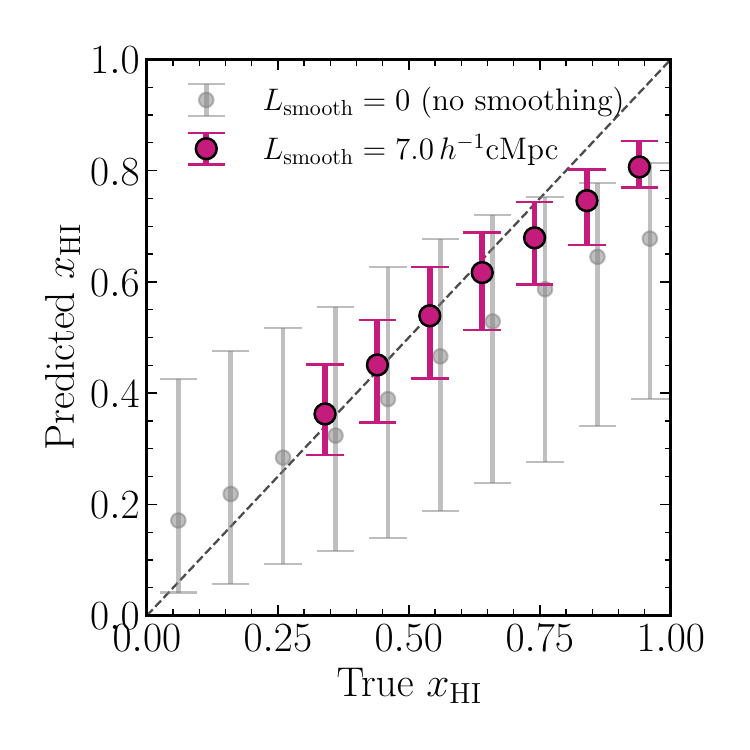}
\end{tcolorbox}

\begin{tcolorbox}[title={\textbf{Early Model: Shallow Survey}}]
\includegraphics[width=12.5cm, trim={3.5cm 0.5cm 3cm 1cm}, clip]{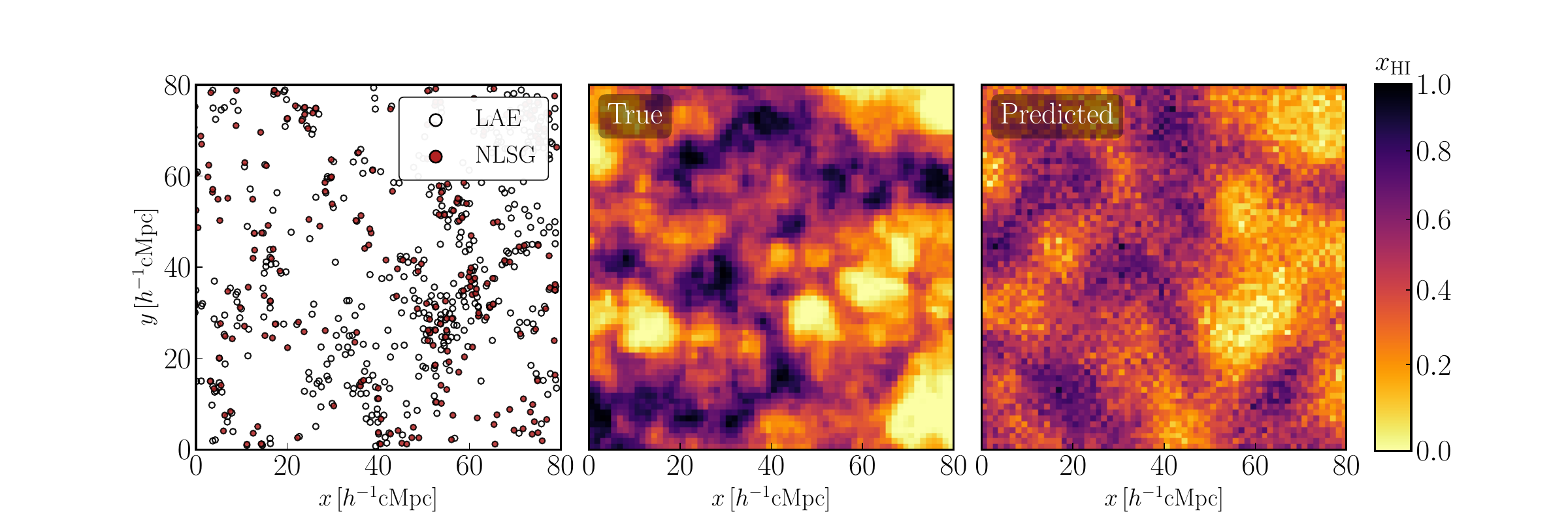}%
\hspace{0.15cm}%
\includegraphics[width=4.4cm, trim={0cm 0.7cm 0.67cm 0.2cm}, clip]{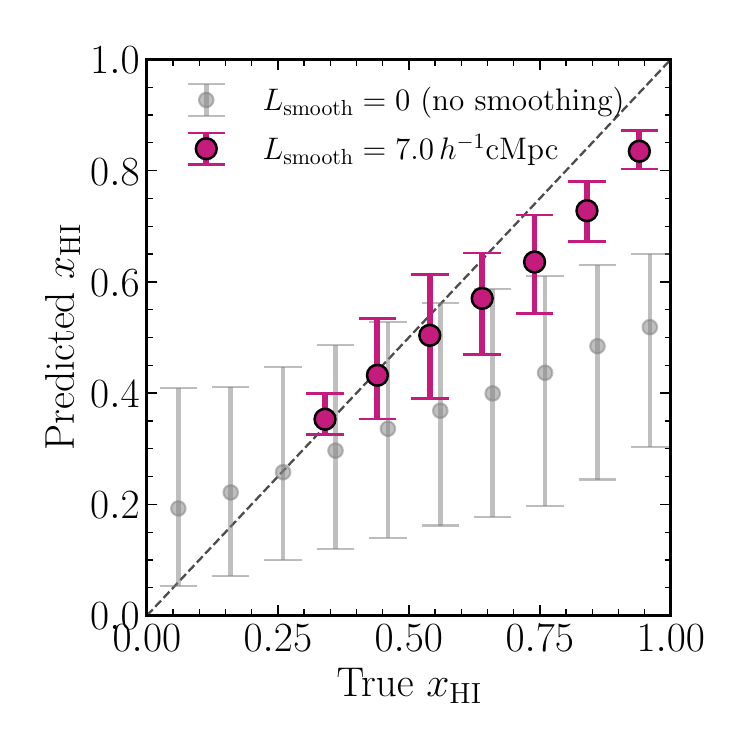}
\end{tcolorbox}

\begin{tcolorbox}[title={\textbf{Early Model: LAE-only Survey} }]
\includegraphics[width=12.5cm, trim={3.5cm 0.5cm 3cm 1cm}, clip]{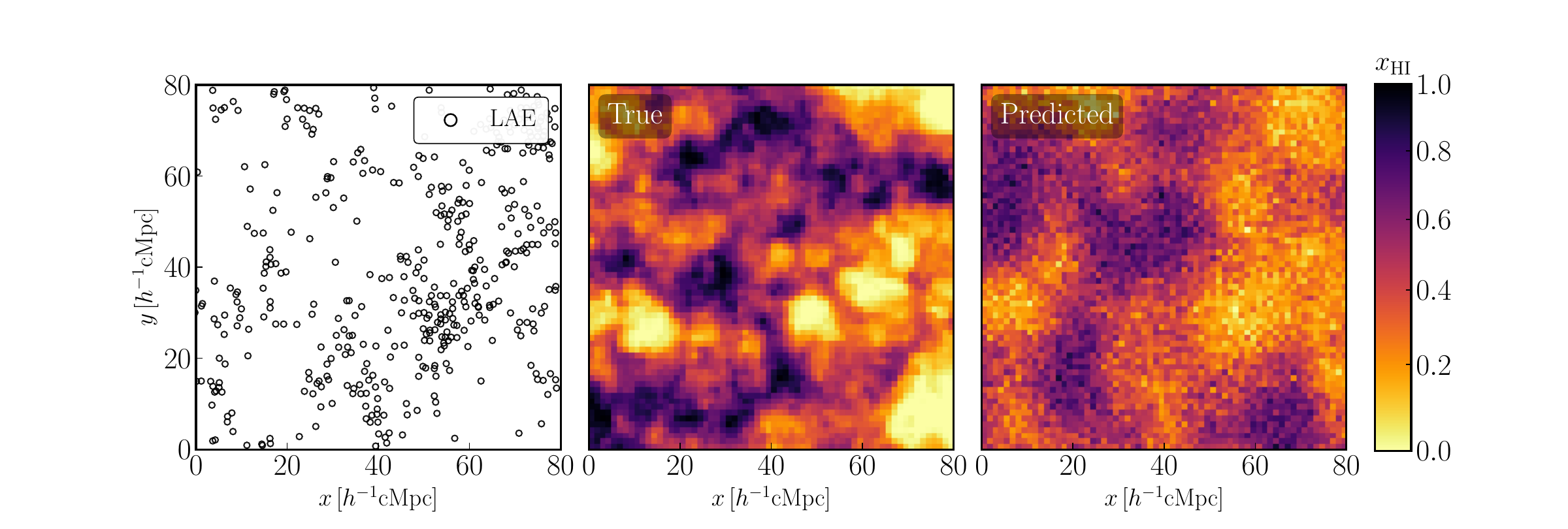}%
\hspace{0.15cm}%
\includegraphics[width=4.4cm, trim={0cm 0.7cm 0.67cm 0.2cm}, clip]{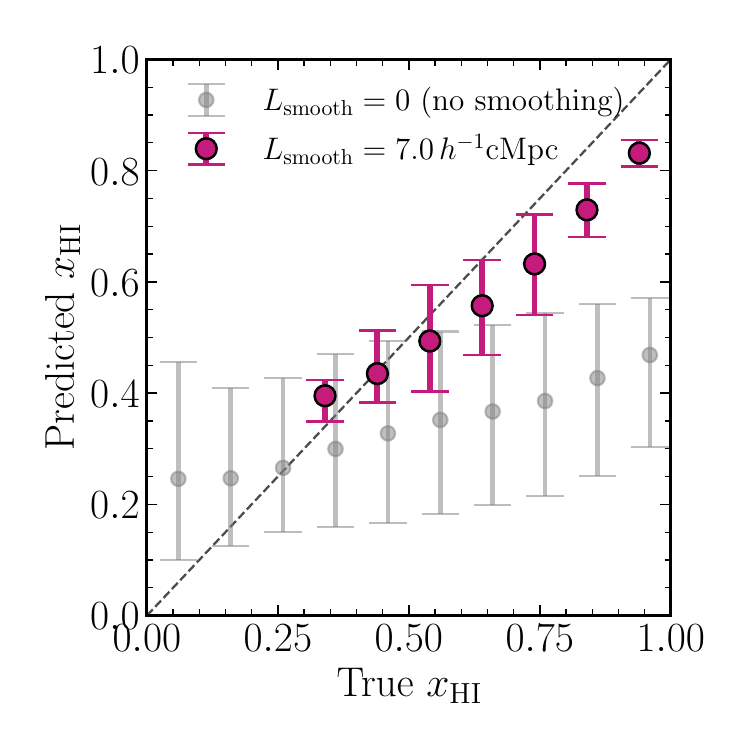}
\end{tcolorbox}

\caption{
Same as Figure \ref{fig:Early_NoSmooth}, but with the line-of-sight positions of the LAEs and NLSGs randomly reassigned within the 30 $h^{-1}$ cMpc thickness of the slice along the $z$-direction to simulate redshift uncertainty along the slice.
}
\label{fig:Early_NoSmooth_random_z}
\end{figure*}

\begin{figure*}
\centering

% ==================== OLIGARCHIC MODEL (NO SMOOTHING) ====================
\begin{tcolorbox}[title={\textbf{Oligarchic Model: Deep Survey} }]
\includegraphics[width=12.5cm, trim={3.5cm 0.5cm 3cm 1cm}, clip]{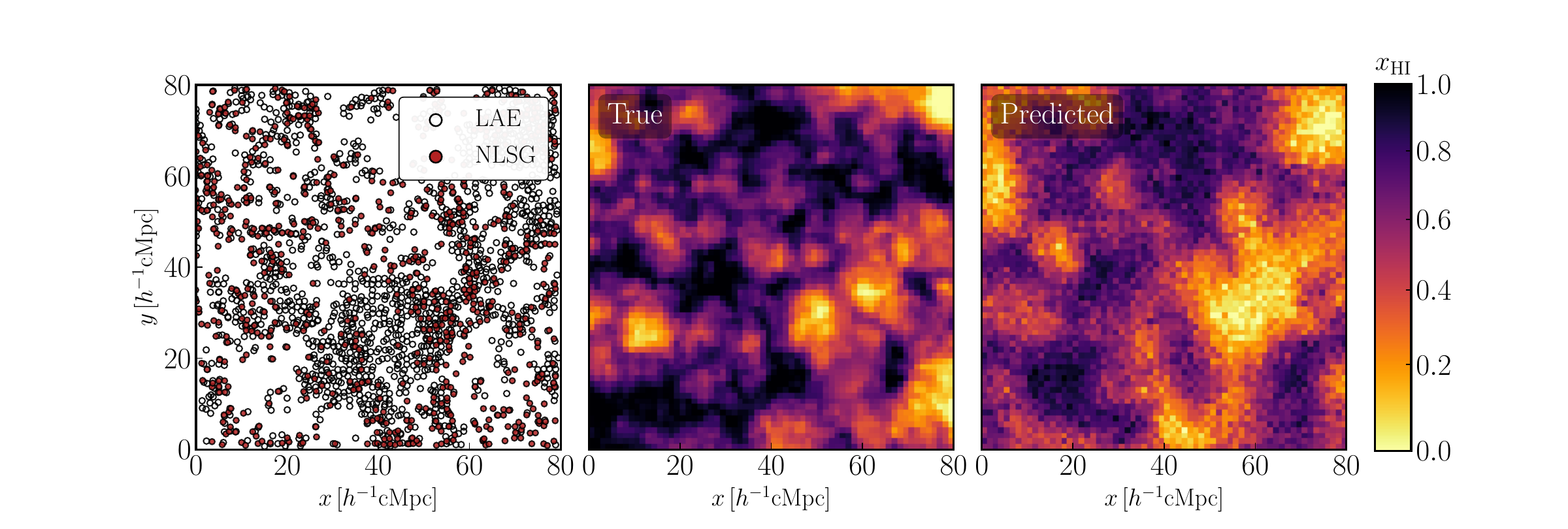}%
\hspace{0.15cm}%
\includegraphics[width=4.4cm, trim={0cm 0.7cm 0.67cm 0.2cm}, clip]{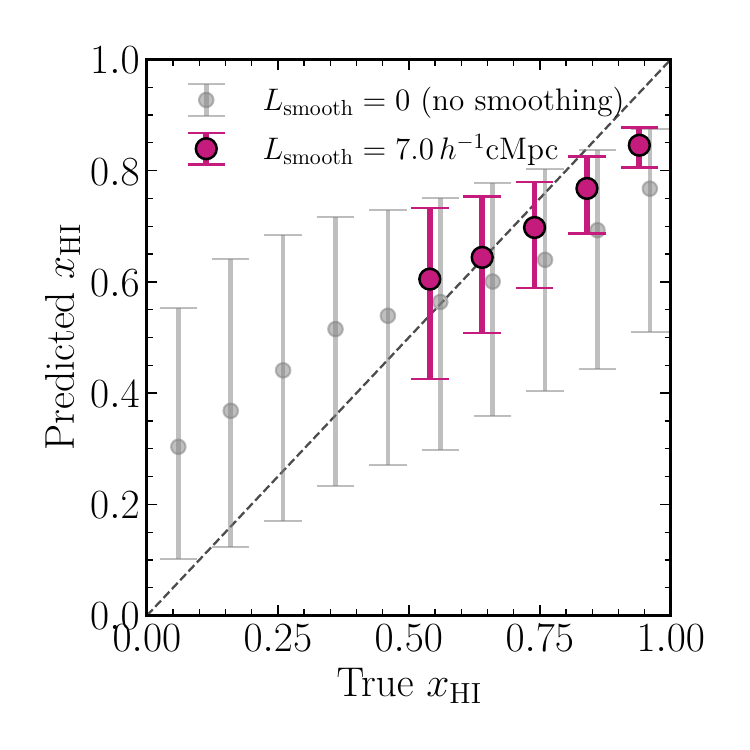}
\end{tcolorbox}

\begin{tcolorbox}[title={\textbf{Oligarchic Model: Shallow Survey} }]
\includegraphics[width=12.5cm, trim={3.5cm 0.5cm 3cm 1cm}, clip]{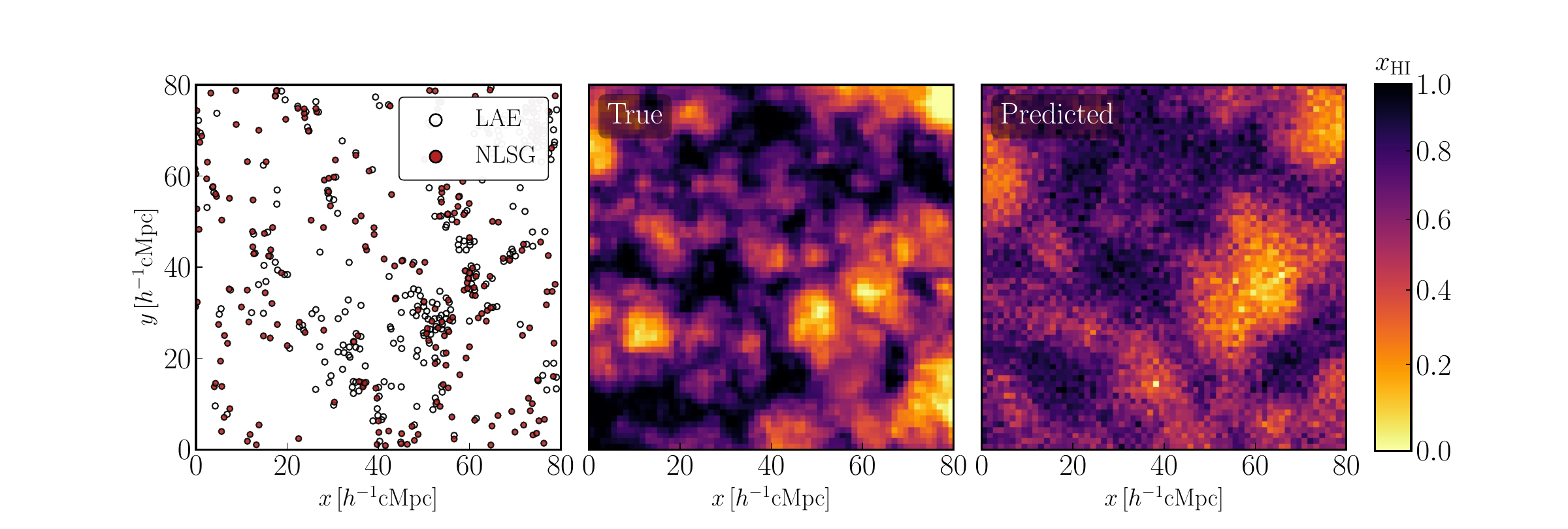}%
\hspace{0.15cm}%
\includegraphics[width=4.4cm, trim={0cm 0.7cm 0.67cm 0.2cm}, clip]{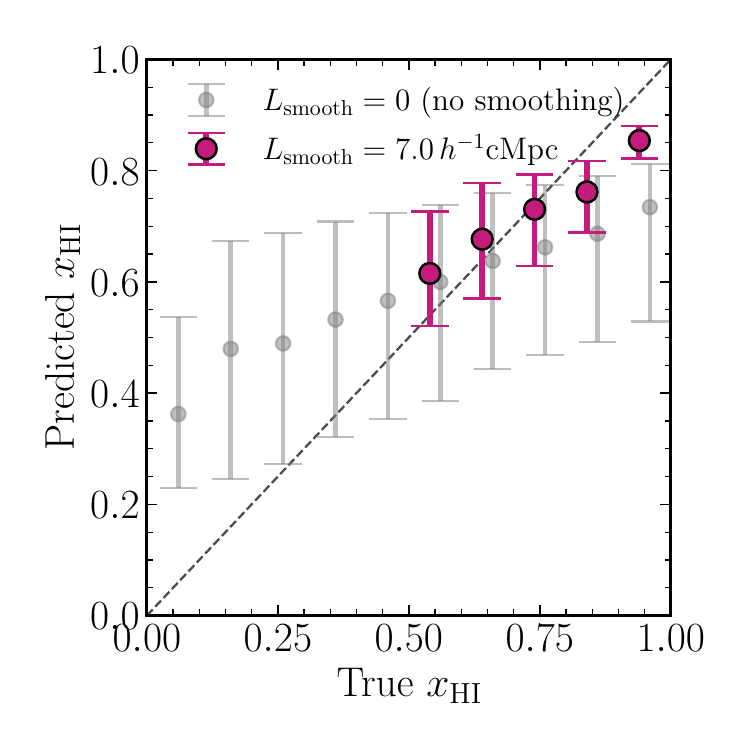}
\end{tcolorbox}

\begin{tcolorbox}[title={\textbf{Oligarchic Model: LAE-only Survey}}]
\includegraphics[width=12.5cm, trim={3.5cm 0.5cm 3cm 1cm}, clip]{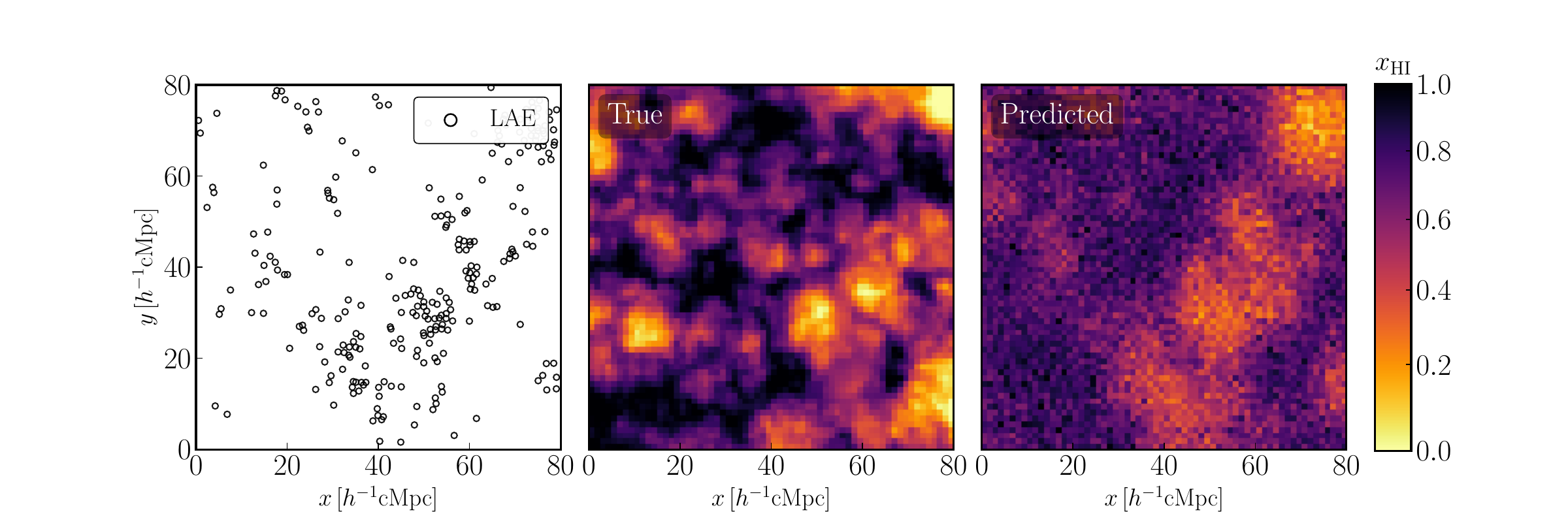}%
\hspace{0.15cm}%
\includegraphics[width=4.4cm, trim={0cm 0.7cm 0.67cm 0.2cm}, clip]{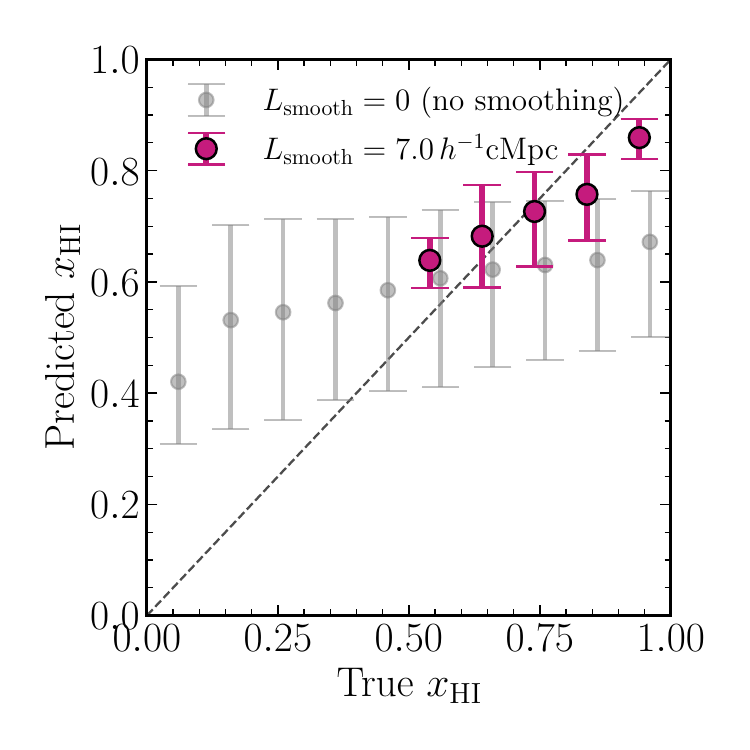}
\end{tcolorbox}

\caption{
Same as Figure \ref{fig:Oligarchic_NoSmooth}, but with the line-of-sight positions of the LAEs and NLSGs randomly reassigned within the 30 $h^{-1}$ cMpc thickness of the slice along the $z$-direction to simulate redshift uncertainty along the slice.
}
\label{fig:Oligarchic_NoSmooth_random_z}
\end{figure*}

\begin{figure*}
    \centering
    \includegraphics[width=0.4\linewidth, trim=15 50 10 15, clip]{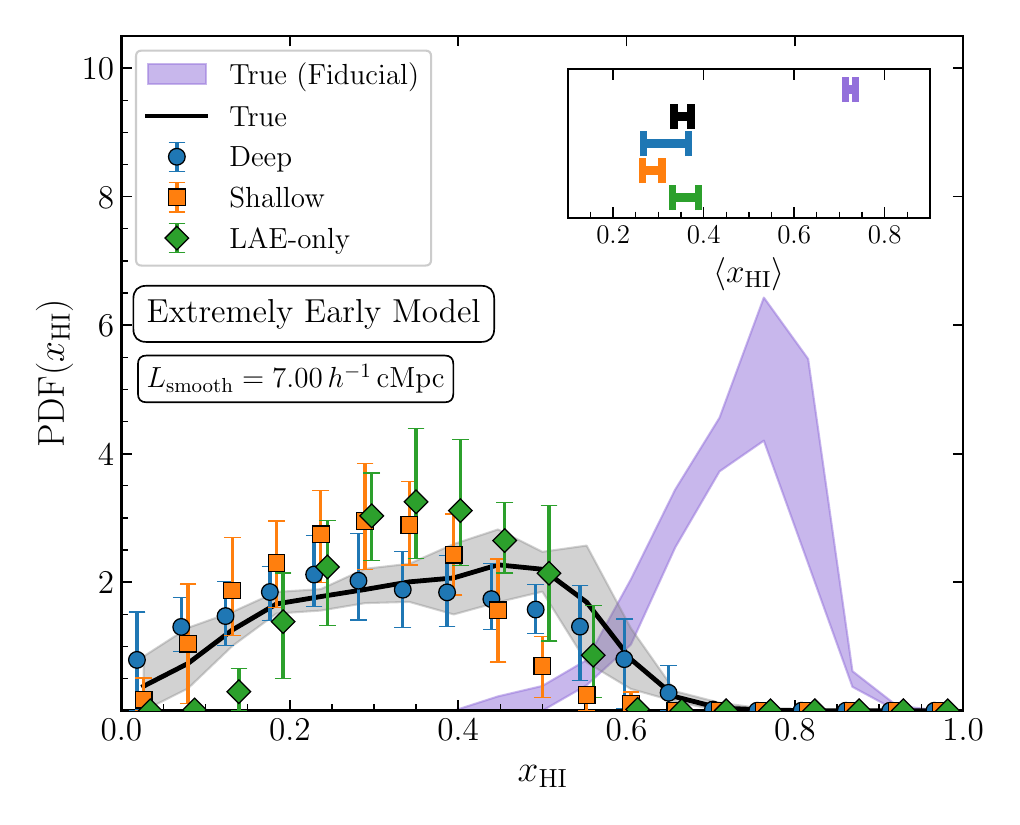}%
    \includegraphics[width=0.365\linewidth, trim=55 50 10 15, clip]{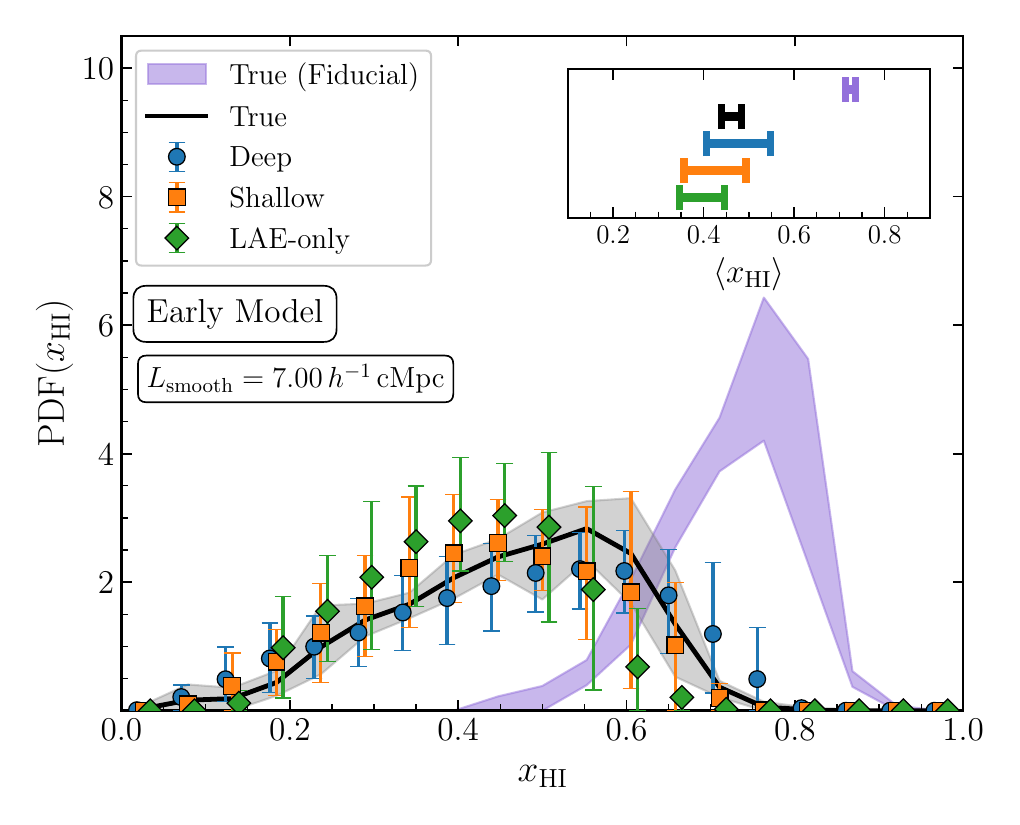}
    
    \includegraphics[width=0.4\linewidth, trim=15 10 10 10, clip]{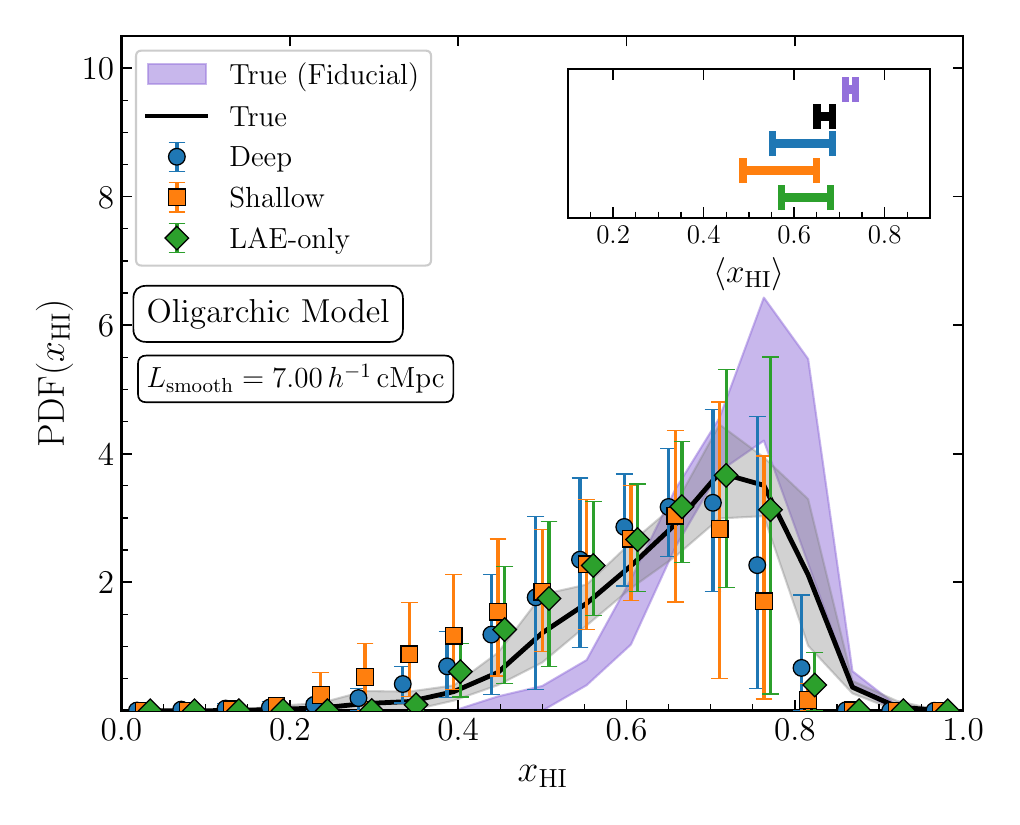}%
    \includegraphics[width=0.365\linewidth, trim=55 10 10 10, clip]{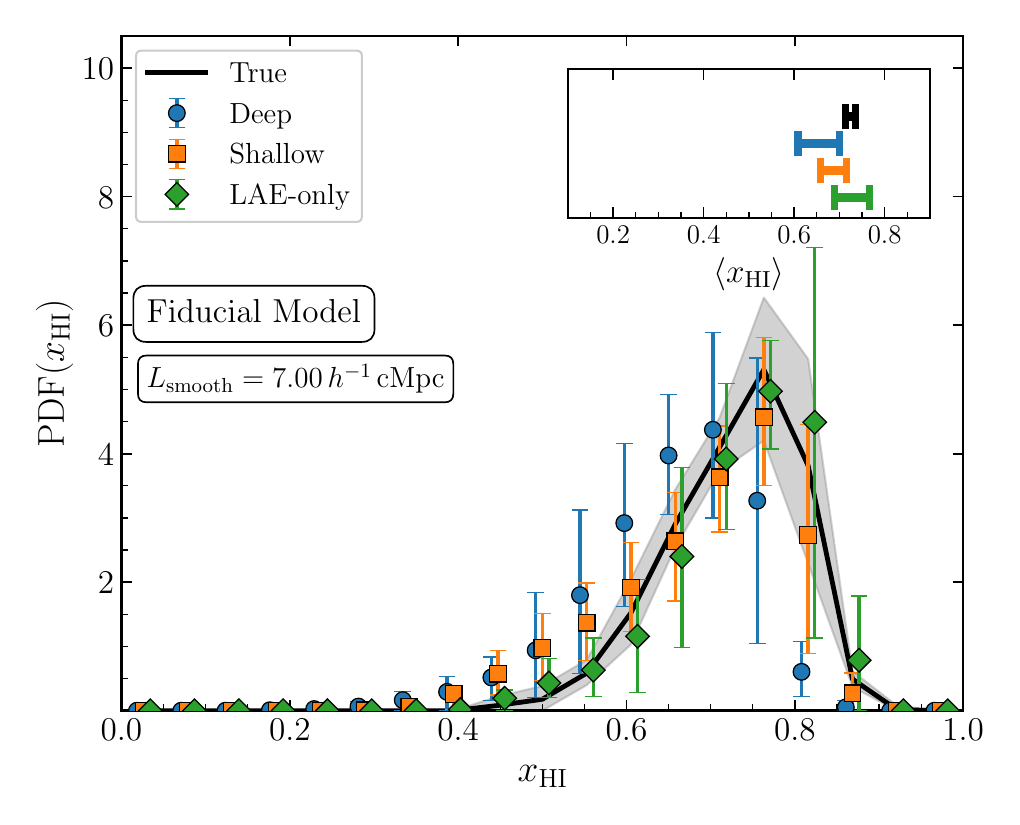}
    
    \caption{
Same as Figure \ref{fig:PDF}, but with the line-of-sight positions of the LAEs and NLSGs randomly reassigned within the 30 $h^{-1}$ cMpc thickness of the slice along the $z$-direction to simulate redshift uncertainty along the slice.
}
\label{fig:PDF_random_z}
\end{figure*}

\begin{figure*}
    \centering
    \includegraphics[width=0.4\linewidth, trim=15 55 10 15, clip]{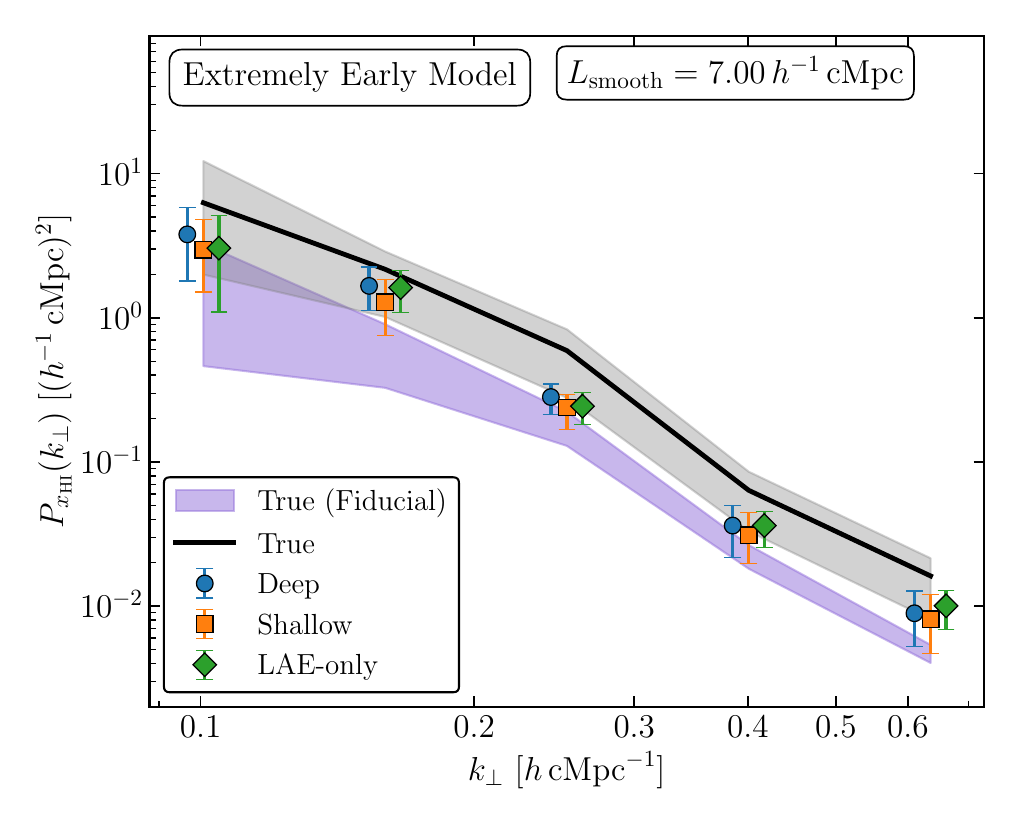}%
    \includegraphics[width=0.353\linewidth, trim=70 55 10 15, clip]{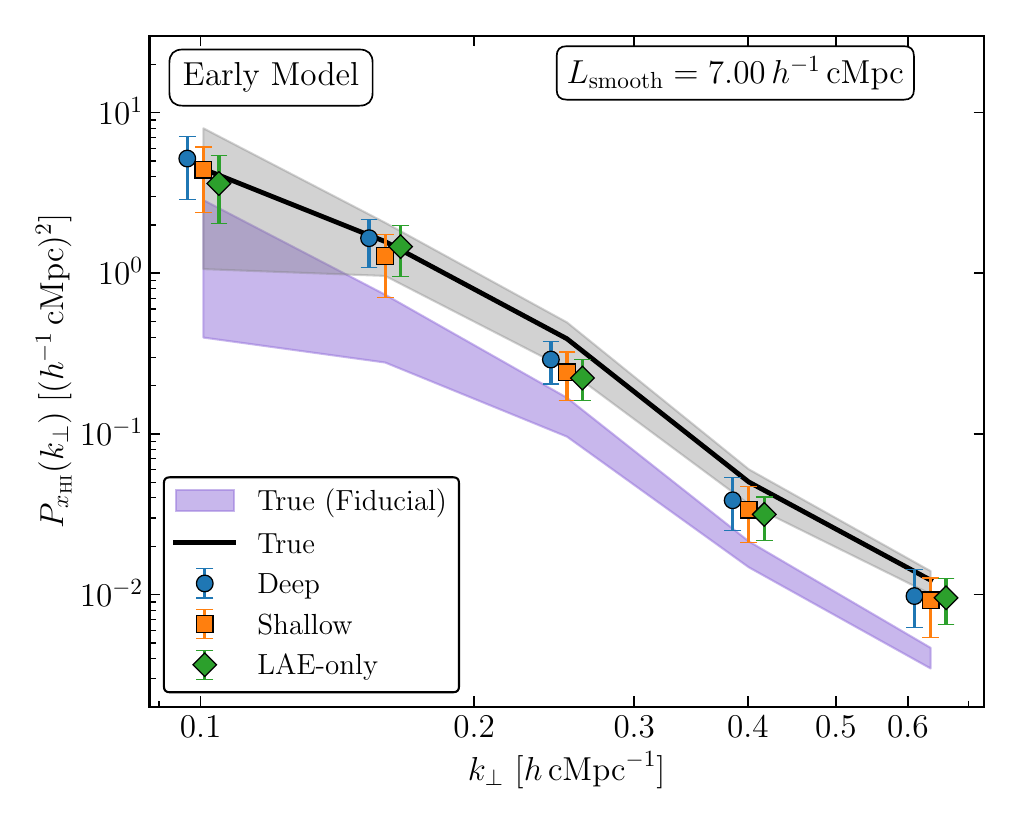}
    
    \includegraphics[width=0.4\linewidth, trim=15 10 10 10, clip]{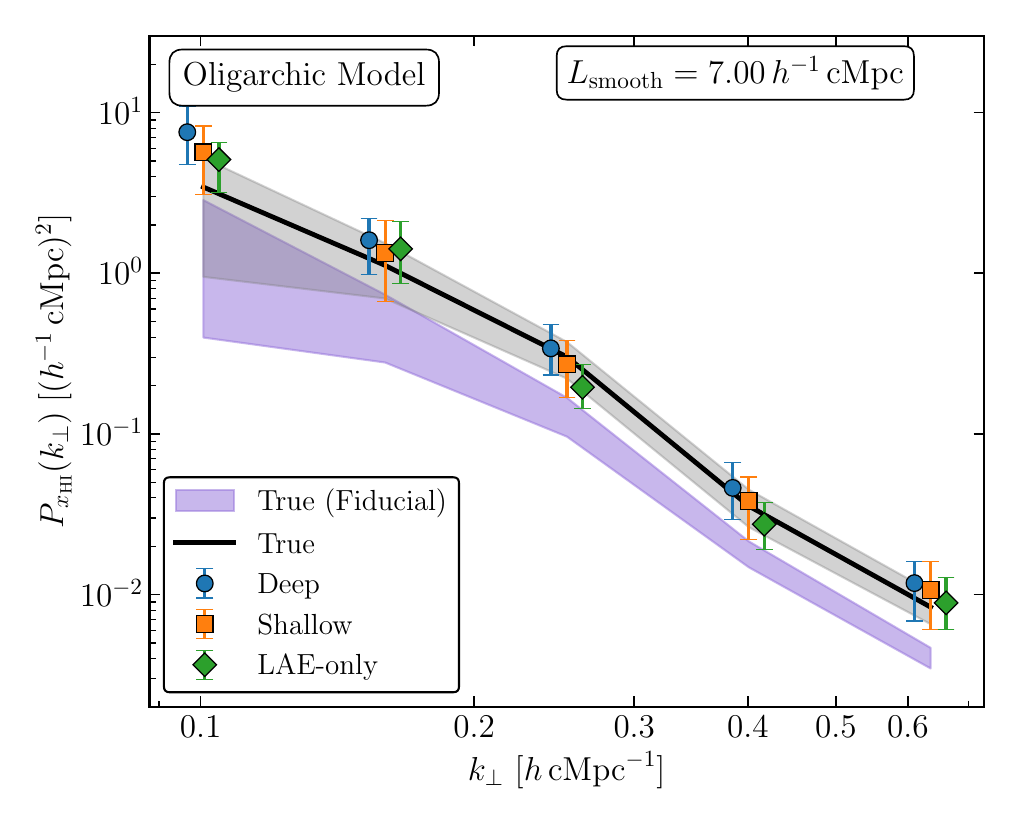}%
    \includegraphics[width=0.353\linewidth, trim=70 10 10 10, clip]{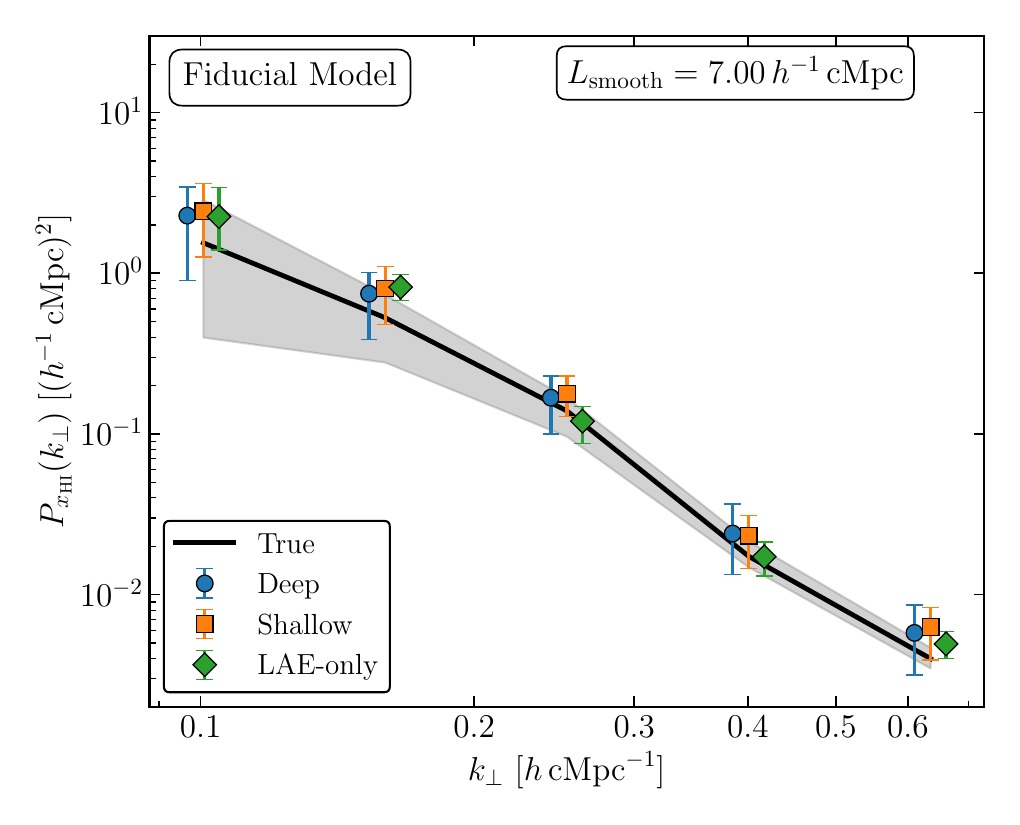}
    \caption{
Same as Figure \ref{fig:PS}, but with the line-of-sight positions of the LAEs and NLSGs randomly reassigned within the 30 $h^{-1}$ cMpc thickness of the slice along the $z$-direction to simulate redshift uncertainty along the slice.
}

    \label{fig:PS_random_z}
\end{figure*}

\section{Robustness checks}
\label{sec:robustness}

In practice, any tomographic reconstruction framework intended for upcoming LAE surveys must remain reliable under a variety of observational and modelling uncertainties. Unlike idealised simulations, where subvolumes are extracted at a single snapshot and tracer positions are specified without error, real surveys probe galaxy populations whose properties and associated ionization conditions evolve, and are affected by additional uncertainties arising from narrow-band selection functions and measurement noise. It is therefore essential to verify that our reconstruction network does not depend sensitively on these idealised assumptions or on the specific ionization conditions present in the training data.

We perform two complementary robustness tests. First, we assess whether a model trained at a single simulation snapshot (here, $z=7.14$) can be successfully applied to tomographic data drawn from a different stage of the same reionization history without retraining. In this test, the change in redshift serves as a controlled way to realise an evolved ionization state, allowing us to examine whether the network generalises beyond the specific ionization conditions encountered during training. Second, we investigate the impact of uncertainties in the line-of-sight positions of the galaxy tracers themselves. Although our fiducial configuration assumes exact spectroscopic redshifts, real LAE and NLSG selections span a finite range in wavelength. To emulate this in a controlled manner, we introduce line-of-sight uncertainty consistent with a narrow-band selection and examine how these propagate through the reconstruction pipeline. Together, these tests establish the degree to which our framework is robust to both evolving ionization conditions and the dominant observational uncertainties expected in real high-redshift surveys.

\subsection{Generalisation across different ionization conditions}
\label{subsec:redshift_robustness}

In earlier sections, the reconstruction network was trained on, and
evaluated against, tomographic data drawn from the same simulation
snapshot, such that the ionization conditions encountered during
testing were statistically identical to those seen during training.
While this provides a controlled baseline for assessing reconstruction
fidelity, a more stringent test is whether the network can generalise
to ionization fields that differ from the specific conditions present
in the training data.

To address this, we perform a targeted robustness experiment using the
Fiducial reionization model, in which the network is trained exclusively
on subvolumes extracted at a single redshift, $z = 7.14$, and is then
applied, without any retraining or fine-tuning, to mock survey
realizations drawn from a later snapshot at $z = 6.6$. By this epoch,
the Fiducial reionization history has evolved to a more advanced
ionization state, with a different neutral-fraction
field. This setup therefore directly tests whether the network can
successfully reconstruct ionization morphologies that were not present
in the training set, rather than merely reproducing the conditions it
was trained on. The resulting reconstructions and associated statistics
are shown in Fig.~\ref{fig:tomo_z=6.6}.

The \textit{top panels} of Fig.~\ref{fig:tomo_z=6.6} compare the true
projected neutral-hydrogen fraction field, $x_{\mathrm{HI}}$, at
$z = 6.6$ with the corresponding reconstructions obtained from the
three survey selections considered throughout this work:
Deep, Shallow and the LAE-only case. Despite the fact that
the network was trained on an earlier stage of reionization, the
reconstructed maps recover the locations of the main ionized regions
and neutral islands, as well as the overall bubble topology, in close
analogy to the results obtained at the training snapshot. As before,
the joint LAE+NLSG survey configurations outperform the LAE-only survey
in recovering the large-scale morphology of the ionization field.

To quantify this behaviour, the \textit{bottom-left} panel shows the
probability distribution function (PDF) of the projected
$x_{\mathrm{HI}}$ field at $z = 6.6$, computed after smoothing with a
Gaussian kernel of $L_{\mathrm{smooth}} = 7\,h^{-1}\,\mathrm{cMpc}$.
The true distribution (black curve with grey $1\sigma$ band) is well
recovered by all three survey selections, with the reconstructed PDFs
closely matching both the peak and width of the true distribution. The
mean neutral fraction $\langle x_{\mathrm{HI}} \rangle$ is reproduced
to within a few per cent for all tracer combinations, indicating that
the network does not rely sensitively on the specific ionization level
present during training.

The \textit{bottom-right} panel presents the dimensionless
two-dimensional power spectrum, $P_{2\mathrm{D}}(k_\perp)$, of the true
and reconstructed $x_{\mathrm{HI}}$ fields at
$L_{\mathrm{smooth}} = 7\,h^{-1}\,\mathrm{cMpc}$. Across the range of
modes probed by our subvolumes, the reconstructions reproduce both the
shape and amplitude of the true power spectrum within the expected
field-to-field scatter, with only mild deviations for the LAE-only
sample on the largest scales.

Taken together, these results demonstrate that the reconstruction
network generalises beyond the specific ionization conditions on which
it was trained. In particular, a model trained at a single snapshot can
successfully recover the ionization morphology at a later stage of the
same reionization history, indicating that it has learned a physically
meaningful mapping between the large-scale LAE/NLSG distribution and
the underlying ionization field, rather than memorising a particular
training configuration.

\subsection{Redshift uncertainty of tracers}
\label{subsec:tracer_redshift_uncertainty}

A second source of observational uncertainty arises from the limited redshift determination of LAEs and NLSGs associated with wide-field high-redshift surveys. In our setup, each tomographic subvolume spans $80 h^{-1}\mathrm{cMpc}\times80 h^{-1}\mathrm{cMpc}$ in the transverse directions and a line-of-sight thickness of $30 h^{-1}\mathrm{cMpc}$, chosen to approximately match the comoving radial extent probed by the narrow-band filters of Subaru/Hyper Suprime-Cam LAE programmes such as SILVERRUSH at $z\sim7$. While the redshift of an individual LAE within such a slice is uncertain at the level of the filter width, NLSGs—typically identified through continuum selection or heterogeneous spectroscopic follow-up are expected to exhibit substantially larger redshift uncertainties. Explicitly modeling these larger, tracer-dependent uncertainties is not currently feasible within our finite simulation volume, as the corresponding line-of-sight smearing would exceed the subvolume depth. We therefore adopt a simplified, forward-looking treatment in which NLSGs are assigned a redshift uncertainty comparable to that expected from future JWST narrow-band imaging of rest-frame optical emission lines at EoR redshifts. Although this approximation does not capture the full complexity of NLSG redshift errors in current ground-based surveys, it provides a controlled and physically motivated baseline for assessing the impact of line-of-sight uncertainty on tomographic reconstruction.

To mimic this effect in a conservative manner, we construct realizations in which the angular positions and sample selection of the tracers are kept fixed, but their line-of-sight coordinates within each subvolume are randomised. Specifically, for every LAE and NLSG we draw a new line-of-sight position from a uniform distribution across the $30\,h^{-1}\mathrm{cMpc}$ thickness of the slab, while leaving all other galaxy properties unchanged. This procedure preserves the projected number densities and clustering of the tracers on the sky, but erases any line-of-sight tracer information relevant for reconstructing the structure of the ionization field within the slab. We emphasize that this approach is not intended to reproduce realistic tracer redshift uncertainties in detail, but rather to isolate and quantify the impact of a loss of positional information along the redshift direction on the tomographic reconstruction. We then feed these randomised tracer distributions through the trained network (without any retraining) and compare the resulting reconstructions to the true ionization maps.

Figs.~\ref{fig:Early_NoSmooth_random_z} and \ref{fig:Oligarchic_NoSmooth_random_z} illustrate the impact of this redshift randomization on
individual slices for the Early and Oligarchic reionization scenarios.
As in the fiducial case, we show the tracer distribution, the true
projected $x_{\mathrm{HI}}$ field, the corresponding reconstruction, and
the predicted-versus-true $x_{\mathrm{HI}}$ relation for the three survey
selections. Qualitatively, the reconstructed maps continue to recover
the major ionized bubbles and neutral islands, with bubble sizes and
overall topology in good agreement with the truth. The degradation
introduced by the line-of-sight uncertainty mainly manifests as extra
small-scale noise in the unsmoothed maps and as a slight increase in
scatter about the one-to-one relation in the cell-by-cell
$x_{\mathrm{HI}}$ comparison. As in our baseline results, the joint
LAE+NLSG samples perform noticeably better than the LAE-only case in
tracing the large-scale morphology of the ionization field.

The statistical impact of tracer redshift uncertainty is quantified in
Figs.~\ref{fig:PDF_random_z} and \ref{fig:PS_random_z}, which present the PDFs and 2D power spectra of the
projected $x_{\mathrm{HI}}$ fields (smoothed with
$L_{\mathrm{smooth}} = 7\,h^{-1}\mathrm{cMpc}$) for all four reionization
models. The PDFs of the reconstructed fields remain in excellent
agreement with the true distributions: for every model and survey
selection, the peak position and overall width of the PDF are
recovered within the quoted $1\sigma$ inter-volume scatter. The mean
neutral fraction $\langle x_{\mathrm{HI}}\rangle$ is essentially unchanged
relative to the no-scrambling case and shows no systematic bias
induced by the redshift uncertainty. Differences between the three
survey selections are larger than the changes induced by scrambling,
indicating that survey depth and tracer abundance remain the dominant
drivers of reconstruction quality.

A similar picture emerges from the power spectra. As shown in
Fig.~\ref{fig:PS_random_z}, the reconstructed $P_{2\mathrm{D}}(k_\perp)$ curves follow both
the shape and normalisation of the true power spectra over the range
of modes probed by our subvolumes, with deviations that are well
within the field-to-field variance. At most, the LAE-only
configuration shows a mild suppression or enhancement of power on the
largest scales, but the effect is comparable to that already present
in the idealised (no-uncertainty) case and does not alter the overall
slope of $P_{2\mathrm{D}}(k_\perp)$.

Overall, this exercise demonstrates that our tomographic framework
is robust to realistic levels of tracer redshift uncertainty expected
in Subaru/Hyper Suprime-Cam narrow-band surveys. Even when the
line-of-sight positions of all galaxies within each
$30 \,h^{-1}\mathrm{cMpc}$ slice are fully scrambled, the reconstructions
retain their ability to recover the large-scale ionization morphology
and associated one-point and two-point statistics. This suggests that
the network relies primarily on the transverse clustering pattern and
global abundance of tracers, rather than on precise radial ordering
within a single narrow-band slice, and that narrow-band-selected LAE
and NLSG samples should suffice for reliable ionization tomography on
$\gtrsim 7\,h^{-1}\mathrm{cMpc}$ scales.

\section{Summary and Discussion}
\label{sec:summary}

In this work, we have presented a tomographic deep-learning framework to reconstruct the neutral-hydrogen fraction field, \(x_{\mathrm{HI}}(\mathbf{r})\), during the epoch of reionization (EoR) from the observed spatial distributions of Ly\(\alpha\) emitters (LAEs) and non-Ly\(\alpha\)-selected galaxies (NLSGs). The method is trained on radiative-transfer post-processed hydrodynamical simulations at \(z=7.14\), combined with a forward model for LAE and NLSG populations, and learns a non-linear mapping from these discrete galaxy tracers to the underlying ionization morphology. Unlike most existing LAE-based analyses, which primarily constrain the luminosity function or the volume-averaged neutral fraction, this framework performs \emph{field-level reconstruction}, enabling direct access to the spatial topology of ionized regions and their coupling to the galaxy population.

We find that 3D U-Nets, inspired from tomographic reconstruction research \citep{Han2016,Jin2017DeepCT,Garcia2023,Du2025}, are well-suited to the problem at hand. By forgoing a global latent bottleneck in favour of hierarchical receptive fields and dense skip connections, the network prioritises the preservation of local and intermediate-scale structure. This choice is particularly appropriate given the limited number of independent large-scale modes available for training in our simulation setup. The heteroscedastic output head provides voxel-level variance estimates on top of the reconstructed field that act as an effective regulariser and highlight regions where the reconstruction is intrinsically uncertain due to sparse sampling or complex topology.
We find that this Bayesian treatment of the output brings improvement in results in comparison to methods treating outputs as point estimates (for example, using Mean-Squared Error). It will be interesting to see how this variance correlates with the underlying uncertainties in the estimates. We leave this investigation to future work.

To assess observational relevance, we considered three survey configurations spanning a wide range of tracer densities. 
The {Deep} configuration (\(\log L_{\mathrm{Ly}\alpha}>41,\ M_{\mathrm{UV}}<-18\)) is representative of {JWST} spectroscopic programmes such as COSMOS-3D \citep{Kakiichi2024}, combining deep NIRCam imaging with NIRSpec spectroscopy over transverse scales of \(\sim100\,h^{-1}\mathrm{cMpc}\). 
The {Shallow} configuration (\(\log L_{\mathrm{Ly}\alpha}>42,\ M_{\mathrm{UV}}<-19\)) corresponds to wide-field narrow-band and multi-object spectroscopic LAE surveys such as SILVERRUSH, supplemented by UV-continuum-selected galaxies from GOLDRUSH \citep{Ono2018} and forthcoming Subaru/PFS-SSP observations. 
Finally, the {LAE-only} configuration (\(\log L_{\mathrm{Ly}\alpha}>42\)) represents pure line-selected samples characteristic of SILVERRUSH and PFS-like spectroscopic catalogues. 
These tiers bracket realistic survey regimes and provide a controlled framework for quantifying how reconstruction fidelity depends on tracer completeness. 

Across all reionization scenarios considered ({Extremely Early}, {Early}, {Oligarchic}, and {Fiducial}), the reconstructions recover the large-scale ionization topology reliably on scales comparable to the characteristic ionized-bubble size. This is quantified through voxel-wise Pearson correlation analysis, which shows that the reconstructed and true fields become strongly correlated once smoothed on scales \(L_{\mathrm{smooth}}\gtrsim 7\,h^{-1}\mathrm{cMpc}\). For the deepest survey, the Pearson correlation coefficient exceeds \(r\simeq 0.5\) on these scales, while intermediate-depth surveys show a modest reduction and LAE-only samples yield noisier but still coherent reconstructions. The characteristic smoothing scale at which predictive power emerges closely matches the typical bubble size at \(z\sim 7\), indicating that the method is most effective in the regime where ionization structure is dominated by coherent, source-driven regions rather than by small-scale residual neutral fluctuations. On smaller scales, the dynamic range of the reconstructed field is systematically compressed, reflecting both the limited information content of the tracer distribution and the regularising effect of the heteroscedastic loss.

One-point and two-point statistics provide complementary insight into which aspects of the ionization field are robustly constrained by the reconstruction. The probability distribution functions of \(x_{\mathrm{HI}}\), evaluated after smoothing on \(7\,h^{-1}\mathrm{cMpc}\) scales, reproduce both the peak position and width of the true distributions across all reionization models and survey configurations considered. Correspondingly, the inferred volume-averaged neutral fraction, \(\langle x_{\mathrm{HI}}\rangle\), agrees well with the true value, despite never being supplied explicitly during training, demonstrating that global constraints emerge naturally from the field-level reconstruction rather than being imposed as priors. The projected galaxy-ionization cross-correlation function, \(\xi_{x_{\mathrm{HI}},\,\mathrm{Galaxies}}(r_\perp)\), is likewise well recovered. The reconstructions capture the characteristic local anti-correlation induced by ionized regions around galaxies and the turnover toward zero at large separations. As expected, the fidelity of the cross-correlation follows the survey-depth hierarchy, with the joint LAE+NLSG samples providing the most faithful recovery.

The power-spectrum analysis further clarifies the scale dependence of the reconstruction. The projected power spectra measured from smoothed reconstructed maps agree well with the true spectra over the range of transverse wavenumbers probed, consistent with the recovery of large-scale ionization morphology. By contrast, in the absence of smoothing, the reconstructed fields exhibit a systematic suppression of power at high \(k\), indicating reduced fidelity on small spatial scales. Taken together, the recovery of large-scale power and galaxy-ionization cross-correlation, alongside the loss of small-scale power, indicates that the reconstruction is primarily constrained by the large-scale clustering and abundance of the galaxy tracers. The method is therefore most reliable on scales comparable to or larger than the characteristic ionized-bubble size, rather than providing a faithful representation of fine-scale structure.

Robustness tests further examine whether the reconstruction remains predictive when applied beyond the specific ionization conditions on which it was trained. 
Throughout the preceding analysis, the network was trained and evaluated on tomographic data drawn from the same set of reionization scenarios, such that the training and test samples share identical ionization statistics while differing only in their underlying density realizations.
To probe generalisation to an ionization state not encountered during training, we apply a network trained exclusively on subvolumes at \(z=7.14\) to mock survey realizations drawn from a later snapshot at \(z=6.6\).
Despite the change in the underlying neutral fraction and ionization morphology, the reconstructed fields preserve the large-scale structure and associated summary statistics, with only mild degradation relative to the matched-redshift case. 
Similarly, randomising galaxy positions within a narrow-band-like slab along the line-of-sight direction has little impact on the recovered smoothed PDFs or power spectra at the scales of interest. 
Taken together, these tests indicate that the reconstruction relies primarily on transverse clustering and overall tracer abundance, rather than on precise radial ordering, supporting its applicability to realistic narrow-band LAE surveys.

Finally, the reconstructed ionization maps provide a natural interface between galaxy surveys and forthcoming 21-cm observations. LAE/NLSG tomography yields a source-anchored view of ionization morphology that is complementary to 21-cm brightness-temperature maps, which are additionally sensitive to the thermal state of the IGM. Joint analyses could exploit these reconstructions to inform priors on ionization structure, guide targeted 21-cm stacking analyses, or interpret cross-correlations between galaxies and the 21-cm signal. In this sense, the framework presented here represents a step toward a multi-tracer, field-level approach to reionization, in which galaxy and 21-cm data are combined to probe the topology and evolution of the ionized IGM.

Despite these strengths, several caveats remain. The reconstructions are trained on a single hydrodynamical simulation and radiative-transfer model, and the LAE/NLSG forward modelling adopts simplified prescriptions for Ly\(\alpha\) radiative transfer and survey selection. The explored reionization scenarios, while diverse, do not span the full space of possible source populations or feedback processes, and residual systematic biases may arise if the real Universe differs significantly from the assumed models. Addressing these limitations will require extending the training set to multiple simulations and source prescriptions, incorporating more realistic survey systematics, and developing more comprehensive uncertainty quantification. 

Nevertheless, the results presented here establish a concrete proof-of-concept for learning the tracer–IGM connection in the EoR regime and demonstrate that LAE-selected information can recover field-level structure beyond summary statistics. In this sense, our framework provides a scalable baseline on which improved forward models, broader simulation ensembles, and rigorous uncertainty propagation can be layered in a controlled manner. Taken together, these developments would enable field-level tomographic mapping of the ionization topology and its correlation with galaxy environments. This paves the way toward EoR tomography with forthcoming wide-field LAE surveys and joint multi-tracer analyses.

\section{Acknowledgements}

It is a pleasure to thank Jatin Batra, Koki Kakiichi, Romain Meyer, Raghunathan Srianand, and Tejaswi Venumadhav for useful discussions.
MV is supported by the INFN PD51 INDARK grant and by the INAF Theory Grant ``Cosmological investigation of the cosmic web''.  MV is partially supported by the Fondazione ICSC, Spoke 3 ``Astrophysics and Cosmos Observations'', Piano Nazionale di Ripresa e Resilienza Project ID CN00000013 ``Italian Research Center on High-Performance Computing, Big Data and Quantum Computing'' funded by MUR Missione 4 Componente 2 Investimento 1.4: Potenziamento strutture di ricerca e creazione di ``campioni nazionali di R\&S (M4C2-19)'' - Next Generation EU (NGEU). 
This work was performed partially using the DiRAC Data Intensive service CSD3 at the University of Cambridge, managed by the University of Cambridge University Information Services on behalf of the STFC DiRAC HPC Facility, and the DiRAC Extreme Scaling service Tursa at the University of Edinburgh, managed by the Edinburgh Parallel Computing Centre on behalf of the STFC DiRAC HPC Facility (www.dirac.ac.uk). The DiRAC components of CSD3 at Cambridge and Tursa at Edinburgh were funded by BEIS, UKRI and STFC capital funding and STFC operations grants. DiRAC is part of the UKRI Digital Research Infrastructure. The project was also supported by a Swiss National Supercomputing Centre (CSCS) grant under project ID s1114. The authors acknowledge the computational resources provided by the Department of Theoretical Physics, Tata Institute of Fundamental Research (TIFR). SA also thanks the Science and Technology Facilities Council for a PhD studentship (STFC grant reference ST/W507362/1) and the University of Cambridge for providing a UKRI International Fees Bursary. The work has been performed as part of the DAE-STFC collaboration `Building Indo-UK collaborations towards the Square Kilometre Array' (STFC grant reference ST/Y004191/1). JSB acknowledges support by STFC consolidated grant ST/X000982/1. LK acknowledges the support of a Royal Society University Research Fellowship (grant number URF$\backslash$R1$\backslash$251793). GK also gratefully acknowledges the KICC Medium-Term Visitor programme. This research used resources of the Oak Ridge Leadership Computing Facility at the Oak Ridge National Laboratory, which is supported by the Office of Science of the U.S. Department of Energy under Contract No. DE-AC05-00OR22725. These resources were granted via INCITE AST206. We also acknowledge the use of OpenAI's ChatGPT for support with code implementation, text refinement, and certain analytical calculations. 

%%%%%%%%%%%%%%%%%%%%%%%%%%%%%%%%%%%%%%%%%%%%%%%%%%
\section{Data Availability}

The data and code underlying this article will be shared upon reasonable request to the corresponding author. We make our LAE model publicly available as the code SiMPLE-Gen, which can be accessed
at https://github.com/soumak-maitra/SiMPLE-Gen.



\bibliographystyle{mnras}
\bibliography{main} % if your bibtex file is called example.bib

@ARTICLE{asthana2024,
       author = {{Asthana}, Shikhar and {Haehnelt}, Martin G. and {Kulkarni}, Girish and {Aubert}, Dominique and {Bolton}, James S. and {Keating}, Laura C.},
        title = "{Late-end reionization with ATON-HE: towards constraints from Ly {\ensuremath{\alpha}} emitters observed with JWST}",
      journal = {\mnras},
         year = 2024,
        month = sep,
       volume = {533},
       number = {3},
        pages = {2843-2866},
          doi = {10.1093/mnras/stae1945},
archivePrefix = {arXiv},
       eprint = {2404.06548},
 primaryClass = {astro-ph.CO},
       adsurl = {https://ui.adsabs.harvard.edu/abs/2024MNRAS.533.2843A}
}

@ARTICLE{bolton17,
       author = {{Bolton}, James S. and {Puchwein}, Ewald and {Sijacki}, Debora and {Haehnelt}, Martin G. and {Kim}, Tae-Sun and {Meiksin}, Avery and {Regan}, John A. and {Viel}, Matteo},
        title = "{The Sherwood simulation suite: overview and data comparisons with the Lyman {\ensuremath{\alpha}} forest at redshifts 2 {\ensuremath{\leq}} z {\ensuremath{\leq}} 5}",
      journal = {\mnras},
         year = 2017,
        month = jan,
       volume = {464},
       number = {1},
        pages = {897-914},
          doi = {10.1093/mnras/stw2397},
archivePrefix = {arXiv},
       eprint = {1605.03462},
 primaryClass = {astro-ph.CO},
       adsurl = {https://ui.adsabs.harvard.edu/abs/2017MNRAS.464..897B}
}

@ARTICLE{aubert2008,
       author = {{Aubert}, Dominique and {Teyssier}, Romain},
        title = "{A radiative transfer scheme for cosmological reionization based on a local Eddington tensor}",
      journal = {\mnras},
         year = 2008,
        month = jun,
       volume = {387},
       number = {1},
        pages = {295-307},
          doi = {10.1111/j.1365-2966.2008.13223.x},
archivePrefix = {arXiv},
       eprint = {0709.1544},
 primaryClass = {astro-ph},
       adsurl = {https://ui.adsabs.harvard.edu/abs/2008MNRAS.387..295A}
}

@ARTICLE{bacon2017,
       author = {{Bacon}, Roland and {Conseil}, Simon and {Mary}, David and {Brinchmann}, Jarle and {Shepherd}, Martin and {Akhlaghi}, Mohammad and {Weilbacher}, Peter M. and {Piqueras}, Laure and {Wisotzki}, Lutz and {Lagattuta}, David and {Epinat}, Benoit and {Guerou}, Adrien and {Inami}, Hanae and {Cantalupo}, Sebastiano and {Courbot}, Jean Baptiste and {Contini}, Thierry and {Richard}, Johan and {Maseda}, Michael and {Bouwens}, Rychard and {Bouch{\'e}}, Nicolas and {Kollatschny}, Wolfram and {Schaye}, Joop and {Marino}, Raffaella Anna and {Pello}, Roser and {Herenz}, Christian and {Guiderdoni}, Bruno and {Carollo}, Marcella},
        title = "{The MUSE Hubble Ultra Deep Field Survey. I. Survey description, data reduction, and source detection}",
      journal = {\aap},
         year = 2017,
        month = dec,
       volume = {608},
          eid = {A1},
        pages = {A1},
          doi = {10.1051/0004-6361/201730833},
archivePrefix = {arXiv},
       eprint = {1710.03002},
 primaryClass = {astro-ph.GA},
       adsurl = {https://ui.adsabs.harvard.edu/abs/2017A&A...608A...1B}
}

@ARTICLE{becker2024,
       author = {{Becker}, George D. and {Bolton}, James S. and {Zhu}, Yongda and {Hashemi}, Seyedazim},
        title = "{Damping wing absorption associated with a giant Ly {\ensuremath{\alpha}} trough at z < 6: direct evidence for late-ending reionization}",
      journal = {\mnras},
         year = 2024,
        month = sep,
       volume = {533},
       number = {2},
        pages = {1525-1540},
          doi = {10.1093/mnras/stae1918},
archivePrefix = {arXiv},
       eprint = {2405.08885},
 primaryClass = {astro-ph.CO},
       adsurl = {https://ui.adsabs.harvard.edu/abs/2024MNRAS.533.1525B}
}

@ARTICLE{bouwens2015,
       author = {{Bouwens}, R.~J. and {Illingworth}, G.~D. and {Oesch}, P.~A. and {Trenti}, M. and {Labb{\'e}}, I. and {Bradley}, L. and {Carollo}, M. and {van Dokkum}, P.~G. and {Gonzalez}, V. and {Holwerda}, B. and {Franx}, M. and {Spitler}, L. and {Smit}, R. and {Magee}, D.},
        title = "{UV Luminosity Functions at Redshifts z {\ensuremath{\sim}} 4 to z {\ensuremath{\sim}} 10: 10,000 Galaxies from HST Legacy Fields}",
      journal = {\apj},
         year = 2015,
        month = apr,
       volume = {803},
       number = {1},
          eid = {34},
        pages = {34},
          doi = {10.1088/0004-637X/803/1/34},
archivePrefix = {arXiv},
       eprint = {1403.4295},
 primaryClass = {astro-ph.CO},
       adsurl = {https://ui.adsabs.harvard.edu/abs/2015ApJ...803...34B}
}

@ARTICLE{casey2023,
       author = {{Casey}, Caitlin M. and {Kartaltepe}, Jeyhan S. and {Drakos}, Nicole E. and {Franco}, Maximilien and {Harish}, Santosh and {Paquereau}, Louise and {Ilbert}, Olivier and {Rose}, Caitlin and {Cox}, Isabella G. and {Nightingale}, James W. and {Robertson}, Brant E. and {Silverman}, John D. and {Koekemoer}, Anton M. and {Massey}, Richard and {McCracken}, Henry Joy and {Rhodes}, Jason and {Akins}, Hollis B. and {Allen}, Natalie and {Amvrosiadis}, Aristeidis and {Arango-Toro}, Rafael C. and {Bagley}, Micaela B. and {Bongiorno}, Angela and {Capak}, Peter L. and {Champagne}, Jaclyn B. and {Chartab}, Nima and {Ch{\'a}vez Ortiz}, {\'O}scar A. and {Chworowsky}, Katherine and {Cooke}, Kevin C. and {Cooper}, Olivia R. and {Darvish}, Behnam and {Ding}, Xuheng and {Faisst}, Andreas L. and {Finkelstein}, Steven L. and {Fujimoto}, Seiji and {Gentile}, Fabrizio and {Gillman}, Steven and {Gould}, Katriona M.~L. and {Gozaliasl}, Ghassem and {Hayward}, Christopher C. and {He}, Qiuhan and {Hemmati}, Shoubaneh and {Hirschmann}, Michaela and {Jahnke}, Knud and {Jin}, Shuowen and {Khostovan}, Ali Ahmad and {Kokorev}, Vasily and {Lambrides}, Erini and {Laigle}, Clotilde and {Larson}, Rebecca L. and {Leung}, Gene C.~K. and {Liu}, Daizhong and {Liaudat}, Tobias and {Long}, Arianna S. and {Magdis}, Georgios and {Mahler}, Guillaume and {Mainieri}, Vincenzo and {Manning}, Sinclaire M. and {Maraston}, Claudia and {Martin}, Crystal L. and {McCleary}, Jacqueline E. and {McKinney}, Jed and {McPartland}, Conor J.~R. and {Mobasher}, Bahram and {Pattnaik}, Rohan and {Renzini}, Alvio and {Rich}, R. Michael and {Sanders}, David B. and {Sattari}, Zahra and {Scognamiglio}, Diana and {Scoville}, Nick and {Sheth}, Kartik and {Shuntov}, Marko and {Sparre}, Martin and {Suzuki}, Tomoko L. and {Talia}, Margherita and {Toft}, Sune and {Trakhtenbrot}, Benny and {Urry}, C. Megan and {Valentino}, Francesco and {Vanderhoof}, Brittany N. and {Vardoulaki}, Eleni and {Weaver}, John R. and {Whitaker}, Katherine E. and {Wilkins}, Stephen M. and {Yang}, Lilan and {Zavala}, Jorge A.},
        title = "{COSMOS-Web: An Overview of the JWST Cosmic Origins Survey}",
      journal = {\apj},
         year = 2023,
        month = sep,
       volume = {954},
       number = {1},
          eid = {31},
        pages = {31},
          doi = {10.3847/1538-4357/acc2bc},
archivePrefix = {arXiv},
       eprint = {2211.07865},
 primaryClass = {astro-ph.GA},
       adsurl = {https://ui.adsabs.harvard.edu/abs/2023ApJ...954...31C}
}

@ARTICLE{davies2018,
       author = {{Davies}, Frederick B. and {Hennawi}, Joseph F. and {Ba{\~n}ados}, Eduardo and {Luki{\'c}}, Zarija and {Decarli}, Roberto and {Fan}, Xiaohui and {Farina}, Emanuele P. and {Mazzucchelli}, Chiara and {Rix}, Hans-Walter and {Venemans}, Bram P. and {Walter}, Fabian and {Wang}, Feige and {Yang}, Jinyi},
        title = "{Quantitative Constraints on the Reionization History from the IGM Damping Wing Signature in Two Quasars at z > 7}",
      journal = {\apj},
         year = 2018,
        month = sep,
       volume = {864},
       number = {2},
          eid = {142},
        pages = {142},
          doi = {10.3847/1538-4357/aad6dc},
archivePrefix = {arXiv},
       eprint = {1802.06066},
 primaryClass = {astro-ph.CO},
       adsurl = {https://ui.adsabs.harvard.edu/abs/2018ApJ...864..142D}
}

@ARTICLE{dijkstra2012,
       author = {{Dijkstra}, Mark and {Wyithe}, J. Stuart B.},
        title = "{An empirical study of the relationship between Ly{\ensuremath{\alpha}} and UV-selected galaxies: do theorists and observers 'select' the same objects?}",
      journal = {\mnras},
         year = 2012,
        month = feb,
       volume = {419},
       number = {4},
        pages = {3181-3193},
          doi = {10.1111/j.1365-2966.2011.19958.x},
archivePrefix = {arXiv},
       eprint = {1108.3840},
 primaryClass = {astro-ph.CO},
       adsurl = {https://ui.adsabs.harvard.edu/abs/2012MNRAS.419.3181D}
}

@ARTICLE{dijkstra2014,
       author = {{Dijkstra}, Mark},
        title = "{Ly{\ensuremath{\alpha}} Emitting Galaxies as a Probe of Reionisation}",
      journal = {\pasa},
         year = 2014,
        month = oct,
       volume = {31},
          eid = {e040},
        pages = {e040},
          doi = {10.1017/pasa.2014.33},
archivePrefix = {arXiv},
       eprint = {1406.7292},
 primaryClass = {astro-ph.CO},
       adsurl = {https://ui.adsabs.harvard.edu/abs/2014PASA...31...40D}
}

@ARTICLE{furlanetto2004,
       author = {{Furlanetto}, Steven R. and {Zaldarriaga}, Matias and {Hernquist}, Lars},
        title = "{The Growth of H II Regions During Reionization}",
      journal = {\apj},
         year = 2004,
        month = sep,
       volume = {613},
       number = {1},
        pages = {1-15},
          doi = {10.1086/423025},
archivePrefix = {arXiv},
       eprint = {astro-ph/0403697},
 primaryClass = {astro-ph},
       adsurl = {https://ui.adsabs.harvard.edu/abs/2004ApJ...613....1F}
}

@ARTICLE{ghara2020,
       author = {{Ghara}, Raghunath and {Mellema}, Garrelt},
        title = "{Impact of Ly {\ensuremath{\alpha}} heating on the global 21-cm signal from the Cosmic Dawn}",
      journal = {\mnras},
         year = 2020,
        month = feb,
       volume = {492},
       number = {1},
        pages = {634-644},
          doi = {10.1093/mnras/stz3513},
archivePrefix = {arXiv},
       eprint = {1904.09999},
 primaryClass = {astro-ph.CO},
       adsurl = {https://ui.adsabs.harvard.edu/abs/2020MNRAS.492..634G}
}

@ARTICLE{hoag2019,
       author = {{Hoag}, A. and {Brada{\v{c}}}, M. and {Huang}, K. and {Mason}, C. and {Treu}, T. and {Schmidt}, K.~B. and {Trenti}, M. and {Strait}, V. and {Lemaux}, B.~C. and {Finney}, E.~Q. and {Paddock}, M.},
        title = "{Constraining the Neutral Fraction of Hydrogen in the IGM at Redshift 7.5}",
      journal = {\apj},
         year = 2019,
        month = jun,
       volume = {878},
       number = {1},
          eid = {12},
        pages = {12},
          doi = {10.3847/1538-4357/ab1de7},
archivePrefix = {arXiv},
       eprint = {1901.09001},
 primaryClass = {astro-ph.GA},
       adsurl = {https://ui.adsabs.harvard.edu/abs/2019ApJ...878...12H}
}

@ARTICLE{keating2020,
       author = {{Keating}, Laura C. and {Kulkarni}, Girish and {Haehnelt}, Martin G. and {Chardin}, Jonathan and {Aubert}, Dominique},
        title = "{Constraining the second half of reionization with the Ly {\ensuremath{\beta}} forest}",
      journal = {\mnras},
         year = 2020,
        month = sep,
       volume = {497},
       number = {1},
        pages = {906-915},
          doi = {10.1093/mnras/staa1909},
archivePrefix = {arXiv},
       eprint = {1912.05582},
 primaryClass = {astro-ph.CO},
       adsurl = {https://ui.adsabs.harvard.edu/abs/2020MNRAS.497..906K}
}

@ARTICLE{konno2014,
       author = {{Konno}, Akira and {Ouchi}, Masami and {Ono}, Yoshiaki and {Shimasaku}, Kazuhiro and {Shibuya}, Takatoshi and {Furusawa}, Hisanori and {Nakajima}, Kimihiko and {Naito}, Yoshiaki and {Momose}, Rieko and {Yuma}, Suraphong and {Iye}, Masanori},
        title = "{Accelerated Evolution of the Ly{\ensuremath{\alpha}} Luminosity Function at z >\raisebox{-0.5ex}\textasciitilde 7 Revealed by the Subaru Ultra-deep Survey for Ly{\ensuremath{\alpha}} Emitters at z = 7.3}",
      journal = {\apj},
         year = 2014,
        month = dec,
       volume = {797},
       number = {1},
          eid = {16},
        pages = {16},
          doi = {10.1088/0004-637X/797/1/16},
archivePrefix = {arXiv},
       eprint = {1404.6066},
 primaryClass = {astro-ph.CO},
       adsurl = {https://ui.adsabs.harvard.edu/abs/2014ApJ...797...16K}
}

@ARTICLE{konno2018,
       author = {{Konno}, Akira and {Ouchi}, Masami and {Shibuya}, Takatoshi and {Ono}, Yoshiaki and {Shimasaku}, Kazuhiro and {Taniguchi}, Yoshiaki and {Nagao}, Tohru and {Kobayashi}, Masakazu A.~R. and {Kajisawa}, Masaru and {Kashikawa}, Nobunari and {Inoue}, Akio K. and {Oguri}, Masamune and {Furusawa}, Hisanori and {Goto}, Tomotsugu and {Harikane}, Yuichi and {Higuchi}, Ryo and {Komiyama}, Yutaka and {Kusakabe}, Haruka and {Miyazaki}, Satoshi and {Nakajima}, Kimihiko and {Wang}, Shiang-Yu},
        title = "{SILVERRUSH. IV. Ly{\ensuremath{\alpha}} luminosity functions at z = 5.7 and 6.6 studied with {\ensuremath{\sim}}1300 Ly{\ensuremath{\alpha}} emitters on the 14-21 deg$^{2}$ sky}",
      journal = {\pasj},
         year = 2018,
        month = jan,
       volume = {70},
          eid = {S16},
        pages = {S16},
          doi = {10.1093/pasj/psx131},
archivePrefix = {arXiv},
       eprint = {1705.01222},
 primaryClass = {astro-ph.GA},
       adsurl = {https://ui.adsabs.harvard.edu/abs/2018PASJ...70S..16K}
}

@ARTICLE{kulkarni2019,
       author = {{Kulkarni}, Girish and {Keating}, Laura C. and {Haehnelt}, Martin G. and {Bosman}, Sarah E.~I. and {Puchwein}, Ewald and {Chardin}, Jonathan and {Aubert}, Dominique},
        title = "{Large Ly {\ensuremath{\alpha}} opacity fluctuations and low CMB {\ensuremath{\tau}} in models of late reionization with large islands of neutral hydrogen extending to z < 5.5}",
      journal = {\mnras},
         year = 2019,
        month = may,
       volume = {485},
       number = {1},
        pages = {L24-L28},
          doi = {10.1093/mnrasl/slz025},
archivePrefix = {arXiv},
       eprint = {1809.06374},
 primaryClass = {astro-ph.CO},
       adsurl = {https://ui.adsabs.harvard.edu/abs/2019MNRAS.485L..24K}
}

@ARTICLE{levermore1984,
       author = {{Levermore}, C.~D.},
        title = "{Relating Eddington factors to flux limiters.}",
      journal = {\jqsrt},
         year = 1984,
        month = feb,
       volume = {31},
       number = {2},
        pages = {149-160},
          doi = {10.1016/0022-4073(84)90112-2},
       adsurl = {https://ui.adsabs.harvard.edu/abs/1984JQSRT..31..149L}
}

@ARTICLE{liu2016,
       author = {{Liu}, Chuanwu and {Mutch}, Simon J. and {Angel}, P.~W. and {Duffy}, Alan R. and {Geil}, Paul M. and {Poole}, Gregory B. and {Mesinger}, Andrei and {Wyithe}, J. Stuart B.},
        title = "{Dark-ages reionization and galaxy formation simulation - IV. UV luminosity functions of high-redshift galaxies}",
      journal = {\mnras},
         year = 2016,
        month = oct,
       volume = {462},
       number = {1},
        pages = {235-249},
          doi = {10.1093/mnras/stw1015},
archivePrefix = {arXiv},
       eprint = {1512.00563},
 primaryClass = {astro-ph.GA},
       adsurl = {https://ui.adsabs.harvard.edu/abs/2016MNRAS.462..235L}
}

@ARTICLE{lu2024,
       author = {{Lu}, Ting-Yi and {Mason}, Charlotte A. and {Hutter}, Anne and {Mesinger}, Andrei and {Qin}, Yuxiang and {Stark}, Daniel P. and {Endsley}, Ryan},
        title = "{The reionizing bubble size distribution around galaxies}",
      journal = {\mnras},
         year = 2024,
        month = mar,
       volume = {528},
       number = {3},
        pages = {4872-4890},
          doi = {10.1093/mnras/stae266},
archivePrefix = {arXiv},
       eprint = {2304.11192},
 primaryClass = {astro-ph.GA},
       adsurl = {https://ui.adsabs.harvard.edu/abs/2024MNRAS.528.4872L}
}

@ARTICLE{majumdar2018,
       author = {{Majumdar}, Suman and {Pritchard}, Jonathan R. and {Mondal}, Rajesh and {Watkinson}, Catherine A. and {Bharadwaj}, Somnath and {Mellema}, Garrelt},
        title = "{Quantifying the non-Gaussianity in the EoR 21-cm signal through bispectrum}",
      journal = {\mnras},
         year = 2018,
        month = may,
       volume = {476},
       number = {3},
        pages = {4007-4024},
          doi = {10.1093/mnras/sty535},
archivePrefix = {arXiv},
       eprint = {1708.08458},
 primaryClass = {astro-ph.CO},
       adsurl = {https://ui.adsabs.harvard.edu/abs/2018MNRAS.476.4007M}
}

@ARTICLE{LiuShaw2020,
       author = {{Liu}, Adrian and {Shaw}, J. Richard},
        title = "{Data Analysis for Precision 21 cm Cosmology}",
      journal = {\pasp},
         year = 2020,
        month = jun,
       volume = {132},
       number = {1012},
          eid = {062001},
        pages = {062001},
          doi = {10.1088/1538-3873/ab5bfd},
archivePrefix = {arXiv},
       eprint = {1907.08211},
 primaryClass = {astro-ph.IM},
       adsurl = {https://ui.adsabs.harvard.edu/abs/2020PASP..132f2001L}
}

@ARTICLE{mason2018,
       author = {{Mason}, Charlotte A. and {Treu}, Tommaso and {Dijkstra}, Mark and {Mesinger}, Andrei and {Trenti}, Michele and {Pentericci}, Laura and {de Barros}, Stephane and {Vanzella}, Eros},
        title = "{The Universe Is Reionizing at z {\ensuremath{\sim}} 7: Bayesian Inference of the IGM Neutral Fraction Using Ly{\ensuremath{\alpha}} Emission from Galaxies}",
      journal = {\apj},
         year = 2018,
        month = mar,
       volume = {856},
       number = {1},
          eid = {2},
        pages = {2},
          doi = {10.3847/1538-4357/aab0a7},
archivePrefix = {arXiv},
       eprint = {1709.05356},
 primaryClass = {astro-ph.CO},
       adsurl = {https://ui.adsabs.harvard.edu/abs/2018ApJ...856....2M}
}

@ARTICLE{mason2019,
       author = {{Mason}, Charlotte A. and {Fontana}, Adriano and {Treu}, Tommaso and {Schmidt}, Kasper B. and {Hoag}, Austin and {Abramson}, Louis and {Amorin}, Ricardo and {Brada{\v{c}}}, Maru{\v{s}}a and {Guaita}, Lucia and {Jones}, Tucker and {Henry}, Alaina and {Malkan}, Matthew A. and {Pentericci}, Laura and {Trenti}, Michele and {Vanzella}, Eros},
        title = "{Inferences on the timeline of reionization at z {\ensuremath{\sim}} 8 from the KMOS Lens-Amplified Spectroscopic Survey}",
      journal = {\mnras},
         year = 2019,
        month = may,
       volume = {485},
       number = {3},
        pages = {3947-3969},
          doi = {10.1093/mnras/stz632},
archivePrefix = {arXiv},
       eprint = {1901.11045},
 primaryClass = {astro-ph.CO},
       adsurl = {https://ui.adsabs.harvard.edu/abs/2019MNRAS.485.3947M}
}

@ARTICLE{mcquinn2007,
       author = {{McQuinn}, Matthew and {Lidz}, Adam and {Zahn}, Oliver and {Dutta}, Suvendra and {Hernquist}, Lars and {Zaldarriaga}, Matias},
        title = "{The morphology of HII regions during reionization}",
      journal = {\mnras},
         year = 2007,
        month = may,
       volume = {377},
       number = {3},
        pages = {1043-1063},
          doi = {10.1111/j.1365-2966.2007.11489.x},
archivePrefix = {arXiv},
       eprint = {astro-ph/0610094},
 primaryClass = {astro-ph},
       adsurl = {https://ui.adsabs.harvard.edu/abs/2007MNRAS.377.1043M}
}

@ARTICLE{mesinger2007,
       author = {{Mesinger}, Andrei and {Furlanetto}, Steven},
        title = "{Efficient Simulations of Early Structure Formation and Reionization}",
      journal = {\apj},
         year = 2007,
        month = nov,
       volume = {669},
       number = {2},
        pages = {663-675},
          doi = {10.1086/521806},
archivePrefix = {arXiv},
       eprint = {0704.0946},
 primaryClass = {astro-ph},
       adsurl = {https://ui.adsabs.harvard.edu/abs/2007ApJ...669..663M}
}

@ARTICLE{mesinger2011,
       author = {{Mesinger}, Andrei and {Furlanetto}, Steven and {Cen}, Renyue},
        title = "{21CMFAST: a fast, seminumerical simulation of the high-redshift 21-cm signal}",
      journal = {\mnras},
         year = 2011,
        month = feb,
       volume = {411},
       number = {2},
        pages = {955-972},
          doi = {10.1111/j.1365-2966.2010.17731.x},
archivePrefix = {arXiv},
       eprint = {1003.3878},
 primaryClass = {astro-ph.CO},
       adsurl = {https://ui.adsabs.harvard.edu/abs/2011MNRAS.411..955M}
}

@ARTICLE{morales2021,
       author = {{Morales}, Alexa M. and {Mason}, Charlotte A. and {Bruton}, Sean and {Gronke}, Max and {Haardt}, Francesco and {Scarlata}, Claudia},
        title = "{The Evolution of the Lyman-alpha Luminosity Function during Reionization}",
      journal = {\apj},
         year = 2021,
        month = oct,
       volume = {919},
       number = {2},
          eid = {120},
        pages = {120},
          doi = {10.3847/1538-4357/ac1104},
archivePrefix = {arXiv},
       eprint = {2101.01205},
 primaryClass = {astro-ph.GA},
       adsurl = {https://ui.adsabs.harvard.edu/abs/2021ApJ...919..120M}
}

@ARTICLE{neyer2024,
       author = {{Neyer}, Meredith and {Smith}, Aaron and {Kannan}, Rahul and {Vogelsberger}, Mark and {Garaldi}, Enrico and {Gal{\'a}rraga-Espinosa}, Daniela and {Borrow}, Josh and {Hernquist}, Lars and {Pakmor}, R{\"u}diger and {Springel}, Volker},
        title = "{The THESAN project: connecting ionized bubble sizes to their local environments during the Epoch of Reionization}",
      journal = {\mnras},
         year = 2024,
        month = jul,
       volume = {531},
       number = {3},
        pages = {2943-2957},
          doi = {10.1093/mnras/stae1325},
archivePrefix = {arXiv},
       eprint = {2310.03783},
 primaryClass = {astro-ph.GA},
       adsurl = {https://ui.adsabs.harvard.edu/abs/2024MNRAS.531.2943N}
}

@ARTICLE{ouchi2008,
       author = {{Ouchi}, Masami and {Shimasaku}, Kazuhiro and {Akiyama}, Masayuki and {Simpson}, Chris and {Saito}, Tomoki and {Ueda}, Yoshihiro and {Furusawa}, Hisanori and {Sekiguchi}, Kazuhiro and {Yamada}, Toru and {Kodama}, Tadayuki and {Kashikawa}, Nobunari and {Okamura}, Sadanori and {Iye}, Masanori and {Takata}, Tadafumi and {Yoshida}, Michitoshi and {Yoshida}, Makiko},
        title = "{The Subaru/XMM-Newton Deep Survey (SXDS). IV. Evolution of Ly{\ensuremath{\alpha}} Emitters from z = 3.1 to 5.7 in the 1 deg$^{2}$ Field: Luminosity Functions and AGN}",
      journal = {\apjs},
         year = 2008,
        month = jun,
       volume = {176},
       number = {2},
        pages = {301-330},
          doi = {10.1086/527673},
archivePrefix = {arXiv},
       eprint = {0707.3161},
 primaryClass = {astro-ph},
       adsurl = {https://ui.adsabs.harvard.edu/abs/2008ApJS..176..301O}
}

@ARTICLE{ouchi2010,
       author = {{Ouchi}, Masami and {Shimasaku}, Kazuhiro and {Furusawa}, Hisanori and {Saito}, Tomoki and {Yoshida}, Makiko and {Akiyama}, Masayuki and {Ono}, Yoshiaki and {Yamada}, Toru and {Ota}, Kazuaki and {Kashikawa}, Nobunari and {Iye}, Masanori and {Kodama}, Tadayuki and {Okamura}, Sadanori and {Simpson}, Chris and {Yoshida}, Michitoshi},
        title = "{Statistics of 207 Ly{\ensuremath{\alpha}} Emitters at a Redshift Near 7: Constraints on Reionization and Galaxy Formation Models}",
      journal = {\apj},
         year = 2010,
        month = nov,
       volume = {723},
       number = {1},
        pages = {869-894},
          doi = {10.1088/0004-637X/723/1/869},
archivePrefix = {arXiv},
       eprint = {1007.2961},
 primaryClass = {astro-ph.CO},
       adsurl = {https://ui.adsabs.harvard.edu/abs/2010ApJ...723..869O}
}

@ARTICLE{ouchi2018,
       author = {{Ouchi}, Masami and {Harikane}, Yuichi and {Shibuya}, Takatoshi and {Shimasaku}, Kazuhiro and {Taniguchi}, Yoshiaki and {Konno}, Akira and {Kobayashi}, Masakazu and {Kajisawa}, Masaru and {Nagao}, Tohru and {Ono}, Yoshiaki and {Inoue}, Akio K. and {Umemura}, Masayuki and {Mori}, Masao and {Hasegawa}, Kenji and {Higuchi}, Ryo and {Komiyama}, Yutaka and {Matsuda}, Yuichi and {Nakajima}, Kimihiko and {Saito}, Tomoki and {Wang}, Shiang-Yu},
        title = "{Systematic Identification of LAEs for Visible Exploration and Reionization Research Using Subaru HSC (SILVERRUSH). I. Program strategy and clustering properties of {\ensuremath{\sim}}2000 Ly{\ensuremath{\alpha}} emitters at z = 6-7 over the 0.3-0.5 Gpc$^{2}$ survey area}",
      journal = {\pasj},
         year = 2018,
        month = jan,
       volume = {70},
          eid = {S13},
        pages = {S13},
          doi = {10.1093/pasj/psx074},
archivePrefix = {arXiv},
       eprint = {1704.07455},
 primaryClass = {astro-ph.GA},
       adsurl = {https://ui.adsabs.harvard.edu/abs/2018PASJ...70S..13O}
}

@ARTICLE{planck2014,
   author = {{Planck Collaboration} and {Ade}, P.~A.~R. and {Aghanim}, N. and 
	{Armitage-Caplan}, C. and {Arnaud}, M. and {Ashdown}, M. and 
	{Atrio-Barandela}, F. and {Aumont}, J. and {Baccigalupi}, C. and 
	{Banday}, A.~J. and et al.},
    title = "{Planck 2013 results. XVI. Cosmological parameters}",
  journal = {\aap},
archivePrefix = "arXiv",
   eprint = {1303.5076},
     year = 2014,
    month = nov,
   volume = 571,
      eid = {A16},
    pages = {A16},
      doi = {10.1051/0004-6361/201321591},
   adsurl = {http://adsabs.harvard.edu/abs/2014A%26A...571A..16P}
}

@ARTICLE{planck2018cp,
       author = {{Planck Collaboration} and {Aghanim}, N. and {Akrami}, Y. and {Ashdown}, M. and {Aumont}, J. and {Baccigalupi}, C. and {Ballardini}, M. and {Banday}, A.~J. and {Barreiro}, R.~B. and {Bartolo}, N. and {Basak}, S. and {Battye}, R. and {Benabed}, K. and {Bernard}, J. -P. and {Bersanelli}, M. and {Bielewicz}, P. and {Bock}, J.~J. and {Bond}, J.~R. and {Borrill}, J. and {Bouchet}, F.~R. and {Boulanger}, F. and {Bucher}, M. and {Burigana}, C. and {Butler}, R.~C. and {Calabrese}, E. and {Cardoso}, J. -F. and {Carron}, J. and {Challinor}, A. and {Chiang}, H.~C. and {Chluba}, J. and {Colombo}, L.~P.~L. and {Combet}, C. and {Contreras}, D. and {Crill}, B.~P. and {Cuttaia}, F. and {de Bernardis}, P. and {de Zotti}, G. and {Delabrouille}, J. and {Delouis}, J. -M. and {Di Valentino}, E. and {Diego}, J.~M. and {Dor{\'e}}, O. and {Douspis}, M. and {Ducout}, A. and {Dupac}, X. and {Dusini}, S. and {Efstathiou}, G. and {Elsner}, F. and {En{\ss}lin}, T.~A. and {Eriksen}, H.~K. and {Fantaye}, Y. and {Farhang}, M. and {Fergusson}, J. and {Fernandez-Cobos}, R. and {Finelli}, F. and {Forastieri}, F. and {Frailis}, M. and {Fraisse}, A.~A. and {Franceschi}, E. and {Frolov}, A. and {Galeotta}, S. and {Galli}, S. and {Ganga}, K. and {G{\'e}nova-Santos}, R.~T. and {Gerbino}, M. and {Ghosh}, T. and {Gonz{\'a}lez-Nuevo}, J. and {G{\'o}rski}, K.~M. and {Gratton}, S. and {Gruppuso}, A. and {Gudmundsson}, J.~E. and {Hamann}, J. and {Handley}, W. and {Hansen}, F.~K. and {Herranz}, D. and {Hildebrandt}, S.~R. and {Hivon}, E. and {Huang}, Z. and {Jaffe}, A.~H. and {Jones}, W.~C. and {Karakci}, A. and {Keih{\"a}nen}, E. and {Keskitalo}, R. and {Kiiveri}, K. and {Kim}, J. and {Kisner}, T.~S. and {Knox}, L. and {Krachmalnicoff}, N. and {Kunz}, M. and {Kurki-Suonio}, H. and {Lagache}, G. and {Lamarre}, J. -M. and {Lasenby}, A. and {Lattanzi}, M. and {Lawrence}, C.~R. and {Le Jeune}, M. and {Lemos}, P. and {Lesgourgues}, J. and {Levrier}, F. and {Lewis}, A. and {Liguori}, M. and {Lilje}, P.~B. and {Lilley}, M. and {Lindholm}, V. and {L{\'o}pez-Caniego}, M. and {Lubin}, P.~M. and {Ma}, Y. -Z. and {Mac{\'\i}as-P{\'e}rez}, J.~F. and {Maggio}, G. and {Maino}, D. and {Mandolesi}, N. and {Mangilli}, A. and {Marcos-Caballero}, A. and {Maris}, M. and {Martin}, P.~G. and {Martinelli}, M. and {Mart{\'\i}nez-Gonz{\'a}lez}, E. and {Matarrese}, S. and {Mauri}, N. and {McEwen}, J.~D. and {Meinhold}, P.~R. and {Melchiorri}, A. and {Mennella}, A. and {Migliaccio}, M. and {Millea}, M. and {Mitra}, S. and {Miville-Desch{\^e}nes}, M. -A. and {Molinari}, D. and {Montier}, L. and {Morgante}, G. and {Moss}, A. and {Natoli}, P. and {N{\o}rgaard-Nielsen}, H.~U. and {Pagano}, L. and {Paoletti}, D. and {Partridge}, B. and {Patanchon}, G. and {Peiris}, H.~V. and {Perrotta}, F. and {Pettorino}, V. and {Piacentini}, F. and {Polastri}, L. and {Polenta}, G. and {Puget}, J. -L. and {Rachen}, J.~P. and {Reinecke}, M. and {Remazeilles}, M. and {Renzi}, A. and {Rocha}, G. and {Rosset}, C. and {Roudier}, G. and {Rubi{\~n}o-Mart{\'\i}n}, J.~A. and {Ruiz-Granados}, B. and {Salvati}, L. and {Sandri}, M. and {Savelainen}, M. and {Scott}, D. and {Shellard}, E.~P.~S. and {Sirignano}, C. and {Sirri}, G. and {Spencer}, L.~D. and {Sunyaev}, R. and {Suur-Uski}, A. -S. and {Tauber}, J.~A. and {Tavagnacco}, D. and {Tenti}, M. and {Toffolatti}, L. and {Tomasi}, M. and {Trombetti}, T. and {Valenziano}, L. and {Valiviita}, J. and {Van Tent}, B. and {Vibert}, L. and {Vielva}, P. and {Villa}, F. and {Vittorio}, N. and {Wandelt}, B.~D. and {Wehus}, I.~K. and {White}, M. and {White}, S.~D.~M. and {Zacchei}, A. and {Zonca}, A.},
        title = "{Planck 2018 results. VI. Cosmological parameters}",
      journal = {\aap},
         year = 2020,
        month = sep,
       volume = {641},
          eid = {A6},
        pages = {A6},
          doi = {10.1051/0004-6361/201833910},
archivePrefix = {arXiv},
       eprint = {1807.06209},
 primaryClass = {astro-ph.CO},
       adsurl = {https://ui.adsabs.harvard.edu/abs/2020A&A...641A...6P}
}

@ARTICLE{puchwein2019,
       author = {{Puchwein}, Ewald and {Haardt}, Francesco and {Haehnelt}, Martin G. and {Madau}, Piero},
        title = "{Consistent modelling of the meta-galactic UV background and the thermal/ionization history of the intergalactic medium}",
      journal = {\mnras},
         year = 2019,
        month = may,
       volume = {485},
       number = {1},
        pages = {47-68},
          doi = {10.1093/mnras/stz222},
archivePrefix = {arXiv},
       eprint = {1801.04931},
 primaryClass = {astro-ph.GA},
       adsurl = {https://ui.adsabs.harvard.edu/abs/2019MNRAS.485...47P}
}

@ARTICLE{puchwein2023,
       author = {{Puchwein}, Ewald and {Bolton}, James S. and {Keating}, Laura C. and {Molaro}, Margherita and {Gaikwad}, Prakash and {Kulkarni}, Girish and {Haehnelt}, Martin G. and {Ir{\v{s}}i{\v{c}}}, Vid and {{\v{S}}oltinsk{\'y}}, Tom{\'a}{\v{s}} and {Viel}, Matteo and {Aubert}, Dominique and {Becker}, George D. and {Meiksin}, Avery},
        title = "{The Sherwood-Relics simulations: overview and impact of patchy reionization and pressure smoothing on the intergalactic medium}",
      journal = {\mnras},
         year = 2023,
        month = mar,
       volume = {519},
       number = {4},
        pages = {6162-6183},
          doi = {10.1093/mnras/stac3761},
archivePrefix = {arXiv},
       eprint = {2207.13098},
 primaryClass = {astro-ph.CO},
       adsurl = {https://ui.adsabs.harvard.edu/abs/2023MNRAS.519.6162P}
}

@ARTICLE{ribli2019,
       author = {{Ribli}, Dezs{\H{o}} and {Pataki}, B{\'a}lint {\'A}rmin and {Csabai}, Istv{\'a}n},
        title = "{An improved cosmological parameter inference scheme motivated by deep learning}",
      journal = {Nature Astronomy},
         year = 2019,
        month = jan,
       volume = {3},
        pages = {93-98},
          doi = {10.1038/s41550-018-0596-8},
archivePrefix = {arXiv},
       eprint = {1806.05995},
 primaryClass = {astro-ph.CO},
       adsurl = {https://ui.adsabs.harvard.edu/abs/2019NatAs...3...93R}
}

@ARTICLE{robertson2022,
       author = {{Robertson}, Brant E.},
        title = "{Galaxy Formation and Reionization: Key Unknowns and Expected Breakthroughs by the James Webb Space Telescope}",
      journal = {\araa},
         year = 2022,
        month = aug,
       volume = {60},
        pages = {121-158},
          doi = {10.1146/annurev-astro-120221-044656},
archivePrefix = {arXiv},
       eprint = {2110.13160},
 primaryClass = {astro-ph.CO},
       adsurl = {https://ui.adsabs.harvard.edu/abs/2022ARA&A..60..121R}
}

@article{sadoun2017,
  author = {Sadoun, Raphael and Zheng, Zheng and Miralda-Escudé, Jordi},
  title = {On the Impact of Large-scale Environments on Ly$\alpha$ Emission from Galaxies},
  journal = {The Astrophysical Journal},
  volume = {839},
  number = {1},
  year = {2017},
  pages = {44},
  doi = {10.3847/1538-4357/aa684d}
}

@ARTICLE{springel2005,
   author = {{Springel}, V.},
    title = "{The cosmological simulation code GADGET-2}",
  journal = {\mnras},
   eprint = {astro-ph/0505010},
     year = 2005,
    month = dec,
   volume = 364,
    pages = {1105-1134},
      doi = {10.1111/j.1365-2966.2005.09655.x},
   adsurl = {http://adsabs.harvard.edu/abs/2005MNRAS.364.1105S}
}

@ARTICLE{takada2014,
       author = {{Takada}, Masahiro and {Ellis}, Richard S. and {Chiba}, Masashi and {Greene}, Jenny E. and {Aihara}, Hiroaki and {Arimoto}, Nobuo and {Bundy}, Kevin and {Cohen}, Judith and {Dor{\'e}}, Olivier and {Graves}, Genevieve and {Gunn}, James E. and {Heckman}, Timothy and {Hirata}, Christopher M. and {Ho}, Paul and {Kneib}, Jean-Paul and {Le F{\`e}vre}, Olivier and {Lin}, Lihwai and {More}, Surhud and {Murayama}, Hitoshi and {Nagao}, Tohru and {Ouchi}, Masami and {Seiffert}, Michael and {Silverman}, John D. and {Sodr{\'e}}, Laerte and {Spergel}, David N. and {Strauss}, Michael A. and {Sugai}, Hajime and {Suto}, Yasushi and {Takami}, Hideki and {Wyse}, Rosemary},
        title = "{Extragalactic science, cosmology, and Galactic archaeology with the Subaru Prime Focus Spectrograph}",
      journal = {\pasj},
         year = 2014,
        month = feb,
       volume = {66},
       number = {1},
          eid = {R1},
        pages = {R1},
          doi = {10.1093/pasj/pst019},
archivePrefix = {arXiv},
       eprint = {1206.0737},
 primaryClass = {astro-ph.CO},
       adsurl = {https://ui.adsabs.harvard.edu/abs/2014PASJ...66R...1T}
}

@INPROCEEDINGS{tamura2016,
       author = {{Tamura}, Naoyuki and {Takato}, Naruhisa and {Shimono}, Atsushi and {Moritani}, Yuki and {Yabe}, Kiyoto and {Ishizuka}, Yuki and {Ueda}, Akitoshi and {Kamata}, Yukiko and {Aghazarian}, Hrand and {Arnouts}, St{\'e}phane and {Barban}, Gabriel and {Barkhouser}, Robert H. and {Borges}, Renato C. and {Braun}, David F. and {Carr}, Michael A. and {Chabaud}, Pierre-Yves and {Chang}, Yin-Chang and {Chen}, Hsin-Yo and {Chiba}, Masashi and {Chou}, Richard C.~Y. and {Chu}, You-Hua and {Cohen}, Judith and {de Almeida}, Rodrigo P. and {de Oliveira}, Antonio C. and {de Oliveira}, Ligia S. and {Dekany}, Richard G. and {Dohlen}, Kjetil and {dos Santos}, Jesulino B. and {dos Santos}, Leandro H. and {Ellis}, Richard and {Fabricius}, Maximilian and {Ferrand}, Didier and {Ferreira}, D{\'e}cio and {Golebiowski}, Mirek and {Greene}, Jenny E. and {Gross}, Johannes and {Gunn}, James E. and {Hammond}, Randolph and {Harding}, Albert and {Hart}, Murdock and {Heckman}, Timothy M. and {Hirata}, Christopher M. and {Ho}, Paul and {Hope}, Stephen C. and {Hovland}, Larry and {Hsu}, Shu-Fu and {Hu}, Yen-Shan and {Huang}, Ping-Jie and {Jaquet}, Marc and {Jing}, Yipeng and {Karr}, Jennifer and {Kimura}, Masahiko and {King}, Matthew E. and {Komatsu}, Eiichiro and {Le Brun}, Vincent and {Le F{\`e}vre}, Olivier and {Le Fur}, Arnaud and {Le Mignant}, David and {Ling}, Hung-Hsu and {Loomis}, Craig P. and {Lupton}, Robert H. and {Madec}, Fabrice and {Mao}, Peter and {Marrara}, Lucas S. and {Mendes de Oliveira}, Claudia and {Minowa}, Yosuke and {Morantz}, Chaz and {Murayama}, Hitoshi and {Murray}, Graham J. and {Ohyama}, Youichi and {Orndorff}, Joseph and {Pascal}, Sandrine and {Pereira}, Jefferson M. and {Reiley}, Daniel and {Reinecke}, Martin and {Ritter}, Andreas and {Roberts}, Mitsuko and {Schwochert}, Mark A. and {Seiffert}, Michael D. and {Smee}, Stephen A. and {Sodre}, Laerte and {Spergel}, David N. and {Steinkraus}, Aaron J. and {Strauss}, Michael A. and {Surace}, Christian and {Suto}, Yasushi and {Suzuki}, Nao and {Swinbank}, John and {Tait}, Philip J. and {Takada}, Masahiro and {Tamura}, Tomonori and {Tanaka}, Yoko and {Tresse}, Laurence and {Verducci}, Orlando and {Vibert}, Didier and {Vidal}, Clement and {Wang}, Shiang-Yu and {Wen}, Chih-Yi and {Yan}, Chi-Hung and {Yasuda}, Naoki},
        title = "{Prime Focus Spectrograph (PFS) for the Subaru telescope: overview, recent progress, and future perspectives}",
    booktitle = {Ground-based and Airborne Instrumentation for Astronomy VI},
         year = 2016,
       editor = {{Evans}, Christopher J. and {Simard}, Luc and {Takami}, Hideki},
       series = {Society of Photo-Optical Instrumentation Engineers (SPIE) Conference Series},
       volume = {9908},
        month = aug,
          eid = {99081M},
        pages = {99081M},
          doi = {10.1117/12.2232103},
archivePrefix = {arXiv},
       eprint = {1608.01075},
 primaryClass = {astro-ph.IM},
       adsurl = {https://ui.adsabs.harvard.edu/abs/2016SPIE.9908E..1MT}
}

@ARTICLE{trenti2010,
       author = {{Trenti}, M. and {Stiavelli}, M. and {Bouwens}, R.~J. and {Oesch}, P. and {Shull}, J.~M. and {Illingworth}, G.~D. and {Bradley}, L.~D. and {Carollo}, C.~M.},
        title = "{The Galaxy Luminosity Function During the Reionization Epoch}",
      journal = {\apjl},
         year = 2010,
        month = may,
       volume = {714},
       number = {2},
        pages = {L202-L207},
          doi = {10.1088/2041-8205/714/2/L202},
archivePrefix = {arXiv},
       eprint = {1004.0384},
 primaryClass = {astro-ph.CO},
       adsurl = {https://ui.adsabs.harvard.edu/abs/2010ApJ...714L.202T}
}

@ARTICLE{viel2004a,
       author = {{Viel}, M. and {Matarrese}, S. and {Heavens}, A. and {Haehnelt}, M.~G. and
         {Kim}, T. -S. and {Springel}, V. and {Hernquist}, L.},
        title = "{The bispectrum of the Lyman {\ensuremath{\alpha}} forest at z\raisebox{-0.5ex}\textasciitilde 2-2.4 from a large sample of UVES QSO absorption spectra (LUQAS)}",
      journal = {\mnras},
         year = "2004",
        month = "Jan",
       volume = {347},
       number = {2},
        pages = {L26-L30},
          doi = {10.1111/j.1365-2966.2004.07404.x},
archivePrefix = {arXiv},
       eprint = {astro-ph/0308151},
 primaryClass = {astro-ph},
       adsurl = {https://ui.adsabs.harvard.edu/abs/2004MNRAS.347L..26V}
}

@ARTICLE{weinberger2019,
       author = {{Weinberger}, Lewis H. and {Haehnelt}, Martin G. and {Kulkarni}, Girish},
        title = "{Modelling the observed luminosity function and clustering evolution of Ly {\ensuremath{\alpha}} emitters: growing evidence for late reionization}",
      journal = {\mnras},
         year = 2019,
        month = may,
       volume = {485},
       number = {1},
        pages = {1350-1366},
          doi = {10.1093/mnras/stz481},
archivePrefix = {arXiv},
       eprint = {1902.05077},
 primaryClass = {astro-ph.GA},
       adsurl = {https://ui.adsabs.harvard.edu/abs/2019MNRAS.485.1350W}
}

@ARTICLE{chardin2017,
   author = {{Chardin}, J. and {Puchwein}, E. and {Haehnelt}, M.~G.},
    title = "{Large-scale opacity fluctuations in the Ly{$\alpha$} forest: evidence for QSOs dominating the ionizing UV background at z {\sim} 5.5-6?}",
  journal = {\mnras},
archivePrefix = "arXiv",
   eprint = {1606.08231},
     year = 2017,
    month = mar,
   volume = 465,
    pages = {3429-3445},
      doi = {10.1093/mnras/stw2943},
   adsurl = {http://adsabs.harvard.edu/abs/2017MNRAS.465.3429C}
}

@article{Harish2022,
  author = {Harish, S. and Zheng, Z. Y. and Hu, W. and Song, M. and Jiang, L. and Malhotra, S. and Rhoads, J. E. and Walker, A. and others},
  title = {LAGER: A Large Sample of Ly$\alpha$ Emitters at $z = 6.9$},
  journal = {The Astrophysical Journal},
  volume = {926},
  number = {2},
  pages = {153},
  year = {2022},
  doi = {10.3847/1538-4357/ac447b}
}

@ARTICLE{Hu2019,
       author = {{Hu}, Weida and {Wang}, Junxian and {Zheng}, Zhen-Ya and {Malhotra}, Sangeeta and {Rhoads}, James E. and {Infante}, Leopoldo and {Barrientos}, L. Felipe and {Yang}, Huan and {Jiang}, Chunyan and {Kang}, Wenyong and {Perez}, Lucia A. and {Wold}, Isak and {Hibon}, Pascale and {Jiang}, Linhua and {Khostovan}, Ali Ahmad and {Valdes}, Francisco and {Walker}, Alistair R. and {Galaz}, Gaspar and {Coughlin}, Alicia and {Harish}, Santosh and {Kong}, Xu and {Pharo}, John and {Zheng}, XianZhong},
        title = "{The Ly{\ensuremath{\alpha}} Luminosity Function and Cosmic Reionization at z {\ensuremath{\sim}} 7.0: A Tale of Two LAGER Fields}",
      journal = {\apj},
         year = 2019,
        month = dec,
       volume = {886},
       number = {2},
          eid = {90},
        pages = {90},
          doi = {10.3847/1538-4357/ab4cf4},
archivePrefix = {arXiv},
       eprint = {1903.09046},
 primaryClass = {astro-ph.GA},
       adsurl = {https://ui.adsabs.harvard.edu/abs/2019ApJ...886...90H}
}

@ARTICLE{Witstok2025,
       author = {{Witstok}, Joris and {Jakobsen}, Peter and {Maiolino}, Roberto and {Helton}, Jakob M. and {Johnson}, Benjamin D. and {Robertson}, Brant E. and {Tacchella}, Sandro and {Cameron}, Alex J. and {Smit}, Renske and {Bunker}, Andrew J. and {Saxena}, Aayush and {Sun}, Fengwu and {Alberts}, Stacey and {Arribas}, Santiago and {Baker}, William M. and {Bhatawdekar}, Rachana and {Boyett}, Kristan and {Cargile}, Phillip A. and {Carniani}, Stefano and {Charlot}, St{\'e}phane and {Chevallard}, Jacopo and {Curti}, Mirko and {Curtis-Lake}, Emma and {D'Eugenio}, Francesco and {Eisenstein}, Daniel J. and {Hainline}, Kevin N. and {Jones}, Gareth C. and {Kumari}, Nimisha and {Maseda}, Michael V. and {P{\'e}rez-Gonz{\'a}lez}, Pablo G. and {Rinaldi}, Pierluigi and {Scholtz}, Jan and {{\"U}bler}, Hannah and {Williams}, Christina C. and {Willmer}, Christopher N.~A. and {Willott}, Chris and {Zhu}, Yongda},
        title = "{Witnessing the onset of reionization through Lyman-{\ensuremath{\alpha}} emission at redshift 13}",
      journal = {\nat},
         year = 2025,
        month = mar,
       volume = {639},
       number = {8056},
        pages = {897-901},
          doi = {10.1038/s41586-025-08779-5},
archivePrefix = {arXiv},
       eprint = {2408.16608},
 primaryClass = {astro-ph.GA},
       adsurl = {https://ui.adsabs.harvard.edu/abs/2025Natur.639..897W}
}

@ARTICLE{Mesinger2016,
       author = {{Mesinger}, Andrei and {Greig}, Bradley and {Sobacchi}, Emanuele},
        title = "{The Evolution Of 21 cm Structure (EOS): public, large-scale simulations of Cosmic Dawn and reionization}",
      journal = {\mnras},
         year = 2016,
        month = jul,
       volume = {459},
       number = {3},
        pages = {2342-2353},
          doi = {10.1093/mnras/stw831},
archivePrefix = {arXiv},
       eprint = {1602.07711},
 primaryClass = {astro-ph.CO},
       adsurl = {https://ui.adsabs.harvard.edu/abs/2016MNRAS.459.2342M}
}

@ARTICLE{Naidu2020,
       author = {{Naidu}, Rohan P. and {Tacchella}, Sandro and {Mason}, Charlotte A. and {Bose}, Sownak and {Oesch}, Pascal A. and {Conroy}, Charlie},
        title = "{Rapid Reionization by the Oligarchs: The Case for Massive, UV-bright, Star-forming Galaxies with High Escape Fractions}",
      journal = {\apj},
         year = 2020,
        month = apr,
       volume = {892},
       number = {2},
          eid = {109},
        pages = {109},
          doi = {10.3847/1538-4357/ab7cc9},
archivePrefix = {arXiv},
       eprint = {1907.13130},
 primaryClass = {astro-ph.GA},
       adsurl = {https://ui.adsabs.harvard.edu/abs/2020ApJ...892..109N}
}

@ARTICLE{Weinberger2018,
       author = {{Weinberger}, Lewis H. and {Kulkarni}, Girish and {Haehnelt}, Martin G. and {Choudhury}, Tirthankar Roy and {Puchwein}, Ewald},
        title = "{Lyman-{\ensuremath{\alpha}} emitters gone missing: the different evolution of the bright and faint populations}",
      journal = {\mnras},
         year = 2018,
        month = sep,
       volume = {479},
       number = {2},
        pages = {2564-2587},
          doi = {10.1093/mnras/sty1563},
archivePrefix = {arXiv},
       eprint = {1803.03789},
 primaryClass = {astro-ph.GA},
       adsurl = {https://ui.adsabs.harvard.edu/abs/2018MNRAS.479.2564W}
}

@ARTICLE{Hutter2015,
       author = {{Hutter}, Anne and {Dayal}, Pratika and {M{\"u}ller}, Volker},
        title = "{Clustering and lifetime of Lyman Alpha Emitters in the Epoch of Reionization}",
      journal = {\mnras},
         year = 2015,
        month = jul,
       volume = {450},
       number = {4},
        pages = {4025-4034},
          doi = {10.1093/mnras/stv929},
archivePrefix = {arXiv},
       eprint = {1503.05201},
 primaryClass = {astro-ph.CO},
       adsurl = {https://ui.adsabs.harvard.edu/abs/2015MNRAS.450.4025H}
}

@ARTICLE{Raste2023,
       author = {{Raste}, Janakee and {Kulkarni}, Girish and {Watkinson}, Catherine A. and {Keating}, Laura C. and {Haehnelt}, Martin G.},
        title = "{The 21-cm bispectrum from neutral hydrogen islands at z < 6}",
      journal = {\mnras},
         year = 2024,
        month = mar,
       volume = {529},
       number = {1},
        pages = {129-140},
          doi = {10.1093/mnras/stae492},
archivePrefix = {arXiv},
       eprint = {2308.09744},
 primaryClass = {astro-ph.CO},
       adsurl = {https://ui.adsabs.harvard.edu/abs/2024MNRAS.529..129R}
}

@ARTICLE{Greig2013,
       author = {{Greig}, Bradley and {Komatsu}, Eiichiro and {Wyithe}, J. Stuart B.},
        title = "{Cosmology from clustering of Ly{\ensuremath{\alpha}} galaxies: breaking non-gravitational Ly{\ensuremath{\alpha}} radiative transfer degeneracies using the bispectrum}",
      journal = {\mnras},
         year = 2013,
        month = may,
       volume = {431},
       number = {2},
        pages = {1777-1794},
          doi = {10.1093/mnras/stt292},
archivePrefix = {arXiv},
       eprint = {1212.0977},
 primaryClass = {astro-ph.CO},
       adsurl = {https://ui.adsabs.harvard.edu/abs/2013MNRAS.431.1777G}
}

@ARTICLE{Matthee2023EIGER,
       author = {{Matthee}, Jorryt and {Mackenzie}, Ruari and {Simcoe}, Robert A. and {Kashino}, Daichi and {Lilly}, Simon J. and {Bordoloi}, Rongmon and {Eilers}, Anna-Christina},
        title = "{EIGER. II. First Spectroscopic Characterization of the Young Stars and Ionized Gas Associated with Strong H{\ensuremath{\beta}} and [O III] Line Emission in Galaxies at z = 5-7 with JWST}",
      journal = {\apj},
         year = 2023,
        month = jun,
       volume = {950},
       number = {1},
          eid = {67},
        pages = {67},
          doi = {10.3847/1538-4357/acc846},
archivePrefix = {arXiv},
       eprint = {2211.08255},
 primaryClass = {astro-ph.GA},
       adsurl = {https://ui.adsabs.harvard.edu/abs/2023ApJ...950...67M}
}

@article{Wang2023ASPIRE,
  author = {Wang, Feige and Yang, Jinyi and Fan, Xiaohui and Bañados, Eduardo and Davies, Frederick B. and Hennawi, Joseph F. and Venemans, Bram and Willott, Chris J. and others},
  title = {A SPectroscopic survey of biased halos In the Reionization Era (ASPIRE): JWST Reveals a Filamentary Structure around a $z = 6.61$ Quasar},
  journal = {The Astrophysical Journal Letters},
  volume = {946},
  number = {1},
  pages = {L10},
  year = {2023},
  doi = {10.3847/2041-8213/accd6f}
}

@ARTICLE{Bosman2022,
       author = {{Bosman}, Sarah E.~I. and {Davies}, Frederick B. and {Becker}, George D. and {Keating}, Laura C. and {Davies}, Rebecca L. and {Zhu}, Yongda and {Eilers}, Anna-Christina and {D'Odorico}, Valentina and {Bian}, Fuyan and {Bischetti}, Manuela and {Cristiani}, Stefano V. and {Fan}, Xiaohui and {Farina}, Emanuele P. and {Haehnelt}, Martin G. and {Hennawi}, Joseph F. and {Kulkarni}, Girish and {Mesinger}, Andrei and {Meyer}, Romain A. and {Onoue}, Masafusa and {Pallottini}, Andrea and {Qin}, Yuxiang and {Ryan-Weber}, Emma and {Schindler}, Jan-Torge and {Walter}, Fabian and {Wang}, Feige and {Yang}, Jinyi},
        title = "{Hydrogen reionization ends by z = 5.3: Lyman-{\ensuremath{\alpha}} optical depth measured by the XQR-30 sample}",
      journal = {\mnras},
         year = 2022,
        month = jul,
       volume = {514},
       number = {1},
        pages = {55-76},
          doi = {10.1093/mnras/stac1046},
archivePrefix = {arXiv},
       eprint = {2108.03699},
 primaryClass = {astro-ph.CO},
       adsurl = {https://ui.adsabs.harvard.edu/abs/2022MNRAS.514...55B}
}

@ARTICLE{Cirasuolo2020,
       author = {{Cirasuolo}, M. and {Fairley}, A. and {Rees}, P. and {Gonzalez}, O.~A. and {Taylor}, W. and {Maiolino}, R. and {Afonso}, J. and {Evans}, C. and {Flores}, H. and {Lilly}, S. and {Oliva}, E. and {Paltani}, S. and {Vanzi}, L. and {Abreu}, M. and {Accardo}, M. and {Adams}, N. and {{\'A}lvarez M{\'e}ndez}, D. and {Amans}, J. -P. and {Amarantidis}, S. and {Atek}, H. and {Atkinson}, D. and {Banerji}, M. and {Barrett}, J. and {Barrientos}, F. and {Bauer}, F. and {Beard}, S. and {B{\'e}chet}, C. and {Belfiore}, A. and {Bellazzini}, M. and {Benoist}, C. and {Best}, P. and {Biazzo}, K. and {Black}, M. and {Boettger}, D. and {Bonifacio}, P. and {Bowler}, R. and {Bragaglia}, A. and {Brierley}, S. and {Brinchmann}, J. and {Brinkmann}, M. and {Buat}, V. and {Buitrago}, F. and {Burgarella}, D. and {Burningham}, B. and {Buscher}, D. and {Cabral}, A. and {Caffau}, E. and {Cardoso}, L. and {Carnall}, A. and {Carollo}, M. and {Castillo}, R. and {Castignani}, G. and {Catelan}, M. and {Cicone}, C. and {Cimatti}, A. and {Cioni}, M. -R.~L. and {Clementini}, G. and {Cochrane}, W. and {Coelho}, J. and {Colling}, M. and {Contini}, T. and {Contreras}, R. and {Conzelmann}, R. and {Cresci}, G. and {Cropper}, M. and {Cucciati}, O. and {Cullen}, F. and {Cumani}, C. and {Curti}, M. and {Da Silva}, A. and {Daddi}, E. and {Dalessandro}, E. and {Dalessio}, F. and {Dauvin}, L. and {Davidson}, G. and {de Laverny}, P. and {Delplancke-Str{\"o}bele}, F. and {De Lucia}, G. and {Del Vecchio}, C. and {Dessauges-Zavadsky}, M. and {Di Matteo}, P. and {Dole}, H. and {Drass}, H. and {Dunlop}, J. and {D{\"u}nner}, R. and {Eales}, S. and {Ellis}, R. and {Enriques}, B. and {Fasola}, G. and {Ferguson}, A. and {Ferruzzi}, D. and {Fisher}, M. and {Flores}, M. and {Fontana}, A. and {Forchi}, V. and {Francois}, P. and {Franzetti}, P. and {Gargiulo}, A. and {Garilli}, B. and {Gaudemard}, J. and {Gieles}, M. and {Gilmore}, G. and {Ginolfi}, M. and {Gomes}, J.~M. and {Guinouard}, I. and {Gutierrez}, P. and {Haigron}, R. and {Hammer}, F. and {Hammersley}, P. and {Haniff}, C. and {Harrison}, C. and {Haywood}, M. and {Hill}, V. and {Hubin}, N. and {Humphrey}, A. and {Ibata}, R. and {Infante}, L. and {Ives}, D. and {Ivison}, R. and {Iwert}, O. and {Jablonka}, P. and {Jakob}, G. and {Jarvis}, M. and {King}, D. and {Kneib}, J. -P. and {Laporte}, P. and {Lawrence}, A. and {Lee}, D. and {Li Causi}, G. and {Lorenzoni}, S. and {Lucatello}, S. and {Luco}, Y. and {Macleod}, A. and {Magliocchetti}, M. and {Magrini}, L. and {Mainieri}, V. and {Maire}, C. and {Mannucci}, F. and {Martin}, N. and {Matute}, I. and {Maurogordato}, S. and {McGee}, S. and {Mcleod}, D. and {McLure}, R. and {McMahon}, R. and {Melse}, B. -T. and {Messias}, H. and {Mucciarelli}, A. and {Nisini}, B. and {Nix}, J. and {Norberg}, P. and {Oesch}, P. and {Oliveira}, A. and {Origlia}, L. and {Padilla}, N. and {Palsa}, R. and {Pancino}, E. and {Papaderos}, P. and {Pappalardo}, C. and {Parry}, I. and {Pasquini}, L. and {Peacock}, J. and {Pedichini}, F. and {Pello}, R. and {Peng}, Y. and {Pentericci}, L. and {Pfuhl}, O. and {Piazzesi}, R. and {Popovic}, D. and {Pozzetti}, L. and {Puech}, M. and {Puzia}, T. and {Raichoor}, A. and {Randich}, S. and {Recio-Blanco}, A. and {Reis}, S. and {Reix}, F. and {Renzini}, A. and {Rodrigues}, M. and {Rojas}, F. and {Rojas-Arriagada}, {\'A}. and {Rota}, S. and {Royer}, F. and {Sacco}, G. and {Sanchez-Janssen}, R. and {Sanna}, N. and {Santos}, P. and {Sarzi}, M. and {Schaerer}, D. and {Schiavon}, R. and {Schnell}, R. and {Schultheis}, M. and {Scodeggio}, M. and {Serjeant}, S. and {Shen}, T. -C. and {Simmonds}, C. and {Smoker}, J. and {Sobral}, D. and {Sordet}, M. and {Sp{\'e}rone}, D.},
        title = "{MOONS: The New Multi-Object Spectrograph for the VLT}",
      journal = {The Messenger},
         year = 2020,
        month = jun,
       volume = {180},
        pages = {10-17},
          doi = {10.18727/0722-6691/5195},
archivePrefix = {arXiv},
       eprint = {2009.00628},
 primaryClass = {astro-ph.IM},
       adsurl = {https://ui.adsabs.harvard.edu/abs/2020Msngr.180...10C}
}

@ARTICLE{2023ApJ...947L..24M,
       author = {{Morishita}, Takahiro and {Roberts-Borsani}, Guido and {Treu}, Tommaso and {Brammer}, Gabriel and {Mason}, Charlotte A. and {Trenti}, Michele and {Vulcani}, Benedetta and {Wang}, Xin and {Acebron}, Ana and {Bah{\'e}}, Yannick and {Bergamini}, Pietro and {Boyett}, Kristan and {Bradac}, Marusa and {Calabr{\`o}}, Antonello and {Castellano}, Marco and {Chen}, Wenlei and {De Lucia}, Gabriella and {Filippenko}, Alexei V. and {Fontana}, Adriano and {Glazebrook}, Karl and {Grillo}, Claudio and {Henry}, Alaina and {Jones}, Tucker and {Kelly}, Patrick L. and {Koekemoer}, Anton M. and {Leethochawalit}, Nicha and {Lu}, Ting-Yi and {Marchesini}, Danilo and {Mascia}, Sara and {Mercurio}, Amata and {Merlin}, Emiliano and {Metha}, Benjamin and {Nanayakkara}, Themiya and {Nonino}, Mario and {Paris}, Diego and {Pentericci}, Laura and {Rosati}, Piero and {Santini}, Paola and {Strait}, Victoria and {Vanzella}, Eros and {Windhorst}, Rogier A. and {Xie}, Lizhi},
        title = "{Early Results from GLASS-JWST. XIV. A Spectroscopically Confirmed Protocluster 650 Million Years after the Big Bang}",
      journal = {\apjl},
         year = 2023,
        month = apr,
       volume = {947},
       number = {2},
          eid = {L24},
        pages = {L24},
          doi = {10.3847/2041-8213/acb99e},
archivePrefix = {arXiv},
       eprint = {2211.09097},
 primaryClass = {astro-ph.GA},
       adsurl = {https://ui.adsabs.harvard.edu/abs/2023ApJ...947L..24M}
}

@ARTICLE{2023ApJ...949L..40B,
       author = {{Bruton}, Sean and {Lin}, Yu-Heng and {Scarlata}, Claudia and {Hayes}, Matthew J.},
        title = "{The Universe is at Most 88\% Neutral at z = 10.6}",
      journal = {\apjl},
         year = 2023,
        month = jun,
       volume = {949},
       number = {2},
          eid = {L40},
        pages = {L40},
          doi = {10.3847/2041-8213/acd5d0},
archivePrefix = {arXiv},
       eprint = {2303.03419},
 primaryClass = {astro-ph.GA},
       adsurl = {https://ui.adsabs.harvard.edu/abs/2023ApJ...949L..40B}
}

@ARTICLE{2015MNRAS.446..566M,
       author = {{Mesinger}, Andrei and {Aykutalp}, Aycin and {Vanzella}, Eros and {Pentericci}, Laura and {Ferrara}, Andrea and {Dijkstra}, Mark},
        title = "{Can the intergalactic medium cause a rapid drop in Ly{\ensuremath{\alpha}} emission at z > 6?}",
      journal = {\mnras},
         year = 2015,
        month = jan,
       volume = {446},
       number = {1},
        pages = {566-577},
          doi = {10.1093/mnras/stu2089},
archivePrefix = {arXiv},
       eprint = {1406.6373},
 primaryClass = {astro-ph.CO},
       adsurl = {https://ui.adsabs.harvard.edu/abs/2015MNRAS.446..566M}
}

@ARTICLE{2021ApJ...923..229G,
       author = {{Goto}, Hinako and {Shimasaku}, Kazuhiro and {Yamanaka}, Satoshi and {Momose}, Rieko and {Ando}, Makoto and {Harikane}, Yuichi and {Hashimoto}, Takuya and {Inoue}, Akio K. and {Ouchi}, Masami},
        title = "{SILVERRUSH. XI. Constraints on the Ly{\ensuremath{\alpha}} Luminosity Function and Cosmic Reionization at z = 7.3 with Subaru/Hyper Suprime-Cam}",
      journal = {\apj},
         year = 2021,
        month = dec,
       volume = {923},
       number = {2},
          eid = {229},
        pages = {229},
          doi = {10.3847/1538-4357/ac308b},
archivePrefix = {arXiv},
       eprint = {2110.14474},
 primaryClass = {astro-ph.GA},
       adsurl = {https://ui.adsabs.harvard.edu/abs/2021ApJ...923..229G}
}

@ARTICLE{2018PASJ...70...55I,
       author = {{Inoue}, Akio K. and {Hasegawa}, Kenji and {Ishiyama}, Tomoaki and {Yajima}, Hidenobu and {Shimizu}, Ikkoh and {Umemura}, Masayuki and {Konno}, Akira and {Harikane}, Yuichi and {Shibuya}, Takatoshi and {Ouchi}, Masami and {Shimasaku}, Kazuhiro and {Ono}, Yoshiaki and {Kusakabe}, Haruka and {Higuchi}, Ryo and {Lee}, Chien-Hsiu},
        title = "{SILVERRUSH. VI. A simulation of Ly{\ensuremath{\alpha}} emitters in the reionization epoch and a comparison with Subaru Hyper Suprime-Cam survey early data}",
      journal = {\pasj},
         year = 2018,
        month = jun,
       volume = {70},
       number = {3},
          eid = {55},
        pages = {55},
          doi = {10.1093/pasj/psy048},
archivePrefix = {arXiv},
       eprint = {1801.00067},
 primaryClass = {astro-ph.GA},
       adsurl = {https://ui.adsabs.harvard.edu/abs/2018PASJ...70...55I}
}

@ARTICLE{2024ApJ...971..124U,
       author = {{Umeda}, Hiroya and {Ouchi}, Masami and {Nakajima}, Kimihiko and {Harikane}, Yuichi and {Ono}, Yoshiaki and {Xu}, Yi and {Isobe}, Yuki and {Zhang}, Yechi},
        title = "{JWST Measurements of Neutral Hydrogen Fractions and Ionized Bubble Sizes at z = 7{\textendash}12 Obtained with Ly{\ensuremath{\alpha}} Damping Wing Absorptions in 27 Bright Continuum Galaxies}",
      journal = {\apj},
         year = 2024,
        month = aug,
       volume = {971},
       number = {2},
          eid = {124},
        pages = {124},
          doi = {10.3847/1538-4357/ad554e},
archivePrefix = {arXiv},
       eprint = {2306.00487},
 primaryClass = {astro-ph.GA},
       adsurl = {https://ui.adsabs.harvard.edu/abs/2024ApJ...971..124U}
}

@ARTICLE{Maitra2025,
       author = {{Maitra}, Soumak and {Kulkarni}, Girish and {Asthana}, Shikhar and {Bolton}, James S. and {Haehnelt}, Martin G. and {Keating}, Laura},
        title = "{The Lyman {\ensuremath{\alpha}} emitter bispectrum as a probe of reionization morphology}",
      journal = {\mnras},
         year = 2025,
        month = sep,
       volume = {542},
       number = {1},
        pages = {486-507},
          doi = {10.1093/mnras/staf1262},
archivePrefix = {arXiv},
       eprint = {2505.17188},
 primaryClass = {astro-ph.CO},
       adsurl = {https://ui.adsabs.harvard.edu/abs/2025MNRAS.542..486M}
}

@INPROCEEDINGS{koopmans2015,
       author = {{Koopmans}, L. and {Pritchard}, J. and {Mellema}, G. and {Aguirre}, J. and {Ahn}, K. and {Barkana}, R. and {van Bemmel}, I. and {Bernardi}, G. and {Bonaldi}, A. and {Briggs}, F. and {de Bruyn}, A.~G. and {Chang}, T.~C. and {Chapman}, E. and {Chen}, X. and {Ciardi}, B. and {Dayal}, P. and {Ferrara}, A. and {Fialkov}, A. and {Fiore}, F. and {Ichiki}, K. and {Illiev}, I.~T. and {Inoue}, S. and {Jelic}, V. and {Jones}, M. and {Lazio}, J. and {Maio}, U. and {Majumdar}, S. and {Mack}, K.~J. and {Mesinger}, A. and {Morales}, M.~F. and {Parsons}, A. and {Pen}, U.~L. and {Santos}, M. and {Schneider}, R. and {Semelin}, B. and {de Souza}, R.~S. and {Subrahmanyan}, R. and {Takeuchi}, T. and {Vedantham}, H. and {Wagg}, J. and {Webster}, R. and {Wyithe}, S. and {Datta}, K.~K. and {Trott}, C.},
        title = "{The Cosmic Dawn and Epoch of Reionisation with SKA}",
    booktitle = {Advancing Astrophysics with the Square Kilometre Array (AASKA14)},
         year = 2015,
        month = apr,
          eid = {1},
        pages = {1},
          doi = {10.22323/1.215.0001},
archivePrefix = {arXiv},
       eprint = {1505.07568},
 primaryClass = {astro-ph.CO},
       adsurl = {https://ui.adsabs.harvard.edu/abs/2015aska.confE...1K}
}

@ARTICLE{Beardsley2016,
       author = {{Beardsley}, A.~P. and {Hazelton}, B.~J. and {Sullivan}, I.~S. and {Carroll}, P. and {Barry}, N. and {Rahimi}, M. and {Pindor}, B. and {Trott}, C.~M. and {Line}, J. and {Jacobs}, Daniel C. and {Morales}, M.~F. and {Pober}, J.~C. and {Bernardi}, G. and {Bowman}, Judd D. and {Busch}, M.~P. and {Briggs}, F. and {Cappallo}, R.~J. and {Corey}, B.~E. and {de Oliveira-Costa}, A. and {Dillon}, Joshua S. and {Emrich}, D. and {Ewall-Wice}, A. and {Feng}, L. and {Gaensler}, B.~M. and {Goeke}, R. and {Greenhill}, L.~J. and {Hewitt}, J.~N. and {Hurley-Walker}, N. and {Johnston-Hollitt}, M. and {Kaplan}, D.~L. and {Kasper}, J.~C. and {Kim}, H.~S. and {Kratzenberg}, E. and {Lenc}, E. and {Loeb}, A. and {Lonsdale}, C.~J. and {Lynch}, M.~J. and {McKinley}, B. and {McWhirter}, S.~R. and {Mitchell}, D.~A. and {Morgan}, E. and {Neben}, A.~R. and {Thyagarajan}, Nithyanandan and {Oberoi}, D. and {Offringa}, A.~R. and {Ord}, S.~M. and {Paul}, S. and {Prabu}, T. and {Procopio}, P. and {Riding}, J. and {Rogers}, A.~E.~E. and {Roshi}, A. and {Udaya Shankar}, N. and {Sethi}, Shiv K. and {Srivani}, K.~S. and {Subrahmanyan}, R. and {Tegmark}, M. and {Tingay}, S.~J. and {Waterson}, M. and {Wayth}, R.~B. and {Webster}, R.~L. and {Whitney}, A.~R. and {Williams}, A. and {Williams}, C.~L. and {Wu}, C. and {Wyithe}, J.~S.~B.},
        title = "{First Season MWA EoR Power spectrum Results at Redshift 7}",
      journal = {\apj},
         year = 2016,
        month = dec,
       volume = {833},
       number = {1},
          eid = {102},
        pages = {102},
          doi = {10.3847/1538-4357/833/1/102},
archivePrefix = {arXiv},
       eprint = {1608.06281},
 primaryClass = {astro-ph.IM},
       adsurl = {https://ui.adsabs.harvard.edu/abs/2016ApJ...833..102B}
}

@ARTICLE{Kubota2018,
       author = {{Kubota}, Kenji and {Yoshiura}, Shintaro and {Takahashi}, Keitaro and {Hasegawa}, Kenji and {Yajima}, Hidenobu and {Ouchi}, Masami and {Pindor}, B. and {Webster}, R.~L.},
        title = "{Detectability of the 21-cm signal during the epoch of reionization with 21-cm Lyman {\ensuremath{\alpha}} emitter cross-correlation - I}",
      journal = {\mnras},
         year = 2018,
        month = sep,
       volume = {479},
       number = {2},
        pages = {2754-2766},
          doi = {10.1093/mnras/sty1471},
archivePrefix = {arXiv},
       eprint = {1708.06291},
 primaryClass = {astro-ph.CO},
       adsurl = {https://ui.adsabs.harvard.edu/abs/2018MNRAS.479.2754K}
}

@ARTICLE{Gillet2019,
       author = {{Gillet}, Nicolas and {Mesinger}, Andrei and {Greig}, Bradley and {Liu}, Adrian and {Ucci}, Graziano},
        title = "{Deep learning from 21-cm tomography of the cosmic dawn and reionization}",
      journal = {\mnras},
         year = 2019,
        month = mar,
       volume = {484},
       number = {1},
        pages = {282-293},
          doi = {10.1093/mnras/stz010},
archivePrefix = {arXiv},
       eprint = {1805.02699},
 primaryClass = {astro-ph.CO},
       adsurl = {https://ui.adsabs.harvard.edu/abs/2019MNRAS.484..282G}
}

@ARTICLE{LaPlante2019,
       author = {{La Plante}, Paul and {Ntampaka}, Michelle},
        title = "{Machine Learning Applied to the Reionization History of the Universe in the 21 cm Signal}",
      journal = {\apj},
         year = 2019,
        month = aug,
       volume = {880},
       number = {2},
          eid = {110},
        pages = {110},
          doi = {10.3847/1538-4357/ab2983},
archivePrefix = {arXiv},
       eprint = {1810.08211},
 primaryClass = {astro-ph.CO},
       adsurl = {https://ui.adsabs.harvard.edu/abs/2019ApJ...880..110L}
}

@ARTICLE{Hassan2020,
       author = {{Hassan}, Sultan and {Andrianomena}, Sambatra and {Doughty}, Caitlin},
        title = "{Constraining the astrophysics and cosmology from 21 cm tomography using deep learning with the SKA}",
      journal = {\mnras},
         year = 2020,
        month = jun,
       volume = {494},
       number = {4},
        pages = {5761-5774},
          doi = {10.1093/mnras/staa1151},
archivePrefix = {arXiv},
       eprint = {1907.07787},
 primaryClass = {astro-ph.CO},
       adsurl = {https://ui.adsabs.harvard.edu/abs/2020MNRAS.494.5761H}
}

@ARTICLE{McCann2017ReviewInverseProblems,
       author = {{McCann}, Michael T. and {Jin}, Kyong Hwan and {Unser}, Michael},
        title = "{Convolutional Neural Networks for Inverse Problems in Imaging: A Review}",
      journal = {IEEE Signal Processing Magazine},
         year = 2017,
        month = nov,
       volume = {34},
       number = {6},
        pages = {85-95},
          doi = {10.1109/MSP.2017.2739299},
archivePrefix = {arXiv},
       eprint = {1710.04011},
 primaryClass = {eess.IV},
       adsurl = {https://ui.adsabs.harvard.edu/abs/2017ISPM...34...85M}
}

@book{Tarantola2005InverseProblemTheory,
  author       = {Tarantola, Albert},
  title        = {Inverse Problem Theory and Methods for Model Parameter Estimation},
  publisher    = {Society for Industrial and Applied Mathematics (SIAM)},
  address      = {Philadelphia, PA, USA},
  year         = {2005},
  isbn         = {978-0-89871-572-9},
  doi          = {10.1137/1.9780898717921}
}

@ARTICLE{Jasche2013BayesianLSS,
       author = {{Jasche}, Jens and {Wandelt}, Benjamin D.},
        title = "{Bayesian physical reconstruction of initial conditions from large-scale structure surveys}",
      journal = {\mnras},
         year = 2013,
        month = jun,
       volume = {432},
       number = {2},
        pages = {894-913},
          doi = {10.1093/mnras/stt449},
archivePrefix = {arXiv},
       eprint = {1203.3639},
 primaryClass = {astro-ph.CO},
       adsurl = {https://ui.adsabs.harvard.edu/abs/2013MNRAS.432..894J}
}

@ARTICLE{Stuart2010Inverse,
       author = {{Stuart}, A.~M.},
        title = "{Inverse problems: A Bayesian perspective}",
      journal = {Acta Numerica},
         year = 2010,
        month = jan,
       volume = {19},
        pages = {451-559},
          doi = {10.1017/S0962492910000061},
       adsurl = {https://ui.adsabs.harvard.edu/abs/2010AcNum..19..451S}
}

@ARTICLE{BuiThanh2013Scale,
       author = {{Bui-Thanh}, Tan and {Ghattas}, Omar and {Martin}, James and {Stadler}, Georg},
        title = "{A Computational Framework for Infinite-Dimensional Bayesian Inverse Problems Part I: The Linearized Case, with Application to Global Seismic Inversion}",
      journal = {SIAM Journal on Scientific Computing},
         year = 2013,
        month = jan,
       volume = {35},
       number = {6},
        pages = {A2494-A2523},
          doi = {10.1137/12089586X},
archivePrefix = {arXiv},
       eprint = {1308.1313},
 primaryClass = {math.NA},
       adsurl = {https://ui.adsabs.harvard.edu/abs/2013SJSC...35A2494B}
}

@ARTICLE{Jin2017DeepCT,
  author={Jin, Kyong Hwan and McCann, Michael T. and Froustey, Emmanuel and Unser, Michael},
  journal={IEEE Transactions on Image Processing}, 
  title={Deep Convolutional Neural Network for Inverse Problems in Imaging}, 
  year={2017},
  volume={26},
  number={9},
  pages={4509-4522},
  doi={10.1109/TIP.2017.2713099}}

@INPROCEEDINGS{Wang2016DeepMRI,
  author={Wang, Shanshan and Su, Zhenghang and Ying, Leslie and Peng, Xi and Zhu, Shun and Liang, Feng and Feng, Dagan and Liang, Dong},
  booktitle={2016 IEEE 13th International Symposium on Biomedical Imaging (ISBI)}, 
  title={Accelerating magnetic resonance imaging via deep learning}, 
  year={2016},
  volume={},
  number={},
  pages={514-517},
  doi={10.1109/ISBI.2016.7493320}}

@article{li2023seismicinversion,
author = {Li, Ming and Yan, Xue-song and Zhang, Ming-zhao},
year = {2023},
month = {08},
pages = {},
title = {A comprehensive review of seismic inversion based on neural networks},
volume = {16},
journal = {Earth Science Informatics},
doi = {10.1007/s12145-023-01079-4}
}

@ARTICLE{ribli2019b,
       author = {{Ribli}, Dezs{\H{o}} and {Pataki}, B{\'a}lint {\'A}rmin and {Zorrilla Matilla}, Jos{\'e} Manuel and {Hsu}, Daniel and {Haiman}, Zolt{\'a}n and {Csabai}, Istv{\'a}n},
        title = "{Weak lensing cosmology with convolutional neural networks on noisy data}",
      journal = {\mnras},
         year = 2019,
        month = dec,
       volume = {490},
       number = {2},
        pages = {1843-1860},
          doi = {10.1093/mnras/stz2610},
archivePrefix = {arXiv},
       eprint = {1902.03663},
 primaryClass = {astro-ph.CO},
       adsurl = {https://ui.adsabs.harvard.edu/abs/2019MNRAS.490.1843R}
}

@ARTICLE{Krachmalnicoff2019,
       author = {{Krachmalnicoff}, N. and {Tomasi}, M.},
        title = "{Convolutional neural networks on the HEALPix sphere: a pixel-based algorithm and its application to CMB data analysis}",
      journal = {\aap},
         year = 2019,
        month = aug,
       volume = {628},
          eid = {A129},
        pages = {A129},
          doi = {10.1051/0004-6361/201935211},
archivePrefix = {arXiv},
       eprint = {1902.04083},
 primaryClass = {astro-ph.IM},
       adsurl = {https://ui.adsabs.harvard.edu/abs/2019A&A...628A.129K}
}

@ARTICLE{Lu2023,
       author = {{Lu}, Tianhuan and {Haiman}, Zolt{\'a}n and {Li}, Xiangchong},
        title = "{Cosmological constraints from HSC survey first-year data using deep learning}",
      journal = {\mnras},
         year = 2023,
        month = may,
       volume = {521},
       number = {2},
        pages = {2050-2066},
          doi = {10.1093/mnras/stad686},
archivePrefix = {arXiv},
       eprint = {2301.01354},
 primaryClass = {astro-ph.CO},
       adsurl = {https://ui.adsabs.harvard.edu/abs/2023MNRAS.521.2050L}
}

@ARTICLE{Ronneberger2015,
       author = {{Ronneberger}, Olaf and {Fischer}, Philipp and {Brox}, Thomas},
        title = "{U-Net: Convolutional Networks for Biomedical Image Segmentation}",
      journal = {arXiv e-prints},
         year = 2015,
        month = may,
          eid = {arXiv:1505.04597},
        pages = {arXiv:1505.04597},
          doi = {10.48550/arXiv.1505.04597},
archivePrefix = {arXiv},
       eprint = {1505.04597},
 primaryClass = {cs.CV},
       adsurl = {https://ui.adsabs.harvard.edu/abs/2015arXiv150504597R}
}

@ARTICLE{Maitra2025b,
       author = {{Maitra}, Soumak and {Viel}, Matteo and {Kulkarni}, Girish},
        title = "{DeepCHART: Mapping the 3D dark matter density field from Ly$α$ forest surveys using deep learning}",
      journal = {arXiv e-prints},
         year = 2025,
        month = jun,
          eid = {arXiv:2507.00135},
        pages = {arXiv:2507.00135},
          doi = {10.48550/arXiv.2507.00135},
archivePrefix = {arXiv},
       eprint = {2507.00135},
 primaryClass = {astro-ph.CO},
       adsurl = {https://ui.adsabs.harvard.edu/abs/2025arXiv250700135M}
}

@ARTICLE{Cicek2016,
       author = {{{\c{C}}i{\c{c}}ek}, {\"O}zg{\"u}n and {Abdulkadir}, Ahmed and {Lienkamp}, Soeren S. and {Brox}, Thomas and {Ronneberger}, Olaf},
        title = "{3D U-Net: Learning Dense Volumetric Segmentation from Sparse Annotation}",
      journal = {arXiv e-prints},
         year = 2016,
        month = jun,
          eid = {arXiv:1606.06650},
        pages = {arXiv:1606.06650},
          doi = {10.48550/arXiv.1606.06650},
archivePrefix = {arXiv},
       eprint = {1606.06650},
 primaryClass = {cs.CV},
       adsurl = {https://ui.adsabs.harvard.edu/abs/2016arXiv160606650C}
}

@ARTICLE{Milletari2016VNet,
       author = {{Milletari}, Fausto and {Navab}, Nassir and {Ahmadi}, Seyed-Ahmad},
        title = "{V-Net: Fully Convolutional Neural Networks for Volumetric Medical Image Segmentation}",
      journal = {arXiv e-prints},
         year = 2016,
        month = jun,
          eid = {arXiv:1606.04797},
        pages = {arXiv:1606.04797},
          doi = {10.48550/arXiv.1606.04797},
archivePrefix = {arXiv},
       eprint = {1606.04797},
 primaryClass = {cs.CV},
       adsurl = {https://ui.adsabs.harvard.edu/abs/2016arXiv160604797M}
}

@article{Isensee2021nnUNet,
author = {Isensee, Fabian and Jaeger, Paul and Kohl, Simon and Petersen, Jens and Maier-Hein, Klaus},
year = {2021},
month = {02},
pages = {1-9},
title = {nnU-Net: a self-configuring method for deep learning-based biomedical image segmentation},
volume = {18},
journal = {Nature Methods},
doi = {10.1038/s41592-020-01008-z}
}

@ARTICLE{Du2025,
       author = {{Du}, Wenying and {Luo}, Xiaolin and {Jiang}, Zhujun and {Xiao}, Xu and {Lin}, Qiufan and {Wang}, Xin and {Wang}, Yang and {Yin}, Fenfen and {Zhang}, Le and {Li}, Xiao-Dong},
        title = "{AI-Driven Reconstruction of Large-Scale Structure from Combined Photometric and Spectroscopic Surveys}",
      journal = {arXiv e-prints},
         year = 2025,
        month = apr,
          eid = {arXiv:2504.06309},
        pages = {arXiv:2504.06309},
          doi = {10.48550/arXiv.2504.06309},
archivePrefix = {arXiv},
       eprint = {2504.06309},
 primaryClass = {astro-ph.IM},
       adsurl = {https://ui.adsabs.harvard.edu/abs/2025arXiv250406309D}
}

@ARTICLE{Aragon-Calvo2019,
       author = {{Aragon-Calvo}, M.~A.},
        title = "{Classifying the large-scale structure of the universe with deep neural networks}",
      journal = {\mnras},
         year = 2019,
        month = apr,
       volume = {484},
       number = {4},
        pages = {5771-5784},
          doi = {10.1093/mnras/stz393},
archivePrefix = {arXiv},
       eprint = {1804.00816},
 primaryClass = {astro-ph.CO},
       adsurl = {https://ui.adsabs.harvard.edu/abs/2019MNRAS.484.5771A}
}

@ARTICLE{Garcia2023,
       author = {{Garc{\'\i}a}, Cristhian and {Santa}, Camilo and {Romano}, Antonio Enea},
        title = "{Deep learning reconstruction of the large-scale structure of the Universe from luminosity distance}",
      journal = {\mnras},
         year = 2023,
        month = jan,
       volume = {518},
       number = {2},
        pages = {2241-2246},
          doi = {10.1093/mnras/stac2916},
archivePrefix = {arXiv},
       eprint = {2107.05771},
 primaryClass = {astro-ph.CO},
       adsurl = {https://ui.adsabs.harvard.edu/abs/2023MNRAS.518.2241G}
}

@ARTICLE{KingmaWelling2013,
       author = {{Kingma}, Diederik P and {Welling}, Max},
        title = "{Auto-Encoding Variational Bayes}",
      journal = {arXiv e-prints},
         year = 2013,
        month = dec,
          eid = {arXiv:1312.6114},
        pages = {arXiv:1312.6114},
          doi = {10.48550/arXiv.1312.6114},
archivePrefix = {arXiv},
       eprint = {1312.6114},
 primaryClass = {stat.ML},
       adsurl = {https://ui.adsabs.harvard.edu/abs/2013arXiv1312.6114K}
}

@InProceedings{Wu2018GroupNorm,
author="Wu, Yuxin
and He, Kaiming",
editor="Ferrari, Vittorio
and Hebert, Martial
and Sminchisescu, Cristian
and Weiss, Yair",
title="Group Normalization",
booktitle="Computer Vision -- ECCV 2018",
year="2018",
publisher="Springer International Publishing",
address="Cham",
pages="3--19",
isbn="978-3-030-01261-8"
}

@article{Srivastava2014Dropout,
  author  = {Nitish Srivastava and Geoffrey Hinton and Alex Krizhevsky and Ilya Sutskever and Ruslan Salakhutdinov},
  title   = {Dropout: A Simple Way to Prevent Neural Networks from Overfitting},
  journal = {Journal of Machine Learning Research},
  year    = {2014},
  volume  = {15},
  number  = {56},
  pages   = {1929--1958},
  url     = {http://jmlr.org/papers/v15/srivastava14a.html}
}

@ARTICLE{Tompson2015SpatialDropout,
       author = {{Tompson}, Jonathan and {Goroshin}, Ross and {Jain}, Arjun and {LeCun}, Yann and {Bregler}, Christopher},
        title = "{Efficient Object Localization Using Convolutional Networks}",
      journal = {arXiv e-prints},
         year = 2014,
        month = nov,
          eid = {arXiv:1411.4280},
        pages = {arXiv:1411.4280},
          doi = {10.48550/arXiv.1411.4280},
archivePrefix = {arXiv},
       eprint = {1411.4280},
 primaryClass = {cs.CV},
       adsurl = {https://ui.adsabs.harvard.edu/abs/2014arXiv1411.4280T}
}

@ARTICLE{Bolan2022,
       author = {{Bolan}, Patricia and {Lemaux}, Brian C. and {Mason}, Charlotte and {Brada{\v{c}}}, Maru{\v{s}}a and {Treu}, Tommaso and {Strait}, Victoria and {Pelliccia}, Debora and {Pentericci}, Laura and {Malkan}, Matthew},
        title = "{Inferring the intergalactic medium neutral fraction at z   6-8 with low-luminosity Lyman break galaxies}",
      journal = {\mnras},
         year = 2022,
        month = dec,
       volume = {517},
       number = {3},
        pages = {3263-3274},
          doi = {10.1093/mnras/stac1963},
archivePrefix = {arXiv},
       eprint = {2111.14912},
 primaryClass = {astro-ph.GA},
       adsurl = {https://ui.adsabs.harvard.edu/abs/2022MNRAS.517.3263B}
}

@ARTICLE{Gnedin2022,
       author = {{Gnedin}, Nickolay Y. and {Madau}, Piero},
        title = "{Modeling cosmic reionization}",
      journal = {Living Reviews in Computational Astrophysics},
         year = 2022,
        month = dec,
       volume = {8},
       number = {1},
          eid = {3},
        pages = {3},
          doi = {10.1007/s41115-022-00015-5},
archivePrefix = {arXiv},
       eprint = {2208.02260},
 primaryClass = {astro-ph.CO},
       adsurl = {https://ui.adsabs.harvard.edu/abs/2022LRCA....8....3G}
}

@ARTICLE{aubert2010,
       author = {{Aubert}, Dominique and {Teyssier}, Romain},
        title = "{Reionization Simulations Powered by Graphics Processing Units. I. On the Structure of the Ultraviolet Radiation Field}",
      journal = {\apj},
         year = 2010,
        month = nov,
       volume = {724},
       number = {1},
        pages = {244-266},
          doi = {10.1088/0004-637X/724/1/244},
archivePrefix = {arXiv},
       eprint = {1004.2503},
 primaryClass = {astro-ph.CO},
       adsurl = {https://ui.adsabs.harvard.edu/abs/2010ApJ...724..244A}
}

@ARTICLE{Kannan2022,
       author = {{Kannan}, R. and {Garaldi}, E. and {Smith}, A. and {Pakmor}, R. and {Springel}, V. and {Vogelsberger}, M. and {Hernquist}, L.},
        title = "{Introducing the THESAN project: radiation-magnetohydrodynamic simulations of the epoch of reionization}",
      journal = {\mnras},
         year = 2022,
        month = apr,
       volume = {511},
       number = {3},
        pages = {4005-4030},
          doi = {10.1093/mnras/stab3710},
archivePrefix = {arXiv},
       eprint = {2110.00584},
 primaryClass = {astro-ph.GA},
       adsurl = {https://ui.adsabs.harvard.edu/abs/2022MNRAS.511.4005K}
}

@ARTICLE{Yeh2023,
       author = {{Yeh}, Jessica Y.-C. and {Smith}, Aaron and {Kannan}, Rahul and {Garaldi}, Enrico and {Vogelsberger}, Mark and {Borrow}, Josh and {Pakmor}, R{\"u}diger and {Springel}, Volker and {Hernquist}, Lars},
        title = "{The THESAN project: ionizing escape fractions of reionization-era galaxies}",
      journal = {\mnras},
         year = 2023,
        month = apr,
       volume = {520},
       number = {2},
        pages = {2757-2780},
          doi = {10.1093/mnras/stad210},
archivePrefix = {arXiv},
       eprint = {2205.02238},
 primaryClass = {astro-ph.GA},
       adsurl = {https://ui.adsabs.harvard.edu/abs/2023MNRAS.520.2757Y}
}

@ARTICLE{Lin2016,
       author = {{Lin}, Yin and {Oh}, S. Peng and {Furlanetto}, Steven R. and {Sutter}, P.~M.},
        title = "{The distribution of bubble sizes during reionization}",
      journal = {\mnras},
         year = 2016,
        month = sep,
       volume = {461},
       number = {3},
        pages = {3361-3374},
          doi = {10.1093/mnras/stw1542},
archivePrefix = {arXiv},
       eprint = {1511.01506},
 primaryClass = {astro-ph.CO},
       adsurl = {https://ui.adsabs.harvard.edu/abs/2016MNRAS.461.3361L}
}

@ARTICLE{Doussot2022,
       author = {{Doussot}, Aristide and {Semelin}, Beno{\^\i}t},
        title = "{A bubble size distribution model for the Epoch of Reionization}",
      journal = {\aap},
         year = 2022,
        month = nov,
       volume = {667},
          eid = {A118},
        pages = {A118},
          doi = {10.1051/0004-6361/202244108},
archivePrefix = {arXiv},
       eprint = {2208.14044},
 primaryClass = {astro-ph.CO},
       adsurl = {https://ui.adsabs.harvard.edu/abs/2022A&A...667A.118D}
}

@ARTICLE{Gorce2019,
       author = {{Gorce}, Ad{\'e}lie and {Pritchard}, Jonathan R.},
        title = "{Studying the morphology of reionization with the triangle correlation function of phases}",
      journal = {\mnras},
         year = 2019,
        month = oct,
       volume = {489},
       number = {1},
        pages = {1321-1337},
          doi = {10.1093/mnras/stz2195},
archivePrefix = {arXiv},
       eprint = {1903.11402},
 primaryClass = {astro-ph.CO},
       adsurl = {https://ui.adsabs.harvard.edu/abs/2019MNRAS.489.1321G}
}

@ARTICLE{Giri2021betti,
       author = {{Giri}, Sambit K. and {Mellema}, Garrelt},
        title = "{Measuring the topology of reionization with Betti numbers}",
      journal = {\mnras},
         year = 2021,
        month = aug,
       volume = {505},
       number = {2},
        pages = {1863-1877},
          doi = {10.1093/mnras/stab1320},
archivePrefix = {arXiv},
       eprint = {2012.12908},
 primaryClass = {astro-ph.CO},
       adsurl = {https://ui.adsabs.harvard.edu/abs/2021MNRAS.505.1863G}
}

@ARTICLE{Shimabukuro2022,
       author = {{Shimabukuro}, Hayato and {Mao}, Yi and {Tan}, Jianrong},
        title = "{Estimation of H II Bubble Size Distribution from 21 cm Power Spectrum with Artificial Neural Networks}",
      journal = {Research in Astronomy and Astrophysics},
         year = 2022,
        month = mar,
       volume = {22},
       number = {3},
          eid = {035027},
        pages = {035027},
          doi = {10.1088/1674-4527/ac4ca3},
archivePrefix = {arXiv},
       eprint = {2002.08238},
 primaryClass = {astro-ph.CO},
       adsurl = {https://ui.adsabs.harvard.edu/abs/2022RAA....22c5027S}
}

@ARTICLE{Giri2019islands,
       author = {{Giri}, Sambit K. and {Mellema}, Garrelt and {Aldheimer}, Thomas and {Dixon}, Keri L. and {Iliev}, Ilian T.},
        title = "{Neutral island statistics during reionization from 21-cm tomography}",
      journal = {\mnras},
         year = 2019,
        month = oct,
       volume = {489},
       number = {2},
        pages = {1590-1605},
          doi = {10.1093/mnras/stz2224},
archivePrefix = {arXiv},
       eprint = {1903.01294},
 primaryClass = {astro-ph.GA},
       adsurl = {https://ui.adsabs.harvard.edu/abs/2019MNRAS.489.1590G}
}

@ARTICLE{Giri2025islands,
       author = {{Giri}, Sambit K. and {Kakiichi}, Koki and {Bianco}, Michele and {Meerburg}, P. Daniel},
        title = "{Mapping neutral islands during end stages of reionization with photometric intergalactic medium tomography}",
      journal = {\mnras},
         year = 2025,
        month = nov,
          doi = {10.1093/mnras/staf1947},
archivePrefix = {arXiv},
       eprint = {2505.06350},
 primaryClass = {astro-ph.GA},
       adsurl = {https://ui.adsabs.harvard.edu/abs/2025MNRAS.tmp.1843G}
}

@ARTICLE{Jennings2020,
       author = {{Jennings}, W.~D. and {Watkinson}, C.~A. and {Abdalla}, F.~B.},
        title = "{Analysing the Epoch of Reionization with three-point correlation functions and machine learning techniques}",
      journal = {\mnras},
         year = 2020,
        month = nov,
       volume = {498},
       number = {3},
        pages = {4518-4532},
          doi = {10.1093/mnras/staa2598},
archivePrefix = {arXiv},
       eprint = {2011.14157},
 primaryClass = {astro-ph.CO},
       adsurl = {https://ui.adsabs.harvard.edu/abs/2020MNRAS.498.4518J}
}

@ARTICLE{Witstok2024env,
       author = {{Witstok}, Joris and {Smit}, Renske and {Saxena}, Aayush and {Jones}, Gareth C. and {Helton}, Jakob M. and {Sun}, Fengwu and {Maiolino}, Roberto and {Kumari}, Nimisha and {Stark}, Daniel P. and {Bunker}, Andrew J. and {Arribas}, Santiago and {Baker}, William M. and {Bhatawdekar}, Rachana and {Boyett}, Kristan and {Cameron}, Alex J. and {Carniani}, Stefano and {Charlot}, Stephane and {Chevallard}, Jacopo and {Curti}, Mirko and {Curtis-Lake}, Emma and {Eisenstein}, Daniel J. and {Endsley}, Ryan and {Hainline}, Kevin and {Ji}, Zhiyuan and {Johnson}, Benjamin D. and {Looser}, Tobias J. and {Nelson}, Erica and {Perna}, Michele and {Rix}, Hans-Walter and {Robertson}, Brant E. and {Sandles}, Lester and {Scholtz}, Jan and {Simmonds}, Charlotte and {Tacchella}, Sandro and {{\"U}bler}, Hannah and {Williams}, Christina C. and {Willmer}, Christopher N.~A. and {Willott}, Chris},
        title = "{Inside the bubble: exploring the environments of reionisation-era Lyman-{\ensuremath{\alpha}} emitting galaxies with JADES and FRESCO}",
      journal = {\aap},
         year = 2024,
        month = feb,
       volume = {682},
          eid = {A40},
        pages = {A40},
          doi = {10.1051/0004-6361/202347176},
archivePrefix = {arXiv},
       eprint = {2306.04627},
 primaryClass = {astro-ph.GA},
       adsurl = {https://ui.adsabs.harvard.edu/abs/2024A&A...682A..40W}
}

@ARTICLE{Saxena2024,
       author = {{Saxena}, Aayush and {Bunker}, Andrew J. and {Jones}, Gareth C. and {Stark}, Daniel P. and {Cameron}, Alex J. and {Witstok}, Joris and {Arribas}, Santiago and {Baker}, William M. and {Baum}, Stefi and {Bhatawdekar}, Rachana and {Bowler}, Rebecca and {Boyett}, Kristan and {Carniani}, Stefano and {Charlot}, Stephane and {Chevallard}, Jacopo and {Curti}, Mirko and {Curtis-Lake}, Emma and {Eisenstein}, Daniel J. and {Endsley}, Ryan and {Hainline}, Kevin and {Helton}, Jakob M. and {Johnson}, Benjamin D. and {Kumari}, Nimisha and {Looser}, Tobias J. and {Maiolino}, Roberto and {Rieke}, Marcia and {Rix}, Hans-Walter and {Robertson}, Brant E. and {Sandles}, Lester and {Simmonds}, Charlotte and {Smit}, Renske and {Tacchella}, Sandro and {Williams}, Christina C. and {Willmer}, Christopher N.~A. and {Willott}, Chris},
        title = "{JADES: The production and escape of ionizing photons from faint Lyman-alpha emitters in the epoch of reionization}",
      journal = {\aap},
         year = 2024,
        month = apr,
       volume = {684},
          eid = {A84},
        pages = {A84},
          doi = {10.1051/0004-6361/202347132},
archivePrefix = {arXiv},
       eprint = {2306.04536},
 primaryClass = {astro-ph.GA},
       adsurl = {https://ui.adsabs.harvard.edu/abs/2024A&A...684A..84S}
}

@ARTICLE{Mesinger2008lae,
       author = {{Mesinger}, Andrei and {Furlanetto}, Steven R.},
        title = "{Ly{\ensuremath{\alpha}} emitters during the early stages of reionization}",
      journal = {\mnras},
         year = 2008,
        month = jun,
       volume = {386},
       number = {4},
        pages = {1990-2002},
          doi = {10.1111/j.1365-2966.2008.13039.x},
archivePrefix = {arXiv},
       eprint = {0708.0006},
 primaryClass = {astro-ph},
       adsurl = {https://ui.adsabs.harvard.edu/abs/2008MNRAS.386.1990M}
}

@ARTICLE{Trapp2023,
       author = {{Trapp}, A.~C. and {Furlanetto}, Steven R. and {Davies}, Frederick B.},
        title = "{Lyman {\ensuremath{\alpha}} emitters in ionized bubbles: constraining the environment and ionized fraction}",
      journal = {\mnras},
         year = 2023,
        month = oct,
       volume = {524},
       number = {4},
        pages = {5891-5903},
          doi = {10.1093/mnras/stad2228},
archivePrefix = {arXiv},
       eprint = {2210.06504},
 primaryClass = {astro-ph.CO},
       adsurl = {https://ui.adsabs.harvard.edu/abs/2023MNRAS.524.5891T}
}

@ARTICLE{NeyerLAE2025,
       author = {{Neyer}, Meredith and {Smith}, Aaron and {Vogelsberger}, Mark and {{\'A}ngela Garc{\'\i}a}, Luz and {Kannan}, Rahul and {Garaldi}, Enrico and {Keating}, Laura},
        title = "{The THESAN project: Lyman-alpha emitters as probes of ionized bubble sizes}",
      journal = {arXiv e-prints},
         year = 2025,
        month = oct,
          eid = {arXiv:2510.18946},
        pages = {arXiv:2510.18946},
          doi = {10.48550/arXiv.2510.18946},
archivePrefix = {arXiv},
       eprint = {2510.18946},
 primaryClass = {astro-ph.GA},
       adsurl = {https://ui.adsabs.harvard.edu/abs/2025arXiv251018946N}
}

@ARTICLE{Giri2018bsd,
       author = {{Giri}, Sambit K. and {Mellema}, Garrelt and {Dixon}, Keri L. and {Iliev}, Ilian T.},
        title = "{Bubble size statistics during reionization from 21-cm tomography}",
      journal = {\mnras},
         year = 2018,
        month = jan,
       volume = {473},
       number = {3},
        pages = {2949-2964},
          doi = {10.1093/mnras/stx2539},
archivePrefix = {arXiv},
       eprint = {1706.00665},
 primaryClass = {astro-ph.CO},
       adsurl = {https://ui.adsabs.harvard.edu/abs/2018MNRAS.473.2949G}
}

@ARTICLE{Wiersma2013,
       author = {{Wiersma}, R.~P.~C. and {Ciardi}, B. and {Thomas}, R.~M. and {Harker}, G.~J.~A. and {Zaroubi}, S. and {Bernardi}, G. and {Brentjens}, M. and {de Bruyn}, A.~G. and {Daiboo}, S. and {Jelic}, V. and {Kazemi}, S. and {Koopmans}, L.~V.~E. and {Labropoulos}, P. and {Martinez}, O. and {Mellema}, G. and {Offringa}, A. and {Pandey}, V.~N. and {Schaye}, J. and {Veligatla}, V. and {Vedantham}, H. and {Yatawatta}, S.},
        title = "{LOFAR insights into the epoch of reionization from the cross-power spectrum of 21 cm emission and galaxies}",
      journal = {\mnras},
         year = 2013,
        month = jul,
       volume = {432},
       number = {3},
        pages = {2615-2624},
          doi = {10.1093/mnras/stt624},
archivePrefix = {arXiv},
       eprint = {1209.5727},
 primaryClass = {astro-ph.CO},
       adsurl = {https://ui.adsabs.harvard.edu/abs/2013MNRAS.432.2615W}
}

@ARTICLE{Sobacchi2016,
       author = {{Sobacchi}, Emanuele and {Mesinger}, Andrei and {Greig}, Bradley},
        title = "{Cross-correlation of the cosmic 21-cm signal and Lyman {\ensuremath{\alpha}} emitters during reionization}",
      journal = {\mnras},
         year = 2016,
        month = jul,
       volume = {459},
       number = {3},
        pages = {2741-2750},
          doi = {10.1093/mnras/stw811},
archivePrefix = {arXiv},
       eprint = {1602.04837},
 primaryClass = {astro-ph.CO},
       adsurl = {https://ui.adsabs.harvard.edu/abs/2016MNRAS.459.2741S}
}

@ARTICLE{Hutter2017,
       author = {{Hutter}, Anne and {Dayal}, Pratika and {M{\"u}ller}, Volker and {Trott}, Cathryn M.},
        title = "{Exploring 21cm-Lyman Alpha Emitter Synergies for SKA}",
      journal = {\apj},
         year = 2017,
        month = feb,
       volume = {836},
       number = {2},
          eid = {176},
        pages = {176},
          doi = {10.3847/1538-4357/836/2/176},
archivePrefix = {arXiv},
       eprint = {1605.01734},
 primaryClass = {astro-ph.CO},
       adsurl = {https://ui.adsabs.harvard.edu/abs/2017ApJ...836..176H}
}

@ARTICLE{Heneka2017,
       author = {{Heneka}, Caroline and {Cooray}, Asantha and {Feng}, Chang},
        title = "{Probing the Intergalactic Medium with Ly{\ensuremath{\alpha}} and 21 cm Fluctuations}",
      journal = {\apj},
         year = 2017,
        month = oct,
       volume = {848},
       number = {1},
          eid = {52},
        pages = {52},
          doi = {10.3847/1538-4357/aa8eed},
archivePrefix = {arXiv},
       eprint = {1611.09682},
 primaryClass = {astro-ph.CO},
       adsurl = {https://ui.adsabs.harvard.edu/abs/2017ApJ...848...52H}
}

@ARTICLE{Vrbanec2020,
       author = {{Vrbanec}, Dijana and {Ciardi}, Benedetta and {Jeli{\'c}}, Vibor and {Jensen}, Hannes and {Iliev}, Ilian T. and {Mellema}, Garrelt and {Zaroubi}, Saleem},
        title = "{Predictions for the 21cm-galaxy cross-power spectrum observable with SKA and future galaxy surveys}",
      journal = {\mnras},
         year = 2020,
        month = mar,
       volume = {492},
       number = {4},
        pages = {4952-4958},
          doi = {10.1093/mnras/staa183},
archivePrefix = {arXiv},
       eprint = {2001.08814},
 primaryClass = {astro-ph.CO},
       adsurl = {https://ui.adsabs.harvard.edu/abs/2020MNRAS.492.4952V}
}

@ARTICLE{Lu2025,
       author = {{Lu}, Ting-Yi and {Mason}, Charlotte A. and {Mesinger}, Andrei and {Prelogovi{\'c}}, David and {Nikoli{\'c}}, Ivan and {Hutter}, Anne and {Gagnon-Hartman}, Samuel and {Tang}, Mengtao and {Qin}, Yuxiang and {Kakiichi}, Koki},
        title = "{Mapping reionization bubbles in JWST era: I. Empirical edge detection with Lyman alpha emission from galaxies}",
      journal = {\aap},
         year = 2025,
        month = may,
       volume = {697},
          eid = {A69},
        pages = {A69},
          doi = {10.1051/0004-6361/202452912},
archivePrefix = {arXiv},
       eprint = {2411.04176},
 primaryClass = {astro-ph.GA},
       adsurl = {https://ui.adsabs.harvard.edu/abs/2025A&A...697A..69L}
}

@ARTICLE{Bianco2021,
       author = {{Bianco}, Michele and {Giri}, Sambit K. and {Iliev}, Ilian T. and {Mellema}, Garrelt},
        title = "{Deep learning approach for identification of H II regions during reionization in 21-cm observations}",
      journal = {\mnras},
         year = 2021,
        month = aug,
       volume = {505},
       number = {3},
        pages = {3982-3997},
          doi = {10.1093/mnras/stab1518},
archivePrefix = {arXiv},
       eprint = {2102.06713},
 primaryClass = {astro-ph.IM},
       adsurl = {https://ui.adsabs.harvard.edu/abs/2021MNRAS.505.3982B}
}

@ARTICLE{Hainline2024,
       author = {{Hainline}, Kevin N. and {Johnson}, Benjamin D. and {Robertson}, Brant and {Tacchella}, Sandro and {Helton}, Jakob M. and {Sun}, Fengwu and {Eisenstein}, Daniel J. and {Simmonds}, Charlotte and {Topping}, Michael W. and {Whitler}, Lily and {Willmer}, Christopher N.~A. and {Rieke}, Marcia and {Suess}, Katherine A. and {Hviding}, Raphael E. and {Cameron}, Alex J. and {Alberts}, Stacey and {Baker}, William M. and {Baum}, Stefi and {Bhatawdekar}, Rachana and {Bonaventura}, Nina and {Boyett}, Kristan and {Bunker}, Andrew J. and {Carniani}, Stefano and {Charlot}, Stephane and {Chevallard}, Jacopo and {Chen}, Zuyi and {Curti}, Mirko and {Curtis-Lake}, Emma and {D'Eugenio}, Francesco and {Egami}, Eiichi and {Endsley}, Ryan and {Hausen}, Ryan and {Ji}, Zhiyuan and {Looser}, Tobias J. and {Lyu}, Jianwei and {Maiolino}, Roberto and {Nelson}, Erica and {Pusk{\'a}s}, D{\'a}vid and {Rawle}, Tim and {Sandles}, Lester and {Saxena}, Aayush and {Smit}, Renske and {Stark}, Daniel P. and {Williams}, Christina C. and {Willott}, Chris and {Witstok}, Joris},
        title = "{The Cosmos in Its Infancy: JADES Galaxy Candidates at z > 8 in GOODS-S and GOODS-N}",
      journal = {\apj},
         year = 2024,
        month = mar,
       volume = {964},
       number = {1},
          eid = {71},
        pages = {71},
          doi = {10.3847/1538-4357/ad1ee4},
archivePrefix = {arXiv},
       eprint = {2306.02468},
 primaryClass = {astro-ph.GA},
       adsurl = {https://ui.adsabs.harvard.edu/abs/2024ApJ...964...71H}
}

@ARTICLE{Runnholm2025,
       author = {{Runnholm}, Axel and {Hayes}, Matthew J. and {Mehta}, Vihang and {Malkan}, Matthew A. and {Scarlata}, Claudia and {Nedkova}, Kalina V. and {Rafelski}, Marc and {Vulcani}, Benedetta and {Huberty}, Mason and {Herenz}, E. Christian and {Hutter}, Anne and {Bruton}, Sean and {Acharyya}, Ayan and {Atek}, Hakim and {Baronchelli}, Ivano and {Battisti}, Andrew J. and {Brada{\v{c}}}, Maru{\v{s}}a and {Bunker}, Andrew J. and {Dai}, Y. Sophia and {Hannahs}, Clea and {Hasan}, Farhanul and {Kim}, Keunho J. and {Leethochawalit}, Nicha and {Lin}, Yu-Heng and {Rutkowski}, Michael J. and {Saldana-Lopez}, Alberto and {Sattari}, Zahra and {Wang}, Xin},
        title = "{The JWST/PASSAGE Survey: Testing Reionization Histories with JWST's First Unbiased Survey for Ly{\ensuremath{\alpha}} Emitters at Redshifts 7.5{\textendash}9.5}",
      journal = {\apj},
         year = 2025,
        month = may,
       volume = {984},
       number = {1},
          eid = {95},
        pages = {95},
          doi = {10.3847/1538-4357/adc008},
archivePrefix = {arXiv},
       eprint = {2502.19174},
 primaryClass = {astro-ph.GA},
       adsurl = {https://ui.adsabs.harvard.edu/abs/2025ApJ...984...95R}
}

@ARTICLE{Seitzer2022,
       author = {{Seitzer}, Maximilian and {Tavakoli}, Arash and {Antic}, Dimitrije and {Martius}, Georg},
        title = "{On the Pitfalls of Heteroscedastic Uncertainty Estimation with Probabilistic Neural Networks}",
      journal = {arXiv e-prints},
         year = 2022,
        month = mar,
          eid = {arXiv:2203.09168},
        pages = {arXiv:2203.09168},
          doi = {10.48550/arXiv.2203.09168},
archivePrefix = {arXiv},
       eprint = {2203.09168},
 primaryClass = {cs.LG},
       adsurl = {https://ui.adsabs.harvard.edu/abs/2022arXiv220309168S}
}

@ARTICLE{Mortlock2011,
       author = {{Mortlock}, Daniel J. and {Warren}, Stephen J. and {Venemans}, Bram P. and {Patel}, Mitesh and {Hewett}, Paul C. and {McMahon}, Richard G. and {Simpson}, Chris and {Theuns}, Tom and {Gonz{\'a}les-Solares}, Eduardo A. and {Adamson}, Andy and {Dye}, Simon and {Hambly}, Nigel C. and {Hirst}, Paul and {Irwin}, Mike J. and {Kuiper}, Ernst and {Lawrence}, Andy and {R{\"o}ttgering}, Huub J.~A.},
        title = "{A luminous quasar at a redshift of z = 7.085}",
      journal = {\nat},
         year = 2011,
        month = jun,
       volume = {474},
       number = {7353},
        pages = {616-619},
          doi = {10.1038/nature10159},
archivePrefix = {arXiv},
       eprint = {1106.6088},
 primaryClass = {astro-ph.CO},
       adsurl = {https://ui.adsabs.harvard.edu/abs/2011Natur.474..616M}
}

@ARTICLE{Banados2018,
       author = {{Ba{\~n}ados}, Eduardo and {Venemans}, Bram P. and {Mazzucchelli}, Chiara and {Farina}, Emanuele P. and {Walter}, Fabian and {Wang}, Feige and {Decarli}, Roberto and {Stern}, Daniel and {Fan}, Xiaohui and {Davies}, Frederick B. and {Hennawi}, Joseph F. and {Simcoe}, Robert A. and {Turner}, Monica L. and {Rix}, Hans-Walter and {Yang}, Jinyi and {Kelson}, Daniel D. and {Rudie}, Gwen C. and {Winters}, Jan Martin},
        title = "{An 800-million-solar-mass black hole in a significantly neutral Universe at a redshift of 7.5}",
      journal = {\nat},
         year = 2018,
        month = jan,
       volume = {553},
       number = {7689},
        pages = {473-476},
          doi = {10.1038/nature25180},
archivePrefix = {arXiv},
       eprint = {1712.01860},
 primaryClass = {astro-ph.GA},
       adsurl = {https://ui.adsabs.harvard.edu/abs/2018Natur.553..473B}
}

@ARTICLE{Kist2025,
       author = {{Kist}, Timo and {Hennawi}, Joseph F. and {Davies}, Frederick B.},
        title = "{Quantifying the precision of IGM damping wing measurements towards quasars}",
      journal = {\mnras},
         year = 2025,
        month = apr,
       volume = {538},
       number = {4},
        pages = {2704-2728},
          doi = {10.1093/mnras/staf460},
archivePrefix = {arXiv},
       eprint = {2406.12071},
 primaryClass = {astro-ph.CO},
       adsurl = {https://ui.adsabs.harvard.edu/abs/2025MNRAS.538.2704K}
}

@ARTICLE{Spina2024,
       author = {{Spina}, Benedetta and {Bosman}, Sarah E.~I. and {Davies}, Frederick B. and {Gaikwad}, Prakash and {Zhu}, Yongda},
        title = "{Damping wings in the Lyman-{\ensuremath{\alpha}} forest: A model-independent measurement of the neutral fraction at 5.4 < z < 6.1}",
      journal = {\aap},
         year = 2024,
        month = aug,
       volume = {688},
          eid = {L26},
        pages = {L26},
          doi = {10.1051/0004-6361/202450798},
archivePrefix = {arXiv},
       eprint = {2405.12273},
 primaryClass = {astro-ph.CO},
       adsurl = {https://ui.adsabs.harvard.edu/abs/2024A&A...688L..26S}
}

@ARTICLE{Zhu2024,
       author = {{Zhu}, Yongda and {Becker}, George D. and {Bosman}, Sarah E.~I. and {Cain}, Christopher and {Keating}, Laura C. and {Nasir}, Fahad and {D'Odorico}, Valentina and {Ba{\~n}ados}, Eduardo and {Bian}, Fuyan and {Bischetti}, Manuela and {Bolton}, James S. and {Chen}, Huanqing and {D'Aloisio}, Anson and {Davies}, Frederick B. and {Davies}, Rebecca L. and {Eilers}, Anna-Christina and {Fan}, Xiaohui and {Gaikwad}, Prakash and {Greig}, Bradley and {Haehnelt}, Martin G. and {Kulkarni}, Girish and {Lai}, Samuel and {Puchwein}, Ewald and {Qin}, Yuxiang and {Ryan-Weber}, Emma V. and {Satyavolu}, Sindhu and {Spina}, Benedetta and {Walter}, Fabian and {Wang}, Feige and {Wolfson}, Molly and {Yang}, Jinyi},
        title = "{Damping wing-like features in the stacked Ly {\ensuremath{\alpha}} forest: Potential neutral hydrogen islands at z < 6}",
      journal = {\mnras},
         year = 2024,
        month = sep,
       volume = {533},
       number = {1},
        pages = {L49-L56},
          doi = {10.1093/mnrasl/slae061},
archivePrefix = {arXiv},
       eprint = {2405.12275},
 primaryClass = {astro-ph.CO},
       adsurl = {https://ui.adsabs.harvard.edu/abs/2024MNRAS.533L..49Z}
}

@ARTICLE{Bosman2018,
       author = {{Bosman}, Sarah E.~I. and {Fan}, Xiaohui and {Jiang}, Linhua and {Reed}, Sophie and {Matsuoka}, Yoshiki and {Becker}, George and {Haehnelt}, Martin},
        title = "{New constraints on Lyman-{\ensuremath{\alpha}} opacity with a sample of 62 quasarsat z > 5.7}",
      journal = {\mnras},
         year = 2018,
        month = sep,
       volume = {479},
       number = {1},
        pages = {1055-1076},
          doi = {10.1093/mnras/sty1344},
archivePrefix = {arXiv},
       eprint = {1802.08177},
 primaryClass = {astro-ph.GA},
       adsurl = {https://ui.adsabs.harvard.edu/abs/2018MNRAS.479.1055B}
}

@ARTICLE{Daloisio2015,
       author = {{D'Aloisio}, Anson and {McQuinn}, Matthew and {Trac}, Hy},
        title = "{Large Opacity Variations in the High-redshift Ly{\ensuremath{\alpha}} Forest: The Signature of Relic Temperature Fluctuations from Patchy Reionization}",
      journal = {\apjl},
         year = 2015,
        month = nov,
       volume = {813},
       number = {2},
          eid = {L38},
        pages = {L38},
          doi = {10.1088/2041-8205/813/2/L38},
archivePrefix = {arXiv},
       eprint = {1509.02523},
 primaryClass = {astro-ph.CO},
       adsurl = {https://ui.adsabs.harvard.edu/abs/2015ApJ...813L..38D}
}

@ARTICLE{Sobacchi2015,
       author = {{Sobacchi}, Emanuele and {Mesinger}, Andrei},
        title = "{The clustering of Lyman {\ensuremath{\alpha}} emitters at z {\ensuremath{\approx}} 7: implications for reionization and host halo masses}",
      journal = {\mnras},
         year = 2015,
        month = oct,
       volume = {453},
       number = {2},
        pages = {1843-1854},
          doi = {10.1093/mnras/stv1751},
archivePrefix = {arXiv},
       eprint = {1505.02787},
 primaryClass = {astro-ph.CO},
       adsurl = {https://ui.adsabs.harvard.edu/abs/2015MNRAS.453.1843S}
}

@ARTICLE{Chen2025,
       author = {{Chen}, Zuyi and {Stark}, Daniel P. and {Mason}, Charlotte A. and {Tang}, Mengtao and {Whitler}, Lily and {Lu}, Ting-Yi and {Topping}, Michael W.},
        title = "{The Impact of Galaxy Overdensities and Ionized Bubbles on Ly$α$ Emission at $z\sim7.0-8.5$}",
      journal = {arXiv e-prints},
         year = 2025,
        month = may,
          eid = {arXiv:2505.24080},
        pages = {arXiv:2505.24080},
          doi = {10.48550/arXiv.2505.24080},
archivePrefix = {arXiv},
       eprint = {2505.24080},
 primaryClass = {astro-ph.GA},
       adsurl = {https://ui.adsabs.harvard.edu/abs/2025arXiv250524080C}
}

@ARTICLE{Barry2016,
       author = {{Barry}, N. and {Hazelton}, B. and {Sullivan}, I. and {Morales}, M.~F. and {Pober}, J.~C.},
        title = "{Calibration requirements for detecting the 21 cm epoch of reionization power spectrum and implications for the SKA}",
      journal = {\mnras},
         year = 2016,
        month = sep,
       volume = {461},
       number = {3},
        pages = {3135-3144},
          doi = {10.1093/mnras/stw1380},
archivePrefix = {arXiv},
       eprint = {1603.00607},
 primaryClass = {astro-ph.IM},
       adsurl = {https://ui.adsabs.harvard.edu/abs/2016MNRAS.461.3135B}
}

@ARTICLE{EwallWice2017,
       author = {{Ewall-Wice}, Aaron and {Dillon}, Joshua S. and {Liu}, Adrian and {Hewitt}, Jacqueline},
        title = "{The impact of modelling errors on interferometer calibration for 21 cm power spectra}",
      journal = {\mnras},
         year = 2017,
        month = sep,
       volume = {470},
       number = {2},
        pages = {1849-1870},
          doi = {10.1093/mnras/stx1221},
archivePrefix = {arXiv},
       eprint = {1610.02689},
 primaryClass = {astro-ph.CO},
       adsurl = {https://ui.adsabs.harvard.edu/abs/2017MNRAS.470.1849E}
}

@ARTICLE{Patil2016,
       author = {{Patil}, Ajinkya H. and {Yatawatta}, Sarod and {Zaroubi}, Saleem and {Koopmans}, L{\'e}on V.~E. and {de Bruyn}, A.~G. and {Jeli{\'c}}, Vibor and {Ciardi}, Benedetta and {Iliev}, Ilian T. and {Mevius}, Maaijke and {Pandey}, Vishambhar N. and {Gehlot}, Bharat K.},
        title = "{Systematic biases in low-frequency radio interferometric data due to calibration: the LOFAR-EoR case}",
      journal = {\mnras},
         year = 2016,
        month = dec,
       volume = {463},
       number = {4},
        pages = {4317-4330},
          doi = {10.1093/mnras/stw2277},
archivePrefix = {arXiv},
       eprint = {1605.07619},
 primaryClass = {astro-ph.IM},
       adsurl = {https://ui.adsabs.harvard.edu/abs/2016MNRAS.463.4317P}
}

@ARTICLE{Mertens2020,
       author = {{Mertens}, F.~G. and {Mevius}, M. and {Koopmans}, L.~V.~E. and {Offringa}, A.~R. and {Mellema}, G. and {Zaroubi}, S. and {Brentjens}, M.~A. and {Gan}, H. and {Gehlot}, B.~K. and {Pandey}, V.~N. and {Sardarabadi}, A.~M. and {Vedantham}, H.~K. and {Yatawatta}, S. and {Asad}, K.~M.~B. and {Ciardi}, B. and {Chapman}, E. and {Gazagnes}, S. and {Ghara}, R. and {Ghosh}, A. and {Giri}, S.~K. and {Iliev}, I.~T. and {Jeli{\'c}}, V. and {Kooistra}, R. and {Mondal}, R. and {Schaye}, J. and {Silva}, M.~B.},
        title = "{Improved upper limits on the 21 cm signal power spectrum of neutral hydrogen at z {\ensuremath{\approx}} 9.1 from LOFAR}",
      journal = {\mnras},
         year = 2020,
        month = apr,
       volume = {493},
       number = {2},
        pages = {1662-1685},
          doi = {10.1093/mnras/staa327},
archivePrefix = {arXiv},
       eprint = {2002.07196},
 primaryClass = {astro-ph.CO},
       adsurl = {https://ui.adsabs.harvard.edu/abs/2020MNRAS.493.1662M}
}

@ARTICLE{Gan2023,
       author = {{Gan}, H. and {Mertens}, F.~G. and {Koopmans}, L.~V.~E. and {Offringa}, A.~R. and {Mevius}, M. and {Pandey}, V.~N. and {Brackenhoff}, S.~A. and {Ceccotti}, E. and {Ciardi}, B. and {Gehlot}, B.~K. and {Ghara}, R. and {Giri}, S.~K. and {Iliev}, I.~T. and {Munshi}, S.},
        title = "{Assessing the impact of two independent direction-dependent calibration algorithms on the LOFAR 21 cm signal power spectrum. And applications to an observation of a field flanking the north celestial pole}",
      journal = {\aap},
         year = 2023,
        month = jan,
       volume = {669},
          eid = {A20},
        pages = {A20},
          doi = {10.1051/0004-6361/202244316},
archivePrefix = {arXiv},
       eprint = {2209.07854},
 primaryClass = {astro-ph.CO},
       adsurl = {https://ui.adsabs.harvard.edu/abs/2023A&A...669A..20G}
}

@ARTICLE{Brackenhoff2024,
       author = {{Brackenhoff}, S.~A. and {Mevius}, M. and {Koopmans}, L.~V.~E. and {Offringa}, A. and {Ceccotti}, E. and {Chege}, J.~K. and {Gehlot}, B.~K. and {Ghosh}, S. and {H{\"o}fer}, C. and {Mertens}, F.~G. and {Munshi}, S. and {Zaroubi}, S.},
        title = "{Ionospheric contributions to the excess power in high-redshift 21-cm power-spectrum observations with LOFAR}",
      journal = {\mnras},
         year = 2024,
        month = sep,
       volume = {533},
       number = {1},
        pages = {632-656},
          doi = {10.1093/mnras/stae1856},
archivePrefix = {arXiv},
       eprint = {2407.20220},
 primaryClass = {astro-ph.CO},
       adsurl = {https://ui.adsabs.harvard.edu/abs/2024MNRAS.533..632B}
}

@ARTICLE{Shibuya2018,
       author = {{Shibuya}, Takatoshi and {Ouchi}, Masami and {Konno}, Akira and {Higuchi}, Ryo and {Harikane}, Yuichi and {Ono}, Yoshiaki and {Shimasaku}, Kazuhiro and {Taniguchi}, Yoshiaki and {Kobayashi}, Masakazu A.~R. and {Kajisawa}, Masaru and {Nagao}, Tohru and {Furusawa}, Hisanori and {Goto}, Tomotsugu and {Kashikawa}, Nobunari and {Komiyama}, Yutaka and {Kusakabe}, Haruka and {Lee}, Chien-Hsiu and {Momose}, Rieko and {Nakajima}, Kimihiko and {Tanaka}, Masayuki and {Wang}, Shiang-Yu and {Yuma}, Suraphong},
        title = "{SILVERRUSH. II. First catalogs and properties of {\ensuremath{\sim}}2000 Ly{\ensuremath{\alpha}} emitters and blobs at z {\ensuremath{\sim}} 6-7 identified over the 14-21 deg$^{2}$ sky$^{*}$}",
      journal = {\pasj},
         year = 2018,
        month = jan,
       volume = {70},
          eid = {S14},
        pages = {S14},
          doi = {10.1093/pasj/psx122},
archivePrefix = {arXiv},
       eprint = {1704.08140},
 primaryClass = {astro-ph.GA},
       adsurl = {https://ui.adsabs.harvard.edu/abs/2018PASJ...70S..14S}
}

@ARTICLE{Khostovan2020,
       author = {{Khostovan}, A.~A. and {Malhotra}, S. and {Rhoads}, J.~E. and {Jiang}, C. and {Wang}, J. and {Wold}, I. and {Zheng}, Z.-Y. and {Barrientos}, L.~F. and {Coughlin}, A. and {Harish}, S. and {Hu}, W. and {Infante}, L. and {Perez}, L.~A. and {Pharo}, J. and {Valdes}, F. and {Walker}, A.~R. and {Yang}, H.},
        title = "{A large, deep 3 deg$^{2}$ survey of H {\ensuremath{\alpha}}, [O III], and [O II] emitters from LAGER: constraining luminosity functions}",
      journal = {\mnras},
         year = 2020,
        month = apr,
       volume = {493},
       number = {3},
        pages = {3966-3984},
          doi = {10.1093/mnras/staa175},
archivePrefix = {arXiv},
       eprint = {2001.04989},
 primaryClass = {astro-ph.GA},
       adsurl = {https://ui.adsabs.harvard.edu/abs/2020MNRAS.493.3966K}
}

@ARTICLE{Bacon2023,
       author = {{Bacon}, Roland and {Brinchmann}, Jarle and {Conseil}, Simon and {Maseda}, Michael and {Nanayakkara}, Themiya and {Wendt}, Martin and {Bacher}, Raphael and {Mary}, David and {Weilbacher}, Peter M. and {Krajnovi{\'c}}, Davor and {Boogaard}, Leindert and {Bouch{\'e}}, Nicolas and {Contini}, Thierry and {Epinat}, Beno{\^\i}t and {Feltre}, Anna and {Guo}, Yucheng and {Herenz}, Christian and {Kollatschny}, Wolfram and {Kusakabe}, Haruka and {Leclercq}, Floriane and {Michel-Dansac}, L{\'e}o and {Pello}, Roser and {Richard}, Johan and {Roth}, Martin and {Salvignol}, Gregory and {Schaye}, Joop and {Steinmetz}, Matthias and {Tresse}, Laurence and {Urrutia}, Tanya and {Verhamme}, Anne and {Vitte}, Eloise and {Wisotzki}, Lutz and {Zoutendijk}, Sebastiaan L.},
        title = "{The MUSE Hubble Ultra Deep Field surveys: Data release II}",
      journal = {\aap},
         year = 2023,
        month = feb,
       volume = {670},
          eid = {A4},
        pages = {A4},
          doi = {10.1051/0004-6361/202244187},
archivePrefix = {arXiv},
       eprint = {2211.08493},
 primaryClass = {astro-ph.GA},
       adsurl = {https://ui.adsabs.harvard.edu/abs/2023A&A...670A...4B}
}

@INPROCEEDINGS{Cirasuolo2014,
       author = {{Cirasuolo}, M. and {Afonso}, J. and {Carollo}, M. and {Flores}, H. and {Maiolino}, R. and {Oliva}, E. and {Paltani}, S. and {Vanzi}, Leonardo and {Evans}, Christopher and {Abreu}, M. and {Atkinson}, David and {Babusiaux}, C. and {Beard}, Steven and {Bauer}, F. and {Bellazzini}, M. and {Bender}, Ralf and {Best}, P. and {Bezawada}, N. and {Bonifacio}, P. and {Bragaglia}, A. and {Bryson}, I. and {Busher}, D. and {Cabral}, A. and {Caputi}, K. and {Centrone}, M. and {Chemla}, F. and {Cimatti}, A. and {Cioni}, M.-R. and {Clementini}, G. and {Coelho}, J. and {Crnojevic}, D. and {Daddi}, E. and {Dunlop}, J. and {Eales}, S. and {Feltzing}, S. and {Ferguson}, A. and {Fisher}, M. and {Fontana}, A. and {Fynbo}, J. and {Garilli}, B. and {Gilmore}, G. and {Glauser}, A. and {Guinouard}, I. and {Hammer}, F. and {Hastings}, P. and {Hess}, A. and {Ivison}, R. and {Jagourel}, P. and {Jarvis}, M. and {Kaper}, L. and {Kauffman}, G. and {Kitching}, A.~T. and {Lawrence}, A. and {Lee}, D. and {Lemasle}, B. and {Licausi}, G. and {Lilly}, S. and {Lorenzetti}, D. and {Lunney}, D. and {Maiolino}, R. and {Mannucci}, F. and {McLure}, R. and {Minniti}, D. and {Montgomery}, D. and {Muschielok}, B. and {Nandra}, K. and {Navarro}, R. and {Norberg}, P. and {Oliver}, S. and {Origlia}, L. and {Padilla}, N. and {Peacock}, J. and {Pedichini}, F. and {Peng}, J. and {Pentericci}, L. and {Pragt}, J. and {Puech}, M. and {Randich}, S. and {Rees}, P. and {Renzini}, A. and {Ryde}, N. and {Rodrigues}, M. and {Roseboom}, I. and {Royer}, F. and {Saglia}, R. and {Sanchez}, A. and {Schiavon}, R. and {Schnetler}, H. and {Sobral}, D. and {Speziali}, R. and {Sun}, D. and {Stuik}, R. and {Taylor}, A. and {Taylor}, W. and {Todd}, S. and {Tolstoy}, E. and {Torres}, M. and {Tosi}, M. and {Vanzella}, E. and {Venema}, L. and {Vitali}, F. and {Wegner}, M. and {Wells}, M. and {Wild}, V. and {Wright}, G. and {Zamorani}, G. and {Zoccali}, M.},
        title = "{MOONS: the Multi-Object Optical and Near-infrared Spectrograph for the VLT}",
    booktitle = {Ground-based and Airborne Instrumentation for Astronomy V},
         year = 2014,
       editor = {{Ramsay}, Suzanne K. and {McLean}, Ian S. and {Takami}, Hideki},
       series = {Society of Photo-Optical Instrumentation Engineers (SPIE) Conference Series},
       volume = {9147},
        month = jul,
          eid = {91470N},
        pages = {91470N},
          doi = {10.1117/12.2056012},
       adsurl = {https://ui.adsabs.harvard.edu/abs/2014SPIE.9147E..0NC}
}

@ARTICLE{Finkelstein2024,
       author = {{Finkelstein}, Steven L. and {Leung}, Gene C.~K. and {Bagley}, Micaela B. and {Dickinson}, Mark and {Ferguson}, Henry C. and {Papovich}, Casey and {Akins}, Hollis B. and {Arrabal Haro}, Pablo and {Dav{\'e}}, Romeel and {Dekel}, Avishai and {Kartaltepe}, Jeyhan S. and {Kocevski}, Dale D. and {Koekemoer}, Anton M. and {Pirzkal}, Nor and {Somerville}, Rachel S. and {Yung}, L.~Y. Aaron and {Amor{\'\i}n}, Ricardo O. and {Backhaus}, Bren E. and {Behroozi}, Peter and {Bisigello}, Laura and {Bromm}, Volker and {Casey}, Caitlin M. and {Ch{\'a}vez Ortiz}, {\'O}scar A. and {Cheng}, Yingjie and {Chworowsky}, Katherine and {Cleri}, Nikko J. and {Cooper}, M.~C. and {Davis}, Kelcey and {de la Vega}, Alexander and {Elbaz}, David and {Franco}, Maximilien and {Fontana}, Adriano and {Fujimoto}, Seiji and {Giavalisco}, Mauro and {Grogin}, Norman A. and {Holwerda}, Benne W. and {Huertas-Company}, Marc and {Hirschmann}, Michaela and {Iyer}, Kartheik G. and {Jogee}, Shardha and {Jung}, Intae and {Larson}, Rebecca L. and {Lucas}, Ray A. and {Mobasher}, Bahram and {Morales}, Alexa M. and {Morley}, Caroline V. and {Mukherjee}, Sagnick and {P{\'e}rez-Gonz{\'a}lez}, Pablo G. and {Ravindranath}, Swara and {Rodighiero}, Giulia and {Rowland}, Melanie J. and {Tacchella}, Sandro and {Taylor}, Anthony J. and {Trump}, Jonathan R. and {Wilkins}, Stephen M.},
        title = "{The Complete CEERS Early Universe Galaxy Sample: A Surprisingly Slow Evolution of the Space Density of Bright Galaxies at z {\ensuremath{\sim}} 8.5{\textendash}14.5}",
      journal = {\apjl},
         year = 2024,
        month = jul,
       volume = {969},
       number = {1},
          eid = {L2},
        pages = {L2},
          doi = {10.3847/2041-8213/ad4495},
archivePrefix = {arXiv},
       eprint = {2311.04279},
 primaryClass = {astro-ph.GA},
       adsurl = {https://ui.adsabs.harvard.edu/abs/2024ApJ...969L...2F}
}

@ARTICLE{Wang2022HLSS,
       author = {{Wang}, Yun and {Zhai}, Zhongxu and {Alavi}, Anahita and {Massara}, Elena and {Pisani}, Alice and {Benson}, Andrew and {Hirata}, Christopher M. and {Samushia}, Lado and {Weinberg}, David H. and {Colbert}, James and {Dor{\'e}}, Olivier and {Eifler}, Tim and {Heinrich}, Chen and {Ho}, Shirley and {Krause}, Elisabeth and {Padmanabhan}, Nikhil and {Spergel}, David and {Teplitz}, Harry I.},
        title = "{The High Latitude Spectroscopic Survey on the Nancy Grace Roman Space Telescope}",
      journal = {\apj},
         year = 2022,
        month = mar,
       volume = {928},
       number = {1},
          eid = {1},
        pages = {1},
          doi = {10.3847/1538-4357/ac4973},
archivePrefix = {arXiv},
       eprint = {2110.01829},
 primaryClass = {astro-ph.CO},
       adsurl = {https://ui.adsabs.harvard.edu/abs/2022ApJ...928....1W}
}

@ARTICLE{Sugai2015,
       author = {{Sugai}, Hajime and {Tamura}, Naoyuki and {Karoji}, Hiroshi and {Shimono}, Atsushi and {Takato}, Naruhisa and {Kimura}, Masahiko and {Ohyama}, Youichi and {Ueda}, Akitoshi and {Aghazarian}, Hrand and {de Arruda}, Marcio Vital and {Barkhouser}, Robert H. and {Bennett}, Charles L. and {Bickerton}, Steve and {Bozier}, Alexandre and {Braun}, David F. and {Bui}, Khanh and {Capocasale}, Christopher M. and {Carr}, Michael A. and {Castilho}, Bruno and {Chang}, Yin-Chang and {Chen}, Hsin-Yo and {Chou}, Richard C.~Y. and {Dawson}, Olivia R. and {Dekany}, Richard G. and {Ek}, Eric M. and {Ellis}, Richard S. and {English}, Robin J. and {Ferrand}, Didier and {Ferreira}, D{\'e}cio and {Fisher}, Charles D. and {Golebiowski}, Mirek and {Gunn}, James E. and {Hart}, Murdock and {Heckman}, Timothy M. and {Ho}, Paul T.~P. and {Hope}, Stephen and {Hovland}, Larry E. and {Hsu}, Shu-Fu and {Hu}, Yen-Shan and {Huang}, Pin Jie and {Jaquet}, Marc and {Karr}, Jennifer E. and {Kempenaar}, Jason G. and {King}, Matthew E. and {le F{\`e}vre}, Olivier and {Mignant}, David Le and {Ling}, Hung-Hsu and {Loomis}, Craig and {Lupton}, Robert H. and {Madec}, Fabrice and {Mao}, Peter and {Souza Marrara}, Lucas and {M{\'e}nard}, Brice and {Morantz}, Chaz and {Murayama}, Hitoshi and {Murray}, Graham J. and {Cesar de Oliveira}, Antonio and {Mendes de Oliveira}, Claudia and {Souza de Oliveira}, Ligia and {Orndorff}, Joe D. and {de Paiva Vila{\c{c}}a}, Rodrigo and {Partos}, Eamon J. and {Pascal}, Sandrine and {Pegot-Ogier}, Thomas and {Reiley}, Daniel J. and {Riddle}, Reed and {Santos}, Leandro and {dos Santos}, Jesulino Bispo and {Schwochert}, Mark A. and {Seiffert}, Michael D. and {Smee}, Stephen A. and {Smith}, Roger M. and {Steinkraus}, Ronald E. and {Sodr{\'e}}, Laerte and {Spergel}, David N. and {Surace}, Christian and {Tresse}, Laurence and {Vidal}, Cl{\'e}ment and {Vives}, Sebastien and {Wang}, Shiang-Yu and {Wen}, Chih-Yi and {Wu}, Amy C. and {Wyse}, Rosie and {Yan}, Chi-Hung},
        title = "{Prime Focus Spectrograph for the Subaru telescope: massively multiplexed optical and near-infrared fiber spectrograph}",
      journal = {Journal of Astronomical Telescopes, Instruments, and Systems},
         year = 2015,
        month = jul,
       volume = {1},
          eid = {035001},
        pages = {035001},
          doi = {10.1117/1.JATIS.1.3.035001},
archivePrefix = {arXiv},
       eprint = {1507.00725},
 primaryClass = {astro-ph.IM},
       adsurl = {https://ui.adsabs.harvard.edu/abs/2015JATIS...1c5001S}
}

@ARTICLE{Inoue2020,
       author = {{Inoue}, Akio K. and {Yamanaka}, Satoshi and {Ouchi}, Masami and {Iwata}, Ikuru and {Shimasaku}, Kazuhiro and {Taniguchi}, Yoshiaki and {Nagao}, Tohru and {Kashikawa}, Nobunari and {Ono}, Yoshiaki and {Mawatari}, Ken and {Shibuya}, Takatoshi and {Hayashi}, Masao and {Ikeda}, Hiroyuki and {Zhang}, Haibin and {Liang}, Yongming and {Lee}, Chien-Hsiu and {Hilmi}, Miftahul and {Kikuta}, Satoshi and {Kusakabe}, Haruka and {Furusawa}, Hisanori and {Hayashino}, Tomoki and {Kajisawa}, Masaru and {Matsuda}, Yuichi and {Nakajima}, Kimihiko and {Momose}, Rieko and {Harikane}, Yuichi and {Saito}, Tomoki and {Kodama}, Tadayuki and {Kikuchihara}, Shotaro and {Iye}, Masanori and {Goto}, Tomotsugu},
        title = "{CHORUS. I. Cosmic HydrOgen Reionization Unveiled with Subaru: Overview}",
      journal = {\pasj},
         year = 2020,
        month = dec,
       volume = {72},
       number = {6},
          eid = {101},
        pages = {101},
          doi = {10.1093/pasj/psaa100},
archivePrefix = {arXiv},
       eprint = {2011.07211},
 primaryClass = {astro-ph.GA},
       adsurl = {https://ui.adsabs.harvard.edu/abs/2020PASJ...72..101I}
}

@MISC{Kakiichi2024,
       author = {{Kakiichi}, Koki and {Egami}, Eiichi and {Fan}, Xiaohui and {Lyu}, Jianwei and {Wang}, Feige and {Yang}, Jinyi and {Bechtel}, Shane and {Behroozi}, Peter and {Bosman}, Sarah E.~I. and {Cai}, Zheng and {Champagne}, Jaclyn and {Davies}, Frederick and {De Rosa}, Gisella and {Decarli}, Roberto and {Eilers}, Anna-Christina and {Ellis}, Richard S. and {Endsley}, Ryan and {Farina}, Emanuele Paolo and {Finkelstein}, Steven L. and {Fujimoto}, Seiji and {Hennawi}, Joseph and {Inoue}, Akio and {Jiang}, Linhua and {Jin}, Xiangyu and {Khusanova}, Yana and {Kirkpatrick}, Allison and {Kocevski}, Dale D. and {Kulkarni}, Girish and {Lee}, Khee-Gan and {Liu}, Weizhe and {Meyer}, Romain Alexis and {Ono}, Yoshiaki and {Onoue}, Masafusa and {Ouchi}, Masami and {Papovich}, Casey and {Satyavolu}, Sindhu and {Schindler}, Jan-Torge and {Sun}, Fengwu and {Tee}, Wei Leong and {Vestergaard}, Marianne and {Zhang}, Haowen and {Zou}, Siwei},
        title = "{COSMOS-3D: A Legacy Spectroscopic/Imaging Survey of the Early Universe}",
 howpublished = {JWST Proposal. Cycle 3, ID. \#5893},
         year = 2024,
        month = feb,
        pages = {5893},
       adsurl = {https://ui.adsabs.harvard.edu/abs/2024jwst.prop.5893K}
}

@ARTICLE{Ono2018,
       author = {{Ono}, Yoshiaki and {Ouchi}, Masami and {Harikane}, Yuichi and {Toshikawa}, Jun and {Rauch}, Michael and {Yuma}, Suraphong and {Sawicki}, Marcin and {Shibuya}, Takatoshi and {Shimasaku}, Kazuhiro and {Oguri}, Masamune and {Willott}, Chris and {Akhlaghi}, Mohammad and {Akiyama}, Masayuki and {Coupon}, Jean and {Kashikawa}, Nobunari and {Komiyama}, Yutaka and {Konno}, Akira and {Lin}, Lihwai and {Matsuoka}, Yoshiki and {Miyazaki}, Satoshi and {Nagao}, Tohru and {Nakajima}, Kimihiko and {Silverman}, John and {Tanaka}, Masayuki and {Taniguchi}, Yoshiaki and {Wang}, Shiang-Yu},
        title = "{Great Optically Luminous Dropout Research Using Subaru HSC (GOLDRUSH). I. UV luminosity functions at z {\ensuremath{\sim}} 4-7 derived with the half-million dropouts on the 100 deg$^{2}$ sky}",
      journal = {\pasj},
         year = 2018,
        month = jan,
       volume = {70},
          eid = {S10},
        pages = {S10},
          doi = {10.1093/pasj/psx103},
archivePrefix = {arXiv},
       eprint = {1704.06004},
 primaryClass = {astro-ph.GA},
       adsurl = {https://ui.adsabs.harvard.edu/abs/2018PASJ...70S..10O}
}

@ARTICLE{Paszke2019,
       author = {{Paszke}, Adam and {Gross}, Sam and {Massa}, Francisco and {Lerer}, Adam and {Bradbury}, James and {Chanan}, Gregory and {Killeen}, Trevor and {Lin}, Zeming and {Gimelshein}, Natalia and {Antiga}, Luca and {Desmaison}, Alban and {K{\"o}pf}, Andreas and {Yang}, Edward and {DeVito}, Zach and {Raison}, Martin and {Tejani}, Alykhan and {Chilamkurthy}, Sasank and {Steiner}, Benoit and {Fang}, Lu and {Bai}, Junjie and {Chintala}, Soumith},
        title = "{PyTorch: An Imperative Style, High-Performance Deep Learning Library}",
      journal = {arXiv e-prints},
         year = 2019,
        month = dec,
          eid = {arXiv:1912.01703},
        pages = {arXiv:1912.01703},
          doi = {10.48550/arXiv.1912.01703},
archivePrefix = {arXiv},
       eprint = {1912.01703},
 primaryClass = {cs.LG},
       adsurl = {https://ui.adsabs.harvard.edu/abs/2019arXiv191201703P}
}

@inproceedings{he2016,
  title={Deep residual learning for image recognition},
  author={He, Kaiming and Zhang, Xiangyu and Ren, Shaoqing and Sun, Jian},
  booktitle={Proceedings of the IEEE conference on computer vision and pattern recognition},
  pages={770--778},
  year={2016}
}

@inproceedings{Nair2010,
  author    = {Nair, Vinod and Hinton, Geoffrey E.},
  title     = {Rectified Linear Units Improve Restricted Boltzmann Machines},
  booktitle = {Proceedings of the 27th International Conference on Machine Learning (ICML)},
  year      = {2010},
  pages     = {807--814}
}

@article{Wu2018,
  author  = {Wu, Yuxin and He, Kaiming},
  title   = {Group Normalization},
  journal = {European Conference on Computer Vision (ECCV)},
  year    = {2018},
  pages   = {3--19}
}

@ARTICLE{Han2016,
       author = {{Han}, Yo Seob and {Yoo}, Jaejun and {Ye}, Jong Chul},
        title = "{Deep Residual Learning for Compressed Sensing CT Reconstruction via Persistent Homology Analysis}",
      journal = {arXiv e-prints},
         year = 2016,
        month = nov,
          eid = {arXiv:1611.06391},
        pages = {arXiv:1611.06391},
          doi = {10.48550/arXiv.1611.06391},
archivePrefix = {arXiv},
       eprint = {1611.06391},
 primaryClass = {cs.CV},
       adsurl = {https://ui.adsabs.harvard.edu/abs/2016arXiv161106391H}
}

%\begin{thebibliography}{99}
%\end{thebibliography}

%%%%%%%%%%%%%%%%%%%%%%%%%%%%%%%%%%%%%%%%%%%%%%%%%%

%%%%%%%%%%%%%%%%% APPENDICES %%%%%%%%%%%%%%%%%%%%%

\appendix

\section{Performance of the reconstruction on other reionization models}

For simplicity, in the main text we highlighted two reionization models. Here we present the corresponding results for additional models considered in this work. Figure~\ref{fig:Crazy_NoSmooth} shows field-level comparisons for the {Extremely Early} model at $z\simeq 7.14$, and Figure~\ref{fig:Fiducial_NoSmooth} shows the analogous comparisons for the {Fiducial} model, each for the three survey selections ({Deep}, {Shallow}, and {LAE-only}). In each case we show the tracer distribution, the true projected $x_{\rm HI}$ field, the reconstructed field, and the voxel-wise true vs. predicted relation. Overall, the reconstructions recover the large-scale ionization morphology across these alternative reionization models, with improved performance for deeper survey selections.

\begin{figure*}
\centering

% ==================== EARLY MODEL (NO SMOOTHING) ====================
\begin{tcolorbox}[title={\textbf{Extremely Early Model: Deep Survey} }]
\includegraphics[width=12.5cm, trim={3.5cm 0.5cm 3cm 1cm}, clip]{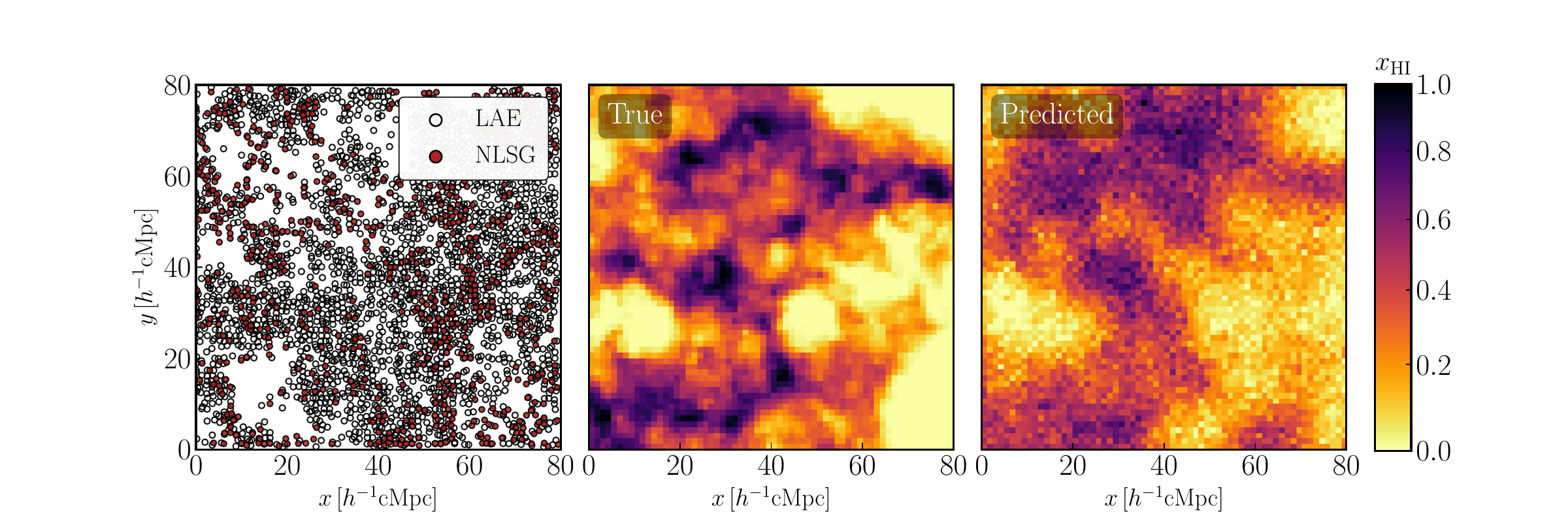}%
\hspace{0.15cm}%
\includegraphics[width=4.4cm, trim={0cm 0.7cm 0.67cm 0.2cm}, clip]{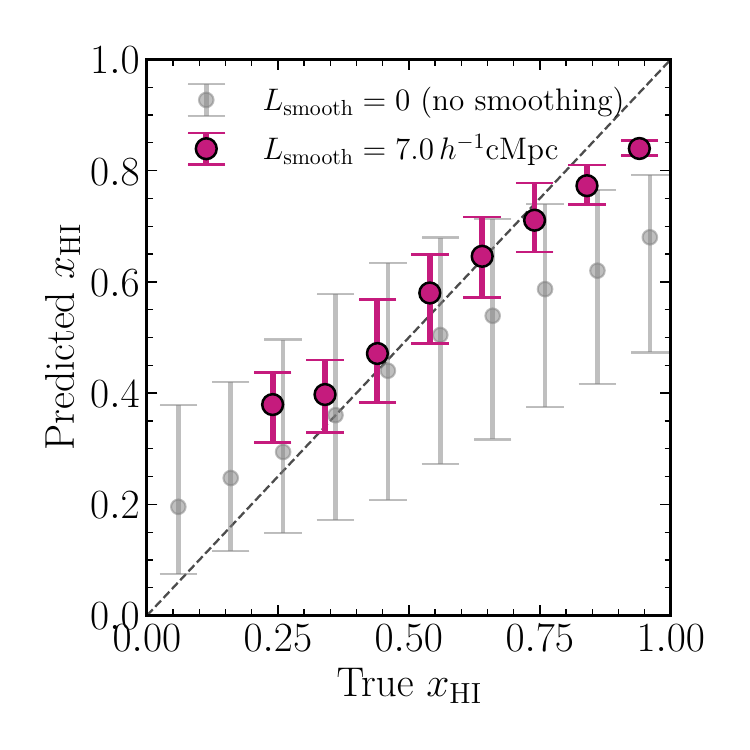}
\end{tcolorbox}

\begin{tcolorbox}[title={\textbf{Extremely Early Model: Shallow Survey} }]
\includegraphics[width=12.5cm, trim={3.5cm 0.5cm 3cm 1cm}, clip]{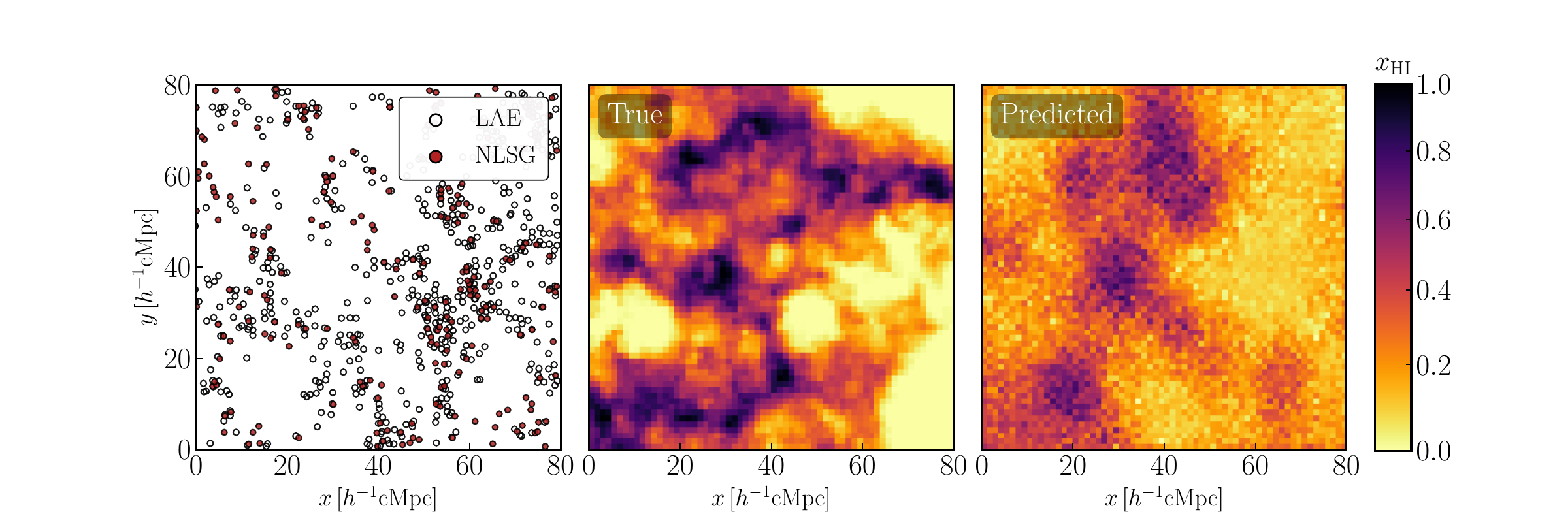}%
\hspace{0.15cm}%
\includegraphics[width=4.4cm, trim={0cm 0.7cm 0.67cm 0.2cm}, clip]{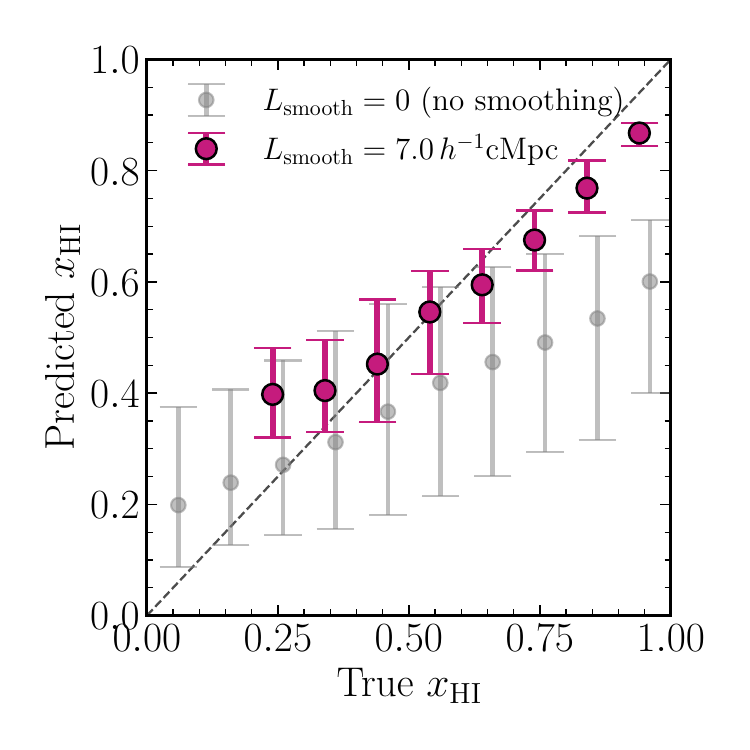}
\end{tcolorbox}

\begin{tcolorbox}[title={\textbf{Extremely Early Model: LAE-only Survey} }]
\includegraphics[width=12.5cm, trim={3.5cm 0.5cm 3cm 1cm}, clip]{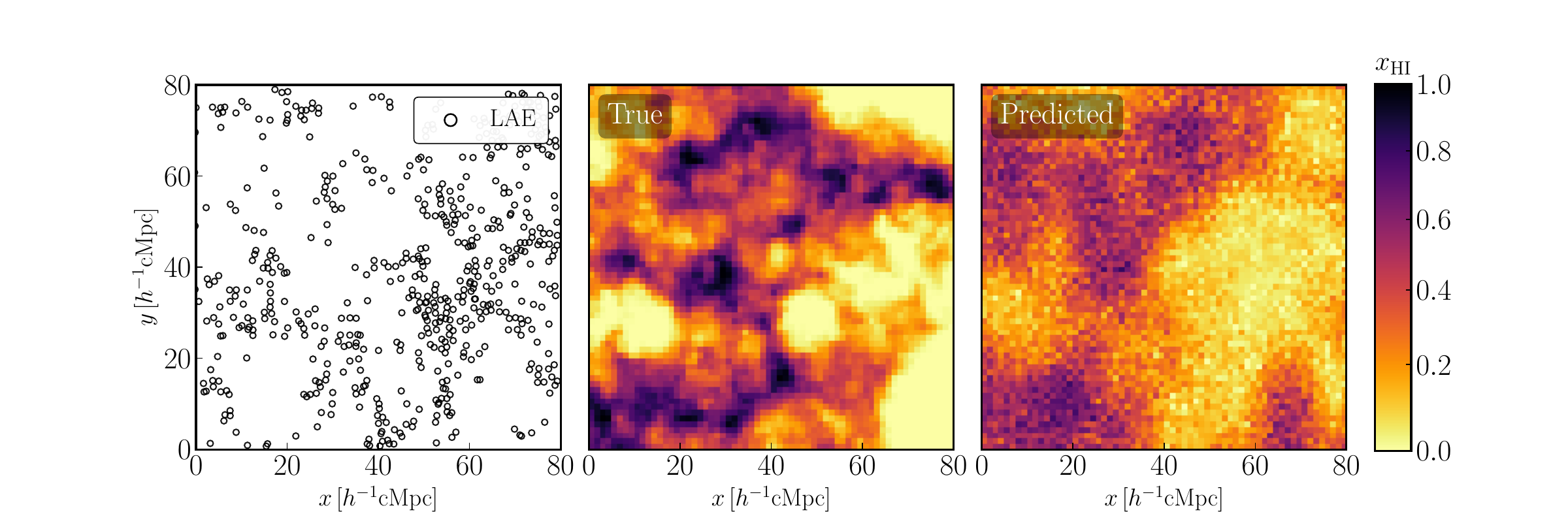}%
\hspace{0.15cm}%
\includegraphics[width=4.4cm, trim={0cm 0.7cm 0.67cm 0.2cm}, clip]{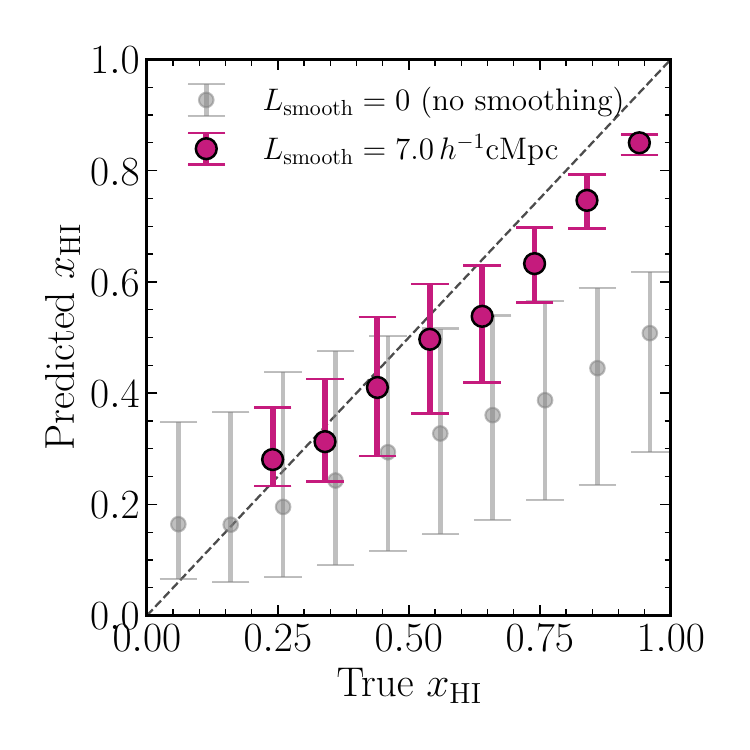}
\end{tcolorbox}

\caption{
Comparison of reconstructed neutral hydrogen fraction fields ($x_{\mathrm{HI}}$) and true–predicted relations for the {Extremely Early reionization model} at $z=7.14$. 
Each row corresponds to a different LAE/NLSG tracer selection: 
{Deep}, {Shallow}, and {LAE-only} (see Table~\ref{tab:survey_configs}). 
For each case, the left three panels show the spatial distribution of LAEs and NLSGs, 
the true ionization map, and the reconstructed map. 
The rightmost panels display the voxel-wise comparison between true and reconstructed 
$x_{\mathrm{HI}}$, showing the median relation and $68\%$ confidence intervals for both the 
unsmoothed ($L_{\mathrm{smooth}}=0$; green) and smoothed 
($L_{\mathrm{smooth}}=7~h^{-1}\,\mathrm{cMpc}$; magenta) fields. 
Smoothing reduces small-scale fluctuations and improves the correlation with the true field, 
while the overall trends remain consistent across different survey selections.
}

\label{fig:Crazy_NoSmooth}
\end{figure*}

\begin{figure*}
\centering

% ==================== Fiducial MODEL (NO SMOOTHING) ====================
\begin{tcolorbox}[title={\textbf{Fiducial Model: Deep Survey} }]
\includegraphics[width=12.5cm, trim={3.5cm 0.5cm 3cm 1cm}, clip]{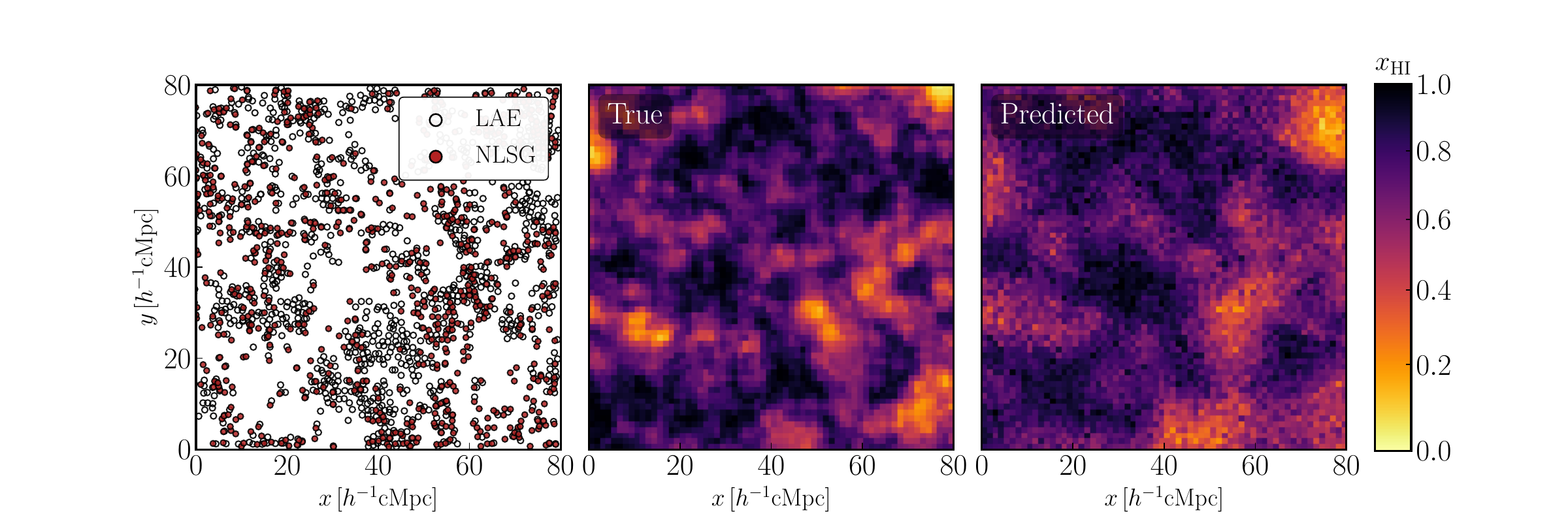}%
\hspace{0.15cm}%
\includegraphics[width=4.4cm, trim={0cm 0.7cm 0.67cm 0.2cm}, clip]{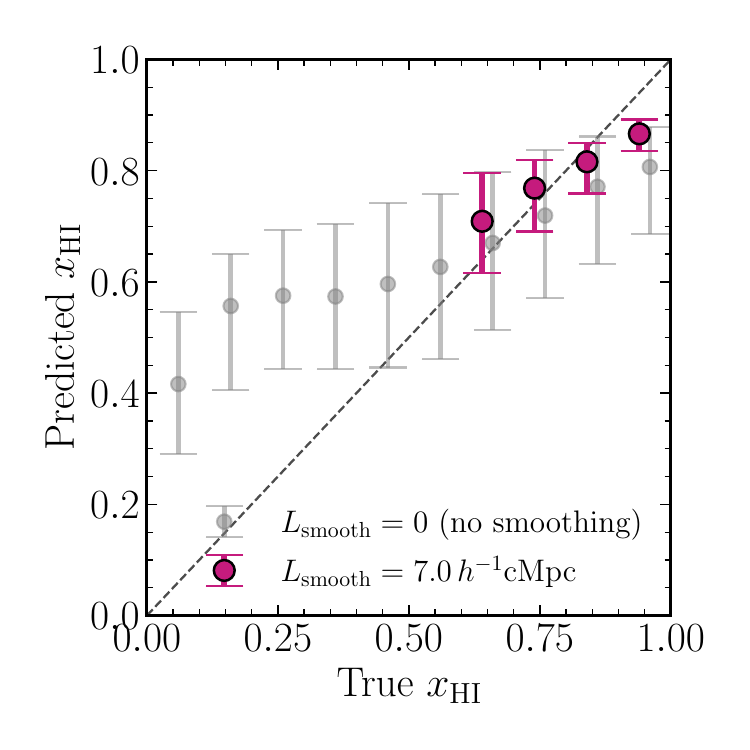}
\end{tcolorbox}

\begin{tcolorbox}[title={\textbf{Fiducial Model: Shallow Survey}}]
\includegraphics[width=12.5cm, trim={3.5cm 0.5cm 3cm 1cm}, clip]{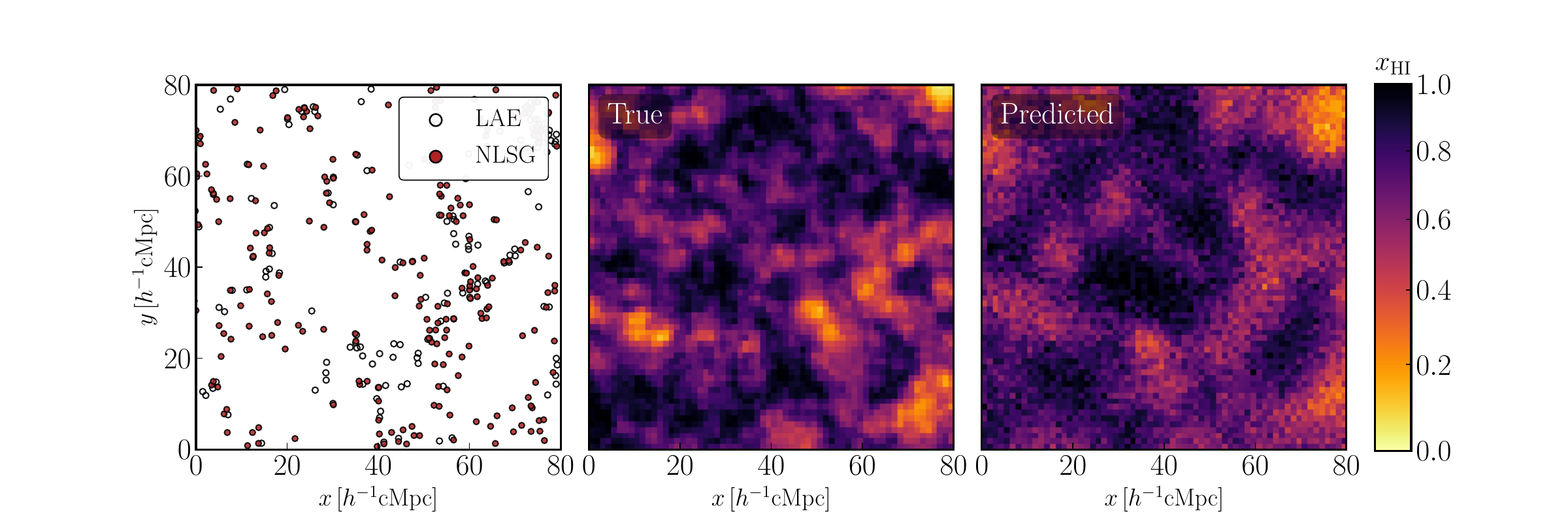}%
\hspace{0.15cm}%
\includegraphics[width=4.4cm, trim={0cm 0.7cm 0.67cm 0.2cm}, clip]{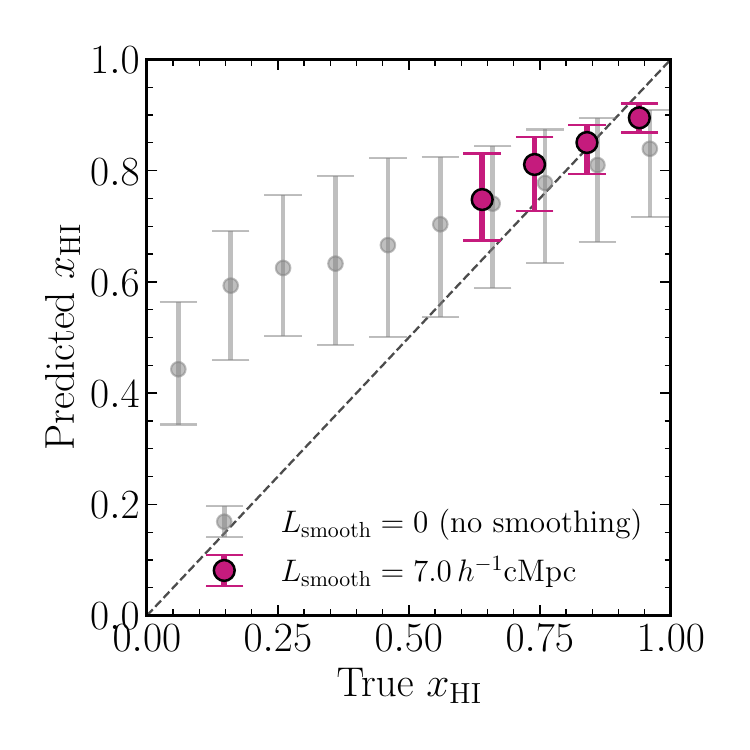}
\end{tcolorbox}

\begin{tcolorbox}[title={\textbf{Fiducial Model: LAE-only Survey} }]
\includegraphics[width=12.5cm, trim={3.5cm 0.5cm 3cm 1cm}, clip]{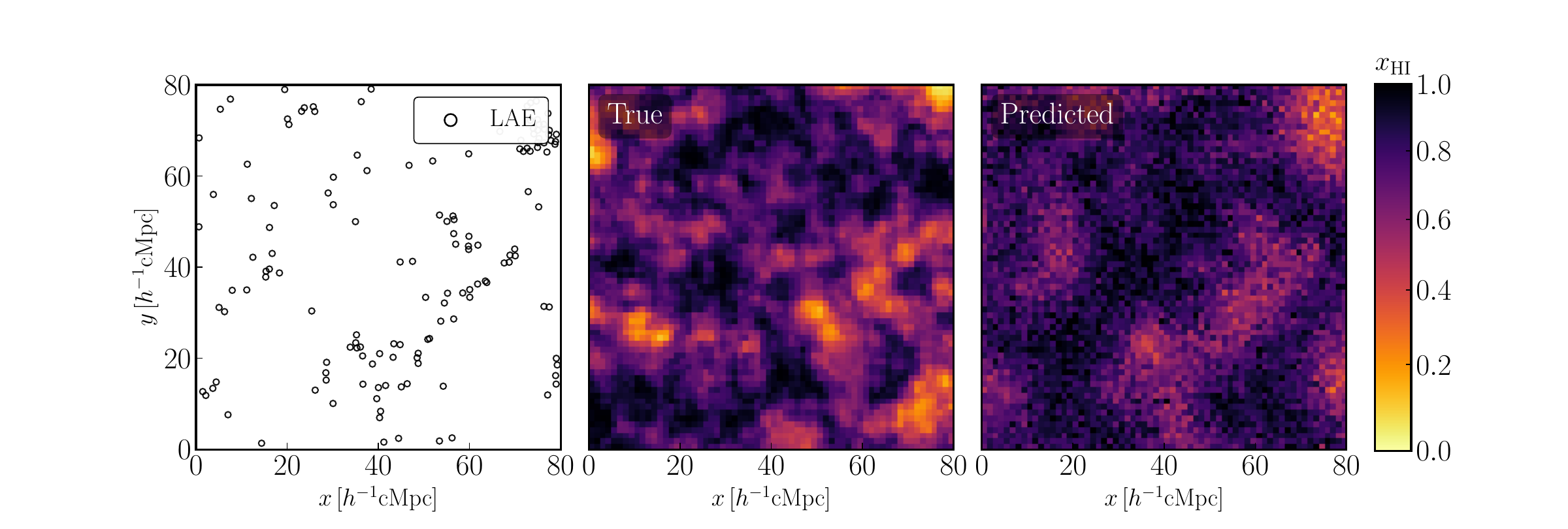}%
\hspace{0.15cm}%
\includegraphics[width=4.4cm, trim={0cm 0.7cm 0.67cm 0.2cm}, clip]{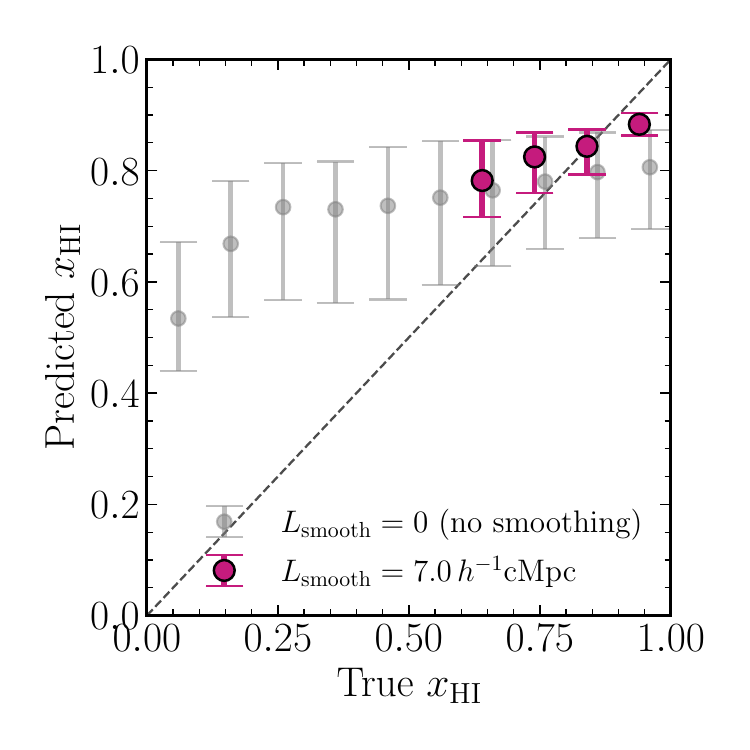}
\end{tcolorbox}

\caption{
Comparison of reconstructed neutral hydrogen fraction fields ($x_{\mathrm{HI}}$) and true–predicted relations for the {Fiducial reionization model} at $z=7.14$. 
Each row corresponds to a different LAE/NLSG tracer selection: 
{Deep}, {Shallow}, and {LAE-only} (see Table~\ref{tab:survey_configs}). 
For each case, the left three panels show the spatial distribution of LAEs and NLSGs, 
the true ionization map, and the reconstructed map. 
The rightmost panels display the voxel-wise comparison between true and reconstructed 
$x_{\mathrm{HI}}$, showing the median relation and $68\%$ confidence intervals for both the 
unsmoothed ($L_{\mathrm{smooth}}=0$; grey) and smoothed 
($L_{\mathrm{smooth}}=7~h^{-1}\,\mathrm{cMpc}$; magenta) fields. 
}
\label{fig:Fiducial_NoSmooth}
\end{figure*}

\section{Effect of smoothing on PDF and Power Spectrum}

For completeness, here we show the \emph{unsmoothed} ($L_{\rm smooth}=0$) summary statistics, whereas the main text focused on smoothed results. Figure~\ref{fig:PDF_nosmooth} presents the unsmoothed PDFs of the reconstructed $x_{\rm HI}$ fields for the different reionization models and survey selections. Although the reconstructed PDFs recover the qualitative trends of the true distributions, they exhibit a somewhat compressed dynamic range, with both low- and high-$x_{\mathrm{HI}}$ tails suppressed.

Figure~\ref{fig:PS_nosmooth} shows the corresponding unsmoothed power spectra. The reconstructed spectra remain consistent with the true large-scale clustering, while small-scale power is underpredicted. Applying smoothing primarily suppresses these small-scale variations and improves the visual and statistical agreement with the true fields.

\begin{figure*}
    \centering
    \includegraphics[width=0.4\linewidth, trim=15 50 10 15, clip]{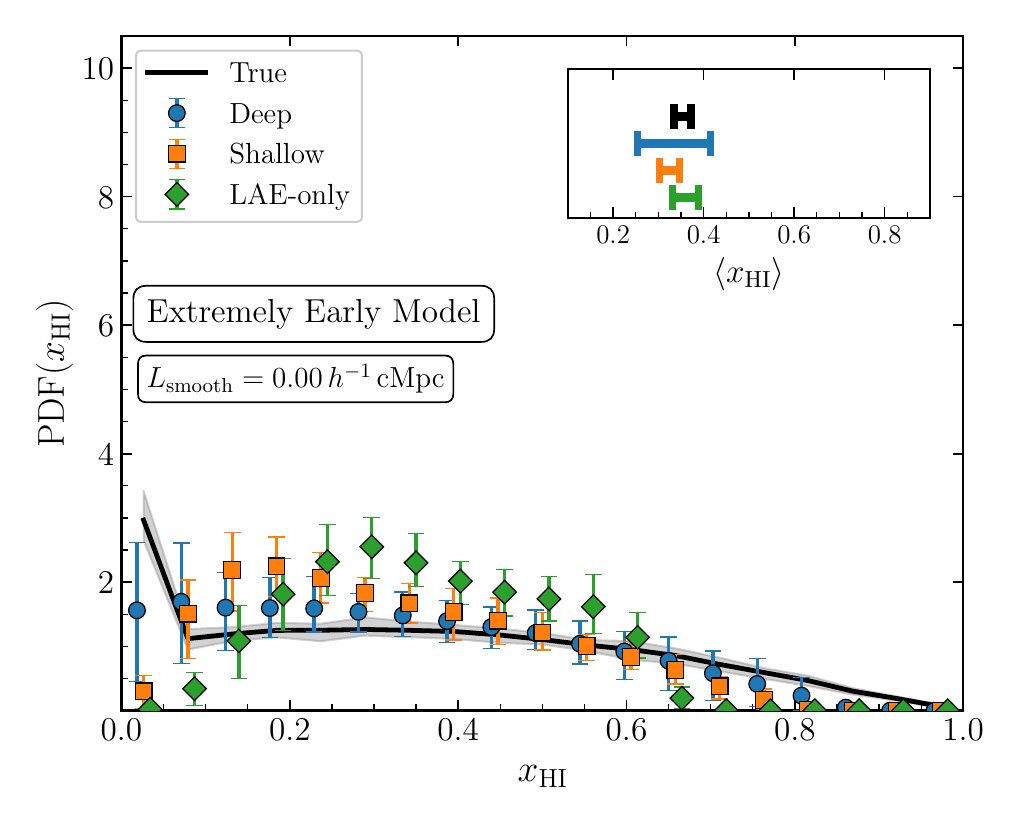}%
    \includegraphics[width=0.365\linewidth, trim=55 50 10 15, clip]{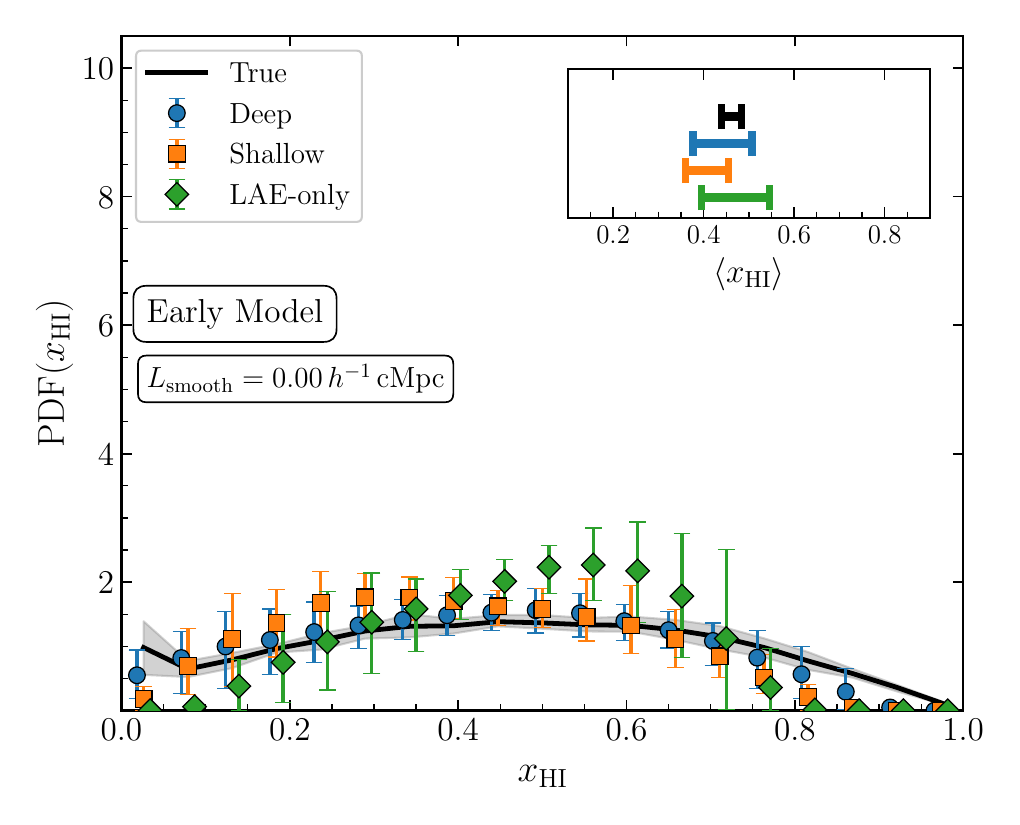}
    
    \includegraphics[width=0.4\linewidth, trim=15 10 10 10, clip]{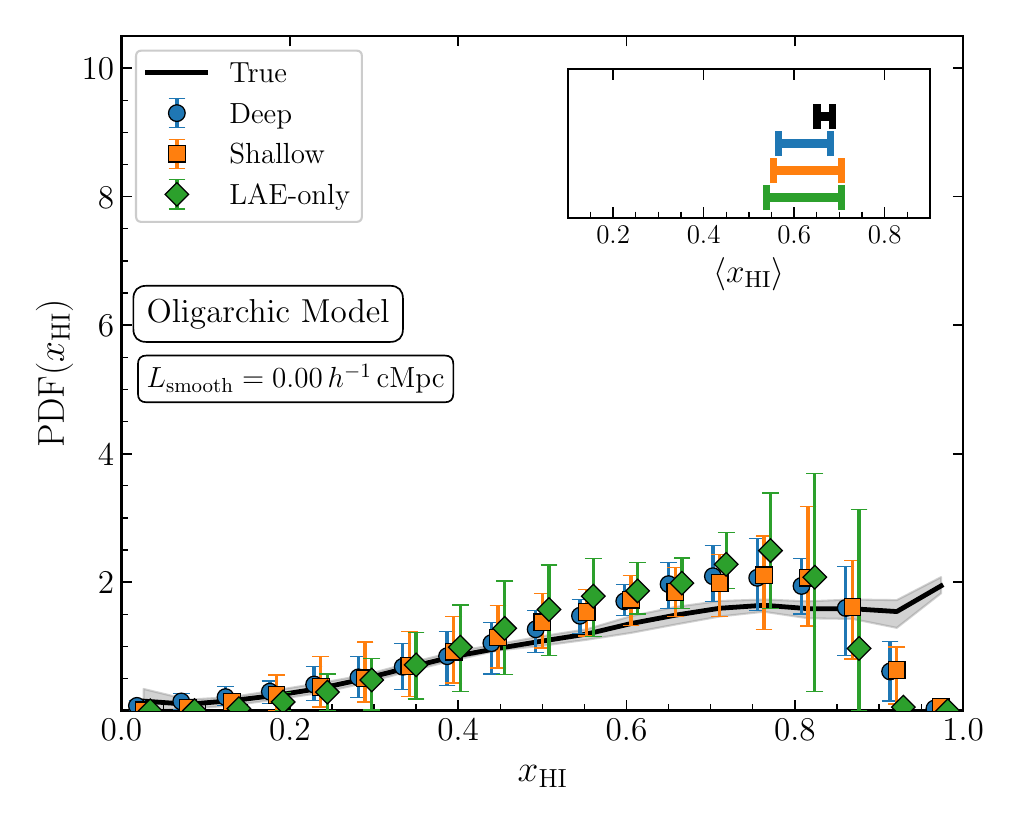}%
    \includegraphics[width=0.365\linewidth, trim=55 10 10 10, clip]{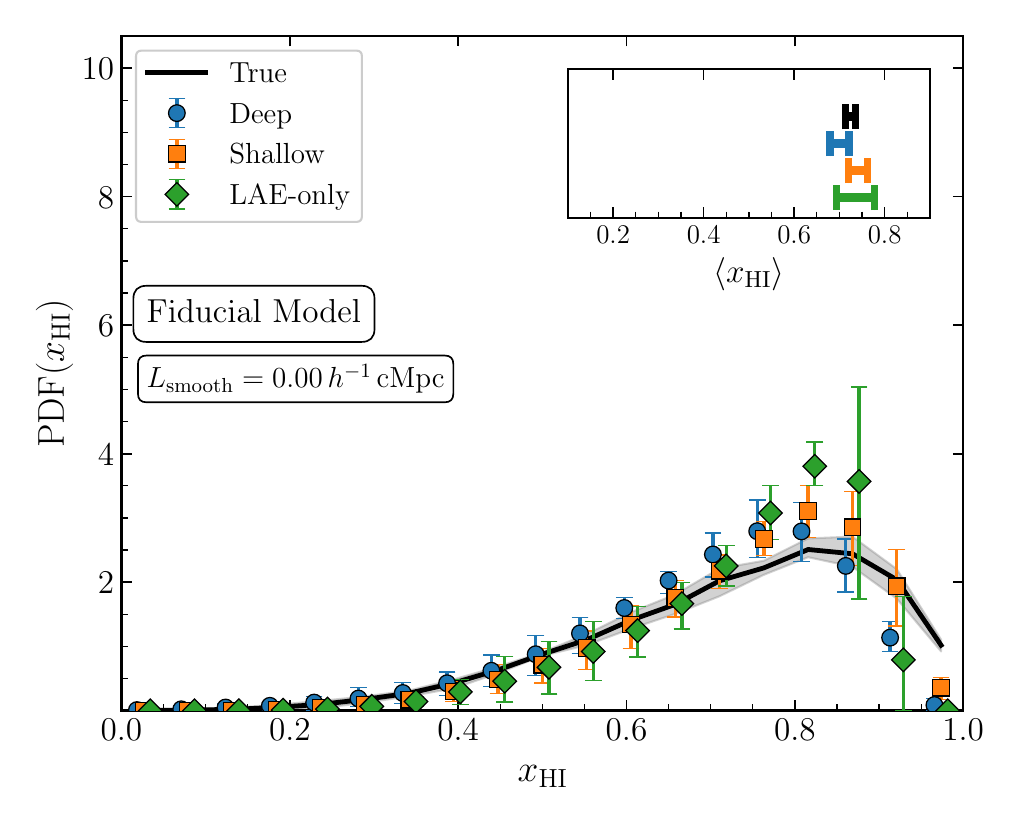}
    \caption{
Probability distribution functions (PDFs) of the unsmoothed projected neutral hydrogen fraction, $x_{\mathrm{HI}}$, reconstructed under four distinct reionization scenarios at $z=7.14$: {Extremely Early}, {Early}, {Oligarchic}, and {Fiducial}.
The solid black curve and grey band indicate the mean and 68\% confidence interval of the true $x_{\mathrm{HI}}$ distribution for the model shown in each panel. Colored symbols with error bars show reconstructed PDFs for the three survey configurations: Deep (blue), Shallow (orange) and LAE-only (green) (see Table~\ref{tab:survey_configs}). The inset panel summarises the mean neutral fraction $\langle x_{\mathrm{HI}}\rangle$ for each case, shown as thick horizontal error bars representing the 68\% confidence intervals.
}

    \label{fig:PDF_nosmooth}
\end{figure*}

\begin{figure*}
    \centering
    \includegraphics[width=0.4\linewidth, trim=15 55 10 15, clip]{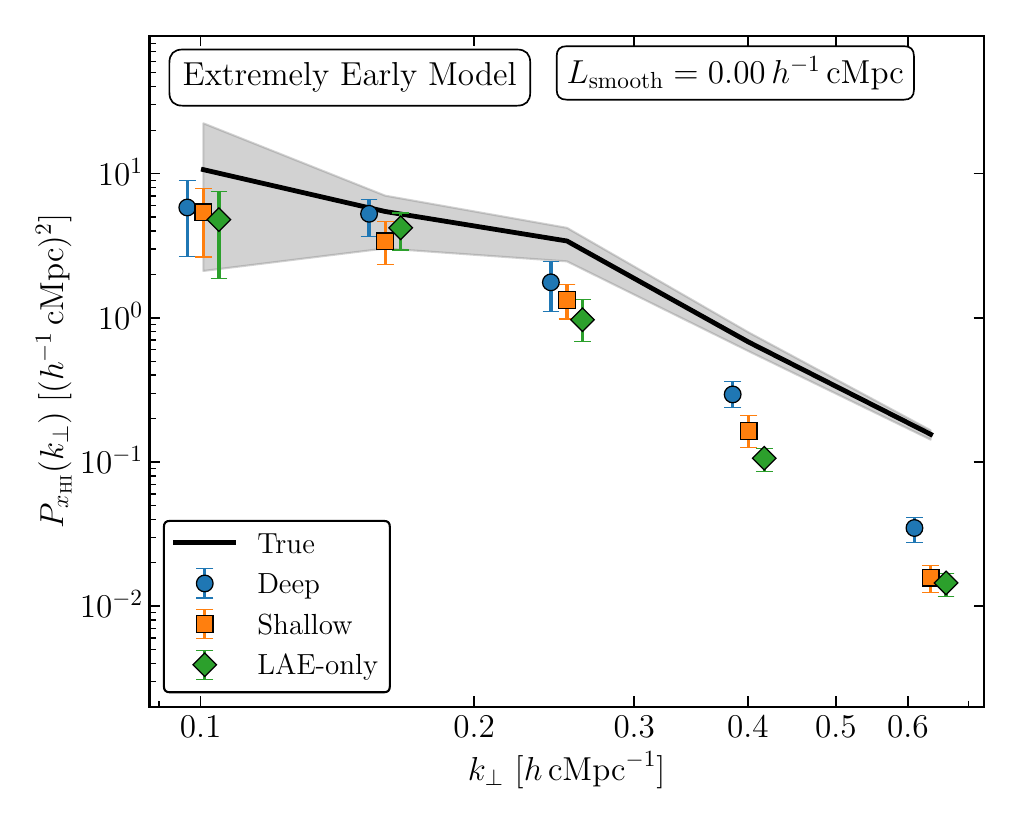}%
    \includegraphics[width=0.353\linewidth, trim=70 55 10 15, clip]{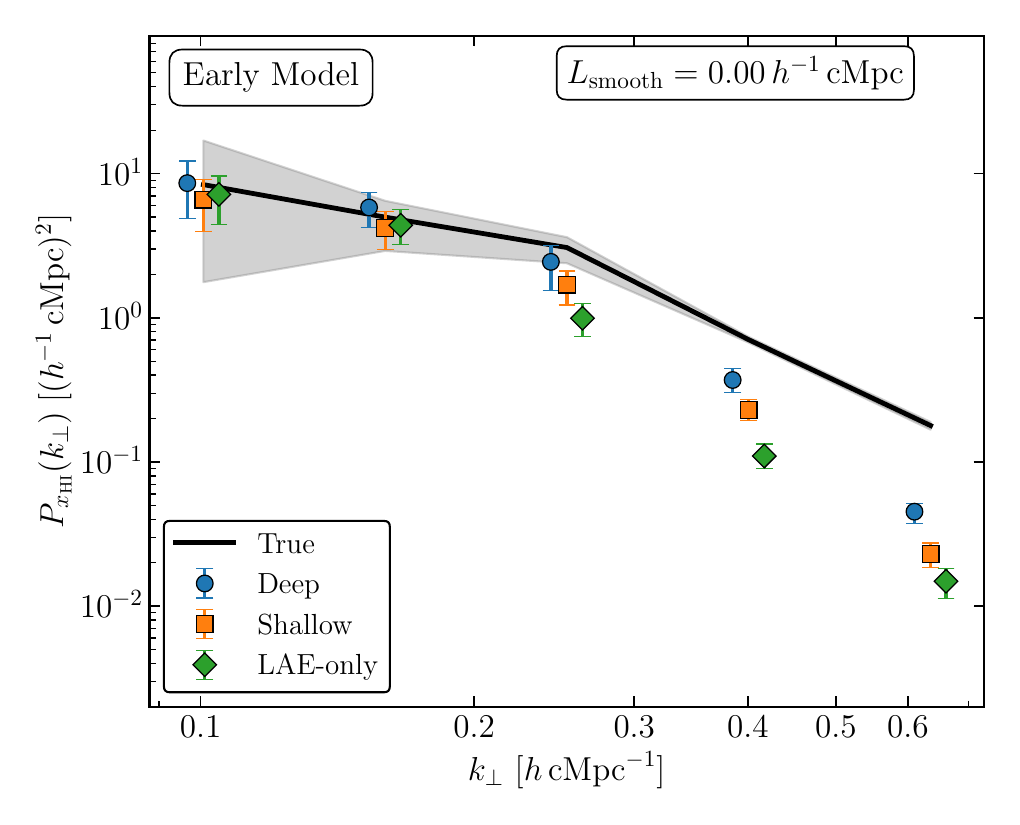}
    
    \includegraphics[width=0.4\linewidth, trim=15 10 10 10, clip]{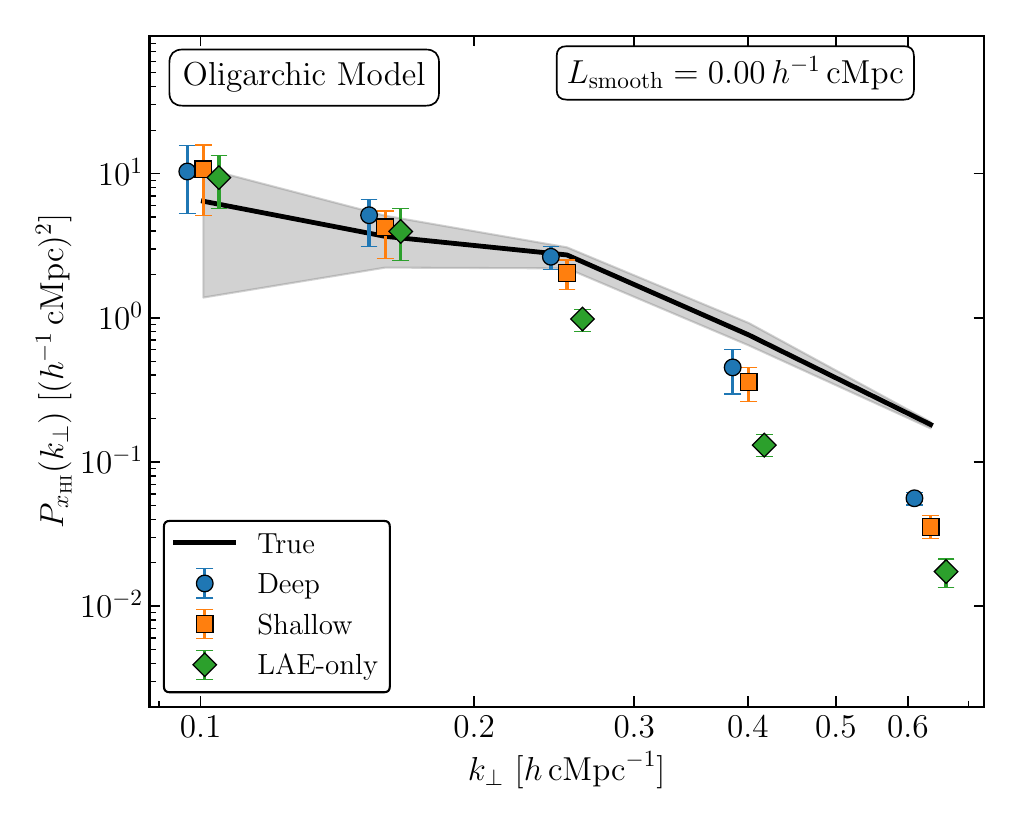}%
    \includegraphics[width=0.353\linewidth, trim=70 10 10 10, clip]{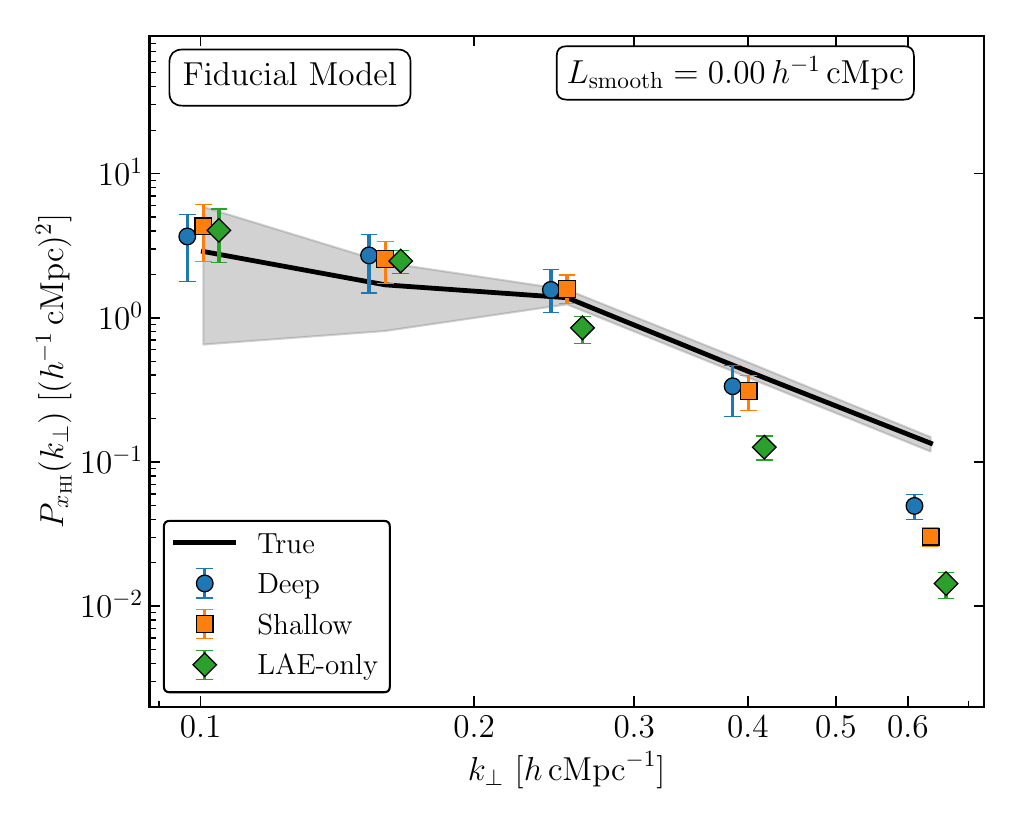}
    \caption{
Projected power spectra of the unsmoothed neutral-hydrogen fraction, 
$P_{x_{\mathrm{HI}}}(k_\perp)$, for true (black curves) and reconstructed 
fields across four reionization scenarios at $z=7.14$: 
{Extremely Early}, {Early}, {Oligarchic}, and {Fiducial}. 
Grey bands denote the $68\%$ scatter of the true spectra across 
the test subvolumes, while colored markers represent reconstructions 
for different tracer selections: 
Deep (blue), Shallow (orange) and LAE-only (green) (see Table~\ref{tab:survey_configs}). 
Reconstructed spectra reproduce the large-scale 
($k_\perp\!\lesssim\!0.3~h\,\mathrm{cMpc}^{-1}$) clustering of 
$x_{\mathrm{HI}}$ across all models, with small-scale power 
slightly suppressed by limited tracer sampling and network regularization. 
Smoothing reduces inter-volume variance and improves overall agreement, 
especially for the denser dual-tracer selections, 
indicating robust recovery of the scale-dependent morphology of the ionization field.
}

    \label{fig:PS_nosmooth}
\end{figure*}

%%%%%%%%%%%%%%%%%%%%%%%%%%%%%%%%%%%%%%%%%%%%%%%%%%

% Don't change these lines
\bsp	% typesetting comment
\label{lastpage}
\end{document}